\documentclass[amsmath,amssymb,11pt,twoside]{book}

\usepackage{graphicx}
\usepackage{dcolumn}
\usepackage{bm}
\usepackage{pstricks}
\usepackage{amssymb}
\usepackage{amsmath}

\newcommand{\bra}[1]{\langle #1|}
\newcommand{\ket}[1]{|#1\rangle}

\newcommand{\proj}[2]{| #1 \rangle \langle #2 |}
\newcommand{\mean}[1]{\langle #1 \rangle} 
\newcommand{\trace}{{\rm Tr}}

\title{Kinetic theory for quantum nanosystems}
\author{Esposito Massimiliano}
\begin{document}

\thispagestyle{empty}
\begin{center}
		
{\large
\begin{tabular}{c}
\bf UNIVERSITE LIBRE DE BRUXELLES \\
\it Facult\'e des Sciences         \\
\it Service de Chimie Physique
\end{tabular}
}
		
\vspace*{6cm}

{\LARGE \bf
\begin{tabular}{p{15cm}}
\hspace{-0.5cm} Kinetic theory for quantum nanosystems
\end{tabular}
}

\vspace*{4.5cm}

{\large
\begin{tabular}{c}
\it Th\`ese pr\'esent\'ee en vue de \\
\it l'obtention du grade acad\'emique de \\
\it Docteur en Sciences \\
\hspace*{1cm} \\
\it R\'ealis\'e sous la direction de \\
{\bf Pierre Gaspard} \\
\vspace*{1.2cm} \hspace*{1cm} \\
{\Large {\bf Massimiliano Esposito}} \\
\hspace*{0.5cm}\\
\it Septembre 2004\\
\end{tabular}
}

\end{center}


\tableofcontents
\pagestyle{headings}
\renewcommand{\chaptermark}[1]{\markboth{\thechapter. #1}{}}
\renewcommand{\sectionmark}[1]{\markright{\thesection \, #1}}

\chapter*{Acknowledgement}

I would like to express my deep gratitude to Gregoire Nicolis for his support and encouragement during this 
thesis and for maintaining a stimulating scientific environment in our department.  \\

This thesis would not have been possible without the intense and very enriching interaction I had with my PhD 
adviser Pierre Gaspard.
His way of approaching a scientific problem will forever exert a deep influence on me. 
I could never thank him enough for his teaching.\\

I would of course like to thank Sebastien Viscardy and Cem Servantie for the fun we had together during these years of PhD.\\

I also thank Ines de Vega for the frequent interesting discussions we had during her one-year visit to Brussels.\\

Furthermore, I wish to thank Doron Cohen who, during the week he spent in Brussels, gave me 
precious advice for the study of the spin-GORM model.   \\

I thank John William Turner who often gave me precious mathematical advices. \\

I gratefully acknowledge everyone I interacted with during this thesis, 
particularly Walter Strunz, Dima Shepelyansky and Francois Bardou, but also many others.\\

I would also like to acknowledge the FRIA for the financial support of my thesis. \\

I finally thank the Universit\'e Libre de Bruxelles (ULB) and all my professors for the high quality 
of their teaching.    \\

A special kind of thanks goes to Carole Linster who shared my life during all these years in Brussels and to
my family who always supported me during my studies.

\chapter{Introduction}

In section \ref{statmec} of this introduction, we place the subject of this thesis in the general context of 
statistical mechanics.
In section \ref{nanosciences}, we discuss the motivations for our contribution to 
statistical mechanics in the emerging field of nanosciences.
In section \ref{QM}, we indicate the link between our approach of quantum statistical mechanics and the 
fundamental problem of quantum measurement.  
In section \ref{goal}, we expose the main goal of this thesis.
Finally in section \ref{plan}, we set out the plan of this thesis. 

\section{Statistical mechanics} \label{statmec}

An intuitive understanding of the natural laws occurs at the spatial scales, temporal scales and energy scales
which are directly accessible to the sensory organs of the human being. 
These typical physical scales, which do not require a sophisticated experimental setup to make the 
measured information accessible to us, were naturally the first historically investigated and broadly define  
what physicists call the macroscopic world.
Thermodynamics is of course the most widely and best defined branch of macroscopic physics which also includes
geometrical optics, fluid dynamics, mechanics and many others.      
Progress in the systematic study of nature came with the improvement of experimental techniques and has 
made possible the investigation of physical scales which are sometimes far less intuitive to the human being. 
For example, the investigation of very small spatial scales has led to the discovery of atoms 
and molecules while the investigation of very large spatial scales has opened the way to the 
discovery of solar systems, galaxies and galaxy clusters.
Each of these new fields of investigation has given rise to physical laws and principles formulated in 
terms of the relevant quantities i.e. the most easily measurable or understandable quantities within each level 
of description.
In astrophysics, the dynamics of planets is, in first approximation, the dynamics of massive points interacting through 
Newtonian gravitational forces. 
In particle physics, the interactions between the smallest matter constituents are often studied through the cross 
sections of colliding particles. \\   
The rational character of the scientific approach motivates a unified understanding of these different physical 
laws occurring at various scales and gives a particularly important status to the physics of small scales, also 
called microscopic physics.
In fact, if matter is made of fundamental elements, say atoms and molecules, it is natural that  
the macroscopic physics describing this matter is the consequence of the microscopic physics describing these 
fundamental elements.
One should be able therefore to reconstruct the macroscopic physics from the microscopic physics instead of building 
it on phenomenological arguments.
This procedure allows a much deeper understanding of the macroscopic physics and the explicit derivation of its 
phenomenological elements.
We know of course today that atoms and molecules are not the fundamental constituents of matter because they have 
an internal structure made of electrons and nucleons, the latter being again made of quarks, etc. 
However, if one can be sure that the internal structure of an atom will not contribute to some macroscopic phenomenon 
that one wishes to describe, the dynamics of the atom can be considered fundamental regarding this macroscopic phenomenon.
This means that the microscopic theory that one uses depends on the nature of the macroscopic theory that one wishes to build. 
It is not clear at present if a definitive fundamental microscopic theory (which cannot be reduced anymore to anything 
else) will ever be found.\\
We can notice that an substential amount of information required for a microscopic description (for example all the 
positions and velocities of the atoms or molecules of a gas) is not relevant to understand a macroscopic phenomenon 
already captured by much fewer variables (such as the volume, pressure and temperature for an ideal gas). 
A huge reduction of the unnecessary information has to be carried out to understand how the macroscopic world can emerge 
from the interaction between the microscopic constituents of matter.
This is precisely the main goal of the branch of physics called statistical mechanics. 
It should be noticed that this body of knowledge is of great importance not only in physics but also in all the other 
fields investigated by the human intelligence such as biology, economy and social behavior, where one wishes to 
understand the behavior of macroscopic variables that consist in a large collection of microscopic variables.
   
In this thesis, we will focus on the specific part of statistical mechanics which deals with the physical systems 
described by non-relativistic quantum mechanics. 
As we know, the microscopic laws of nature are described by quantum mechanics. 
However, for many studies in statistical mechanics, one can suppose that the equations of motion of the microscopic 
constituents of a physical system are classical.
This approach has a well-defined range of applicability (the de Broglie wavelength has to be shorter than the 
mean free path of the microscopic particle) but has the advantage of being more easily tractable.
An important step to construct statistical mechanics is to take the so-called thermodynamic limit of a system.
This means that one supposes that the number of microscopic variables in the system becomes very large so that the
law of large numbers of probability theory applies.
Classical statistical mechanics is concerned with systems which can be described in the classical limit before taking
the thermodynamic limit. 
When the quantum effects manifest themselves, statistical mechanics has to use a quantum mechanical description.
Quantum statistical mechanics therefore formally consists in taking the thermodynamic limit without any classical limit.

The field of statistical mechanics can be roughly subdivided into equilibrium and nonequilibrium statistical mechanics.\\
In equilibrium statistical mechanics all the thermodynamic fluxes in the system are vanishing and the macroscopic 
quantities describing the system are time independent. 
It is well known that the equilibrium probability distribution of the microscopic degrees of freedom of the system depends 
on the type of system one considers (microcanonical distribution for an isolated system, canonical distribution for a system
exchanging energy with its environment and grand-canonical distribution for a system exchanging energy and particles with
its environment \cite{DiouB97,KuboB1,McQuarrieB00}). \\
In nonequilibrium statistical mechanics, there are three main ways of putting a system out of equilibrium.  
On can impose nonzero thermodynamic fluxes through the system using nonequilibrium boundary conditions 
(i.e. escape-rate formalism \cite{GaspardB98}).
This method is essentially used to study the nonequilibrium steady states of the system.
One can also apply time-dependent external perturbations to a system which induce local nonzero thermodynamic fluxes 
in this system (i.e. linear response theory \cite{KuboB2}).
This method is often used to study the response of a system to perturbations of its equilibrium state.
Finally, one can choose an initial condition putting the system out of equilibrium 
(i.e. Boltzmann's kinetic theory \cite{BoltzmannB95}).
This induces transient nonzero thermodynamic fluxes in the system which relax at long times to equilibrium. 
The nonequilibrium probability distribution of the microscopic degrees of freedom of a system is more difficult
to construct than its equilibrium probability distribution and there is unfortunately no general method for this construction. 
Furthermore, the nonequilibrium distributions obeys kinetic equations which manifest irreversible properties.  
The origin of this irreversibility is an important problem at the core of nonequilibrium statistical mechanics. 

The work presented in this thesis deals with nonequilibrium quantum statistical mechanics. 
The kind of systems on which we focus our study of nonequilibrium quantum dynamics consist in a quantum subsystem which 
is in contact with a quantum environment. 
In most cases, the initial conditions of the total system composed of the subsystem plus the environment 
are initially out of equilibrium and one studies the relaxation of the total system toward equilibrium 
through the subsystem dynamics.  
Because of the interaction with the environment, the subsystem dynamics is then expected to become irreversible, 
relaxing toward some equilibrium distribution. 
The kinetic equations we will establish for the description of the irreversible dynamics rely on the weak-coupling assumption
between the subsystem and its environment.

\section{Emerging irreversible processes in nanosystems} \label{nanosciences}

The fundamental assumptions of statistical mechanics are based on probability theory. 
They rely on the concept of statistical ensemble.
A statistical ensemble is a collection of identical systems submitted to the same constraints. 
However, the initial condition of each member of the ensemble is slightly different and a given probability 
is associated with each of these different initial conditions.
The fundamental assumption necessary to make the statistical description of a system useful is that an individual 
member of the ensemble has, with a very high probability, the same behavior as the one averaged over all 
members of the statistical ensemble.  
The use of such arguments requires either a large number of degrees of freedom in the system in order to exploit the laws 
of large numbers or a chaotic dynamics which introduces the dynamical randomness necessary for the probabilistic 
description even if the system only possesses a few degrees of freedom. 
With the recent progress in experimental techniques, one can study in great detail single systems ranging from
macroscopic scales down to the scale of the nanometer.
This ability to study individual systems of size varying from one or a few atoms to thousand or more atoms allows us to 
investigate the crossover between a dynamical behavior very far from the one expected from kinetic theory and a 
statistical behavior well described in terms of the concepts of nonequilibrium statistical mechanics.
It is a great challenge to refine the statistical mechanical description in order to be able to study systems 
in these intermediate regimes. 
It is precisely this domain that nanotechnologies are nowadays reaching.
Notice that such systems are also very useful to investigate the importance of chaos (classical or quantum) for the 
emergence of statistical behavior in few-degree-of-freedom systems.    

We will now give some examples of typical systems in which such questions can be studied.
It is known that semiconducting devices can range from macroscopic scales (standard electronics) down to the scale of micrometer  
(microelectronics) or even to the nanoscopic scales (artificial atoms, quantum dots).
These devices are characterized by their electronic conduction. 
Thanks to nanotechnology, it is nowadays possible to give different shapes to such electronic devices at the nanoscale 
\cite{Alhassid00,Beenakker97,NakamuraB04,Reimann02}.  
Because they can have different shapes, giving rise to either classically integrable dynamics or classically chaotic dynamics,
these quantum dots are ideal candidates to study the emerging statistical behaviors due to the dynamical complexity.  
In such systems one can also focus on the spin of the electron instead of its charge.  
This new emerging field studying the spin dynamics inside quantum dots is called spintronics \cite{Gupta99,Gupta01,Wolf01,Zutic04}.
The study of spin decoherence, which is a statistical phenomenon, in such devices is of particular importance for potential 
future applications (i.e. quantum computers).
Another example of quantum systems which can be studied today are single atoms in microwave cavities \cite{Raimond01}. 
These spectroscopic experiments provide a detailed study of the atom-cavity entanglement and of the decoherence effect induced 
by the interaction of the atom-cavity with its surrounding. 
The femtochemistry study of IVR (intramolecular vibrational relaxation) in small molecules is also a very rich topic for
statistical mechanics. 
Depending on the vibrational dynamical complexity (chaoticity), the energy initially localized by a laser on a given chemical
bond can flow or not in a statistical way through the other bonds and reach equipartition \cite{Callegari99}. 
Besides, the different microscopy techniques (FIM, STM, AFM, $\dots$) \cite{Giessibl03,Hofer03} play a fundamental role to  
allow the nanoscale investigations. 
Let us give some examples of nanometric systems in which emerging statistical behaviors can be studied by such techniques. 
FIM can allow the real time study with atomic resolution of chemical reactions over catalytic surfaces which can give rise to
nonlinear collective statistical phenomena such as surface wave propagation, concentration pattern formation and explosions 
\cite{DeDecker02}.
STM allows the study of electronic transport properties of single molecules (nanotubes, conducting polymers and other molecules)
\cite{Qui04,NanotubeConduction,MoleculeConduction}. 
Besides, mechanical properties of a single macromolecule can be studied by AFM in experiments which require statistical arguments
for their interpretation \cite{Kreuzer01}. \\
These few examples from the nanoscience world show us the great number of possible applications of statistical-mechanical
theories in this expanding field. 
There are numerous theoretical as well as technological challenges opened in the field of nanosciences that need to be investigated.\\

\section{Quantum measurement as a nonequilibrium process} \label{QM}

The measurement postulate of quantum mechanics, which tells us that the state of a measured system is projected with a given 
probability in one of the eigenstates of its measured observable, if applied to single quantum trajectories, is in apparent 
contradiction with the unitary evolution of quantum mechanics.
It is natural, especially in the context of statistical mechanics, to consider that the measurement apparatus as well as the 
measured system can be described by the laws of quantum mechanics without introducing any dichotomy in the physical 
description between macroscopic and microscopic objects. 
Therefore, the unitary quantum evolution has to apply to the total system (measurement apparatus and measured system).
The measurement process in this scheme automatically introduces a statistical description of the system dynamics.
In fact, due to the non-separability of quantum mechanics, even if the system and the measurement apparatus are described
by pure states before the measurement, once the measurement process has started the system cannot be described anymore by a 
pure state but rather by a statistical mixture. 
This means that, according to the measurement postulate in its actual formulation, entropy is produced when a measure is 
performed on a quantum system.   
We see here an important analogy between the case of a subsystem interacting with its environment in nonequilibrium quantum 
statistical mechanics and the case of a measured system interacting with a measurement apparatus in quantum mechanics.
This analogy already appears at the level of the total Hamiltonian, which is the sum of the subsystem Hamiltonian, 
the environment or the measurement apparatus Hamiltonian, and an interaction term describing the coupling between these two systems.  
Adopting this view of the problem, the delicate question to solve becomes to understand how the total unitary evolution 
can statistically induce a measurement-like effect on the dynamics of the subsystem.   
We want here to point out the importance that nonequilibrium quantum statistical mechanics has for the fundamental 
understanding of quantum mechanics and its measurement postulate. 

\section{Goal of the thesis} \label{goal}

The general goal of this thesis is to contribute to the theoretical description of irreversible processes occurring in 
quantum systems.
We would like more specifically to investigate three major questions.\\
The first one concerns the application of the standard description of irreversible quantum processes to nanosystems. 
In fact, in the standard approaches, one studies the dynamics of a small quantum subsystem in contact with an infinitely large 
environment \footnote{System sizes refer here to the energy scales rather than to spacial scales.}.  
However, in nanosystems, the environment cannot always be considered infinite.
We therefore would like to elaborate a new theory permitting the description of the irreversible dynamics of a subsystem 
interacting with a finite environment by taking into account the mutual effects of the environment on the subsystem 
and vice-versa.    
The statistical ensemble relevant to study such problems is no longer the canonical ensemble as for infinite environments, but 
the microcanonical ensemble.\\
The second important question we want to answer in this thesis is related to the necessity of having a system with a continuous 
spectrum to manifest true irreversible quantum dynamics.
We would like to understand how and under which conditions an almost-kinetic process can take place in a system with a discrete but 
dense spectrum. 
Understanding the transition between purely reversible quantum dynamics and kinetic quantum dynamics is also an important issue 
in nanosciences.\\
Finally, the third major question to be solved in this thesis concerns quantum diffusion in spatially extended systems.
We would like to understand how, and under which conditions, a diffusive transport process can take place, due to the interaction 
with the environment, in spatially extended quantum systems.
This problem of quantum diffusion is once again of great importance in nanosciences where transport processes can be studied 
in many different spatially extended nanosystems.  

\section{Plan of the thesis} \label{plan}

In this thesis, after having introduced the basic concepts of quantum statistical mechanics (in chapter \ref{ch1}) 
and after having derived the most important weak-coupling quantum kinetic equations known at present (in chapter \ref{ch2}), 
we will mainly focus on three important aspects of the emergence of statistical behaviors in nanosystems. \\
The first aspect (in chapter \ref{ch3}), is the derivation of a kinetic equation for the relaxation dynamics of a 
quantum subsystem weakly interacting with a quantum environment of finite size whose energy is modified by the 
subsystem-environment energy transfer.
In this way, we generalize the Redfield equation, which under its actual form only applies to quantum subsystems 
interacting with an infinite environment (heat bath) which is not affected by the subsystem. 
This is an important issue in nanosystems, where the non-relevant part of the degrees of freedom 
of the total system, playing the role of an environment, cannot always be considered infinite. 
The application of this kinetic equation to a two-level subsystem interacting in a non-diagonal way with a general 
environment, allows a better understanding of the physical processes behind the general formalism and the explicit 
derivation of the subsystem relaxation rates as well as of the subsystem equilibrium distribution.
The second aspect of the emergence of statistical behaviors in nanosystems is studied (in chapter \ref{ch4})
by specifying the environment operators of the precedent model in terms of random matrices. 
Doing this, one completely specifies the total system and allows a detailed comparison between the predictions of the kinetic 
equation and the exact dynamics of the total system. 
Such an analysis permits the important discussion of the applicability of the kinetic equation to finite quantum systems. 
This specific model, we call spin-GORM model, allows us to derive an analytical solution for the subsystem relaxation in the 
strong-coupling regime.\\
Finally, the third aspect of the emergence of statistical behavior in nanosystems investigated in this thesis (in chapter \ref{ch5}) 
concerns the study of quantum transport on a quantum unidimensional chain due to the interaction of the chain with its environment. 
We show in this model that the coherent ballistic transport on the isolated chain disappears, due to the interaction with 
the environment, and is replaced by a diffusive transport.
This study is an important step towards the understanding of the emergence of transport phenomena in quantum nanosystems.

\chapter{Basic concepts in nonequilibrium quantum statistical mechanics} \label{ch1}

We establish in this chapter the main ideas necessary for the derivation of quantum kinetic equations. 
We start in section \ref{IQD} by defining the concept of statistical ensemble in quantum mechanics in terms 
of density matrices and by deriving the von Neumann equation which rules the dynamics of these ensembles.
On this ground, we discuss the main properties of the time evolution of a quantum statistical ensemble.
In section \ref{Projdyn}, we formally derive a dynamical equation for the relevant part of a statistical 
ensemble using projection operators. 
We show that the main concepts that we need for the derivation of quantum kinetic equations in the following 
chapters, are already present at this formal level.
In section \ref{Posevol}, we discuss the problem of the positivity preservation of quantum kinetic equations
and present the most general form of a quantum master equation preserving positivity (Lindblad equation).    

\section{Quantum dynamics of isolated systems} \label{IQD}

We introduce in this section the basic concepts necessary for the statistical description of an 
ensemble of isolated quantum systems ruled by a Hamiltonian $\hat{H}$.

\subsection{Single quantum system dynamics}

In quantum mechanics, the state of a single system is described by the time-dependent state 
$\vert \psi (t) \rangle$ evolving in the state space $\cal E$ according to the Schr\"odinger equation 
\begin{eqnarray}
\frac{d \vert \psi (t) \rangle}{dt} = - \frac{i}{\hbar} \hat{H} \vert \psi (t) \rangle ,                    \label{1Aaaaa}
\end{eqnarray}
where the Hamiltonian $\hat{H}$ is a Hermitian operator in $\cal E$. 
The formal solution of the Schr\"odinger equation is
\begin{eqnarray}
\vert \psi (t) \rangle = e^{ -\frac{i}{\hbar} \hat{H} t} \vert \psi (0) \rangle ,                           \label{1Aaaab}
\end{eqnarray}
where $e^{ -\frac{i}{\hbar} \hat{H} t}$ is the evolution operator.
The eigenvalue equation of the Hamiltonian reads 
\begin{eqnarray}
\hat{H} \vert n \rangle = E_n \vert n \rangle,                                                              \label{1Aaaac}
\end{eqnarray}
where $\{ E_n \}$ are the real eigenvalues of the Hamiltonian and $\{ \vert n \rangle \}$ its eigenvectors.
The eigenvectors constitute an orthogonal basis in the state space $\cal E$: $\langle n \vert n' \rangle = \delta_{n n'}$.
The closure equation reads $\sum_n \vert n \rangle \langle n \vert = \hat{I}$, where $\hat{I}$ is the identity operator.
The spectral decomposition of the evolution operator in $\cal E$ is given by   
\begin{eqnarray}
\vert \psi (t) \rangle = \sum_{n} e^{ - \frac{i}{\hbar} E_n t} c_n (0) \vert n \rangle .                    \label{1Aaaad}
\end{eqnarray}

\subsection{Quantum statistical ensemble dynamics}

The fundamental concept of statistical ensemble has been introduced by Gibbs in classical statistical mechanics.
A statistical ensemble is a collection of noninteracting identical copies of a system, the copies being in different 
states at a given time. 
A density matrix describes the ensemble by assigning a probability $P^{\mu}$ to the different states found in the 
ensemble (the probability corresponds to the fraction of copies in the ensemble which have the same state and therefore 
$\sum_{\mu} P^{\mu}=1$). 
The density matrix is the central quantity in quantum statistical mechanics and reads
\begin{eqnarray}
\hat{\rho} (t) = \sum_{\mu} P^{\mu} \vert \psi^{\mu} (t) \rangle \langle \psi^{\mu} (t) \vert.             \label{1Aaaae}
\end{eqnarray}
By construction, the density matrix is a Hermitian operator: $\hat{\rho}^{\dagger}=\hat{\rho}$. 
The normalization implies ${\rm Tr} \hat{\rho}=1$.  
Furthermore, a density matrix describing a pure state ($P^{\mu}=\delta_{1 \mu}$) has the property: 
$\hat{\rho}^2=\hat{\rho}$, which is not the case for a density matrix describing a statistical ensemble: 
$\hat{\rho}^2 \neq \hat{\rho}$.
In a given basis, the diagonal elements of the density matrix, which are always positive real numbers, are 
called the populations and are interpreted as the probability of being in a given state. 
The non-diagonal elements of the density matrix are called the coherences and have no classical interpretation.
A detailed presentation of quantum statistical ensembles can be found in Ref. \cite{Fano57}.
The density matrix and therefore the quantum statistical ensemble evolves according to the von Neumann equation
\begin{eqnarray}
\frac{d \hat{\rho} (t)}{dt} = -\frac{i}{\hbar} [\hat{H},\hat{\rho}(t)] = \hat{\hat{{\cal L}}} \hat{\rho}(t), \label{1Aaaaf}
\end{eqnarray}
which is a direct consequence of the definition of the density matrix and of the Schr\"odinger equation. 
The von Neumann equation is the fundamental equation of nonequilibrium statistical mechanics.
All the derivations of kinetic equations start from the von Neumann equation which contains the full microscopic
information of the statistical ensemble.\\
A superoperator is a mathematical object which maps an operator into another operator. 
The superoperator $\hat{\hat{{\cal L}}}$ which generates the evolution in the von Neumann equation is called 
the Liouvillian. 
The formal solution of the von Neumann equation is given by 
\begin{eqnarray}
\hat{\rho} (t) = e^{- \frac{i}{\hbar} \hat{H} t} \hat{\rho} (0) e^{\frac{i}{\hbar} \hat{H} t} 
= e^{ \hat{\hat{{\cal L}}} t} \hat{\rho} (0),                                                                \label{1Aaaag}
\end{eqnarray}
where $e^{ \hat{\hat{{\cal L}}} t}$ is the evolution superoperator.
The Liouvillian eigenvalue equation reads
\begin{eqnarray}
\hat{\hat{{\cal L}}} \hat{\rho}^{\nu} = s_{\nu} \hat{\rho}^{\nu} ,                                           \label{1Aaaah}
\end{eqnarray}
where $\{ s_{\nu} \}$ are the Liouvillian eigenvalues and $\{ \hat{\rho}^{\nu} \}$ the Liouvillian eigenvectors.
The spectral decomposition of the evolution superoperator is given by
\begin{eqnarray}
\hat{\rho} (t) = \sum_{\nu} a_{\nu}(0) e^{ s_{\nu} t} \hat{\rho}^{\nu}.                                      \label{1Aaaai}
\end{eqnarray}
Using the density matrix (\ref{1Aaaae}) with the spectral decomposition (\ref{1Aaaad}), we get
\begin{eqnarray}
\hat{\rho} (t) = \sum_{n,n'} \sum_{\mu} P^{\mu} c^{\mu}_n (0) c^{\mu}_{n'} (0) 
e^{ - \frac{i}{\hbar} (E_n-E_{n'}) t} \vert n \rangle \langle n' \vert .                                      \label{1Aaaaj}
\end{eqnarray}
By identifying (\ref{1Aaaai}) with (\ref{1Aaaaj}) we conclude that the coefficients of the spectral 
decomposition (\ref{1Aaaai}) are given by
\begin{eqnarray}
a_{\nu}(0) = a_{nn'}(0) = \sum_{\mu} P^{\mu} c^{\mu}_n (0) c^{\mu}_{n'} (0),                                 \label{1Aaaak}
\end{eqnarray}
that the Liouvillian eigenvectors are given by
\begin{eqnarray}
\hat{\rho}^{\nu} = \hat{\rho}^{nn'} = \vert n \rangle \langle n' \vert ,                                     \label{1Aaaal}
\end{eqnarray}
and that the Liouvillian eigenvalues are pure complex numbers given by 
\begin{eqnarray}
s_{\nu} = - i \omega_{\nu},                                                                                   \label{1Aaaam}
\end{eqnarray}
where $\omega_{\nu}$ are the Bohr frequencies of the system given by
\begin{eqnarray}
\omega_{\nu} = \omega_{nn'} = \frac{E_n-E_{n'}}{\hbar}.                                                      \label{1Aaaan}
\end{eqnarray}
Notice that if the Hamiltonian operator of the system, and therefore also the evolution operator, 
is represented by a $N \times N$ matrix (two-index object) in $\cal E$, the Liouvillian superoperator 
and the evolution superoperator are represented by a $N \times N \times N \times N$ tetradic 
(four-index object) in $\cal E$. \\

Let consider $\hat{A}$ an observable (Hermitian operator) in $\cal E$. 
Its evolution is given by
\begin{eqnarray}
\langle \hat{A} (t) \rangle = {\rm Tr} \hat{\rho} (t) \hat{A} = 
\sum_{n,n'} e^{ - \frac{i}{\hbar} (E_n-E_{n'}) t} \langle n \vert \hat{\rho} (0) \vert n' \rangle 
\langle n' \vert \hat{A} \vert n \rangle.                                                                     \label{1Aaaao}
\end{eqnarray}
Such an observable is an almost-periodic function of time. 
If the energy eigenvalues $E_n$ are rationally dependent, it is even a periodic function of time. 
The almost-periodicity implies a recurrence theorem (which is the quantum analogue of the Poincar\'e 
recurrence theorem).
The theorem states that there always exists a recurrence time around which the observable reaches a 
value arbitrarily close to its initial condition \cite{Bocchieri}.  
This recurrence theorem, which reflects the reversibility of the full dynamics, limits 
the applicability of kinetic equations to quantum finite systems.
In fact, kinetic behaviors will only be observable on time scales shorter than the Heisenberg time, which 
is defined as $t_H = \hbar n_{av}(\epsilon)$, where $n_{av}(\epsilon)$ is the mean density of states at 
total energy $\epsilon$. 
However, the Heisenberg time grows very fast with the size of the quantum system. 
The greater the quantum system, the denser its density of states and the longer its Heisenberg time.
This discussion shows that the kinetic behavior in a finite quantum system is intimately related to its 
density of states.
Mathematically, of course, a true irreversible quantum dynamics only exist in the thermodynamical limit 
where the size of the system goes to infinity and therefore its spectrum becomes continuous. 
The quasi-continuous limit is a mathematical tool which sends the Heisenberg time to infinity. 
This limit can be performed by assuming that the eigenvalues of the spectrum are so densely distributed 
that the energy can be considered continuous $E_n = \epsilon$.
Furthermore, one has to assume that the Hamiltonian eigenbasis representation of the relevant operators 
is a continuous function of the energy $\bra{n} \hat{A} \ket{n'} = A(\epsilon,\epsilon')$ with $\epsilon=E_n$ and 
$\epsilon'=E_{n'}$.   
Under such circumstances, the sum over the quantum states in the relevant quantities can be replaced 
by integrals over the continuous energy $\sum_{n} \to \int d\epsilon \; n(\epsilon)$ where $n(\epsilon)$ 
is the density of states of the system at energy $\epsilon$.
The quasi-continuous limit of (\ref{1Aaaao}) would give
\begin{eqnarray}
\langle \hat{A} (t) \rangle = {\rm Tr} \hat{\rho} (t) \hat{A} = 
\int d\epsilon \int d\epsilon' n(\epsilon) n(\epsilon') e^{ - \frac{i}{\hbar} (\epsilon-\epsilon') t} 
\rho_0 (\epsilon,\epsilon') A (\epsilon',\epsilon).                                                             \label{1Aaabo}
\end{eqnarray}
Now, the calculation can exploit the power of infinitesimal calculus, the Heisenberg time is infinite, 
and a pure irreversible dynamics can occur.\\
The time average of an evolving observable $\hat{A}(t)$, which defines the equilibrium value of the 
observable, is an important quantity in nonequilibrium quantum statistical mechanics.
It is defined by
\begin{eqnarray}
\langle \hat{A} \rangle_{\infty} = \lim_{T \to \infty} \frac{1}{T} \int_{0}^{T} dt 
\langle \hat{A} (t) \rangle = \sum_{n} \langle n \vert \hat{\rho} (0) \vert n \rangle 
\langle n \vert \hat{A} \vert n \rangle,                                                                         \label{1Aaaap}
\end{eqnarray}
where we used the property 
\begin{eqnarray}
\lim_{T \to \infty} \frac{1}{T} \int_{0}^{T} dt e^{ - \frac{i}{\hbar} (E_n-E_{n'}) t} = \delta_{nn'}.           \label{1Aaaaq}
\end{eqnarray}
By rewriting (\ref{1Aaaap}) in the following way 
\begin{eqnarray}
\langle \hat{A} \rangle_{\infty} = {\rm Tr} \hat{\rho}_{\infty} \hat{A},                                         \label{1Aaaar}
\end{eqnarray}
we define the equilibrium density matrix
\begin{eqnarray}
\hat{\rho}_{\infty} = \vert n \rangle \langle n \vert \; \langle n \vert \hat{\rho} (0) \vert n \rangle .        \label{1Aaaas}
\end{eqnarray}
We notice that the time average of the observable is equivalent to its ensemble average over the invariant 
equilibrium density matrix $\hat{\rho}_{\infty}$. 
Therefore, according to O. Penrose \cite{Penrose75}, one could assert that all finite quantum systems with 
a non-degenerate spectrum are ergodic. 
Unfortunately, this concept of ergodicity is so trivially satisfied, that it is far less important 
for quantum statistical mechanics than for classical statistical mechanics.
Notice also that the equilibrium density matrix depends on the initial conditions.\\
With the definition of the equilibrium density matrix and with the recurrence theorem we have the keys to 
understand the dynamics of statistical ensembles of finite quantum systems.
In a small quantum system containing only a few levels, the dynamics of an observable will strongly  
oscillate and the equilibrium value of the observable has no relevance to characterize the system. 
For such systems a statistical description in terms of kinetic equations has no relevance.
But when the spectrum of the considered system gets denser and the number of levels implied in the dynamics 
increases, the observable dynamics can display a relaxation behavior which reaches the equilibrium 
distribution on time scales shorter than the Heisenberg time. 
On this time scale, a kinetic description in terms of reduced statistical ensembles is relevant. 
On larger time scales however, recurrences appear so that the kinetic description fails. 
The observable may sometimes leave its equilibrium value, even if the observable spends most of its time 
close to its equilibrium value.

\subsection{Liouville space}

We will end this section by a brief description of quantum statistical mechanics in
the Liouville space. 
Because it is rarely used, the notation that we now introduce will not be used further in this thesis 
calculation. 
However, we want to introduce this formalism because we think it gives a nice intuition of the way in which
superoperators act.\\   
The isomorphism between the matrix algebra for operators in Hilbert space and the tetradic 
algebra for superoperators in Hilbert space allows us to represent the operators as vectors 
and the superoperators as operators in a new space called the Liouville space. 
A quantum operator is represented by a matrix in the Hilbert space 
\begin{eqnarray}
\hat{A}= \hat{I} \hat{A}  \hat{I} = \sum_{n,n'} A_{n n'} \vert n \rangle \langle n' \vert,             \label{1Daaaa}
\end{eqnarray}
where $A_{n n'}=\langle n \vert \hat{A} \vert n' \rangle$.
We can represent this operator by a ket $\vert \hat{A} \gg$ in the Liouville space. The adjoint of this
operator is represented by the bra $\ll \hat{A} \vert$. 
The scalar product between two operators in the Liouville space is defined by
\begin{eqnarray}
\ll \hat{B} \vert \hat{A} \gg = {\rm Tr} \hat{B}^{\dagger} \hat{A}.                                    \label{1Daaab}
\end{eqnarray}
The norm in the Liouville space is given by 
$\vert \vert \hat{A} \vert \vert = \sqrt{\ll \hat{A} \vert \hat{A} \gg}$. 
A projector and its adjoint, which are superoperators in the Hilbert space, are respectively represented by the ket 
\begin{eqnarray}
\vert n,n' \gg = \vert n \rangle \langle n' \vert                                                      \label{1Daaac}
\end{eqnarray}
and the bra 
\begin{eqnarray}
\ll n,n' \vert = \vert n' \rangle \langle n \vert                                                      \label{1Daaad}
\end{eqnarray}
in the Liouville space.  
The ensemble of kets $\{ \vert n,n' \gg \}$ constitute an orthonormal basis in the Liouville space
\begin{eqnarray}
\ll n,n' \vert m,m' \gg = \delta_{n m} \delta_{n' m'}.                                                 \label{1Daabd}
\end{eqnarray}
The identity in the Liouville space is given by
\begin{eqnarray}
\hat{\hat{I}} = \sum_{n,n'} \vert n,n' \gg \ll n,n' \vert  .                                           \label{1Daaae}
\end{eqnarray}
An operator $\hat{A}$ is therefore represented by
\begin{eqnarray}
\hat{A} = \sum_{n,n'} \ll n,n' \vert \hat{A} \gg \vert n,n' \gg = \sum_{n,n'} A_{n n'} \vert n,n' \gg , \label{1Daaaf}
\end{eqnarray}
and a superoperator $\hat{\hat{S}}$ by 
\begin{eqnarray}
\hat{\hat{S}} &=& 
\sum_{n,n'} \sum_{m,m'} \ll n,n' \vert \hat{\hat{S}} \vert m,m' \gg \vert n,n' \gg \ll m,m' \vert \nonumber \\
&=& \sum_{n,n'} \sum_{m,m'} S_{n n',m m'} \vert n,n' \gg \ll m,m' \vert    ,                            \label{1Daaag}
\end{eqnarray}
in the Liouville space.
The adjoint of a superoperator in the Liouville space is defined as
\begin{eqnarray}
\ll n,n' \vert \hat{\hat{S}}^{\dagger} \vert m,m' \gg = \ll m,m' \vert \hat{\hat{S}} \vert n,n' \gg^{*} .   \label{1Daaah}     
\end{eqnarray}
A Hermitian superoperator is therefore defined by $\hat{\hat{S}}^{\dagger}=\hat{\hat{S}}$.\\
The Liouvillian can be represented in the Liouville space by
\begin{eqnarray}
\hat{\hat{{\cal L}}} &=& 
\sum_{n,n'} \sum_{m,m'} \ll n,n' \vert \hat{\hat{{\cal L}}} \vert m,m' \gg \vert n,n' \gg \ll m,m' \vert, \label{1Daabh}
\end{eqnarray} 
where
\begin{eqnarray}
\ll n,n' \vert \hat{\hat{{\cal L}}} \vert m,m' \gg = {\cal L}_{n n',m m'}
=- \frac{i}{\hbar} (H_{n m} \delta_{n' m'} - H_{n' m'} \delta_{n m}).                                    \label{1Daach}
\end{eqnarray} 
We see that the Liouville superoperator is anti-Hermitian 
$\hat{\hat{{\cal L}}}^{\dagger}=-\hat{\hat{{\cal L}}}$.\\
If the $n$ and $m$ indices correspond to the eigenstates of the Hamiltonian, we have
\begin{eqnarray}
{\cal L}_{n n',m m'} = - \frac{i}{\hbar} (E_n-E_{n'}) \delta_{n m} \delta_{n' m'}
= - i \omega_{n n'} \delta_{n m} \delta_{n' m'} = s_{n n'} \delta_{n m} \delta_{n' m'}. \nonumber \\    \label{1Daadh}
\end{eqnarray} 
The von Neumann equation in the Liouville space can be written
\begin{eqnarray}
\frac{d \vert \hat{\rho}(t) \gg}{dt} = \hat{\hat{{\cal L}}} \vert \hat{\rho}(t) \gg ,                    \label{1Daaai}
\end{eqnarray}
where the formal solution takes the form
\begin{eqnarray}
\vert \hat{\rho}(t) \gg = e^{\hat{\hat{{\cal L}}} t} \vert \hat{\rho}(0) \gg .                           \label{1Daaaj}
\end{eqnarray}
The spectral decomposition of the evolution superoperator now reads
\begin{eqnarray}
\vert \hat{\rho}(t) \gg = \sum_{n,n'} \sum_{m,m'} \ll n,n' \vert e^{\hat{\hat{{\cal L}}} t} 
\vert m,m' \gg \ll m,m' \vert \hat{\rho}(0) \gg \vert n,n' \gg.                                          \label{1Daaak}
\end{eqnarray}
If the $n$ and $m$ indices again correspond to the eigenstates of the Hamiltonian, we have
\begin{eqnarray}
\vert \hat{\rho}(t) \gg = \sum_{n,n'} e^{s_{n n'} t} \ll n,n' \vert \hat{\rho}(0) \gg \vert n,n' \gg.    \label{1Daaal}
\end{eqnarray}

\section{Projected dynamics} \label{Projdyn}

We will introduce in this section the Nakajima-Zwanzig projection operator formalism 
and derive from it the generalized quantum master equation. 
This method is a relatively general method to construct the relevant statistical operator to describe 
a kinetic process by starting from the full density matrix describing a system.
We notice that this formalism is restricted to linear transformations for the construction of the relevant 
statistical operator. 
In fact, non-linear transformations could also be considered, but we shall not use such ones in this thesis.
By applying the projection technique to the von Neumann equation, one is able to obtain a general master 
equation which is exact and is the starting point of many quantum derivations of kinetic equations 
\cite{KuboB2,ZubarevB96,Zwanzig61}.
We derive this generalized master equation because, even if it is a formal derivation, it shows how
non-Markovian effects arise as a natural consequence of the dynamical description.    
We would also like to mention that the Nakajima-Zwanzig projection operator formalism shares many common
points with the method of nonequilibrium statistical operators by Zubarev and coworkers \cite{ZubarevB96}.
In a sense, the Nakajima-Zwanzig projection operator formalism can be seen as a specific case of the
non-equilibrium statistical operator method \cite{ZubarevB96}.
However, they are very similar for our purpose and the Nakajima-Zwanzig formalism has the advantage of 
being simpler and more compact.

\subsection{Generalized master equation}

We start by defining the projection superoperators. $\hat{\hat{{\cal P}}}$ is the projection superoperator
which selects the relevant part of the dynamics when acting on the full density matrix and 
$\hat{\hat{{\cal Q}}}$ the projection superoperator which selects the non-relevant part of the dynamics
when acting on the full density matrix.\\  
The basic properties of these projection superoperators are
\begin{eqnarray}
&&\hat{\hat{{\cal P}}}+\hat{\hat{{\cal Q}}}=\hat{\hat{I}}                                                 \label{1Baaaa} \\
&&\hat{\hat{{\cal P}}}^2=\hat{\hat{{\cal P}}}                                                             \label{1Baaab} \\
&&\hat{\hat{{\cal Q}}}^2=\hat{\hat{{\cal Q}}}                                                             \label{1Baaac} \\
&&\hat{\hat{{\cal P}}} \hat{\hat{{\cal Q}}} = \hat{\hat{{\cal Q}}} \hat{\hat{{\cal P}}} = 0.              \label{1Baaad}
\end{eqnarray}
We want now to obtain an equation for the relevant part of the dynamics.
To do this, we apply the projection operators to the von Neumann equation (\ref{1Aaaaf})
\begin{eqnarray}
\frac{d}{dt} \hat{\hat{{\cal P}}} \hat{\rho} (t) &=& 
\hat{\hat{{\cal P}}} \hat{\hat{{\cal L}}} \hat{\hat{{\cal P}}} \hat{\rho}(t) + 
\hat{\hat{{\cal P}}} \hat{\hat{{\cal L}}} \hat{\hat{{\cal Q}}} \hat{\rho}(t)                              \label{1Baaae} \\
\frac{d}{dt} \hat{\hat{{\cal Q}}} \hat{\rho} (t) &=& 
\hat{\hat{{\cal Q}}} \hat{\hat{{\cal L}}} \hat{\hat{{\cal Q}}} \hat{\rho}(t) + 
\hat{\hat{{\cal Q}}} \hat{\hat{{\cal L}}} \hat{\hat{{\cal P}}} \hat{\rho}(t).                             \label{1Baaaf}
\end{eqnarray}
The equations for the relevant part and for the non-relevant part of the density matrix are 
coupled together.
Integrating the evolution equation for the non-relevant part (\ref{1Baaaf}), we get
\begin{eqnarray}
\hat{\hat{{\cal Q}}} \hat{\rho} (t) &=& 
\int_{0}^{t} d\tau' e^{(t-\tau') \hat{\hat{{\cal Q}}} \hat{\hat{{\cal L}}} }
\hat{\hat{{\cal Q}}} \hat{\hat{{\cal L}}} \hat{\hat{{\cal P}}} \hat{\rho}(\tau') + 
e^{t \hat{\hat{{\cal Q}}} \hat{\hat{{\cal L}}} } \hat{\hat{{\cal Q}}} \hat{\rho}(0).                      \label{1Baaag}
\end{eqnarray}
Making the change of variable $\tau \to t-\tau'$ in (\ref{1Baaag}) and inserting the result in the 
evolution equation of the relevant part (\ref{1Baaae}), we get the generalized master equation 
\begin{eqnarray}
\frac{d}{dt} \hat{\hat{{\cal P}}} \hat{\rho} (t) &=& 
\hat{\hat{{\cal P}}} \hat{\hat{{\cal L}}} \hat{\hat{{\cal P}}} \hat{\rho}(t) + 
\hat{\hat{{\cal P}}} \hat{\hat{{\cal L}}} \int_{0}^{t} d\tau e^{\tau \hat{\hat{{\cal Q}}} 
\hat{\hat{{\cal L}}} } \hat{\hat{{\cal Q}}} \hat{\hat{{\cal L}}} \hat{\hat{{\cal P}}} \hat{\rho}(t-\tau) + 
\hat{\hat{{\cal P}}} \hat{\hat{{\cal L}}} e^{t \hat{\hat{{\cal Q}}} \hat{\hat{{\cal L}}} } 
\hat{\hat{{\cal Q}}} \hat{\rho}(0) . \nonumber \\                                                        \label{1Baaah} 
\end{eqnarray}
This equation is exact and is simply a way of rewriting the von Neumann equation.
The second term of the right-hand side of (\ref{1Baaah}) is called the memory term. 
Because of this term, the evolution of the relevant part of the density matrix at a given time depends 
on its past evolution.
The dynamics of a quantity having an evolution which depends on its history is called a non-Markovian 
dynamics.
We can therefore make the very important observation that the evolution of the projected part of a density 
matrix following a von Neumann equation is non-Markovian. 
This plays a fundamental role in nonequilibrium statistical mechanics. 
The third term in the right-hand side of (\ref{1Baaah}) is inhomogeneous and depends on the initial 
condition of the non-relevant part. 
In many cases, the initial condition of the non-relevant part is assumed to be zero so 
that this third term vanishes.\\
The generalized master equation (\ref{1Baaah}) has been obtained independently by Nakajima
\cite{Nakajima58} in 1958 and by Zwanzig \cite{Zwanzig60,Zwanzig61} in 1960. 
It should be mentioned that the generalized master equations, obtained respectively by R\'esibois
\cite{Resibois61} and by Montroll \cite{Montroll61}, both in 1961, where shown in Ref. \cite{Zwanzig64}
to be equivalent to the Nakajima-Zwanzig master equation.

\subsection{Markovian approximation}

The generalized master equation can take a much simpler form if one performs the so-called Markovian 
approximation.
This approximation is justified if the memory kernel $e^{\tau \hat{\hat{{\cal Q}}} \hat{\hat{{\cal L}}}} 
\hat{\hat{{\cal Q}}} \hat{\hat{{\cal L}}}$ decays to zero on a short time scale $\tau_c$ compared to 
the rate of change in time of the relevant distribution $\hat{\hat{{\cal P}}} \hat{\rho} (t)$.
The Markovian approximation consists in the following assumption
\begin{eqnarray}
\int_{0}^{t} d\tau e^{\tau \hat{\hat{{\cal Q}}} \hat{\hat{{\cal L}}} } \hat{\hat{{\cal Q}}} 
\hat{\hat{{\cal L}}} \hat{\hat{{\cal P}}} \hat{\rho}(t-\tau) \approx 
\{ \int_{0}^{\infty} d\tau e^{\tau \hat{\hat{{\cal Q}}} \hat{\hat{{\cal L}}} } \hat{\hat{{\cal Q}}} 
\hat{\hat{{\cal L}}} \} \hat{\hat{{\cal P}}} \hat{\rho}(t)    .                                           \label{1Baaai} 
\end{eqnarray}
If one furthermore neglects the third term of (\ref{1Baaah}) representing the memory of the
non-relevant initial distribution $\hat{\hat{{\cal Q}}} \hat{\rho}(0)$, the generalized master
equation (\ref{1Baaah}) takes the simple Markovian form
\begin{eqnarray}
\frac{d}{dt} \hat{\hat{{\cal P}}} \hat{\rho} (t) = 
\hat{\hat{M}} \hat{\hat{{\cal P}}} \hat{\rho} (t) ,                                                       \label{1Baaaj} 
\end{eqnarray}
where the Markovian dynamical generator is given by
\begin{eqnarray}
\hat{\hat{M}} = \hat{\hat{{\cal P}}} \hat{\hat{{\cal L}}} +
\int_{0}^{\infty} d\tau e^{\tau \hat{\hat{{\cal Q}}} \hat{\hat{{\cal L}}} } 
\hat{\hat{{\cal Q}}} \hat{\hat{{\cal L}}} .                                                               \label{1Baaak}   
\end{eqnarray}
A system for which a relevant distribution can be found such that it obeys a Markovian master equation like
(\ref{1Baaak}), is a system which can successfully be described by statistical mechanics and whose relevant 
distribution undergoes an irreversible dynamics to equilibrium.

\subsection{Slippage of initial conditions} \label{slippagegen}

An important remark concerns the Markovian approximation. 
As afore mentioned, this approximation is good if the memory kernel decays fast enough (on time scales of 
the order of the correlation time $\tau_c$) compared to the evolution of the relevant part. 
However, on time scales shorter than $\tau_c$, the non-Markovian evolution is very different 
from the Markovian one.
Therefore the Markovian equation, even if correct on the long-time scale, contains a bias coming from the
wrong initial dynamics.
This problem can be solved by the so-called slippage of initial conditions. 
This slippage is the rapid evolution during the time scale $\tau_c$.
We can take this slippage into account by modifying the initial conditions of the relevant density matrix 
used for the Markovian evolution. 
This modification can be expressed by a slippage superoperator applied to these initial conditions. 
This slippage superoperator takes into account the non-Markovian effects occurring on short time scales 
smaller than $\tau_c$ and maps them on the appropriate initial conditions of the Markovian equation.
This procedure eliminates the bias coming from the initial non-Markovian dynamics on the long-time 
scale of the Markovian evolution.    
If we write the solutions of the generalized master equations for the non-Markovian (NM) and for the 
Markovian (M) dynamics as follows 
\begin{eqnarray}
\hat{\hat{{\cal P}}} \hat{\rho}^{NM} (t)&=& 
{\cal F}_t^{NM} \lbrack \hat{\hat{{\cal P}}} \hat{\rho}^{NM} (0) \rbrack                                 \label{1Baaal}\\  
\hat{\hat{{\cal P}}} \hat{\rho}^{M} (t) &=& 
{\cal F}_t^{M} \lbrack \hat{\hat{{\cal P}}} \hat{\rho}^{M} (0) \rbrack  \nonumber ,
\end{eqnarray}
the slippage superoperator is defined by imposing that, on a long time scale (much longer than $\tau_c$),
we have that $\hat{\hat{{\cal P}}} \hat{\rho}^{NM} (t)=\hat{\hat{{\cal P}}} \hat{\rho}^{M} (t)$. 
This implies that the initial condition of the Markovian equation is "slipped" by the 
superoperator $\hat{\hat{S}}$ satisfying
\begin{eqnarray}
{\cal F}_t^{NM} \lbrack \hat{\hat{{\cal P}}} \hat{\rho}^{NM} (0) \rbrack 
={\cal F}_t^{M} \lbrack \hat{\hat{S}} \hat{\hat{{\cal P}}} \hat{\rho}^{M} (0) \rbrack  .                 \label{1Baaam}
\end{eqnarray}
If a reasonably simple form of the slippage superoperator can be found, it is useful to apply it 
to the real initial condition of the system before implementing the Markovian scheme. 
In this way, one keeps a Markovian evolution equation which is much easier to solve 
than a non-Markovian equation and one eliminates the bias coming from the Markovian short time 
scale evolution on longer time scales.
  
\section{Positivity-preserving evolution} \label{Posevol}

We present in this section the restriction imposed on the form of a quantum kinetic equation if the 
evolving quantity (the relevant projected density matrix) has to be interpreted as a density matrix.\\

The density matrix operator, defined by equation (\ref{1Aaaae}), is a Hermitian operator which satisfies
positivity (the diagonal elements in any representation of the density matrix operator are real positive numbers)
and which is normalized to unity (the trace of the density matrix operator is equal to one). 
Quantum kinetic equations are dynamical equations for the relevant part of a density matrix operator.
If the relevant part of a density matrix has to be interpreted as a density matrix, the kinetic equation has
to preserve, during the dynamics, the basic properties of the density matrix: normalization, Hermiticity and
positivity.   
This is for example the case if one considers the reduced density matrix of a subsystem interacting with an 
environment because the reduced density matrix is the projection of the total density matrix on the subsystem 
state space.
It is therefore natural to wonder if the master equation of the reduced dynamics still preserves positivity, 
Hermiticity and the norm.
This question has been discussed by Lindblad in 1976 \cite{Lindblad76} who found that the more general form of 
a quantum master equation preserving these three properties as well as Markovianity is given by 
the so-called Lindblad equation
\begin{eqnarray}
\frac{d \hat{\hat{{\cal P}}} \hat{\rho} (t)}{dt}&=& 
-i \lbrack \hat{H}, \hat{\hat{{\cal P}}} \hat{\rho} (t) \rbrack                                            \label{1Faaaa}\\
&&+ \sum_{\kappa} (2 \hat{L_{\kappa}} \hat{\hat{{\cal P}}} \hat{\rho} (t) \hat{L_{\kappa}}^{\dagger}
- \hat{L_{\kappa}}^{\dagger} \hat{L_{\kappa}} \hat{\hat{{\cal P}}} \hat{\rho} (t)
- \hat{\hat{{\cal P}}} \hat{\rho} (t) \hat{L_{\kappa}}^{\dagger} \hat{L_{\kappa}}) \nonumber ,
\end{eqnarray}
where $\hat{H}$ is an Hermitian operator. \\
The mathematical formulation of this problem relies on the concept of quantum dynamical semi-groups 
which goes beyond the scope of this thesis. 
For further information, a mathematically rigorous discussion on this subject can be found in 
\cite{DaviesB76,Lindblad76,Spohn78,Spohn80}.

\chapter{Kinetic equations for weakly perturbed systems} \label{ch2}

After having established the main concepts necessary to construct quantum statistical mechanics, we will
derive in this chapter some of the most important equations of nonequilibrium quantum statistical mechanics.  
These equations are presented in this chapter before the presentation of our new kinetic equation
in chapter \ref{ch3} where we shall need the material given here. 
A common point to all the theories discussed in the present chapter is the fact that they start from the same 
perturbative expansion of the von Neumann equation.
Therefore, in section \ref{pertVN} we start by performing the perturbation expansion of the von Neumann equation.
In section \ref{linearresponse}, we present the linear response theory describing the response of a quantum system to 
a time-dependent external perturbation.  
In section \ref{FermiGR}, the Fermi golden rule is derived because of its important role in the interpretation of the 
kinetic equations. 
In section \ref{Pauli}, we derive the Pauli equation describing the dynamics of transitions between the unperturbed states
of a quantum system due to a constant perturbation term in its Hamiltonian. 
Finally, in section \ref{Redfield}, we derive the Redfield equation ruling the dynamics of a small quantum
system weakly interacting with an infinitely large environment.

\section{Perturbative expansion of the von Neumann equation} \label{pertVN}

In this section, we develop the perturbative expansion of the von Neumann equation. 
This expansion will be the staring point of all the kinetic equations derived in this thesis.\\ 
The weak-coupling assumption, even if restrictive, is one of the most powerful tools to study nonequilibrium phenomena.
However, some qualitative arguments indicate that such an approach is not as restrictive as one could first think.
In fact, if one wishes to put a system out of equilibrium, this has to be done by imposing constraints on the system
which very often act at the surface of this system.
Therefore, for reasonably large systems, because the surface/volume ratio is low, the interaction can often be 
considered as being weak.    
Another argument is that even for reasonably strong interactions it is sometimes possible to incorporate
the "simple" component of the interaction in the non-perturbed part of the Hamiltonian and to keep a possibly weaker
"complicated" component of the interaction in the interaction term.
Finally, if the interaction is very strong, one can always renormalize the full Hamiltonian in such a way that the old 
non-perturbed part of the Hamiltonian becomes the new weak interaction term and that the old perturbation term becomes 
the the new non-perturbed part of the Hamiltonian.
Of course, such a qualitative justification to extend the application range of the weak-coupling approximation cannot
claim to be always valid and there are still many situations for which non-perturbative schemes have to be used. 
However, in this thesis almost all the results are restricted to weak-coupling.\\

Let us consider a Hamiltonian which is given by
\begin{eqnarray}
\hat{H}=\hat{H}_{0}+\lambda \hat{V}(t).                                                                    \label{1Eaaaa}
\end{eqnarray}
$\hat{H}_{0}$ is generally considered to be an easily diagonalizable Hamiltonian, $\hat{V}(t)$
is an interaction term which does not commute with $\hat{H}_{0}$ and $\lambda$ is
the interaction parameter which measures the intensity of the interaction term.
The dynamics of the total system is described by the following von Neumann equation
\begin{eqnarray}
\frac{d \hat{\rho} (t)}{dt} &=& 
\hat{\hat{{\cal L}}} \hat{\rho}(t) = -\frac{i}{\hbar} [\hat{H},\hat{\rho}(t)]                          \label{1Eaaab} \\
&=& (\hat{\hat{{\cal L}}}_0 + \lambda \hat{\hat{{\cal L}}}_I) \hat{\rho}(t) = 
-\frac{i}{\hbar} [\hat{H}_0,\hat{\rho}(t)] - \lambda \frac{i}{\hbar} [\hat{V}(t),\hat{\rho}(t)]  .          \label{1Eaaac}
\end{eqnarray}
Let us define the interaction representation of the operators as
\begin{eqnarray}
\hat{\rho}_{I}(t) &=& 
e^{-\hat{\hat{{\cal L}}}_0 t} \hat{\rho}(t)=
e^{\frac{i}{\hbar} \hat{H}_{0}t} \hat{\rho}(t) e^{-\frac{i}{\hbar} \hat{H}_{0}t} ,\nonumber \\
\hat{V}_I(t) &=& 
e^{-\hat{\hat{{\cal L}}}_0 t} \hat{V}(t) =
e^{\frac{i}{\hbar} \hat{H}_{0}t} \hat{V}(t) e^{-\frac{i}{\hbar} \hat{H}_{0}t} .                            \label{1Eaaad}
\end{eqnarray}
Let us also define the interaction representation of the Liouvillian as 
\begin{equation}
\hat{\hat{{\cal L}}}_I(t)=e^{-\hat{\hat{{\cal L}}}_0 t}\hat{\hat{{\cal L}}}_I 
e^{\hat{\hat{{\cal L}}}_0 t}     .                                                                         \label{1Eaabe}
\end{equation}
In the interaction representation, the von Neumann equation (\ref{1Eaaab}) takes the simple form
\begin{equation}
\frac{d \hat{\rho}_I(t)}{dt} = \lambda \hat{\hat{{\cal L}}}_I(t)\hat{\rho}_I(t) 
= - \lambda \frac{i}{\hbar} \lbrack \hat{V}_I(t),\hat{\rho}_I(t) \rbrack  .                                 \label{1Eaaae}
\end{equation}
By integrating (\ref{1Eaaae}), we get
\begin{eqnarray}
\hat{\rho}_{I}(t)
&=& \hat{\rho}(0) + \lambda \int^{t}_{0} dt_1 \hat{\hat{{\cal L}}}_I(t_1)\hat{\rho}(0)                  \label{1Eaabf} \\
& &+ \lambda^2 \int^{t}_{0} dt_1 \int^{t_1}_{0} dt_2 \hat{\hat{{\cal L}}}_I(t_1) 
\hat{\hat{{\cal L}}}_I(t_2)\hat\rho(t_2) \nonumber \\
&=& \hat{\rho}(0) -\lambda \frac{i}{\hbar} \int^{t}_{0} dt_1 
\lbrack \hat{V}_I(t_1),\hat{\rho}(0)\rbrack \nonumber \\
& &- \frac{\lambda^2}{\hbar^2} \int^{t}_{0} dt_1 \int^{t_1}_{0} dt_2 
\lbrack \hat{V}_I(t_1),\lbrack \hat{V}_I(t_2)  ,\hat{\rho}(t_2) \rbrack \rbrack  \nonumber  .                                                                   
\end{eqnarray}
The perturbative expression of the von Neumann equation in the interaction representation is now obtained 
by closing (\ref{1Eaabf}) to order $\lambda^2$.
\begin{eqnarray}
\hat{\rho}_{I}(t)
&=& \hat{\rho}(0) + \lambda \int^{t}_{0} dt_1 \hat{\hat{{\cal L}}}_I(t_1)\hat{\rho}(0)                  \label{1Eaaaf} \\
& &+ \lambda^2 \int^{t}_{0} dt_1 \int^{t_1}_{0} dt_2 \hat{\hat{{\cal L}}}_I(t_1) 
\hat{\hat{{\cal L}}}_I(t_2)\hat\rho(0) + {\cal O}(\lambda^3)  \nonumber \\
&=& \hat{\rho}(0) -\lambda \frac{i}{\hbar} \int^{t}_{0} dt_1 
\lbrack \hat{V}_I(t_1),\hat{\rho}(0)\rbrack \nonumber \\
& &- \frac{\lambda^2}{\hbar^2} \int^{t}_{0} dt_1 \int^{t_1}_{0} dt_2 
\lbrack \hat{V}_I(t_1),\lbrack \hat{V}_I(t_2) ,\hat{\rho}(0) \rbrack \rbrack 
+ {\cal O}(\lambda^3) \nonumber 
\end{eqnarray}
With the change of variable $T=t_1$ and $\tau=t_1-t_2$, we get
\begin{eqnarray}
\hat{\rho}_{I}(t)
&=&\hat{\rho}(0) + \lambda \int^{t}_{0} dT \hat{\hat{{\cal L}}}_I (T) \hat{\rho}(0)                     \label{1Eaaah} \\
& &+ \lambda^2 \int^{t}_{0} dT \int^{T}_{0} d\tau \hat{\hat{{\cal L}}}_I (T)
\hat{\hat{{\cal L}}}_I (T-\tau) \hat{\rho}(0)  + {\cal O}(\lambda^3) \nonumber   \\
&=& \hat{\rho}(0) -\lambda \frac{i}{\hbar} \int^{t}_{0} dT \lbrack 
\hat{V}_I(T),\hat{\rho}(0)\rbrack \nonumber \\
& &- \frac{\lambda^2}{\hbar^2} \int^{t}_{0} dT \int^{T}_{0} 
d\tau \lbrack \hat{V}_I(T),\lbrack \hat{V}_I(T-\tau) 
, \hat{\rho}(0) \rbrack \rbrack + {\cal O}(\lambda^3) \nonumber.
\end{eqnarray}
Going back to the original representation, we get the perturbative expansion at order two for the von Neumann 
equation of a system with the Hamiltonian (\ref{1Eaaaa}): 
\begin{eqnarray}
\hat{\rho}(t)
&=& e^{\hat{\hat{{\cal L}}}_0 t} \hat{\rho}(0) 
+ \lambda e^{\hat{\hat{{\cal L}}}_0 t} \int^{t}_{0} dT \hat{\hat{{\cal L}}}_I (T) \hat{\rho}(0)        \label{1Eaaai} \\
& &+ \lambda^2 e^{\hat{\hat{{\cal L}}}_0 t} \int^{t}_{0} dT \int^{T}_{0} d\tau \hat{\hat{{\cal L}}}_I (T)
\hat{\hat{{\cal L}}}_I (T-\tau) \hat{\rho}(0)  + {\cal O}(\lambda^3) \nonumber   \\
&=&e^{-\frac{i}{\hbar} \hat{H}_{0}t} \hat{\rho}(0) e^{\frac{i}{\hbar} \hat{H}_{0}t} 
-\lambda \frac{i}{\hbar} \int^{t}_{0} dT e^{-\frac{i}{\hbar} \hat{H}_{0}t} 
\lbrack \hat{V}_I(T),\hat{\rho}(0)\rbrack e^{\frac{i}{\hbar} \hat{H}_{0}t} \nonumber \\
& &- \frac{\lambda^2}{\hbar^2} \int^{t}_{0} dT \int^{T}_{0} d\tau e^{-\frac{i}{\hbar} \hat{H}_{0}t}
\lbrack \hat{V}_I(T),\lbrack \hat{V}_I(T-\tau) 
, \hat{\rho}(0) \rbrack \rbrack e^{\frac{i}{\hbar} \hat{H}_{0}t} + {\cal O}(\lambda^3) \nonumber.
\end{eqnarray}
A further formal step can be achieved by applying the relevant superoperator projector defined in 
section \ref{Projdyn} to equation (\ref{1Eaaai}). 
Using the property (\ref{1Baaaa}), we get
\begin{eqnarray}
\hat{\hat{{\cal P}}} \hat{\rho}(t)
&=&\hat{\hat{{\cal P}}} e^{\hat{\hat{{\cal L}}}_0 t} 
(\hat{\hat{{\cal P}}}+\hat{\hat{{\cal Q}}}) \hat{\rho}(0)
+\lambda \hat{\hat{{\cal P}}} e^{\hat{\hat{{\cal L}}}_0 t} \int^{t}_{0} dT 
\hat{\hat{{\cal L}}}_I (T) (\hat{\hat{{\cal P}}}+\hat{\hat{{\cal Q}}}) \hat{\rho}(0)                       \label{1Eaaak} \\    
& &+ \lambda^2 \hat{\hat{{\cal P}}} e^{\hat{\hat{{\cal L}}}_0 t} 
\int^{t}_{0} dT \int^{T}_{0} d\tau \hat{\hat{{\cal L}}}_I (T) \hat{\hat{{\cal L}}}_I (T-\tau) 
(\hat{\hat{{\cal P}}}+\hat{\hat{{\cal Q}}}) \hat{\rho}(0)  + {\cal O}(\lambda^3) \nonumber .
\end{eqnarray}
This equation will be the starting point of all our kinetic equation derivations.
The strategy is to obtain a closed equation for the relevant projected density matrix from equation (\ref{1Eaaak}).
This requires, depending on the type of system one studies, a careful choice of the projector.
Notice that in all our applications, the non-relevant projected density matrix is chosen in such a way that
$\hat{\hat{{\cal Q}}} \hat{\rho}(0)=0$.
This is a standard assumption simplifying (\ref{1Eaaak}). 

\section{Linear response theory} \label{linearresponse}

Weakly perturbing a system at equilibrium in order to study its response to the perturbation and its return 
to equilibrium, is one of the main approaches of nonequilibrium statistical mechanics. 
This theory, often called the linear response theory, has essentially been developed by Kubo in $1957$ \cite{Kubo57}. 
General reviews of this theory can be found in \cite{KuboB2} and \cite{ZubarevB96}.
This approach to study nonequilibrium processes is often used to derive general fluctuation-dissipation relations,
to compute transport coefficients and to prove the Onsager reciprocity relation of the phenomenological coefficients
relating the thermodynamical forces to the fluxes in irreversible thermodynamics.  
We present in this section the general ideas of this theory and derive the fluctuation-dissipation 
relation that relates the response of an externally perturbed system to the correlation 
functions characterizing its equilibrium fluctuations.
Frequent reference will be made to appendix \ref{AppA}, where we discuss the property of the correlation functions 
and their Fourier transforms.
These results are gathered in appendix \ref{AppA} because they will play an important role 
in our later presentation of the quantum kinetic equations and should be easily accessible.\\

Let us consider a non-perturbed system described by a Hamiltonian $\hat{H}_{0}$ at equilibrium in the state 
$\hat{\rho}^{eq}$.
Suppose that at time $t=0$, a time-dependent perturbation is applied to the system.
This time-dependent perturbation takes the system out of equilibrium and is assumed to have the form of a 
time-independent operator $\hat{A}$ multiplied by a time-dependent function $f(t)$. 
Such a perturbation is called a mechanical perturbation.
A coupling parameter $\lambda$ measures the strength of the perturbation.
The total Hamiltonian of the system therefore reads  
\begin{eqnarray}
\hat{H}(t) = \hat{H}_{0} - \lambda \hat{A} f(t),                                                                 \label{1Haaaa}
\end{eqnarray}
where $f(t)=0$ for $t<0$.
This Hamiltonian has the form (\ref{1Eaaaa}) with $\hat{V}(t)=-\hat{A} f(t)$.
Using the fact that the initial condition is an equilibrium density matrix $\hat{\rho}(0)=\hat{\rho}^{eq}$, 
the expansion (\ref{1Eaaak}) reads
\begin{eqnarray}
\hat{\rho}(t)
&=& \hat{\rho}^{eq} +\lambda \frac{i}{\hbar} \int^{t}_{0} dT f(T) \lbrack 
\hat{A}(T-t),\hat{\rho}^{eq} \rbrack + \frac{\lambda^2}{\hbar^2} \int^{t}_{0} dT \int^{T}_{0} d\tau                 \label{1Haaab}\\
&& \hspace*{1cm} f(T) f(T-\tau) e^{-\frac{i}{\hbar} \hat{H}_0 t} \lbrack \hat{A}(T),\lbrack \hat{A}(T-\tau) 
, \hat{\rho}^{eq} \rbrack \rbrack e^{\frac{i}{\hbar} \hat{H}_0 t} + {\cal O}(\lambda^3) \nonumber
\end{eqnarray}
where
\begin{eqnarray}
\hat{A}(t)&=& e^{\frac{i}{\hbar} \hat{H}_0 t}  \hat{A} e^{-\frac{i}{\hbar} \hat{H}_0 t}.                             \label{1Haaac}
\end{eqnarray}
Keeping in (\ref{1Haaab}) only the first-order term in $\lambda$, we get 
\begin{eqnarray}
\Delta \hat{\rho} (t) = \hat{\rho} (t) - \hat{\rho}^{eq} = \frac{i}{\hbar} \lambda \int_{-\infty}^{t} dT f(T) 
\lbrack \hat{A}(T-t) , \hat{\rho}^{eq} \rbrack + {\cal O}(\lambda^2).                                               \label{1Haaad}
\end{eqnarray}
Suppose now that we want to study the effect of the perturbation on a system operator $\hat{B}$.
Its deviation from equilibrium is given by 
\begin{eqnarray}
\mean{\Delta \hat{B} (t)} = \mean{\hat{B} (t)} - \mean{\hat{B}}_{eq}
=  \trace \Delta \hat{\rho} (t) \hat{B}.                                                                             \label{1Haaae}
\end{eqnarray}
Using (\ref{1Haaad}), equation (\ref{1Haaae}) becomes
\begin{eqnarray}
\mean{\Delta \hat{B} (t)} =  \lambda \int_{-\infty}^{t} dT f(T) \phi_{BA}(t-T)
=  \lambda \int_{-\infty}^{\infty} dT f(T) \chi_{BA}(t-T),                                                           \label{1Haaaf}
\end{eqnarray}
where we have defined the response function $\phi_{BA}(t)$
\begin{eqnarray}
\phi_{BA}(t)                                                                 
=  \frac{i}{\hbar}  \trace \lbrack \hat{A}(-t) , \hat{\rho}^{eq} \rbrack \hat{B}                             
=  \frac{i}{\hbar}  \trace \hat{\rho}^{eq} \lbrack \hat{B}(t) , \hat{A} \rbrack ,                                    \label{1Haabk}                     
\end{eqnarray}
and the quantity
\begin{eqnarray}
\chi_{BA}(t) = \Theta(t) \phi_{BA}(t),                                                                               \label{1Haaak}                     
\end{eqnarray}
where $\Theta(t)$ is the Heaviside function.\\
Notice that, for a periodic forcing $f(t)=f_0 e^{-i \omega t}$, equation (\ref{1Haaaf}) becomes
\begin{eqnarray}
\mean{\Delta \hat{B} (t)} = 2 \pi \lambda  f_0   e^{-i \omega t}  \tilde{\chi}_{BA}(\omega) ,                        \label{1Haack}
\end{eqnarray}
where we have defined the susceptibility 
\begin{eqnarray}
\tilde{\chi}_{BA}(\omega)=\int_{-\infty}^{\infty} \frac{dt}{2\pi} e^{i \omega t} \chi_{BA}(t)
=\int_{0}^{\infty} \frac{dt}{2\pi} e^{i \omega t} \phi_{BA}(t).                                                      \label{1Haadk}
\end{eqnarray}
We define now the correlation function of operators $\hat{B}$ and $\hat{A}$ which characterizes the 
equilibrium fluctuations of one operator with respect to the other:
\begin{eqnarray}
\alpha_{BA}(t) = \trace \hat{\rho}^{eq}  \hat{B}(t) \hat{A} = C_{BA}(t) + i D_{BA}(t).                               \label{1Haaal}
\end{eqnarray}
The important properties of these correlation functions are detailed in the appendix \ref{AppA}.
It is remarkable that the susceptibility (\ref{1Haaak}) and the response function (\ref{1Haabk}) can now be 
related to the imaginary part of the correlation function [see equation (\ref{AppAaadb})]
\begin{eqnarray}
\chi_{BA}(t) = \Theta(t) \phi_{BA}(t) = - \frac{2}{\hbar} \Theta(t) D_{BA}(t)  .                                     \label{1Haaaq}                 
\end{eqnarray}
The Fourier transform of the susceptibility is related to the Fourier transform of the response function by
\begin{eqnarray}
\tilde{\chi}_{BA}(\omega)= \frac{\tilde{\phi}_{BA}(\omega)}{2} - \frac{i}{2 \pi} \int_{-\infty}^{\infty} d\omega' 
{{\cal P}} \frac{\tilde{\phi}_{BA}(\omega')}{\omega'-\omega}.                                                          \label{1Haaes}
\end{eqnarray}
Because of the relation (\ref{1Haaaq}), $\tilde{\phi}_{BA}(\omega)$ has the same properties 
as $\tilde{D}_{BA}(\omega)$ in equation (\ref{AppAaagb}).
Using these properties, one can show that
\begin{eqnarray}
\tilde{\phi}_{BA}(\omega)=\tilde{\chi}_{BA}(\omega) - \tilde{\chi}^{*}_{AB}(\omega).                                 \label{1Haaat}
\end{eqnarray}
By defining the symmetric and antisymmetric parts of the response function and of the susceptibility as
\begin{eqnarray}
\tilde{\phi}^{s}_{BA}(\omega)&\equiv&(\tilde{\phi}_{BA}(\omega)+\tilde{\phi}_{AB}(\omega))/2                         \label{1Haaau}\\
\tilde{\phi}^{a}_{BA}(\omega)&\equiv&(\tilde{\phi}_{BA}(\omega)-\tilde{\phi}_{AB}(\omega))/2 \nonumber\\
\tilde{\chi}^{s}_{BA}(\omega)&\equiv&(\tilde{\chi}_{BA}(\omega)+\tilde{\chi}_{AB}(\omega))/2  \nonumber\\
\tilde{\chi}^{a}_{BA}(\omega)&\equiv&(\tilde{\chi}_{BA}(\omega)-\tilde{\chi}_{AB}(\omega))/2, \nonumber                                   
\end{eqnarray}
and by using equation (\ref{1Haaat}), one finds that 
\begin{eqnarray}
\tilde{\phi}^{s}_{BA}(\omega) = 2i {\cal I}m \lbrack \tilde{\chi}^{s}_{BA}(\omega)  \rbrack                           \label{1Haaav}\\
\tilde{\phi}^{a}_{BA}(\omega) = 2 {\cal R}e  \lbrack \tilde{\chi}^{a}_{BA}(\omega) \rbrack \nonumber ,
\end{eqnarray}
and that
\begin{eqnarray}
\tilde{\chi}_{BA}(\omega)=
{\cal R}e \lbrack \tilde{\chi}_{BA}(\omega) \rbrack + i {\cal I}m \lbrack \tilde{\chi}_{BA}(\omega) \rbrack  ,        \label{1Haabv}
\end{eqnarray}
where
\begin{eqnarray}
{\cal R}e \lbrack \tilde{\chi}_{BA}(\omega) \rbrack &=&
\frac{\tilde{\phi}^{a}_{BA}(\omega)}{2} - \frac{i}{2 \pi} \int_{-\infty}^{\infty} d\omega' 
{{\cal P}} \frac{\tilde{\phi}^{s}_{BA}(\omega')}{\omega'-\omega}                                                        \label{1Haabv}\\
{\cal I}m \lbrack \tilde{\chi}_{BA}(\omega) \rbrack &=&
\frac{\tilde{\phi}^{s}_{BA}(\omega)}{2} - \frac{i}{2 \pi} \int_{-\infty}^{\infty} d\omega' 
{{\cal P}} \frac{\tilde{\phi}^{a}_{BA}(\omega')}{\omega'-\omega} \nonumber
\end{eqnarray}

If the equilibrium density matrix is canonical ($\hat{\rho}^{eq}=e^{-\beta \hat{H}_0}/Z$, where 
$Z=\trace e^{-\beta \hat{H}_0}$), the KMS (Kubo-Martin-Schwinger) property of the correlation function 
[see equations (\ref{AppAaaat})-(\ref{AppAaabu})] implies that
\begin{eqnarray}
\tilde{C}_{BA}(\beta;\omega)= 2i \frac{E_{\beta}(\omega)}{\hbar \omega} \tilde{D}_{BA}(\beta;\omega)
= -i \frac{E_{\beta}(\omega)}{\omega} \tilde{\phi}_{BA}(\beta;\omega),                                                \label{1Haaay}
\end{eqnarray}
where
\begin{eqnarray}
E_{\beta}(\omega)=\frac{\hbar \omega}{2} \coth{\frac{\beta \hbar \omega}{2}}   .                                      \label{1Haaaz}
\end{eqnarray}
Using equations (\ref{1Haaav}), (\ref{1Haaay}) and (\ref{AppAaahb}), we also find
\begin{eqnarray}
\tilde{C}^{s}_{BA}(\beta;\omega)
&=& 2 \frac{E_{\beta}(\omega)}{\omega} {\cal I}m \lbrack \tilde{\chi}^{s}_{AB}(\beta,\omega) \rbrack                  \label{1Hbaaa}\\
\tilde{C}^{a}_{BA}(\beta;\omega)
&=& -2i \frac{E_{\beta}(\omega)}{\omega} {\cal R}e \lbrack \tilde{\chi}^{a}_{AB}(\beta,\omega) \rbrack. \nonumber
\end{eqnarray}
Equation (\ref{1Haaay}) and (\ref{1Hbaaa}) are manifestations of the so-called fluctuation dissipation theorem.
In fact, these equations connect the response or the susceptibility of the system to the fluctuation properties of 
the equilibrium state. 
This connection is very important and occurs in many different ways in nonequilibrium statistical mechanics.  

\section{Fermi golden rule} \label{FermiGR}

In this section, we derive the Fermi golden rule which will give us an important intuitive tool for the further studies.
The derivation is made starting from the second-order perturbative expansion of the von Neumann equation, in order to insist
on the similarity between the different weak-coupling theories. We also show that, for a system submitted to a periodic
forcing, the dissipated power can be related to the imaginary part of the susceptibility using the Fermi golden rule.

The systems we consider here are the same ones as in the linear response theory.
We have a non-perturbed system with a Hamiltonian $\hat{H}_{0}$ at equilibrium in the state $\hat{\rho}^{eq}$,
to which a time-dependent perturbation is applied at time $t=0$. 
This time-dependent perturbation has again the form of a time-independent operator $\hat{A}$ multiplied by a 
time-dependent function $f(t)$ and the coupling parameter $\lambda$ measures the strength of the perturbation.
The total Hamiltonian therefore reads  
\begin{eqnarray}
\hat{H}(t) = \hat{H}_{0} - \lambda \hat{A} f(t),                                                                     \label{1Hcaba}
\end{eqnarray}
where $f(t)=0$ for $t<0$.
The von Neumann equation ruling the dynamics of the system is the same as in linear response [see equation (\ref{1Haaab})].
However, when the operator $\hat{B}$, that we defined in linear response theory to study the response of a system to an 
external perturbation [see equation (\ref{1Haaae})], commutes with $\hat{H}_0$, the first-order expansion of the von Neumann 
equation (\ref{1Haaab}) and therefore also the response function (\ref{1Haabk}) of linear response theory are zero.
It is then necessary to go to the second order of the perturbative expansion of the von Neumann equation (\ref{1Haaab}).
The response of $\hat{B}$ will in this case have the form 
\begin{eqnarray}
\mean{\Delta \hat{B}(t)}= 
\trace \Delta \hat{\rho}(t) \hat{B}=
\sum_n \bra{n} \Delta \hat{\rho}(t) \ket{n}  \bra{n} \hat{B} \ket{n} = 
\sum_n \Delta \rho_{nn}(t) B_{nn},                                                                        \label{1Hcaaa}
\end{eqnarray}
where
\begin{eqnarray}
\Delta \rho_{nn}(t)
&=&\frac{\lambda^2}{\hbar^2} \int^{t}_{0} dT \int^{T}_{0} d\tau              
f(T) f(T-\tau) \bra{n} \lbrack \hat{A}(T),\lbrack \hat{A}(T-\tau) 
, \hat{\rho}^{eq} \rbrack \rbrack \ket{n} . \nonumber\\                                                               \label{1Hcaab}
\end{eqnarray}
We choose a stationary state of $\hat{H}_{0}$ as equilibrium state 
\begin{eqnarray}
\hat{\rho}^{eq} = \ket{n} \bra{n} ,                                                                                   \label{1Hcabb}
\end{eqnarray}
and focus on a time-periodic forcing of the form 
\begin{eqnarray}
f(t) = \cos \omega t = \frac{e^{i \omega t}+e^{-i \omega t}}{2}.                                                      \label{1Hcaac}
\end{eqnarray}

\subsection{The derivation of the transition rates}

\scriptsize
Using the fact that
\begin{eqnarray}
f(T) f(T-\tau) = \frac{1}{4} \{ (e^{i \omega 2T}+1) e^{-i \omega \tau} + CC \},                                       \label{1Hcaad}
\end{eqnarray}
where $CC$ stands for complex conjugate, and that
\begin{eqnarray}
\bra{n} \lbrack \hat{A}(T) , \lbrack \hat{A}(T-\tau) , \hat{\rho}^{eq} \rbrack \rbrack \ket{n} &=&
+ \{ \sum_{n''}  e^{-\frac{i}{\hbar} (E_{n''}-E_{n}) \tau} 
\vert A_{nn''} \vert^2 \delta_{nn'} + CC \} \nonumber \\
&&- \{ e^{-\frac{i}{\hbar} (E_{n'}-E_{n}) \tau} \vert A_{nn'} \vert^2 + CC \} ,                                       \label{1Hcaae}
\end{eqnarray}
equation (\ref{1Hcaab}) becomes
\begin{eqnarray}
\Delta \rho_{nn}(t)
&=& {\cal R}e \; \frac{\lambda^2}{2 \hbar^2} 
\int^{t}_{0} dT (e^{i \omega 2T}+1) \int^{T}_{0} d\tau \{ \nonumber \\            
&&+ \sum_{n''} (e^{-\frac{i}{\hbar} (\hbar \omega+E_{n''}-E_{n}) \tau}+e^{-i (\omega+E_{n}-E_{n''}) \tau}) 
\vert A_{nn''} \vert^2 \delta_{nn'}   \nonumber \\  
&&- (e^{-i (\omega+E_{n'}-E_{n}) \tau}+e^{-\frac{i}{\hbar} 
(\hbar \omega+E_{n}-E_{n'}) \tau}) \vert A_{nn'} \vert^2 \; \} .                                                     \label{1Hcaaf}
\end{eqnarray}
Now we make the long-time assumption which consists in taking the upper bound $T$ of the integral over $\tau$ 
in (\ref{1Hcaaf}) to infinity, so that one can use the property
\begin{eqnarray}
\int^{\infty}_{0} d\tau (e^{-\frac{i}{\hbar} (\hbar \omega+E_{n'}-E_{n}) \tau}+
e^{-\frac{i}{\hbar} (\hbar \omega+E_{n}-E_{n'}) \tau}) &=&                                                           \label{1Hcaag}\\
&&\hspace*{-6cm} \pi (\delta(\omega-\omega_{nn'})+\delta(\omega+\omega_{nn'})) 
- i {{\cal P}} (\frac{1}{\omega-\omega_{nn'}}+\frac{1}{\omega+\omega_{nn'}})     \nonumber             
\end{eqnarray}
where $\hbar \omega_{nn'}=E_{n}-E_{n'}$.
Therefore, we can evaluate the time-dependent quantities entering in (\ref{1Hcaaf}) as follows
\begin{eqnarray}
{\cal R}e \int^{t}_{0} dT (e^{i \omega 2T}+1) \int^{\infty}_{0} d\tau 
(e^{-\frac{i}{\hbar} (\hbar \omega+E_{n'}-E_{n}) \tau}+
e^{-\frac{i}{\hbar} (\hbar \omega+E_{n}-E_{n'}) \tau})&&                                                             \label{1Hcaah}\\ 
&&\hspace*{-10cm}=+\pi (\delta(\omega-\omega_{nn'})+\delta(\omega+\omega_{nn'})) 
\int^{t}_{0} dT (1+\cos 2 \omega T) \nonumber\\  
&&\hspace*{-8cm}-i {{\cal P}} (\frac{1}{\omega-\omega_{nn'}}
+\frac{1}{\omega+\omega_{nn'}}) \int^{t}_{0} dT \sin 2 \omega T  \nonumber \\
&&\hspace*{-10cm} = +\pi (\delta(\omega-\omega_{nn'})+\delta(\omega+\omega_{nn'})) 
(t+\frac{\sin 2 \omega t}{2 \omega}) \nonumber\\  
&&\hspace*{-8cm}-i {{\cal P}} (\frac{1}{\omega-\omega_{nn'}}
+\frac{1}{\omega+\omega_{nn'}}) \frac{1-\cos 2 \omega t}{2 \omega}  \nonumber \\ 
&&\hspace*{-10cm} \stackrel{t \to \infty}{=} +\pi (\delta(\omega-\omega_{nn'})+\delta(\omega+\omega_{nn'})) 
(t+\frac{\pi}{2} \delta(\omega)) \nonumber\\  
&&\hspace*{-8cm}+i {{\cal P}} (\frac{1}{\omega-\omega_{nn'}}
+\frac{1}{\omega+\omega_{nn'}}) {{\cal P}} (\frac{1}{2 \omega}) \nonumber \\  
&&\hspace*{-10cm} \approx +\pi (\delta(\omega-\omega_{nn'})+\delta(\omega+\omega_{nn'})) t .
\end{eqnarray}
\normalsize
The long-time behavior of equation (\ref{1Hcaaf}) is given by
\begin{eqnarray}
\Delta \rho_{nn}(t)&\approx& 
+ \frac{\pi \lambda^2}{2 \hbar^2} (\delta(\omega - \omega_{nn'}) 
+ \delta(\omega + \omega_{nn'}) ) \vert A_{nn'} \vert^2 t                                                           \label{1Hcaai}\\
&&- \delta_{nn'} \frac{\pi \lambda^2 }{2 \hbar^2} \sum_{n''} ( \delta(\omega - \omega_{nn''}) 
+ \delta(\omega + \omega_{nn''}) ) \vert A_{nn''} \vert^2 t \nonumber  .                                       
\end{eqnarray}
This expression has to be used carefully. In fact, it is valid on sufficiently long time scales to
perform the preceding calculation, but it is not valid anymore on too long time scales because then 
the probability goes to infinity.
It provides however a simple intuitive picture. $\Delta \rho_{nn}(t)$ is of course the probability
to find the system in the state $\ket{n}$ at time $t$, having started to perturb the system at time zero in
state $\ket{n'}$ with a cosine periodic forcing. 
It is therefore quite natural to define the transition probability rate $W_{n' \to n}(\omega)$ as follows
\begin{eqnarray}
\Delta \rho_{nn}(t)&=& W_{n' \to n}(\omega) t.                                                                       \label{1Hcabj} 
\end{eqnarray}
If $n \neq n'$, the transition rate is given by the Fermi golden rule
\begin{eqnarray}
W_{n' \to n}(\omega)&=& 
+ \frac{\pi \lambda^2}{\hbar^2} (\delta(\omega - \omega_{nn'}) 
+ \delta(\omega + \omega_{nn'}) ) \vert A_{nn'} \vert^2 .                                                            \label{1Hcaaj}                                  
\end{eqnarray}       
This rate gives the probability per unit time to jump from state $\ket{n'}$ to $\ket{n}$ under
the cosine time-dependent perturbation at frequency $\omega$. 
We identify two possible kinds of transitions, the first coming from the absorption of an energy 
$\hbar \omega$ by the system jumping from the initial state at energy $E_{n'}$ to a state at energy 
$E_{n'}+\hbar \omega$, and the second coming from the stimulated emission of an energy 
$-\hbar \omega$ by the system jumping now from the initial state at energy $E_{n'}$ to a state at energy 
$E_{n'}-\hbar \omega$.\\
We can also define the total transition rate, which is the total probability per unit time to leave the initial 
state $\ket{n'}$, whatever the arrival state is, under the cosine time-dependent perturbation. 
By summing the transition rate from state $\ket{n'}$ to $\ket{n}$ over all arrival states $\ket{n}$, 
the total transition rate is therefore given by
\begin{eqnarray}
W_{n'}(\omega) = \sum_{n} W_{n' \to n}(\omega) .                                                                     \label{1Hcacj}
\end{eqnarray} 
If the spectrum is dense enough to consider that the energies vary in an almost continuous way in the spectrum 
$E_n = \epsilon$ and that $\vert A_{nn'} \vert^2 = \vert A(\epsilon,\epsilon') \vert^2$ is a 
smooth function of energy (the initial energy is denoted by $\epsilon'$), we can write that the 
total transition rate given by the Fermi golden rule becomes $W_{n'}(\omega) \to W(\epsilon',\omega)$, where
\begin{eqnarray}
W(\epsilon',\omega)&=& 
\frac{\pi \lambda^2}{\hbar} \int d\epsilon \; n(\epsilon) \large[ \delta(\hbar \omega - \epsilon + \epsilon') 
+ \delta(\hbar \omega + \epsilon - \epsilon') \large] \; \vert A(\epsilon,\epsilon') \vert^2 , \nonumber \\         \label{1Hcadj}
\end{eqnarray}
$n(\epsilon)$ being the density of states at energy $\epsilon$.
This expression can be rewritten as
\begin{center} \fbox{\parbox{12.5cm}{
\begin{eqnarray}                                          
W(\epsilon',\omega)&=& \frac{\pi \lambda^2}{\hbar^2} 
\{ \tilde{\alpha}_{AA}(\epsilon',\omega) + \tilde{\alpha}_{AA}(\epsilon',-\omega) \},                                \label{1Hcaak}
\end{eqnarray}
}} \end{center}
where
\begin{eqnarray}
\tilde{\alpha}_{AA}(\epsilon,\omega)=
\hbar n(\epsilon+\hbar \omega) \vert A(\epsilon+\hbar \omega,\epsilon) \vert^2                                       \label{1Hcaal}
\end{eqnarray}
is the Fourier transform of the microcanonical correlation function defined in equation (\ref{AppAaaaq}).
This is a very interesting connection between the equilibrium fluctuations and the response of the system 
to an external perturbation.
Equation (\ref{1Hcaal}) again shows the absorption and the induced emission contributions to the total 
transition rate.\\
We now have a very nice interpretation of the Fourier transform of a microcanonical correlation function:
\begin{itemize}
\item $\tilde{\alpha}_{AA}(\epsilon,\omega)$ is proportional by a factor $\hbar^2/(\pi \lambda^2)$ to the 
transition rate for a system at energy $\epsilon$ to absorb an energy $\hbar \omega$ under a periodic 
perturbation of frequency $\omega$.
\item $\tilde{\alpha}_{AA}(\epsilon,-\omega)$ is proportional by a factor $\hbar^2/(\pi \lambda^2)$ to the 
transition rate for a system at energy $\epsilon$ to emit an energy $\hbar \omega$ under a periodic 
perturbation of frequency $\omega$.
\end{itemize}
If the system is in a thermal state described by a canonical distribution at temperature $\beta$ before being 
submitted to the periodic forcing, the total transition rate is the thermal average of the total rate 
(\ref{1Hcacj}) and reads 
\begin{eqnarray}
W(\beta,\omega) = \sum_{n'} \frac{e^{-\beta E_{n'}}}{Z} W_{n'}(\omega)                                                 \label{1Hcabl}
\end{eqnarray} 
for a discrete spectrum and 
\begin{eqnarray}
W(\beta,\omega) = \int d\epsilon \frac{n(\epsilon) e^{-\beta \epsilon}}{Z} W(\epsilon,\omega)                          \label{1Hcacl}
\end{eqnarray} 
for a dense spectrum.\\
Using the fact that the canonical correlation function is the thermal average of the microcanonical correlation 
function [see equation (\ref{AppAaaas}) in appendix \ref{AppA}], the total transition rate can be written as
\begin{center} \fbox{\parbox{12.5cm}{
\begin{eqnarray}                                          
W(\beta,\omega)&=& \frac{\pi \lambda^2}{\hbar^2} 
\{ \tilde{\alpha}_{AA}(\beta,\omega) + \tilde{\alpha}_{AA}(\beta,-\omega) \}.                                          \label{1Hcadl}
\end{eqnarray}
}} \end{center}
Here we have the interpretation of the Fourier transform of the canonical correlation function.
\begin{itemize}
\item $\tilde{\alpha}_{AA}(\beta,\omega)$ is proportional by a factor $\hbar^2/(\pi \lambda^2)$ to the 
transition rate for a system at temperature $\beta$ to absorb an energy $\hbar \omega$ under a periodic 
perturbation of frequency $\omega$.
\item $\tilde{\alpha}_{AA}(\beta,-\omega)$ is proportional by a factor $\hbar^2/(\pi \lambda^2)$ to the 
transition rate for a system at temperature $\beta$ to emit an energy $\hbar \omega$ under a periodic 
perturbation of frequency $\omega$.
\end{itemize}
\begin{figure}[h]
\centering
\rotatebox{0}{\scalebox{0.8}{\includegraphics{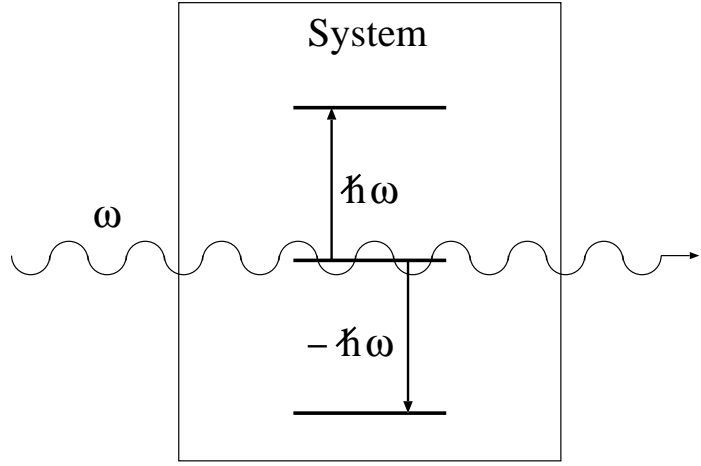}}}
\caption{The system submitted to a periodic forcing of frequency $\omega$ can absorb 
or emit an energy $\hbar \omega$ at a rate given by the Fermi golden rule (\ref{1Hcaak}).} \label{imfermigold}
\end{figure}

\subsection{Dissipated power}

As in linear response theory, a connection can be made between the equilibrium fluctuations of a system, 
its response to an external perturbation and the dissipation to which it is submitted. 
This shows again the many ways in which the fluctuation-dissipation theorem occurs.\\
The power dissipated by the system submitted to a periodic external forcing is defined as
\begin{eqnarray}
P = \lim_{t \to \infty} \frac{\mean{\Delta \hat{H}_0(t)}}{t}                                                        \label{1Hcaam}
\end{eqnarray}
where
\begin{eqnarray}
\mean{\Delta \hat{H}_0(t)} &=& \trace \Delta \hat{\rho}(t) \hat{H}_0 = \sum_{n} \Delta \rho_{nn}(t) E_n            \label{1Hcaan}\\
&=&\sum_{n \neq n'} \Delta \rho_{nn}(t) E_n + \Delta \rho_{n'n'}(t) E_{n'} \nonumber .                                  
\end{eqnarray}
This means that using (\ref{1Hcaai}), we can write
\begin{eqnarray}
P = \sum_{n \neq n'} \frac{\pi \lambda^2}{2 \hbar^2} (\delta(\omega - \omega_{nn'}) 
+ \delta(\omega + \omega_{nn'}) ) \vert A_{nn'} \vert^2 (E_n-E_{n'})  .                                        \label{1Hcaao} 
\end{eqnarray}
Assuming as afore mentioned that the spectrum is dense enough, we finally have that
\begin{eqnarray}
P &=& \int d\epsilon \frac{\pi \lambda^2}{2 \hbar} (\delta(\hbar \omega - \epsilon + \epsilon') 
+ \delta(\hbar \omega + \epsilon - \epsilon') ) 
\vert A(\epsilon,\epsilon') \vert^2 (\epsilon - \epsilon') \nonumber \\
&=& \frac{\pi \lambda^2}{2 \hbar} \omega \{ \tilde{\alpha}_{AA}(\epsilon',\omega) 
- \tilde{\alpha}_{AA}(\epsilon',-\omega) \}   \nonumber \\
&=& - \pi \lambda^2 \omega \; {\cal I}m \lbrack \tilde{\chi}_{AA}(\epsilon',\omega) \rbrack ,                        \label{1Hcaap}
\end{eqnarray}
where we used equations (\ref{1Haaav}), (\ref{1Haaaq}), (\ref{AppAaagb}) and the property (\ref{AppAaahb}) 
to obtain the last line.\\
We see that the power dissipated by a system submitted to a periodic external forcing is directly 
connected with the imaginary part of the susceptibility. 
As we have seen in the linear response theory, the imaginary part of the susceptibility of the system is related to 
the response function [see equation (\ref{1Haaav})], which in turn is related to the real part of the correlation 
function [see equation (\ref{1Haaaq})].
All these connections are manifestations of the fluctuation-dissipation theorem.
 
\section{Pauli equation} \label{Pauli}

We will derive in this section the first quantum kinetic equation derived by Pauli in $1928$ \cite{Pauli28}.
In $1955$, this equation has been studied in great detail by Van Hove \cite{VanHove55}.
General results on this topic can be found in Refs. \cite{KuboB2,ZubarevB96,Zwanzig61}.
In this chapter we will derive the Pauli equation starting from the perturbative expansion (\ref{1Eaaah}) 
of the von Neumann equation in order to insist on the similarity between the Pauli equation and the other 
weak-coupling kinetic equations derived in this chapter.\\ 

The Pauli equation applies to systems whose Hamiltonian is given by equation 
(\ref{1Eaaaa}), where the perturbation term is time-independent:
\begin{eqnarray}
\hat{H}=\hat{H}_{0}+\lambda \hat{V}.                                                     \label{1Caaba}
\end{eqnarray}
Let us first define the notation by writing the eigenvalue equations of the non-perturbed 
Hamiltonian $\hat{H}_{0}$ and the full Hamiltonian $\hat{H}$, respectively, as
\begin{eqnarray}
\hat{H}_{0} \vert n \rangle &=& E^{(0)}_{n} \vert n \rangle     \\                           \label{1Caaaa}
\hat{H} \vert \alpha \rangle &=& E_{\alpha} \vert \alpha \rangle  .                          \label{1Caaab}
\end{eqnarray}

\subsection{The projector}

We define now the form of the projection operators needed 
\begin{eqnarray}
\langle n \vert \hat{\hat{{\cal P}}} \hat{\rho}(t) \vert n' \rangle &=& 
\langle n \vert \hat{\rho}(t) \vert n \rangle \delta_{n n'}       \\                          \label{1Caaac}
\langle n \vert \hat{\hat{{\cal Q}}} \hat{\rho}(t) \vert n' \rangle &=& 
\langle n \vert \hat{\rho}(t) \vert n' \rangle (1-\delta_{n n'})        .                     \label{1Caaad}
\end{eqnarray}
This means that the relevant density operator and the non-relevant density operator 
can be written as
\begin{eqnarray}
\hat{\hat{{\cal P}}} \hat{\rho}(t) &=& 
\sum_{n} \bra{n} \hat{\rho}(t) \ket{n} \vert n \rangle \langle n \vert                        \label{1Caaae}\\
\hat{\hat{{\cal Q}}} \hat{\rho}(t) &=& 
\sum_{n,n'\neq n} \bra{n} \hat{\rho}(t) \ket{n'} \vert n \rangle \langle n' \vert.            \label{1Caaaf}
\end{eqnarray}
The relevant projector has the effect of keeping only the populations of the density matrix 
while loosing the information regarding the coherences. 
The Pauli equation is obtained using this projection operators to close the projected 
perturbative expansion of the von Neumann equation (\ref{1Eaaak}) where the terms
containing the projector $\hat{\hat{{\cal Q}}}$ are neglected.
Notice that this last assumption is exact if the initial coherences are zero: 
$\hat{\hat{{\cal Q}}} \hat{\rho}(0)=0$.

\subsection{The derivation of the Pauli equation}

\scriptsize
Writing the projected perturbative expansion of the von Neumann equation (\ref{1Eaaak}) 
where $\hat{\hat{{\cal Q}}} \hat{\rho}(0)=0$ in the eigenbasis of $\hat{H}_{0}$, we get
\begin{eqnarray}
\langle n \vert \hat{\hat{{\cal P}}} \hat{\rho}(t) \vert n' \rangle 
&=& \langle n \vert \hat{\hat{{\cal P}}} e^{\hat{\hat{{\cal L}}}_0 t} \hat{\hat{{\cal P}}} 
\hat{\rho}(0) \vert n' \rangle
+\lambda \int^{t}_{0} dT \langle n \vert \hat{\hat{{\cal P}}} e^{\hat{\hat{{\cal L}}}_0 t} 
\hat{\hat{{\cal L}}}_I (T) \hat{\hat{{\cal P}}} \hat{\rho}(0) \vert n' \rangle                   \label{1Caaah}\\             
& &+ \lambda^2 \int^{t}_{0} dT \int^{T}_{0} d\tau \langle n \vert \hat{\hat{{\cal P}}} 
e^{\hat{\hat{{\cal L}}}_0 t} \hat{\hat{{\cal L}}}_I (T) \hat{\hat{{\cal L}}}_I (T-\tau) 
\hat{\hat{{\cal P}}} \hat{\rho}(0) \vert n' \rangle + {\cal O}(\lambda^3) \nonumber .
\end{eqnarray}
Using (\ref{1Caaaa}), we have
\begin{eqnarray}
\langle n \vert \hat{\rho}(t) \vert n \rangle 
&=& \langle n \vert \hat{\rho}(0) \vert n \rangle
+\lambda \int^{t}_{0} dT \langle n \vert \hat{\hat{{\cal L}}}_I (T) 
\hat{\hat{{\cal P}}} \hat{\rho}(0) \vert n \rangle                                               \label{1Caaai}\\             
& &+ \lambda^2 \int^{t}_{0} dT \int^{T}_{0} d\tau \langle n \vert  
\hat{\hat{{\cal L}}}_I (T) \hat{\hat{{\cal L}}}_I (T-\tau) 
\hat{\hat{{\cal P}}} \hat{\rho}(0) \vert n \rangle + {\cal O}(\lambda^3) \nonumber .
\end{eqnarray}
We now need to evaluate 
$\langle n \vert \hat{\hat{{\cal L}}}_I (T) \hat{\hat{{\cal P}}} \hat{\rho}(0) \vert n \rangle$ and 
$\langle n \vert  \hat{\hat{{\cal L}}}_I (T) \hat{\hat{{\cal L}}}_I (T-\tau) \hat{\hat{{\cal P}}} 
\hat{\rho}(0) \vert n \rangle$.
\begin{eqnarray}
\langle n \vert \hat{\hat{{\cal L}}}_I (T) \hat{\hat{{\cal P}}} \hat{\rho}(0) \vert n \rangle 
&=& - \frac{i}{\hbar} \langle n \vert \lbrack \hat{V}(T), \hat{\hat{{\cal P}}} \hat{\rho}(0) 
\rbrack \vert n \rangle                                                                         \label{1Caaaj}\\
&=& - \frac{i}{\hbar} \sum_{n'} \langle n \vert \hat{V}(T) \vert n' \rangle \langle n' \vert 
\hat{\hat{{\cal P}}} \hat{\rho}(0) \vert n \rangle \nonumber \\
&&+ \frac{i}{\hbar} \sum_{n'} \langle n \vert \hat{\hat{{\cal P}}} \hat{\rho}(0) \vert n' \rangle 
\langle n' \vert \hat{V}(T) \vert n \rangle \nonumber \\
&=& 0 \nonumber
\end{eqnarray}
\begin{eqnarray}
\langle n \vert  \hat{\hat{{\cal L}}}_I (T) \hat{\hat{{\cal L}}}_I (T-\tau) \hat{\hat{{\cal P}}} 
\hat{\rho}(0) \vert n \rangle 
&=& - \frac{1}{\hbar^2} \langle n \vert \lbrack \hat{V}(T),\lbrack \hat{V}(T-\tau) , 
\hat{\hat{{\cal P}}}  \hat{\rho}(0) \rbrack \rbrack \vert n \rangle                                \label{1Caaak}\\
&=& - \frac{1}{\hbar^2} \sum_{n'} \langle n \vert \hat{V}(T) \vert n' \rangle 
\langle n' \vert \hat{V}(T-\tau) \vert n \rangle 
\langle n \vert \hat{\rho}(0) \vert n \rangle \nonumber\\
&& + \frac{1}{\hbar^2} \sum_{n'} \langle n \vert \hat{V}(T) \vert n' \rangle 
\langle n' \vert \hat{\rho}(0) \vert n' \rangle 
\langle n' \vert \hat{V}(T-\tau) \vert n \rangle \nonumber\\
&& + \frac{1}{\hbar^2} \sum_{n'} \langle n \vert \hat{V}(T-\tau) \vert n' \rangle 
\langle n' \vert \hat{\rho}(0) \vert n' \rangle 
\langle n' \vert \hat{V}(T) \vert n \rangle \nonumber\\
&& - \frac{1}{\hbar^2} \sum_{n'} \langle n \vert \hat{\rho}(0) \vert n \rangle 
\langle n \vert \hat{V}(T-\tau) \vert n' \rangle 
\langle n' \vert \hat{V}(T) \vert n \rangle \nonumber\\
&=& - \frac{2}{\hbar^2} \sum_{n'} 
\vert \langle n \vert \hat{V} \vert n' \rangle \vert^2 
\langle n \vert \hat{\rho}(0) \vert n \rangle 
\cos \frac{E_n - E_{n'}}{\hbar} \tau \nonumber\\
&& + \frac{2}{\hbar^2} \sum_{n'} 
\vert \langle n \vert \hat{V} \vert n' \rangle \vert^2 
\langle n' \vert \hat{\rho}(0) \vert n' \rangle 
\cos \frac{E_n - E_{n'}}{\hbar} \tau \nonumber
\end{eqnarray}
Equation (\ref{1Caaai}) then becomes
\begin{eqnarray}
\langle n \vert \hat{\rho}(t) \vert n \rangle                                                    
&=& \langle n \vert \hat{\rho}(0) \vert n \rangle                                                \label{1Caaal}\\
&&- 2 \frac{\lambda^2}{\hbar} \sum_{n'} 
\vert \langle n \vert \hat{V} \vert n' \rangle \vert^2
\{ \langle n \vert \hat{\rho}(0) \vert n \rangle - 
\langle n' \vert \hat{\rho}(0) \vert n' \rangle \} \nonumber \\
&& \hspace*{4cm} \int^{t}_{0} dT \frac{\sin (E_n - E_{n'}) T / \hbar}{E_n - E_{n'}} 
+ {\cal O}(\lambda^3). \nonumber
\end{eqnarray}
Differentiating this expression with respect to time, we get
\begin{eqnarray}
\frac{d \langle n \vert \hat{\rho}(t) \vert n \rangle}{dt} &=&       
- 2 \frac{\lambda^2}{\hbar} \sum_{n'} 
\vert \langle n \vert \hat{V} \vert n' \rangle \vert^2
\{ \langle n \vert \hat{\rho}(0) \vert n \rangle - 
\langle n' \vert \hat{\rho}(0) \vert n' \rangle \}                                                \label{1Caabl}\\
&& \hspace*{5cm} \frac{\sin (E_n - E_{n'}) t / \hbar}{E_n - E_{n'}} 
+ {\cal O}(\lambda^3). \nonumber
\end{eqnarray}
In order to close the equation, we will have to assume that $\langle n \vert \hat{\rho}(t) 
\vert n \rangle = \langle n \vert \hat{\rho}(0) \vert n \rangle + {\cal O}(\lambda^2)$.
$\langle n \vert \hat{\rho}(0) \vert n \rangle$ can be replaced by 
$\langle n \vert \hat{\rho}(t) \vert n \rangle$ in the right-hand side of (\ref{1Caabl})
without affecting the equation on lower order than ${\cal O}(\lambda^3)$. 
This assumption is the main assumption of this derivation and is common to many derivations of
weak-coupling kinetic equations.
Simplifying the notation and defining $\hbar \omega_{nn'}=E_n - E_{n'}$, equation (\ref{1Caabl}) 
becomes equation (\ref{1Caaam}).\\
\normalsize
The non-Markovian version of the Pauli equation for a discrete spectrum reads
\begin{eqnarray}
\frac{d \rho_{nn}(t)}{dt} 
&=& - 2 \frac{\lambda^2}{\hbar^2} \sum_{n'} \vert \hat{V}_{nn'} \vert^2 
\frac{\sin \omega_{nn} t}{\omega_{nn}} \{ \rho_{nn}(t) - \rho_{n'n'}(t) \}.                       \label{1Caaam}  
\end{eqnarray}
On long times scales, because $\lim_{t \to \infty} \frac{\sin \omega_{nn'} t}{\omega_{nn'}} 
= \pi \delta(\omega_{nn'})$, equation (\ref{1Caaam}) can be simplified to
\begin{eqnarray}
\frac{d \rho_{nn}(t)}{dt}  
= - 2 \pi \frac{\lambda^2}{\hbar^2} \sum_{n'} \vert V_{nn'} \vert^2 
\delta(\omega_{nn'})  \{ \rho_{nn}(t) - \rho_{n'n'}(t) \}      .                                 \label{1Caaan}
\end{eqnarray} 
Equation (\ref{1Caaan}) is the Markovian version of the Pauli equation for a discrete spectrum. 
The Pauli equation is however typically written for a system with a continuous spectrum. 
If $\alpha$ denotes the eigenvalues of all the operators that commute with $\hat{H}_{0}$
(all quantum numbers of the unperturbed system except energy), the continuous version 
of (\ref{1Caaam}) and (\ref{1Caaan}) is obtained in the following way 
\begin{eqnarray}
\rho_{nn}(t) &\to& P(\epsilon,\alpha,t) ,                                                        \label{1Caabn}\\
\vert V_{nn'} \vert^2 &\to& 
\vert \langle \epsilon,\alpha \vert \hat{V} \vert \epsilon',\alpha' \rangle \vert^2 \nonumber \\ 
\sum_{n'} &\to& \int d \epsilon' \int d \alpha' n(\epsilon',\alpha') \nonumber .
\end{eqnarray}
The correlation function $\tilde{\alpha}_{VV}(\epsilon,\omega)$ defined in equation (\ref{AppAaaaq}) 
becomes $\tilde{\alpha}_{VV}(\epsilon,\alpha,\alpha',\omega)$.
Similarly, the total transition rate given by the Fermi golden rule $W(\epsilon',\omega)$ defined 
in equation (\ref{1Hcaak}) becomes $W(\epsilon',\alpha,\alpha',\omega)$.
We now have
\begin{eqnarray}
\tilde{\alpha}_{VV}(\epsilon,\alpha,\alpha',\omega)= \hbar n(\epsilon+\hbar \omega,\alpha') 
\vert \langle \epsilon+\hbar \omega,\alpha' \vert \hat{V} \vert \epsilon,\alpha \rangle \vert^2   \label{1Caacn}
\end{eqnarray} 
and 
\begin{eqnarray}                                          
W(\epsilon',\alpha,\alpha',\omega) &=& \frac{\pi \lambda^2}{\hbar^2} 
\{ \tilde{\alpha}_{VV}(\epsilon',\alpha,\alpha',\omega) 
+ \tilde{\alpha}_{VV}(\epsilon',\alpha,\alpha',-\omega) \}.                                       \label{1Caadn}
\end{eqnarray}
Taking the continuous limit in equation (\ref{1Caaam}), we get the non-Markovian version of 
the Pauli equation for a continuous spectrum
\begin{eqnarray}
\frac{d P(\epsilon,\alpha,t)}{dt} 
&=& - 2 \frac{\lambda^2}{\hbar} \int d \epsilon' \int d\alpha' n(\epsilon',\alpha') 
\vert \langle \epsilon,\alpha \vert \hat{V} \vert \epsilon',\alpha' \rangle \vert^2               \label{1Caaao}\\
&& \hspace*{3cm} \{ P(\epsilon,\alpha,t) - P(\epsilon',\alpha',t) \} 
\frac{\sin (\epsilon - \epsilon') t / \hbar}{\epsilon - \epsilon'} .\nonumber    
\end{eqnarray} 
On long time scales, the sum in (\ref{1Caaam}) is limited to states for which 
$\vert \epsilon - \epsilon \vert = {\cal O}(\hbar / t)$. 
The number of states in this domain is roughly $n(\epsilon) \vert \epsilon - \epsilon' \vert$. 
This number has to be larger than unit to convert the sum in an integral 
$t/(n(\epsilon) \hbar) \ll 1$.
We therefore conclude that the continuous limit taken to go from (\ref{1Caaam}) to (\ref{1Caaao}) 
is valid for time scales shorter than the Heisenberg time $t_H=n(\epsilon) \hbar$.
This result applies to all the other kinetic equations derived in this thesis.\\
Equation (\ref{1Caaao}) can be rewritten in terms of the Fourier transform of the correlation function 
using the change of variable $\hbar \omega =\epsilon-\epsilon'$ as follows 
\begin{eqnarray}
\frac{d P(\epsilon,\alpha,t)}{dt} 
&=& - 2 \frac{\lambda^2}{\hbar^2} \int d \omega \int d\alpha' 
\tilde{\alpha}_{VV}(\epsilon,\alpha,\alpha',-\omega)                                              \label{1Caabo}\\
&& \hspace*{3cm} \{ P(\epsilon,\alpha,t) - P(\epsilon-\hbar \omega,\alpha',t) \} 
\frac{\sin \omega t}{\omega} .\nonumber    
\end{eqnarray} 
Using the long-time property $\lim_{t \to \infty} \frac{\sin \omega t}
{\omega} = \pi \delta(\omega)$, we get the continuous version of equation 
(\ref{1Caaan}), which is the Markovian version of the Pauli equation for a continuous spectrum. 
This equation is the original equation derived by Pauli in $1928$ \cite{Pauli28}:
\begin{center} \fbox{\parbox{12.5cm}{
\begin{eqnarray}
\frac{d P(\epsilon,\alpha,t)}{dt} 
= - \int d\alpha' W(\epsilon,\alpha,\alpha',0)
\{ P(\epsilon,\alpha,t) - P(\epsilon,\alpha',t) \} ,                                               \label{1Caaap}
\end{eqnarray} 
}} \end{center}
where 
\begin{eqnarray}
W(\epsilon,\alpha,\alpha',0)= 
2 \pi \frac{\lambda^2}{\hbar^2} \tilde{\alpha}_{VV}(\epsilon,\alpha,\alpha',0)
=2 \pi \frac{\lambda^2}{\hbar} n(\epsilon,\alpha')
\vert \langle \epsilon,\alpha \vert \hat{V} \vert \epsilon,\alpha' \rangle \vert^2                \label{1Caaaq}
\end{eqnarray} 
is the total transition probability given by the Fermi golden rule at frequency
zero [see equation (\ref{1Hcaak})].\\
Notice that the long-time limit can be justified when $t$ is large enough to assume that 
$W(\epsilon,\alpha,\alpha',{\cal O}(\hbar / t)) 
\approx W(\epsilon,\alpha,\alpha',0)$.
Let us call $t_c$ the minimal value of $t$ which satisfies this relation.\\
The Van Hove limit \cite{VanHove55} is often cited in the literature because it is the
limit for which the Pauli equation is exact.
This limit consists in successively taking the following limits:
\begin{itemize}
\item $N \to \infty$
\item $\lambda \to 0$ and $t \to \infty$ with $\lambda^2 t$ fixed.
\end{itemize}
The Van Hove limit is physically not very useful.
When this limit is not taken, the Pauli equation is valid for times $t$ such that $t_c < t < t_H$.\\
 
The Pauli equation is an important equation. 
It is the simplest quantum kinetic equation and a simple interpretation in terms of stochastic dynamics is possible.
The dynamics can be seen as resulting from random transitions (given by the Fermi golden rule) between eigenstates 
of the non-perturbed Hamiltonian, caused by the presence of a perturbation term in the total Hamiltonian.
The stochasticity of the Pauli equation is a consequence of the reduction of the entire density matrix describing 
the deterministic dynamics to the population dynamics.
The Pauli equation is an irreversible equation which allows to describe quantum non-equilibrium processes.
This equation has however a limited range of applicability.
In fact, in order to be able to describe coherent processes, the coherences of the system have to be included in 
the description.
This is what the Redfield equation achieves as we will see in the next subsection.    
\begin{figure}[h]
\centering
\rotatebox{0}{\scalebox{0.8}{\includegraphics{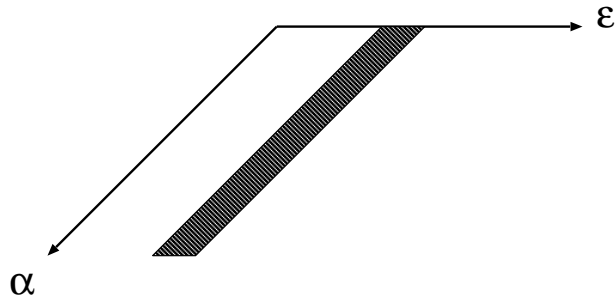}}}
\caption{The Pauli equation describes transitions occurring between levels of different quantum numbers 
$\alpha$ belonging to the same narrow energy shell.} \label{imeqpauli}
\end{figure}

\section{Redfield equation} \label{Redfield}

As already mentioned in the introduction, one of the common ways to study the relaxation to equilibrium of
a system is to let it interact with its environment.
We will derive in this section the Redfield equation which is the best known quantum kinetic equation describing 
the dynamics of a small subsystem interacting with an infinitely large environment.
In this scheme, the irreversible subsystem dynamics arises as a consequence of the fact that the description 
only focuses on the reduced dynamics of the subsystem which is a projection of the full dynamics on the 
degrees of freedom of the subsystem.
The Redfield equation has been first derived semi-empirically by Redfield in $1957$ in the context of NMR \cite{Redfield57}.
Many systematic derivations have been performed since then \cite{GardinerB00,GaspardRed99,Haake73,KampenB97,KuboB2,ZubarevB96} 
although we find equations that not always bear the name of Redfield in these different references. 
This equation has often been used in various fields ranging from NMR \cite{MehringB83,SlichterB90}
to quantum chemistry \cite{SchatzRatner93}, electron transfer theories \cite{Jean94},  
and condensed-phase reaction dynamics \cite{Kuhn03}.\\

The systems we consider in this section are composed of two parts, a small quantum subsystem 
with a Hamiltonian $\hat{H}_{S}$ and an environment with a Hamiltonian $\hat{H}_{B}$. 
These two parts are interacting with each other through an interaction term. 
The magnitude of this term is measured by the coupling parameter $\lambda$. 
The interaction term can be written as a sum of products of Hermitian operators of 
the subsystem $\hat{S}_{\kappa}$ with Hermitian operators of the environment $\hat{B}_{\kappa}$. 
The total Hamiltonian of such systems reads  
\begin{eqnarray}
\hat{H}=\hat{H}_{S}+\hat{H}_{B}+\lambda \sum_{\kappa} \hat{S}_{\kappa} \hat{B}_{\kappa} .                  \label{1Gaaaa}
\end{eqnarray}
This Hamiltonian is equivalent to the Hamiltonian (\ref{1Eaaaa}), where
$\hat{H}_{0}=\hat{H}_{S}+\hat{H}_{B}$ and $\hat{V}=\sum_{\kappa} \hat{S}_{\kappa} \hat{B}_{\kappa}$.
We use the notation
\begin{eqnarray}
\hat{H}_{S} \vert s \rangle &=& E_{s} \vert s \rangle \\
\hat{H}_{B} \vert b \rangle &=& E_{b} \vert b \rangle .                                                        \label{1Gaaba}
\end{eqnarray}
The dynamics of the total system is given by
\begin{eqnarray}
\frac{d \hat{\rho} (t)}{dt} &=& 
\hat{\hat{{\cal L}}} \hat{\rho}(t) =  
(\hat{\hat{{\cal L}}}_S + \hat{\hat{{\cal L}}}_B + \lambda \hat{\hat{{\cal L}}}_I) \hat{\rho}(t)               \label{1Gaaab} \\
&=& -\frac{i}{\hbar} [\hat{H},\hat{\rho}(t)] = 
-\frac{i}{\hbar} [\hat{H}_S,\hat{\rho}(t)] -\frac{i}{\hbar} [\hat{H}_B,\hat{\rho}(t)] 
- \lambda \frac{i}{\hbar} \sum_{\kappa} [\hat{S}_{\kappa} \hat{B}_{\kappa},\hat{\rho}(t)]  . \nonumber     
\end{eqnarray}
In the interaction representation the coupling operators become
\begin{eqnarray}
\hat{S}_{\kappa}(t) &=& 
e^{-\hat{\hat{{\cal L}}}_S t} \hat{S}_{\kappa} =
e^{\frac{i}{\hbar} \hat{H}_{S}t} \hat{S}_{\kappa} e^{-\frac{i}{\hbar} \hat{H}_{S}t}, \nonumber \\ 
\hat{B}_{\kappa}(t) &=& 
e^{-\hat{\hat{{\cal L}}}_B t} \hat{B}_{\kappa} =
e^{\frac{i}{\hbar} \hat{H}_{B}t} \hat{B}_{\kappa} e^{-\frac{i}{\hbar} \hat{H}_{B}t} ,                           \label{1Gaaac}
\end{eqnarray}
and the interaction representation of the Liouvillian reads 
\begin{equation}
\hat{\hat{{\cal L}}}_I(t)=e^{-(\hat{\hat{{\cal L}}}_S + \hat{\hat{{\cal L}}}_B)t} 
\hat{\hat{{\cal L}}}_I e^{(\hat{\hat{{\cal L}}}_S + \hat{\hat{{\cal L}}}_B) t} .                                 \label{1Gaaaf}
\end{equation}

\subsection{The projector}

The subsystem dynamics is described by the reduced density matrix of the subsystem 
\begin{equation}
\hat{\rho}_{S}(t) = \trace_{B} \hat{\rho}(t) = 
\sum_{s,s'} \proj{s}{s'} \sum_b \bra{s b} \hat{\rho}(t) \ket{s'b} .                                              \label{1Gaaag}
\end{equation}
Let us define the following projector over the relevant part of the density matrix 
\begin{equation}
\hat{\hat{{\cal P}}} \hat{\rho}(t) 
=\sum_{s,s'} \sum_{b} \bra{s b} \hat{\rho}(t) \ket{s' b} \proj{s b}{s' b} ,                                      \label{1Gaaah}
\end{equation}
and the projector on the non-relevant part of the density matrix as 
\begin{equation}
\hat{\hat{{\cal Q}}} \hat{\rho}(t) 
=\sum_{s,s'} \sum_{b,b'\neq b} \bra{s b} \hat{\rho}(t) \ket{s' b'} \proj{sb}{s'b'} .                             \label{1Gaaai}
\end{equation}
Equivalently, one can write that
\begin{equation}
\bra{s b} \hat{\hat{{\cal P}}} \hat{\rho}(t) \ket{s' b'}
= \bra{s b}  \hat{\rho}(t) \ket{s' b} \delta_{b b'},                                                             \label{1Gaabi}
\end{equation}
and
\begin{equation}
\bra{s b} \hat{\hat{{\cal Q}}} \hat{\rho}(t) \ket{s' b'} 
= \bra{s b} \hat{\rho}(t) \ket{s' b'}  (1-\delta_{b b'}).                                                        \label{1Gaaci}
\end{equation}
The reduced density matrix of the subsystem is obtained from the relevant-part projector as follows
\begin{equation}
\trace_{B} \hat{\hat{{\cal P}}} \hat{\rho}(t) 
= \trace_{B} \hat{\rho}(t) = \hat{\rho}_{S}(t) .                                                                \label{1Gaadi}
\end{equation}
This reduced density matrix contains all the information about the system, 
whereas all the information about the environment has been traced out and is therefore lost.
The reduced density matrix of the subsystem is the central quantity in the Redfield theory and the 
standard derivation of the Redfield equation consists in getting a closed equation for the reduced 
density matrix of the subsystem from the von Neumann equation.
All our derivations start from the projected weak-coupling expansion of the von Neumann equation (\ref{1Eaaak}). 
As in the Pauli equation derivation, in order to eliminate the $\hat{\hat{{\cal Q}}} \hat{\rho}(0)$
terms in the equation (\ref{1Eaaak}), we always choose an initial condition for which this quantity is zero.
In the Redfield theory, this corresponds to take an initial condition of the total system which is the tensor 
product of a general initial condition for the subsystem with an invariant density matrix for the environment
(i.e. a density matrix which is diagonal in the eigenbasis of $\hat{H}_B$ 
and therefore with the property $\hat{\hat{{\cal L}}}_B \hat{\rho}^{eq}_{B}=0$).
This initial condition reads 
\begin{equation}
\hat{\rho}(0) = \hat{\rho}_{S}(0) \hat{\rho}^{eq}_{B}.                                                            \label{1Gaaaj}
\end{equation}
Therefore, $\hat{\hat{{\cal P}}} \hat{\rho}(0) =\hat{\rho}_{S}(0) \hat{\rho}^{eq}_{B}$ and
$\hat{\hat{{\cal Q}}} \hat{\rho}(0) =0$.

\subsection{The derivation of the Redfield equation}

We now explicitly start the derivation.\\
\scriptsize
The projected von Neumann equation (\ref{1Eaaak}) becomes
\begin{eqnarray}
\hat{\hat{{\cal P}}} \hat{\rho}(t)
&=&\hat{\hat{{\cal P}}} e^{(\hat{\hat{{\cal L}}}_S+\hat{\hat{{\cal L}}}_B) t} \hat{\hat{{\cal P}}} \hat{\rho}(0)
+\lambda \hat{\hat{{\cal P}}} e^{(\hat{\hat{{\cal L}}}_S+\hat{\hat{{\cal L}}}_B) t} \int^{t}_{0} dT 
\hat{\hat{{\cal L}}}_I (T) \hat{\hat{{\cal P}}} \hat{\rho}(0)                                                      \label{1Gaaak} \\    
& &+ \lambda^2 \hat{\hat{{\cal P}}} e^{(\hat{\hat{{\cal L}}}_S+\hat{\hat{{\cal L}}}_B) t} 
\int^{t}_{0} dT \int^{T}_{0} d\tau \hat{\hat{{\cal L}}}_I (T) \hat{\hat{{\cal L}}}_I (T-\tau) 
\hat{\hat{{\cal P}}} \hat{\rho}(0)  + {\cal O}(\lambda^3) \nonumber .
\end{eqnarray}
We first need to evaluate $\hat{\hat{{\cal L}}}_I (T) \hat{\hat{{\cal P}}} \hat{\rho}(0)$ 
and $\hat{\hat{{\cal L}}}_I (T) \hat{\hat{{\cal L}}}_I (T-\tau) \hat{\hat{{\cal P}}} \hat{\rho}(0)$.
We have that
\begin{eqnarray}
\hat{\hat{{\cal L}}}_I (T) \hat{\hat{{\cal P}}} \hat{\rho}(0)
&=& - \frac{i}{\hbar} \sum_{\kappa} \lbrack \hat{S}_{\kappa}(T) \hat{B}_{\kappa}(T) 
, \hat{\hat{{\cal P}}} \hat{\rho}(0) \rbrack                                                                       \label{1Gaaal}\\
&=& - \frac{i}{\hbar} \sum_{\kappa} \hat{S}_{\kappa}(T) \hat{\rho}_{S}(0) 
\hat{B}_{\kappa}(T) \hat{\rho}^{eq}_{B} 
+ \frac{i}{\hbar} \sum_{\kappa}  \hat{\rho}_{S}(0) \hat{S}_{\kappa}(T) 
\hat{\rho}^{eq}_{B} \hat{B}_{\kappa}(T)  \nonumber 
\end{eqnarray}
and
\begin{eqnarray}
\hat{\hat{{\cal L}}}_I (T) \hat{\hat{{\cal L}}}_I (T-\tau) \hat{\hat{{\cal P}}} \hat{\rho}(0) 
&=& - \frac{1}{\hbar^2} \sum_{\kappa,\kappa'} \lbrack \hat{S}_{\kappa}(T) \hat{B}_{\kappa}(T) 
, \lbrack \hat{S}_{\kappa'}(T-\tau) \hat{B}_{\kappa'}(T-\tau), 
\hat{\hat{{\cal P}}}  \hat{\rho}(0) \rbrack \rbrack \nonumber\\
&=& - \frac{1}{\hbar^2} \sum_{\kappa,\kappa'} \hat{S}_{\kappa}(T) \hat{S}_{\kappa'}(T-\tau) \hat{\rho}_{S}(0) 
\hat{B}_{\kappa}(T) \hat{B}_{\kappa'}(T-\tau) \hat{\rho}^{eq}_{B} \nonumber\\
&& + \frac{1}{\hbar^2} \sum_{\kappa,\kappa'} \hat{S}_{\kappa}(T) \hat{\rho}_{S}(0) \hat{S}_{\kappa'}(T-\tau) 
\hat{B}_{\kappa}(T) \hat{\rho}^{eq}_{B} \hat{B}_{\kappa'}(T-\tau) \nonumber\\
&& + \frac{1}{\hbar^2} \sum_{\kappa,\kappa'} \hat{S}_{\kappa'}(T-\tau) \hat{\rho}_{S}(0) \hat{S}_{\kappa}(T) 
\hat{B}_{\kappa'}(T-\tau) \hat{\rho}^{eq}_{B} \hat{B}_{\kappa}(T)  \nonumber\\
&& - \frac{1}{\hbar^2} \sum_{\kappa,\kappa'} \hat{\rho}_{S}(0) \hat{S}_{\kappa'}(T-\tau) \hat{S}_{\kappa}(T) 
\hat{\rho}^{eq}_{B} \hat{B}_{\kappa'}(T-\tau) \hat{B}_{\kappa}(T)    ,                                          \label{1Gaaam}
\end{eqnarray}
so that equation (\ref{1Gaaak}) becomes
\begin{eqnarray}
\hat{\rho}_{S}(t)
&=& e^{\hat{\hat{{\cal L}}}_S t} \hat{\rho}_{S}(0) \trace_{B} \hat{\rho}^{eq}_{B}                                \label{1Gaaan}\\ 
&&- i \frac{\lambda}{\hbar} \sum_{\kappa} e^{\hat{\hat{{\cal L}}}_S t} \int^{t}_{0} dT 
\hat{S}_{\kappa}(T) \hat{\rho}_{S}(0) \trace_{B} \hat{B}_{\kappa}(T) \hat{\rho}^{eq}_{B} \nonumber \\
&&+ i \frac{\lambda}{\hbar} \sum_{\kappa} e^{\hat{\hat{{\cal L}}}_S t} \int^{t}_{0} dT 
\hat{\rho}_{S}(0) \hat{S}_{\kappa}(T) \trace_{B} \hat{\rho}^{eq}_{B} \hat{B}_{\kappa}(T) \nonumber \\
&&- \frac{\lambda^2}{\hbar^2} \sum_{\kappa,\kappa'} e^{\hat{\hat{{\cal L}}}_S t} \int^{t}_{0} dT \int^{T}_{0} d\tau 
\hat{S}_{\kappa}(T) \hat{S}_{\kappa'}(T-\tau) \hat{\rho}_{S}(0) 
\trace_{B} \hat{B}_{\kappa}(T) \hat{B}_{\kappa'}(T-\tau) \hat{\rho}^{eq}_{B} \nonumber \\
&&+ \frac{\lambda^2}{\hbar^2} \sum_{\kappa,\kappa'} e^{\hat{\hat{{\cal L}}}_S t} \int^{t}_{0} dT \int^{T}_{0} d\tau 
\hat{S}_{\kappa}(T) \hat{\rho}_{S}(0) \hat{S}_{\kappa'}(T-\tau) 
\trace_{B} \hat{B}_{\kappa}(T) \hat{\rho}^{eq}_{B} \hat{B}_{\kappa'}(T-\tau) \nonumber \\
&&+ \frac{\lambda^2}{\hbar^2} \sum_{\kappa,\kappa'} e^{\hat{\hat{{\cal L}}}_S t} \int^{t}_{0} dT \int^{T}_{0} d\tau 
\hat{S}_{\kappa'}(T-\tau) \hat{\rho}_{S}(0) \hat{S}_{\kappa}(T) 
\trace_{B} \hat{B}_{\kappa'}(T-\tau) \hat{\rho}^{eq}_{B} \hat{B}_{\kappa}(T) \nonumber \\
&&- \frac{\lambda^2}{\hbar^2} \sum_{\kappa,\kappa'} e^{\hat{\hat{{\cal L}}}_S t} \int^{t}_{0} dT \int^{T}_{0} d\tau 
\hat{\rho}_{S}(0) \hat{S}_{\kappa'}(T-\tau) \hat{S}_{\kappa}(T) 
\trace_{B} \hat{\rho}^{eq}_{B} \hat{B}_{\kappa'}(T-\tau) \hat{B}_{\kappa}(T) 
+ {\cal O}(\lambda^3) \nonumber .
\end{eqnarray}
Differentiating equation (\ref{1Gaaan}) with respect to time, we get
\begin{eqnarray}
\frac{d \hat{\rho}_{S}(t)}{dt}
&=& \hat{\hat{{\cal L}}}_S \hat{\rho}_{S}(t)                                                                     \label{1Gaaao}\\ 
&&- i \frac{\lambda}{\hbar} \sum_{\kappa} \hat{S}_{\kappa} \{e^{\hat{\hat{{\cal L}}}_S t} \hat{\rho}_{S}(0) \} 
\trace_{B} \hat{\rho}^{eq}_{B} \hat{B}_{\kappa} 
+ i \frac{\lambda}{\hbar} \sum_{\kappa} \{ e^{\hat{\hat{{\cal L}}}_S t} \hat{\rho}_{S}(0) \} \hat{S}_{\kappa} 
\trace_{B} \hat{\rho}^{eq}_{B} \hat{B}_{\kappa} \nonumber \\
&&- \frac{\lambda^2}{\hbar^2} \sum_{\kappa,\kappa'} \int^{t}_{0} d\tau 
\hat{S}_{\kappa} \hat{S}_{\kappa'}(-\tau) \{ e^{\hat{\hat{{\cal L}}}_S t} \hat{\rho}_{S}(0) \} 
\trace_{B} \hat{\rho}^{eq}_{B} \hat{B}_{\kappa}(t) \hat{B}_{\kappa'}(t-\tau) \nonumber \\
&&+ \frac{\lambda^2}{\hbar^2} \sum_{\kappa,\kappa'} \int^{t}_{0} d\tau 
\hat{S}_{\kappa} \{ e^{\hat{\hat{{\cal L}}}_S t} \hat{\rho}_{S}(0) \} \hat{S}_{\kappa'}(-\tau) 
\trace_{B} \hat{\rho}^{eq}_{B} \hat{B}_{\kappa'}(t-\tau) \hat{B}_{\kappa}(t) \nonumber \\
&&+ \frac{\lambda^2}{\hbar^2} \sum_{\kappa,\kappa'} \int^{t}_{0} d\tau 
\hat{S}_{\kappa'}(-\tau) \{ e^{\hat{\hat{{\cal L}}}_S t} \hat{\rho}_{S}(0) \} \hat{S}_{\kappa} 
\trace_{B} \hat{\rho}^{eq}_{B} \hat{B}_{\kappa}(t) \hat{B}_{\kappa'}(t-\tau) \nonumber \\
&&- \frac{\lambda^2}{\hbar^2} \sum_{\kappa,\kappa'} \int^{t}_{0} d\tau 
\{ e^{\hat{\hat{{\cal L}}}_S t} \hat{\rho}_{S}(0) \} \hat{S}_{\kappa'}(-\tau) \hat{S}_{\kappa}
\trace_{B} \hat{\rho}^{eq}_{B} \hat{B}_{\kappa'}(t-\tau) \hat{B}_{\kappa}(t) 
+ {\cal O}(\lambda^3) \nonumber .
\end{eqnarray}
In order to close the equation, we use the fact that $\hat{\rho}_{S}(t)= e^{\hat{\hat{{\cal L}}}_S t} 
\hat{\rho}_{S}(0) + {\cal O}(\lambda^2)$. 
This does not affect equation (\ref{1Gaaao}) to orders lower than ${\cal O}(\lambda^3)$. \\
\normalsize
The final result is the Redfield equation which reads
\begin{center} \fbox{\parbox{12.5cm}{
\begin{eqnarray}
\frac{d \hat{\rho}_{S}(t)}{dt}
&=& \hat{\hat{{\cal L}}}_S \hat{\rho}_{S}(t)                                                                     
- i \frac{\lambda}{\hbar} \sum_{\kappa} \lbrack \hat{S}_{\kappa} 
, \hat{\rho}_{S}(t) \rbrack \mean{\hat{B}_{\kappa}}                                                           
+ \frac{\lambda^2}{\hbar^2} \sum_{\kappa,\kappa'} \int^{t}_{0} d\tau \{ \;                                        \label{1Gaaap}\\
&&\hspace*{1cm} - \alpha_{\kappa \kappa'}(\tau) \hat{S}_{\kappa} \hat{S}_{\kappa'}(-\tau) \hat{\rho}_{S}(t) 
- \alpha^{*}_{\kappa \kappa'}(\tau) \hat{\rho}_{S}(t) \hat{S}_{\kappa'}(-\tau) \hat{S}_{\kappa}  \nonumber \\
&&\hspace*{1cm} + \alpha^{*}_{\kappa \kappa'}(\tau) \hat{S}_{\kappa} \hat{\rho}_{S}(t) \hat{S}_{\kappa'}(-\tau) 
+ \alpha_{\kappa \kappa'}(\tau) \hat{S}_{\kappa'}(-\tau) \hat{\rho}_{S}(t) \hat{S}_{\kappa} \; \} \nonumber ,
\end{eqnarray} 
}} \end{center}
where the expectation of the environment coupling operator is given by
\begin{eqnarray}
\mean{\hat{B}_{\kappa}}=\trace_{B} \hat{\rho}^{eq}_{B} \hat{B}_{\kappa} ,                                       \label{1Gaaaq}
\end{eqnarray}
and where the environment correlation function by 
\begin{eqnarray}
\alpha_{\kappa \kappa'}(\tau)=\trace_{B} \hat{\rho}^{eq}_{B} \hat{B}_{\kappa'}(\tau) \hat{B}_{\kappa}
=\alpha^{*}_{\kappa \kappa'}(-\tau) .                                                                             \label{1Gaaar}
\end{eqnarray}
Equation (\ref{1Gaaap}) is the non-Markovian version of the Redfield equation.
This equation preserves trace and Hermiticity.
The Markovian version of the Redfield equation (as originally derived by Redfield) is simply obtained 
by replacing $\int_{0}^{t} d\tau$ by $\int_{0}^{\infty} d\tau$ in equation (\ref{1Gaaap}).\\
The correlation function appearing in (\ref{1Gaaap}) has the same form as the one 
we found in linear response theory [section \ref{linearresponse}, equation (\ref{1Haaal})]. 
The general properties of this correlation function are detailed in appendix \ref{AppA}.
This correlation function characterizes the equilibrium fluctuations of the environment. \\
Notice that if $\mean{\hat{B}_{\kappa}(t)} \neq 0$, with the redefinition
\begin{eqnarray}
\hat{H}_S &\to& \hat{H}_S + \lambda \sum_{\kappa} \hat{S}_{\kappa} \mean{\hat{B}_{\kappa}}                       \label{1Gaaas}\\
\sum_{\kappa} \hat{S}_{\kappa} \hat{B}_{\kappa} &\to& \sum_{\kappa} \hat{S}_{\kappa} \hat{B}_{\kappa} 
- \lambda \sum_{\kappa} \hat{S}_{\kappa} \mean{\hat{B}_{\kappa}} ,                                               \label{1Gaabs}
\end{eqnarray}
the ${\cal O}(\lambda)$-term can be eliminated from equation (\ref{1Gaaap}).

\subsection{Slippage of initial conditions}

An important remark is that the Markovian version of the Redfield equation does not in general have 
the Lindblad form (see section \ref{Posevol}) and is known to violate positivity if the initial 
condition is near the border of the set of physically admissible density matrices
\cite{AlickiB87,GaspardRed99,Gnutzmann96,Pechukas94,Suarez92}. 
We notice however that the trace and Hermiticity are preserved by the the Markovian Redfield equation.
The non-Markovian version of the Redfield equation is believed to preserve positivity.
There is no general proof for this statement, but we do not know any work having infirmed this statement.
Therefore, the non-preservation of positivity in the Markovian Redfield equation is meant to 
arise as a consequence of not having taken into account the initial non-Markovian dynamics which 
corresponds to the stage of the dynamics where correlations are established between the subsystem and its 
environment from the uncorrelated initial condition (\ref{1Gaaaj}).
This highly non-Markovian initial stage of the dynamics occurs on time scales of the order of the 
environment correlation time, which is the typical time necessary for the correlation function of the
environment to decay to zero.
A slippage (displacement) of initial conditions has been proposed by various authors \cite{GaspardRed99,Suarez92} 
to correct this initial highly non-Markovian dynamics which is not taken into account by the Markovian equation.
The general method of the slippage of initial conditions has been mentioned in section \ref{slippagegen}.
In this way, it can be shown that when the Markovian evolution starts from an initial condition modified by the 
slippage superoperator, it preserves the positivity and gives the same results as the non-Markovian equation 
on time scales longer than the correlation time of the environment \cite{GaspardRed99,Suarez92}.

\subsection{Delta correlated environments}

An important observation can be made when the environment has a very fast dynamics 
compared to the subsystem dynamics.
In this case, the environment correlation function decays on a very short time scale compared to 
the subsystem dynamics and can therefore be considered proportional to a delta distribution in time
\begin{eqnarray}
\alpha_{\kappa \kappa'}(\tau) = 2 D_{\kappa \kappa'} \delta(\tau).                                                \label{1Gaaav}
\end{eqnarray}
In this case, the Markovian Redfield equation takes the Lindblad form and therefore always preserves positivity.\\
This can be understood by the following argument.
$D_{\kappa \kappa'}$ is Hermitian and can be diagonalized as $\hat{D}=\hat{U}^{\dagger} \hat{d} \hat{U}$ 
where $\hat{d}$ is a diagonal matrix with eigenvalues $d_{\kappa}$.
By introducing the Lindblad operator defined as follows
\begin{eqnarray}
\hat{L}_{\kappa}= \sqrt{d_{\kappa}} \sum_{\kappa'} U_{\kappa \kappa'} \hat{S}_{\kappa'} ,                          \label{1Gaaaw}
\end{eqnarray}
the Markovian Redfield equation becomes of Lindblad type (\ref{1Faaaa})
\begin{eqnarray}
\frac{d \hat{\rho}_{S} (t)}{dt}&=& 
-\frac{i}{\hbar} \lbrack \hat{H}_{S}, \hat{\rho} (t) \rbrack                                                       \label{1Gaaax}\\
&&+ \frac{\lambda^2}{\hbar^2} \sum_{\kappa} (2 \hat{L}_{\kappa} \hat{\rho}_{S} (t) \hat{L}_{\kappa}^{\dagger}
- \hat{L}_{\kappa}^{\dagger} \hat{L}_{\kappa} \hat{\rho}_{S} (t)
- \hat{\rho}_{S} (t) \hat{L}_{\kappa}^{\dagger} \hat{L}_{\kappa}) \nonumber .
\end{eqnarray}

\subsection{Markovian approximation}

To simplify the notation, we shall eliminate the summation over $\kappa$ and consider a single coupling term.
This does not affect the discussion and the summation is easy to restore. 
The Redfield equation (\ref{1Gaaap}) becomes
\begin{eqnarray}
\dot{\hat{\rho}}_{S}(t)
&=& \hat{\hat{{\cal L}}}_S \hat{\rho}_{S}(t)                                                                     
- i \frac{\lambda}{\hbar} \lbrack \hat{S}, \hat{\rho}_{S}(t) \rbrack \mean{\hat{B}}                                                           
+ \frac{\lambda^2}{\hbar^2} \int^{t}_{0} d\tau \{ \;                                                           \label{1Gbabb}\\
&&\hspace*{1cm} - \alpha(\tau) \hat{S} \hat{S}(-\tau) \hat{\rho}_{S}(t) 
- \alpha^{*}(\tau) \hat{\rho}_{S}(t) \hat{S}(-\tau) \hat{S} \nonumber \\
&&\hspace*{1cm} + \alpha^{*}(\tau) \hat{S} \hat{\rho}_{S}(t) \hat{S}(-\tau) 
+ \alpha(\tau) \hat{S}(-\tau) \hat{\rho}_{S}(t) \hat{S} \; \} \nonumber ,
\end{eqnarray}
To study the population and the coherence dynamics, we will project the Redfield equation 
in the eigenbasis of the subsystem.\\
\scriptsize
Using the notation $\bra{s} \hat{\rho}_{S}(t) \ket{s'}=\rho_{ss'}(t)$, equation (\ref{1Gbabb}) becomes
\begin{eqnarray}
\dot{\rho}_{ss'}(t)
&=& - \frac{i}{\hbar} (E_s-E_{s'}) \rho_{ss'}(t)                                                                     
- i \frac{\lambda}{\hbar} \sum_{\bar{s}} ( S_{s\bar{s}} \rho_{\bar{s}s'}(t) - 
\rho_{s\bar{s}}(t) S_{\bar{s}s'} )  \mean{\hat{B}_{\kappa}} \nonumber\\                                              
&&+ \frac{\lambda^2}{\hbar^2} \sum_{\bar{s}\bar{s}'} \int^{t}_{0} \{ \; \nonumber \\                                   
&&\hspace*{2cm} - S_{s\bar{s}} S_{\bar{s}\bar{s}'} 
\rho_{\bar{s}'s'}(t) \alpha(\tau) 
e^{-\frac{i}{\hbar}(E_{\bar{s}}-E_{\bar{s}'})\tau} \nonumber \\
&&\hspace*{2cm} - \rho_{s\bar{s}}(t) S_{\bar{s}\bar{s}'} S_{\bar{s}'s'} 
\alpha^{*}(\tau) e^{-\frac{i}{\hbar}(E_{\bar{s}}-E_{\bar{s}'})\tau} \nonumber \\
&&\hspace*{2cm} + S_{s\bar{s}} \rho_{\bar{s}\bar{s}'}(t) S_{\bar{s}'s'} 
\alpha^{*}(\tau) e^{-\frac{i}{\hbar}(E_{\bar{s}'}-E_{s'})\tau} \nonumber \\
&&\hspace*{2cm} + S_{s\bar{s}} \rho_{\bar{s}\bar{s}'}(t) S_{\bar{s}'s'} 
\alpha(\tau) e^{-\frac{i}{\hbar}(E_{s}-E_{\bar{s}})\tau} \; \} .                                              \label{1Gbaac}
\end{eqnarray}
Using $\alpha(t)=\int d\omega \tilde{\alpha}(\omega) e^{-i\omega t}$, the property
$\tilde{\alpha}(\omega)=\tilde{\alpha}^{*}(\omega)$, and defining $\hbar \omega_{s,s'}=(E_s-E_{s'})$, we get
\begin{eqnarray}
\dot{\rho}_{ss'}(t)
&=& - i \omega_{ss'} \rho_{ss'}(t)                                                                     
- i \frac{\lambda}{\hbar} \sum_{\bar{s}} ( S_{s\bar{s}} \rho_{\bar{s}s'}(t) - 
\rho_{s\bar{s}}(t) S_{\bar{s}s'} )  \mean{\hat{B}_{\kappa}} \nonumber\\                                              
&&+ \frac{\lambda^2}{\hbar^2} \sum_{\bar{s}\bar{s}'} \int^{t}_{0} d\tau \int d\omega \{ \; \nonumber \\                                   
&&\hspace*{3cm} - S_{s\bar{s}} S_{\bar{s}\bar{s}'} \rho_{\bar{s}'s'}(t) \tilde{\alpha}(\omega) 
e^{i(\omega_{\bar{s}'\bar{s}}-\omega)\tau} \nonumber \\
&&\hspace*{3cm} - \rho_{s\bar{s}}(t) S_{\bar{s}\bar{s}'} S_{\bar{s}'s'} 
\tilde{\alpha}(\omega) e^{-i(\omega_{\bar{s}\bar{s}'}- \omega)\tau} \nonumber \\
&&\hspace*{3cm} + S_{s\bar{s}} \rho_{\bar{s}\bar{s}'}(t) S_{\bar{s}'s'}
\tilde{\alpha}(\omega) e^{-i(\omega_{\bar{s}'s'}-\omega)\tau} \nonumber \\
&&\hspace*{3cm} + S_{s\bar{s}} \rho_{\bar{s}\bar{s}'}(t) S_{\bar{s}'s'} 
\tilde{\alpha}(\omega) e^{i(\omega_{\bar{s}s}-\omega)\tau} \; \}.                                             \label{1Gbaad}
\end{eqnarray}
Taking the Markovian limit and using
\begin{eqnarray}
\int_{0}^{\infty} d\tau e^{\pm i \omega \tau} = \pm i{{\cal P}} \frac{1}{\omega} + \pi \delta(\omega) ,         \label{1Gbaae}
\end{eqnarray}
we get equation (\ref{1Gbaaf}).\\
\normalsize
The Markovian version of the Redfield equation in the eigenbasis of the subsystem reads
\begin{eqnarray}
\dot{\rho}_{ss'}(t)
&=& -i \omega_{ss'} \rho_{ss'}(t)                                                                     
- i \frac{\lambda}{\hbar} \sum_{\bar{s}} ( S_{s\bar{s}} \rho_{\bar{s}s'}(t) - 
\rho_{s\bar{s}}(t) S_{\bar{s}s'} )  \mean{\hat{B}}                                               \label{1Gbaaf}\\   
&&+ \frac{\lambda^2}{\hbar^2} \sum_{\bar{s}\bar{s}'}                                                                                        
\{ \; S_{s\bar{s}} S_{\bar{s}\bar{s}'} \rho_{\bar{s}'s'}(t) 
(\; -\pi \tilde{\alpha}(\omega_{\bar{s}'\bar{s}}) 
-i \int d\omega {{\cal P}} \frac{\tilde{\alpha}(\omega)}{\omega_{\bar{s}'\bar{s}}-\omega} \;)\nonumber \\
&&\hspace*{1.5cm} + \rho_{s\bar{s}}(t) S_{\bar{s}\bar{s}'} S_{\bar{s}'s'} 
(\; -\pi \tilde{\alpha}(\omega_{\bar{s}\bar{s}'}) 
+i \int d\omega {{\cal P}} \frac{\tilde{\alpha}(\omega)}{\omega_{\bar{s}\bar{s}'}- \omega} \;)\nonumber \\
&&\hspace*{1.5cm} + S_{s\bar{s}} \rho_{\bar{s}\bar{s}'}(t) S_{\bar{s}'s'}
(\; \pi \tilde{\alpha}(\omega_{\bar{s}'s'})
-i \int d\omega {{\cal P}} \frac{\tilde{\alpha}(\omega)}{\omega_{\bar{s}'s'}-\omega} \;)\nonumber \\
&&\hspace*{1.5cm} + S_{s\bar{s}} \rho_{\bar{s}\bar{s}'}(t) S_{\bar{s}'s'} 
(\; \pi \tilde{\alpha}(\omega_{\bar{s}s}) 
+i \int d\omega {{\cal P}} \frac{\tilde{\alpha}(\omega)}{\omega_{\bar{s}s}-\omega} \;) \; \}. \nonumber                   
\end{eqnarray}
No further approximation have been done with respect to the Markovian Redfield equation.

\subsubsection{Population dynamics}

To study the population dynamics we will now neglect the contribution of the coherences to the population dynamics.
We get
\begin{eqnarray} 
\dot{\rho}_{ss}(t)
&=&- 2 \pi \frac{\lambda^2}{\hbar^2} \sum_{\bar{s}} \vert S_{\bar{s}s} \vert^2 
\tilde{\alpha}(-\omega_{\bar{s}s}) \rho_{ss}(t)                                                       \label{1Gbaag} \\
&&+ 2 \pi \frac{\lambda^2}{\hbar^2} \sum_{\bar{s}} \vert S_{\bar{s}s} \vert^2 
\tilde{\alpha}(\omega_{\bar{s}s}) \rho_{\bar{s}\bar{s}}(t)  \nonumber   .                                               
\end{eqnarray}
This master equation describes the population dynamics of the subsystem. 
This equation is commonly used is quantum optics \cite{CohenTannoudjiB96}.
Its simple structure allows us to easily understand the physical processes it describes. 
According to our results of section \ref{FermiGR}, the quantity $\tilde{\alpha}(-\omega_{\bar{s}s})$ 
is proportional to the Fermi golden rule transition rate for the emission of an energy 
$\hbar \omega_{\bar{s}s}$ by the environment. 
This energy is therefore absorbed by the subsystem which undergoes a transition from the state 
$s$ to the state $\bar{s}$.
Since this transition depopulates the state $s$, this term appears as a negative contribution to the 
$\rho_{ss}(t)$ dynamics.
$\tilde{\alpha}(\omega_{\bar{s}s})$ is proportional to the transition rate for 
the absorption of an energy $\hbar \omega_{\bar{s}s}$ by the environment. 
This energy is emitted by the subsystem in the state $\bar{s}$ which jumps to the state $s$.
This process contributes positively to the $\rho_{ss}(t)$ dynamics.\\
An important remark, which will become fundamental in chapter \ref{ch3}, is that the
state of the environment is not affected by the emission or by the absorption of a quantum 
in the Redfield scheme.
This becomes a problem if the environment is finite and of small size.
\begin{figure}[h]
\centering
\rotatebox{0}{\scalebox{0.6}{\includegraphics{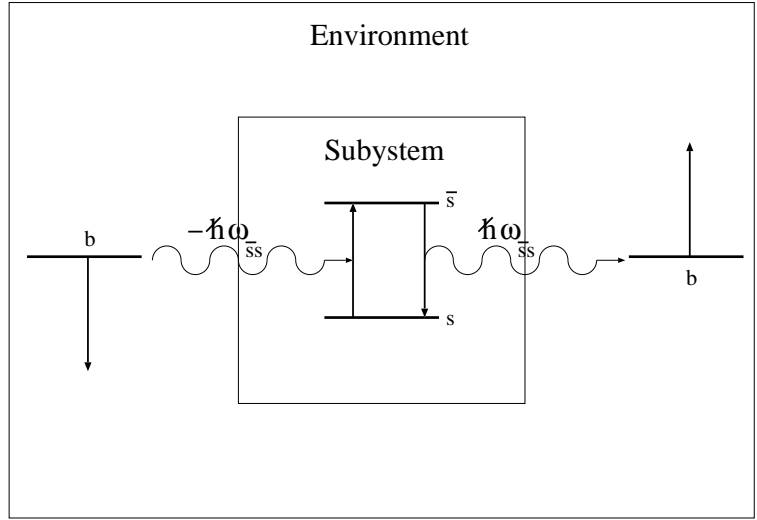}}}
\caption{The two processes of emission (on the right) and of absorption (on the left) of energy from the 
environment to the subsystem are schematically represented. $b$ schematically represent the state of the 
environment which is either of given energy in the microcanonical case or of given temperature in the 
canonical case.}                                                                                             \label{imRedfield}
\end{figure}
We finally notice that equation (\ref{1Gbaag}) can satisfy detailed balance. 
If $\rho_{ss}^{eq}$ denotes the stationary density matrix, the detailed balance condition reads
\begin{eqnarray} 
\tilde{\alpha}(-\omega_{\bar{s}s}) \rho_{ss}^{eq}=\tilde{\alpha}(\omega_{\bar{s}s}) 
\rho_{\bar{s}\bar{s}}^{eq}.                                                                                  \label{1Gbabg}
\end{eqnarray}
The environment correlation functions generally used in the Redfield equation are canonical. 
In this case, if the temperature is $1/\beta$ and using the KMS property (\ref{AppAaaau}), 
the detailed balance condition allows us to find a Boltzmann distribution for the subsystem 
equilibrium distribution 
\begin{eqnarray} 
\frac{\rho_{ss}^{eq}}{\rho_{\bar{s}\bar{s}}^{eq}}= 
\frac{\tilde{\alpha}(\beta,\omega_{\bar{s}s})}{\tilde{\alpha}(\beta,-\omega_{\bar{s}s})}=
e^{\beta \hbar \omega_{\bar{s}s}} .                                                                          \label{1Gbadg}
\end{eqnarray}
If one applies the detailed balance condition to the Redfield equation with microcanonical 
environment correlation functions (at energy $\epsilon$), we get the following equilibrium 
distribution for the subsystem
\begin{eqnarray} 
\frac{\rho_{ss}^{eq}}{\hat{\rho}_{\bar{s}\bar{s}}^{eq}}= 
\frac{\tilde{\alpha}(\epsilon,\omega_{\bar{s}s})}{\tilde{\alpha}(\epsilon,-\omega_{\bar{s}s})}.              \label{1Gbacg}
\end{eqnarray}
When the environment is large, the equivalence between the microcanonical and the canonical ensemble 
should allow us to understand how the Boltzmann subsystem canonical distribution (\ref{1Gbadg}) can emerge 
from the microcanonical one (\ref{1Gbacg}).  
However, the Redfield theory provides no answer to this question, indicating that the Redfield theory, 
when applied to environments in microcanonical states, may contain some non-physical features. 
In chapter \ref{ch3} we will see how our new kinetic equation solves this problem.

\subsubsection{Coherence dynamics}

To study the coherence dynamics, we isolate one coherence and neglect its coupling to the other coherences 
and populations. 
We get an equation of the form
\begin{eqnarray}  
\dot{\rho}_{ss'}(t) &\approx& (-i \tilde{\omega}_{ss'} - \gamma_{ss'} ) \rho_{ss'}(t),                        \label{1Gbaah}
\end{eqnarray}
where
\begin{eqnarray} 
\tilde{\omega}_{ss'} = \frac{\tilde{E}_s-\tilde{E}_{s'}}{\hbar} ,                                              \label{1Gbaai}
\end{eqnarray}
where
\begin{eqnarray} 
\tilde{E}_s = E_s + \frac{\lambda}{\hbar} \mean{\hat{B}(t)} S_{ss}
+ \frac{\lambda^2}{\hbar^2} \sum_{\bar{s}} \vert S_{s\bar{s}} \vert^2 
\int d\omega {{\cal P}} \frac{\tilde{\alpha}(\omega)}{\omega_{s\bar{s}}-\omega}                               \label{1Gbaaj}
\end{eqnarray}
is the Bohr frequency modified by the so-called Lamb shifts and
\begin{eqnarray} 
\gamma_{s,s'} &=&- \pi \frac{\lambda^2}{\hbar^2} (S_{ss}-S_{s's'})^2 \tilde{\alpha}(0)                       \label{1Gbaak}\\ 
&&+ \pi \frac{\lambda^2}{\hbar^2} \sum_{\bar{s} \neq s} 
( \vert S_{s\bar{s}} \vert^2 \tilde{\alpha}(\omega_{s\bar{s}})                                               
+ \vert S_{s'\bar{s}} \vert^2 \tilde{\alpha}(\omega_{s'\bar{s}}) ) \nonumber
\end{eqnarray}
the damping rate.\\
We thus have two kind of effects on the dynamics of the coherence of the subsystem induced by the environment interaction.
The first is a Lamb shift which modifies the frequency of the free oscillation of the subsystem coherence and
the second effect is the irreversible damping of the coherence, this last effect being at the origin of 
decoherence.

\chapter{New kinetic equation for finite environment} \label{ch3}

In the present chapter, we derive one of the main results of this thesis.
We have seen in the introduction that the advances in nanotechnologies raise the question of the validity
of the standard methods of statistical mechanics at small sizes.
We here present a new kinetic theory to describe the dynamics of a subsystem interacting with an environment.
The standard theory to describe such systems is the Redfield theory. 
However, in this theory, the environment is implicitly assumed to be infinite and therefore not affected 
by the subsystem dynamics.
This assumption is probably not valid for many nanosystems for which the environment of the subsystem 
is itself of finite size. 
Examples of such nanosystems are atomic or molecular clusters in which a two-level impurity is enclosed.
Our new formalism permits to derive a more general kinetic equation than the Redfield equation, which takes 
into account the possible finiteness of the environment.
This equation has been derived recently by Esposito and Gaspard in Ref. \cite{Esposito1}.\\

Our new equation is derived in section \ref{derivation}.
The population and the coherence dynamics are studied in section \ref{popcoh}. 
In section \ref{infinitelimit}, we show that our new equation transforms into the Redfield equation 
in the infinite environment limit.
 
\section{Our new kinetic equation} \label{derivation}

As in the Redfield theory, we consider a system composed of two parts, a quantum subsystem 
whose Hamiltonian is $\hat{H}_{S}$ and an environment whose Hamiltonian is $\hat{H}_{B}$. 
These two parts are interacting with each other via an interaction term. 
The magnitude of the interaction is characterized by the coupling parameter $\lambda$.
The interaction term is made of a sum of products of Hermitian operators of the subsystem
$\hat{S}_{\kappa}$ with Hermitian operators of the environment $\hat{B}_{\kappa}$. 
The Hamiltonian of the total system reads  
\begin{eqnarray}
\hat{H}=\hat{H}_{S}+\hat{H}_{B}+\lambda \sum_{\kappa} \hat{S}_{\kappa} \hat{B}_{\kappa} .                  \label{2Aaaaa}
\end{eqnarray}
This Hamiltonian has the form (\ref{1Eaaaa}), where $\hat{H}_{0}=\hat{H}_{S}+\hat{H}_{S}$ 
and $\hat{V}=\sum_{\kappa} \hat{S}_{\kappa} \hat{B}_{\kappa}$.
We use the notation
\begin{eqnarray}
\hat{H}_{S} \vert s \rangle &=& E_{s} \vert s \rangle \\
\hat{H}_{B} \vert b \rangle &=& E_{b} \vert b \rangle .                                                        \label{2Aaaba}
\end{eqnarray}
The dynamics of the total system is ruled by the von Neumann equation
\begin{eqnarray}
\frac{d \hat{\rho} (t)}{dt} &=& 
\hat{\hat{{\cal L}}} \hat{\rho}(t) =  
(\hat{\hat{{\cal L}}}_S + \hat{\hat{{\cal L}}}_B + \lambda \hat{\hat{{\cal L}}}_I) \hat{\rho}(t)               \label{2Aaaab} \\
&=& -\frac{i}{\hbar} [\hat{H},\hat{\rho}(t)] = 
-\frac{i}{\hbar} [\hat{H}_S,\hat{\rho}(t)] -\frac{i}{\hbar} [\hat{H}_B,\hat{\rho}(t)] 
- \lambda \frac{i}{\hbar} \sum_{\kappa} [\hat{S}_{\kappa} \hat{B}_{\kappa},\hat{\rho}(t)]  . \nonumber     
\end{eqnarray}
In the interaction representation, the operators become
\begin{eqnarray}
\hat{\rho}_{I}(t) &=& 
e^{-(\hat{\hat{{\cal L}}}_S+\hat{\hat{{\cal L}}}_B) t} \hat{\rho}(t)=
e^{\frac{i}{\hbar} (\hat{H}_{S}+\hat{H}_{B}) t} \hat{\rho}(t) 
e^{-\frac{i}{\hbar} (\hat{H}_{S}+\hat{H}_{B}) t} ,\nonumber \\
\hat{S}_{\kappa}(t) &=& 
e^{-\hat{\hat{{\cal L}}}_S t} \hat{S}_{\kappa} =
e^{\frac{i}{\hbar} \hat{H}_{S}t} \hat{S}_{\kappa} e^{-\frac{i}{\hbar} \hat{H}_{S}t}, \nonumber \\ 
\hat{B}_{\kappa}(t) &=& 
e^{-\hat{\hat{{\cal L}}}_B t} \hat{B}_{\kappa} =
e^{\frac{i}{\hbar} \hat{H}_{B}t} \hat{B}_{\kappa} e^{-\frac{i}{\hbar} \hat{H}_{B}t} ,                           \label{2Aaaac}
\end{eqnarray}
and the von Neumann equation (\ref{2Aaaab}) reads
\begin{equation}
\frac{d \hat{\rho}_I(t)}{dt} = \lambda \hat{\hat{{\cal L}}}_I(t)\hat{\rho}_I(t) 
= - \lambda \frac{i}{\hbar} \sum_{\kappa} \lbrack \hat{S}_{\kappa}(t) \hat{B}_{\kappa}(t) 
,\hat{\rho}_I(t) \rbrack  ,                                                                                     \label{2Aaaae}
\end{equation}
where we have defined the interaction representation of the Liouvillian 
\begin{equation}
\hat{\hat{{\cal L}}}_I(t)=e^{-(\hat{\hat{{\cal L}}}_S + \hat{\hat{{\cal L}}}_B)t} 
\hat{\hat{{\cal L}}}_I e^{(\hat{\hat{{\cal L}}}_S + \hat{\hat{{\cal L}}}_B) t} .                                \label{2Aaaaf}
\end{equation}

\subsection{The new projector}

The central quantity $\hat{P}(\hat{H}_B,t)$ of our new theory is defined by the following  
projection operator 
\begin{equation}
\hat{\hat{{\cal P}}} \hat{\rho}(t) 
=\sum_{s,s'} \sum_{b} \bra{s b}  \hat{\rho}(t) \ket{s' b} \proj{s b}{s' b} 
\equiv \frac{\hat{P}(\hat{H}_B,t)}{n(\hat{H}_B)},                                                               \label{2Aaaah}
\end{equation}
where $n(\hat{H}_B)$ is the density of states of the environment.
The projector on the non-relevant part of the density matrix therefore reads
\begin{equation}
\hat{\hat{{\cal Q}}} \hat{\rho}(t) 
=\sum_{s,s'} \sum_{b,b'\neq b} \bra{s b}  \hat{\rho}(t) \ket{s' b'} \proj{sb}{s'b'} .                           \label{2Aaaai}
\end{equation}
Equivalently, we can write that
\begin{equation}
\bra{s b} \hat{\hat{{\cal P}}} \hat{\rho}(t) \ket{s' b'}
= \bra{s b} \hat{\rho}(t) \ket{s' b} \delta_{b b'}
\equiv \frac{\hat{P}_{s,s'}(E_b,t)}{n(E_b)} \delta_{b b'},                                                      \label{2Aaabi}
\end{equation}
and
\begin{equation}
\bra{s b} \hat{\hat{{\cal Q}}} \hat{\rho}(t) \ket{s' b'} 
= \bra{s b}  \hat{\rho}(t) \ket{s' b'}  (1-\delta_{b b'}).                                                      \label{2Aaaci}
\end{equation}
Our new kinetic equation is an evolution equation for the quantity $\hat{P}(\hat{H}_B,t)$.
The only information lost by describing the total system dynamics in term of the quantity $\hat{P}(\hat{H}_B,t)$ 
is the information regarding the environment coherences.
However, the information on the subsystem populations and coherences as well as the information on the 
environment energy is kept.  
The reduced density matrix of the subsystem is obtained from $\hat{P}(\hat{H}_B,t)$ by the tracing 
over the environment states
\begin{equation}
\hat{\rho}_{S}(t)=\trace_{B} \hat{\hat{{\cal P}}} \hat{\rho}(t) 
= \trace_{B} \frac{\hat{P}(\hat{H}_B,t)}{n(\hat{H}_B)}.                                                         \label{2Aaadi}
\end{equation}
Similarly as for the Redfield equation, the initial conditions which 
eliminate the terms containing $\hat{\hat{{\cal Q}}} \hat{\rho}(0)$ in equation (\ref{1Eaaak}) 
are initial conditions which are diagonal in the eigenbasis of $\hat{H}_B$.
These initial conditions read 
\begin{equation}
\hat{\rho}(0) = \frac{\hat{P}(\hat{H}_B,0)}{n(\hat{H}_B)}.                                                     \label{2Baaaj}
\end{equation}
We therefore have $\hat{\hat{{\cal P}}} \hat{\rho}(0) = \hat{P}(\hat{H}_B,0)/n(\hat{H}_B)$ and
$\hat{\hat{{\cal Q}}} \hat{\rho}(0) =0$.

\subsection{The derivation of our kinetic equation}
 
We start now the explicit derivation of the new kinetic equation. \\
\scriptsize
The projected weak-coupling expansion of the von Neumann equation (\ref{1Eaaak}) becomes
\begin{eqnarray}
\frac{\hat{P}(\hat{H}_B,t)}{n(\hat{H}_B)}
&=&\hat{\hat{{\cal P}}} e^{(\hat{\hat{{\cal L}}}_S+\hat{\hat{{\cal L}}}_B) t} 
\frac{\hat{P}(\hat{H}_B,0)}{n(\hat{H}_B)}
+\lambda \hat{\hat{{\cal P}}} e^{(\hat{\hat{{\cal L}}}_S+\hat{\hat{{\cal L}}}_B) t} \int^{t}_{0} dT 
\hat{\hat{{\cal L}}}_I (T) \frac{\hat{P}(\hat{H}_B,0)}{n(\hat{H}_B)}                                           \label{2Baaak} \\    
& &+ \lambda^2 \hat{\hat{{\cal P}}} e^{(\hat{\hat{{\cal L}}}_S+\hat{\hat{{\cal L}}}_B) t} 
\int^{t}_{0} dT \int^{T}_{0} d\tau \hat{\hat{{\cal L}}}_I (T) \hat{\hat{{\cal L}}}_I (T-\tau) 
\frac{\hat{P}(\hat{H}_B,0)}{n(\hat{H}_B)} + {\cal O}(\lambda^3) \nonumber .
\end{eqnarray}
Projecting this equation in the eigenbasis of the environment we get
\begin{eqnarray}
\frac{\hat{P}(E_b,t)}{n(E_b)}
&=&e^{\hat{\hat{{\cal L}}}_S t} \frac{\hat{P}(E_b,0)}{n(E_b)}
+\lambda e^{\hat{\hat{{\cal L}}}_S t} \int^{t}_{0} dT 
\bra{b} \hat{\hat{{\cal L}}}_I (T) \frac{\hat{P}(\hat{H}_B,0)}{n(\hat{H}_B)} \ket{b}                            \label{2Baabk} \\    
& &+ \lambda^2 e^{\hat{\hat{{\cal L}}}_S t} 
\int^{t}_{0} dT \int^{T}_{0} d\tau \bra{b} \hat{\hat{{\cal L}}}_I (T) \hat{\hat{{\cal L}}}_I (T-\tau) 
\frac{\hat{P}(\hat{H}_B,0)}{n(\hat{H}_B)} \ket{b}+ {\cal O}(\lambda^3) \nonumber .
\end{eqnarray}
We now need to evaluate $\bra{b} \hat{\hat{{\cal L}}}_I (T) (\hat{P}(\hat{H}_B,0)/n(\hat{H}_B)) \ket{b}$ 
and $\bra{b} \hat{\hat{{\cal L}}}_I (T) \hat{\hat{{\cal L}}}_I (T-\tau) (\hat{P}(\hat{H}_B,0)/n(\hat{H}_B)) \ket{b}$.
We have that
\begin{eqnarray}
\bra{b} \hat{\hat{{\cal L}}}_I (T) \frac{\hat{P}(\hat{H}_B,0)}{n(\hat{H}_B)} \ket{b}
&=& - \frac{i}{\hbar} \sum_{\kappa} \bra{b} \lbrack \hat{S}_{\kappa}(T) \hat{B}_{\kappa}(T) 
, \frac{\hat{P}(\hat{H}_B,0)}{n(\hat{H}_B)} \rbrack \ket{b}                                                    \label{1Kaaal}\\
&=& - \frac{i}{\hbar} \sum_{\kappa} \hat{S}_{\kappa}(T) \frac{\hat{P}(E_b,0)}{n(E_b)}
\bra{b} \hat{B}_{\kappa} \ket{b} 
+ \frac{i}{\hbar} \sum_{\kappa} \frac{\hat{P}(E_b,0) }{n(E_b)} \hat{S}_{\kappa}(T) 
\bra{b} \hat{B}_{\kappa} \ket{b}  \nonumber 
\end{eqnarray}
and
\begin{eqnarray}
\bra{b} \hat{\hat{{\cal L}}}_I (T) \hat{\hat{{\cal L}}}_I (T-\tau) 
\frac{\hat{P}(\hat{H}_B,0)}{n(\hat{H}_B)} \ket{b}                                                               \label{1CCaaam}\\
&&\hspace*{-4cm} = - \frac{1}{\hbar^2} \sum_{\kappa,\kappa'} \bra{b} 
\lbrack \hat{S}_{\kappa}(T) \hat{B}_{\kappa}(T) 
, \lbrack \hat{S}_{\kappa'}(T-\tau) \hat{B}_{\kappa'}(T-\tau), 
\frac{\hat{P}(\hat{H}_B,0)}{n(\hat{H}_B)} \rbrack \rbrack \ket{b} \nonumber\\
&&\hspace*{-4cm} = - \frac{1}{\hbar^2} \sum_{\kappa,\kappa'} \sum_{b'} \hat{S}_{\kappa}(T) 
\hat{S}_{\kappa'}(T-\tau) \frac{\hat{P}(E_b,0)}{n(E_b)}
\bra{b} \hat{B}_{\kappa} \ket{b'} \bra{b'} \hat{B}_{\kappa'} \ket{b} e^{\frac{i}{\hbar}(E_b-E_{b'})\tau} \nonumber\\
&&\hspace*{-3.5cm} + \frac{1}{\hbar^2} \sum_{\kappa,\kappa'} \sum_{b'} \hat{S}_{\kappa}(T) 
\frac{\hat{P}(E_{b'},0)}{n(E_{b'})} \hat{S}_{\kappa'}(T-\tau) 
\bra{b} \hat{B}_{\kappa} \ket{b'} \bra{b'} \hat{B}_{\kappa'} \ket{b} e^{\frac{i}{\hbar}(E_b-E_{b'})\tau} \nonumber\\
&&\hspace*{-3.5cm} + \frac{1}{\hbar^2} \sum_{\kappa,\kappa'} \sum_{b'} \hat{S}_{\kappa'}(T-\tau) 
\frac{\hat{P}(E_{b'},0)}{n(E_{b'})} \hat{S}_{\kappa}(T) 
\bra{b} \hat{B}_{\kappa'} \ket{b'} \bra{b'} \hat{B}_{\kappa} \ket{b} e^{-\frac{i}{\hbar}(E_b-E_{b'})\tau} \nonumber\\
&&\hspace*{-3.5cm} - \frac{1}{\hbar^2} \sum_{\kappa,\kappa'} \sum_{b'} \frac{\hat{P}(E_{b},0)}{n(E_{b})} 
\hat{S}_{\kappa'}(T-\tau) \hat{S}_{\kappa}(T) 
\bra{b} \hat{B}_{\kappa'} \ket{b'} \bra{b'} \hat{B}_{\kappa} \ket{b} e^{-\frac{i}{\hbar}(E_b-E_{b'})\tau} \nonumber   .          
\end{eqnarray}
Equation (\ref{2Baabk}) becomes
\begin{eqnarray}
\frac{\hat{P}(E_b,t)}{n(E_{b})}
&=&e^{\hat{\hat{{\cal L}}}_S t} \frac{\hat{P}(E_b,0)}{n(E_{b})}                                              
- \frac{i}{\hbar} \lambda e^{\hat{\hat{{\cal L}}}_S t} \sum_{\kappa}  
\int^{t}_{0} dT \hat{S}_{\kappa}(T) \frac{\hat{P}(E_b,0)}{n(E_{b})} \bra{b} \hat{B}_{\kappa} \ket{b}             \label{2Baaan} \\
&&+ \frac{i}{\hbar} \lambda e^{\hat{\hat{{\cal L}}}_S t} \sum_{\kappa}  
\int^{t}_{0} dT \frac{\hat{P}(E_b,0)}{n(E_{b})} \hat{S}_{\kappa}(T) \bra{b} \hat{B}_{\kappa} \ket{b}  \nonumber\\                                         
&&+ \frac{\lambda^2}{\hbar^2} e^{\hat{\hat{{\cal L}}}_S t} \sum_{\kappa,\kappa'} \sum_{b'}  
\int^{t}_{0} dT \int^{T}_{0} d\tau \; \{ \nonumber\\
&&\hspace*{1cm}- \hat{S}_{\kappa}(T) \hat{S}_{\kappa'}(T-\tau) \frac{\hat{P}(E_b,0)}{n(E_{b})}
\bra{b} \hat{B}_{\kappa} \ket{b'} \bra{b'} \hat{B}_{\kappa'} \ket{b} e^{\frac{i}{\hbar}(E_b-E_{b'})\tau} \nonumber\\
&&\hspace*{1cm}+ \hat{S}_{\kappa}(T) \frac{\hat{P}(E_{b'},0)}{n(E_{b'})} \hat{S}_{\kappa'}(T-\tau) 
\bra{b} \hat{B}_{\kappa} \ket{b'} \bra{b'} \hat{B}_{\kappa'} \ket{b} e^{\frac{i}{\hbar}(E_b-E_{b'})\tau} \nonumber\\
&&\hspace*{1cm}+ \hat{S}_{\kappa'}(T-\tau) \frac{\hat{P}(E_{b'},0)}{n(E_{b'})} \hat{S}_{\kappa}(T) 
\bra{b} \hat{B}_{\kappa'} \ket{b'} \bra{b'} \hat{B}_{\kappa} \ket{b} e^{-\frac{i}{\hbar}(E_b-E_{b'})\tau} \nonumber\\
&&\hspace*{1cm}- \frac{\hat{P}(E_b,0)}{n(E_{b})} \hat{S}_{\kappa'}(T-\tau) \hat{S}_{\kappa}(T) 
\bra{b} \hat{B}_{\kappa'} \ket{b'} \bra{b'} \hat{B}_{\kappa} \ket{b} e^{-\frac{i}{\hbar}(E_b-E_{b'})\tau} \; \} 
+ {\cal O}(\lambda^3) \nonumber .
\end{eqnarray}
Differentiating equation (\ref{2Baaan}) with respect to time, we get
\begin{eqnarray}
\frac{d}{dt} \frac{\hat{P}(E_b,t)}{n(E_{b})}
&=&\hat{\hat{{\cal L}}}_S \frac{\hat{P}(E_b,t)}{n(E_{b})}                                                     
- \frac{i}{\hbar} \lambda e^{\hat{\hat{{\cal L}}}_S t} \sum_{\kappa}  
\hat{S}_{\kappa}(t) \frac{\hat{P}(E_b,0)}{n(E_{b})} \bra{b} \hat{B}_{\kappa} \ket{b}                             \label{2Baaao} \\
&&+ \frac{i}{\hbar} \lambda e^{\hat{\hat{{\cal L}}}_S t} \sum_{\kappa}  
\frac{\hat{P}(E_b,0)}{n(E_{b})} \hat{S}_{\kappa}(t) \bra{b} \hat{B}_{\kappa} \ket{b}  \nonumber\\                                         
&&+ \frac{\lambda^2}{\hbar^2} e^{\hat{\hat{{\cal L}}}_S t} \sum_{\kappa,\kappa'} \sum_{b'}  
\int^{t}_{0} d\tau \; \{ \nonumber\\
&&\hspace*{1cm}- \hat{S}_{\kappa}(t) \hat{S}_{\kappa'}(t-\tau) \frac{\hat{P}(E_b,0)}{n(E_{b})}
\bra{b} \hat{B}_{\kappa} \ket{b'} \bra{b'} \hat{B}_{\kappa'} \ket{b} e^{\frac{i}{\hbar}(E_b-E_{b'})\tau} \nonumber\\
&&\hspace*{1cm}+ \hat{S}_{\kappa}(t) \frac{\hat{P}(E_{b'},0)}{n(E_{b'})} \hat{S}_{\kappa'}(t-\tau) 
\bra{b} \hat{B}_{\kappa} \ket{b'} \bra{b'} \hat{B}_{\kappa'} \ket{b} e^{\frac{i}{\hbar}(E_b-E_{b'})\tau} \nonumber\\
&&\hspace*{1cm}+ \hat{S}_{\kappa'}(t-\tau) \frac{\hat{P}(E_{b'},0)}{n(E_{b'})} \hat{S}_{\kappa}(t) 
\bra{b} \hat{B}_{\kappa'} \ket{b'} \bra{b'} \hat{B}_{\kappa} \ket{b} e^{-\frac{i}{\hbar}(E_b-E_{b'})\tau} \nonumber\\
&&\hspace*{1cm}- \frac{\hat{P}(E_b,0)}{n(E_{b})} \hat{S}_{\kappa'}(t-\tau) \hat{S}_{\kappa}(t) 
\bra{b} \hat{B}_{\kappa'} \ket{b'} \bra{b'} \hat{B}_{\kappa} \ket{b} e^{-\frac{i}{\hbar}(E_b-E_{b'})\tau} \; \} 
+ {\cal O}(\lambda^3) \nonumber .
\end{eqnarray}
In order to close the equation, we use the fact that $\hat{P}(\hat{H}_B,t)= e^{\hat{\hat{{\cal L}}}_S t} 
\hat{P}(\hat{H}_B,0) + {\cal O}(\lambda^2)$. 
This does not affect equation (\ref{2Baaao}) to lower order than ${\cal O}(\lambda^3)$.\\
\normalsize
Finally, our new kinetic equation reads 
\begin{eqnarray}
\frac{d}{dt} \frac{\hat{P}(E_b,t)}{n(E_{b})}
&=&\hat{\hat{{\cal L}}}_S \frac{\hat{P}(E_b,t)}{n(E_{b})}                                                                    
- \frac{i}{\hbar} \lambda \sum_{\kappa}  \lbrack \hat{S}_{\kappa} , \frac{\hat{P}(E_b,t)}{n(E_{b})} \rbrack                        
\bra{b} \hat{B}_{\kappa} \ket{b}                                                                                \label{2Baaap} \\                                 
&&+ \frac{\lambda^2}{\hbar^2} \sum_{\kappa,\kappa'} \sum_{b'}  
\int^{t}_{0} d\tau \; \{ \nonumber\\
&&\hspace*{1cm}- \hat{S}_{\kappa} \hat{S}_{\kappa'}(-\tau) \frac{\hat{P}(E_b,t)}{n(E_{b})}
\bra{b} \hat{B}_{\kappa} \ket{b'} \bra{b'} \hat{B}_{\kappa'} \ket{b} e^{\frac{i}{\hbar}(E_b-E_{b'})\tau} \nonumber\\
&&\hspace*{1cm}+ \hat{S}_{\kappa} \frac{\hat{P}(E_{b'},t)}{n(E_{b'})} \hat{S}_{\kappa'}(-\tau) 
\bra{b} \hat{B}_{\kappa} \ket{b'} \bra{b'} \hat{B}_{\kappa'} \ket{b} e^{\frac{i}{\hbar}(E_b-E_{b'})\tau} \nonumber\\
&&\hspace*{1cm}+ \hat{S}_{\kappa'}(-\tau) \frac{\hat{P}(E_{b'},t)}{n(E_{b'})} \hat{S}_{\kappa} 
\bra{b} \hat{B}_{\kappa'} \ket{b'} \bra{b'} \hat{B}_{\kappa} \ket{b} e^{-\frac{i}{\hbar}(E_b-E_{b'})\tau} \nonumber\\
&&\hspace*{1cm}- \frac{\hat{P}(E_b,t)}{n(E_{b})} \hat{S}_{\kappa'}(-\tau) \hat{S}_{\kappa} 
\bra{b} \hat{B}_{\kappa'} \ket{b'} \bra{b'} \hat{B}_{\kappa} \ket{b} e^{-\frac{i}{\hbar}(E_b-E_{b'})\tau} \; \} \nonumber .
\end{eqnarray}
If the spectrum of the environment is dense enough to be considered quasi-continuous, 
performing the quasi-continuous limit as follows
\begin{eqnarray}
n(E_{b}) &\to& n(\epsilon)                                                                                      \label{2Baaaq}\\ 
\hat{P}(E_b,t) &\to& \hat{P}(\epsilon,t)  \nonumber \\                                               
\sum_{b'} &\to& \int d\epsilon' n(\epsilon') \nonumber \\
\bra{b} \hat{B}_{\kappa} \ket{b'} &\to& B_{\kappa} (\epsilon,\epsilon'), \nonumber
\end{eqnarray}
our kinetic equation for quasi-continuous environments becomes
\begin{eqnarray}
\dot{\hat{P}}(\epsilon,t)
&=&\hat{\hat{{\cal L}}}_S \hat{P}(\epsilon,t)                                                                    
- \frac{i}{\hbar} \lambda \sum_{\kappa}  \lbrack \hat{S}_{\kappa} , \hat{P}(\epsilon,t) \rbrack                        
\bra{\epsilon} \hat{B}_{\kappa} \ket{\epsilon}                                                                   \label{2Baaar} \\                                 
&&+ \frac{\lambda^2}{\hbar^2} \sum_{\kappa,\kappa'} \int d\epsilon' \int^{t}_{0} d\tau \; \{ \nonumber\\
&&\hspace*{1cm}- \hat{S}_{\kappa} \hat{S}_{\kappa'}(-\tau) \hat{P}(\epsilon,t) n(\epsilon')
B_{\kappa} (\epsilon,\epsilon') B_{\kappa'} (\epsilon',\epsilon) 
e^{\frac{i}{\hbar}(\epsilon-\epsilon')\tau} \nonumber\\
&&\hspace*{1cm}+ \hat{S}_{\kappa} \hat{P}(\epsilon',t) \hat{S}_{\kappa'}(-\tau) n(\epsilon)
B_{\kappa} (\epsilon,\epsilon') B_{\kappa'} (\epsilon',\epsilon)
e^{\frac{i}{\hbar}(\epsilon-\epsilon')\tau} \nonumber\\
&&\hspace*{1cm}+ \hat{S}_{\kappa'}(-\tau) \hat{P}(\epsilon',t) \hat{S}_{\kappa} n(\epsilon)
B_{\kappa'} (\epsilon,\epsilon') B_{\kappa} (\epsilon',\epsilon) 
e^{-\frac{i}{\hbar}(\epsilon-\epsilon')\tau} \nonumber\\
&&\hspace*{1cm}- \hat{P}(\epsilon,t) \hat{S}_{\kappa'}(-\tau) \hat{S}_{\kappa} n(\epsilon')
B_{\kappa'} (\epsilon,\epsilon') B_{\kappa} (\epsilon',\epsilon) 
e^{-\frac{i}{\hbar}(\epsilon-\epsilon')\tau} \; \} \nonumber .
\end{eqnarray}
Using the correlation function [defined by equation (\ref{AppAaaap})]
\begin{eqnarray}
\alpha_{\kappa \kappa'}(\epsilon,t) &=& 
\int d\epsilon' n(\epsilon) B_{\kappa} (\epsilon,\epsilon') B_{\kappa'}(\epsilon',\epsilon)                      
e^{\frac{i}{\hbar}(\epsilon-\epsilon')t}                                                                        \label{2Baaas}
\end{eqnarray}
and its Fourier transform [defined by equation (\ref{AppAaaaq})]
\begin{eqnarray}
\tilde{\alpha}_{\kappa \kappa'}(\epsilon,\omega)&=&
\hbar n(\epsilon+\hbar \omega) B_{\kappa}(\epsilon,\epsilon+\hbar \omega) 
B_{\kappa'}(\epsilon+\hbar \omega,\epsilon) ,                                                                    \label{2Baaat}
\end{eqnarray}
we can rewrite equation (\ref{2Baaar}) as
\begin{eqnarray}
\dot{\hat{P}}(\epsilon,t)
&=&\hat{\hat{{\cal L}}}_S \hat{P}(\epsilon,t)                                                                    
- \frac{i}{\hbar} \lambda \sum_{\kappa}  \lbrack \hat{S}_{\kappa} , \hat{P}(\epsilon,t) \rbrack                        
\bra{\epsilon} \hat{B}_{\kappa} \ket{\epsilon}                                                                  \label{2Baaau} \\                                 
&&+ \frac{\lambda^2}{\hbar^3} \sum_{\kappa,\kappa'} 
\int d\epsilon' \int^{t}_{0} d\tau \; \{ \nonumber\\
&&\hspace*{1cm}- \tilde{\alpha}_{\kappa \kappa'}(\epsilon,\frac{\epsilon'-\epsilon}{\hbar})
e^{\frac{i}{\hbar}(\epsilon-\epsilon')\tau} 
\hat{S}_{\kappa} \hat{S}_{\kappa'}(-\tau) \hat{P}(\epsilon,t) \nonumber\\
&&\hspace*{1cm}+ \tilde{\alpha}^{*}_{\kappa \kappa'}(\epsilon',\frac{\epsilon-\epsilon'}{\hbar})
e^{\frac{i}{\hbar}(\epsilon-\epsilon')\tau} 
\hat{S}_{\kappa} \hat{P}(\epsilon',t) \hat{S}_{\kappa'}(-\tau) \nonumber\\
&&\hspace*{1cm}+ \tilde{\alpha}_{\kappa \kappa'}(\epsilon',\frac{\epsilon-\epsilon'}{\hbar})
e^{-\frac{i}{\hbar}(\epsilon-\epsilon')\tau} 
\hat{S}_{\kappa'}(-\tau) \hat{P}(\epsilon',t) \hat{S}_{\kappa} \nonumber\\
&&\hspace*{1cm}- \tilde{\alpha}^{*}_{\kappa \kappa'}(\epsilon,\frac{\epsilon'-\epsilon}{\hbar})
e^{-\frac{i}{\hbar}(\epsilon-\epsilon')\tau} 
\hat{P}(\epsilon,t) \hat{S}_{\kappa'}(-\tau) \hat{S}_{\kappa} \; \} \nonumber .
\end{eqnarray}
With a change of variable in the energy integral, our kinetic equation for a quasi-continuous 
environment becomes
\begin{center} \fbox{\parbox{12.5cm}{
\begin{eqnarray}
\dot{\hat{P}}(\epsilon,t)
&=&\hat{\hat{{\cal L}}}_S \hat{P}(\epsilon,t)                                                                    
- \frac{i}{\hbar} \lambda \sum_{\kappa}  \lbrack \hat{S}_{\kappa} , \hat{P}(\epsilon,t) \rbrack                        
\bra{\epsilon} \hat{B}_{\kappa} \ket{\epsilon}                                                                                                  
+ \frac{\lambda^2}{\hbar^2} \sum_{\kappa,\kappa'} \int^{t}_{0} d\tau \int d\omega \; \{ \nonumber \\ 
&&\hspace*{1cm}- \tilde{\alpha}_{\kappa \kappa'}(\epsilon,\omega) e^{-i \omega \tau} 
\hat{S}_{\kappa} \hat{S}_{\kappa'}(-\tau) \hat{P}(\epsilon,t) \nonumber\\
&&\hspace*{1cm}- \tilde{\alpha}^{*}_{\kappa \kappa'}(\epsilon,\omega) e^{i \omega \tau} 
\hat{P}(\epsilon,t) \hat{S}_{\kappa'}(-\tau) \hat{S}_{\kappa} \nonumber\\
&&\hspace*{1cm}+ \tilde{\alpha}^{*}_{\kappa \kappa'}(\epsilon-\hbar \omega,\omega)
e^{i \omega \tau} \hat{S}_{\kappa} \hat{P}(\epsilon-\hbar \omega,t) \hat{S}_{\kappa'}(-\tau) \nonumber\\
&&\hspace*{1cm}+ \tilde{\alpha}_{\kappa \kappa'}(\epsilon-\hbar \omega,\omega) e^{-i \omega \tau}  
\hat{S}_{\kappa'}(-\tau) \hat{P}(\epsilon-\hbar \omega,t) \hat{S}_{\kappa} \; \} \label{2Baaav}.
\end{eqnarray}
}} \end{center}
This equation preserves trace and Hermiticity.
The Markovian version of this equation is obtained by replacing the time integral 
$\int_{0}^{t} d\tau$ by $\int_{0}^{\infty} d\tau$ in equation (\ref{2Baaav}).\\
In the case of a quasi-continuous environment, using equation (\ref{2Aaadi}), the reduced density matrix of 
the subsystem can be constructed from the solution of equation (\ref{2Baaav}) as follows
\begin{equation}
\hat{\rho}_S(t)=\int d\epsilon
\left(\begin{array}{cccc}
P_{11}(\epsilon,t) & P_{12}(\epsilon,t) & \ldots & P_{1N_S}(\epsilon,t) \\
P_{21}(\epsilon,t) & P_{22}(\epsilon,t) & \ldots & P_{2N_S}(\epsilon,t) \\
\vdots & \vdots & \ddots & \vdots \\
P_{N_S1}(\epsilon,t) & P_{N_S 2}(\epsilon,t) & \ldots &
P_{N_SN_S}(\epsilon,t) \\
\end{array} \right) \; ,                                                                                         \label{2Baaaw}
\end{equation}
where $N_S$ is the number of levels of the subsystem.\\

Notice that our new kinetic equation is a closed evolution equation for the quantity $\hat{P}(\hat{H}_B,t)$. 
However, no closed evolution equation can be obtained for the reduced density matrix of the subsystem. 
Stronger assumptions are needed to derive a closed equation for the reduced density matrix of the subsystem, 
as in the Redfield theory.\\
We see that the dynamic of our new kinetic equation implies different environment energies.
To better understand the energy transfers between the subsystem and the environment, we will project 
our kinetic equation in the eigenbasis of the subsystem in the next section.
  
\section{Markovian approximation} \label{popcoh}

To simplify the notation, we will eliminate the summation over $\kappa$ in our kinetic equation.
This does not affect the discussion and the summation is easy to restore. 
Equation (\ref{2Baaav}) becomes
\begin{eqnarray}
\dot{\hat{P}}(\epsilon,t)
&=&\hat{\hat{{\cal L}}}_S \hat{P}(\epsilon,t)                                                                    
- \frac{i}{\hbar} \lambda  \lbrack \hat{S} , \hat{P}(\epsilon,t) \rbrack                        
\bra{\epsilon} \hat{B} \ket{\epsilon}                                                                                                  
+ \frac{\lambda^2}{\hbar^2} \int^{t}_{0} d\tau \int d\omega \; \{ \; \;                                          \label{2Baaaz} \\ 
&&\hspace*{1cm}- \tilde{\alpha}(\epsilon,\omega) e^{-i \omega \tau}
\hat{S} \hat{S}(-\tau) \hat{P}(\epsilon,t) \nonumber\\
&&\hspace*{1cm}- \tilde{\alpha}(\epsilon,\omega) e^{i \omega \tau} 
\hat{P}(\epsilon,t) \hat{S}(-\tau) \hat{S} \nonumber\\
&&\hspace*{1cm}+ \tilde{\alpha}(\epsilon-\hbar \omega,\omega) e^{i \omega \tau} 
\hat{S} \hat{P}(\epsilon-\hbar \omega,t) \hat{S}(-\tau) \nonumber\\
&&\hspace*{1cm}+ \tilde{\alpha}(\epsilon-\hbar \omega,\omega) e^{-i \omega \tau} 
\hat{S}(-\tau) \hat{P}(\epsilon-\hbar \omega,t) \hat{S} \; \} \nonumber.
\end{eqnarray}
Now, we will project this equation in the eigenbasis of the subsystem and perform the 
Markovian approximation.\\
\scriptsize
Projecting equation (\ref{2Baaaz}) in the eigenbasis of the subsystem, we get
\begin{eqnarray}
\dot{P}_{ss'}(\epsilon,t)
&=&-i \omega_{ss'} P_{ss'}(\epsilon,t)             
- \frac{i}{\hbar} \lambda  \bra{\epsilon} \hat{B} \ket{\epsilon} 
\sum_{\bar{s}} ( S_{s\bar{s}} P_{\bar{s}s'}(\epsilon,t) 
- P_{s\bar{s}}(\epsilon,t) S_{\bar{s}s'})   \nonumber\\                                                                                                                                  
&&+ \frac{\lambda^2}{\hbar^2} \sum_{\bar{s}\bar{s}'}\int^{t}_{0} d\tau \int d\omega \; \{ \; \;                \label{2Bbaaa} \\                              
&&\hspace*{1cm}- \tilde{\alpha}(\epsilon,\omega) e^{i (\omega_{\bar{s}'\bar{s}}-\omega) \tau}
S_{s\bar{s}} S_{\bar{s}\bar{s}'} P_{\bar{s}'s'}(\epsilon,t) \nonumber\\
&&\hspace*{1cm}- \tilde{\alpha}(\epsilon,\omega) e^{-i (\omega_{\bar{s}\bar{s}'}-\omega) \tau} 
P_{s\bar{s}}(\epsilon,t) S_{\bar{s}\bar{s}'} S_{\bar{s}'s'} \nonumber\\
&&\hspace*{1cm}+ \tilde{\alpha}(\epsilon-\hbar \omega,\omega) e^{-i (\omega_{\bar{s}'s'}-\omega) \tau} 
S_{s\bar{s}} P_{\bar{s}\bar{s}'}(\epsilon-\hbar \omega,t) S_{\bar{s}'s'} \nonumber\\
&&\hspace*{1cm}+ \tilde{\alpha}(\epsilon-\hbar \omega,\omega) e^{i (\omega_{\bar{s}s}-\omega) \tau} 
S_{s\bar{s}} P_{\bar{s}\bar{s}'}(\epsilon-\hbar \omega,t) S_{\bar{s}'s'} \; \} \nonumber,
\end{eqnarray}
where $\omega_{ss'}=(E_s-E_{s'})/\hbar$.
Remembering the property 
\begin{eqnarray}
\int_{0}^{\infty} d\tau e^{\pm i \omega \tau} = \pm i{\cal P} \frac{1}{\omega} + \pi \delta(\omega) ,          \label{2Bbaba}
\end{eqnarray}
we can now perform the Markovian approximation in (\ref{2Bbaaa}). \\
\normalsize
The Markovian version of our kinetic equation in the eigenbasis of the subsystem reads
\begin{eqnarray}
\dot{P}_{ss'}(\epsilon,t)
&=&-i \omega_{ss'} P_{ss'}(\epsilon,t)             
- \frac{i}{\hbar} \lambda  \bra{\epsilon} \hat{B} \ket{\epsilon} 
\sum_{\bar{s}} ( S_{s\bar{s}} P_{\bar{s}s'}(\epsilon,t) 
- P_{s\bar{s}}(\epsilon,t) S_{\bar{s}s'}) 
+ \frac{\lambda^2}{\hbar^2} \sum_{\bar{s}\bar{s}'} \; \{ \nonumber \\                              
&&\hspace*{-2.2cm}+ S_{s\bar{s}} S_{\bar{s}\bar{s}'}  
P_{\bar{s}'s'}(\epsilon,t) \left( -\pi \tilde{\alpha}(\epsilon,\omega_{\bar{s}'\bar{s}}) 
-i \int d\omega {\cal P} \frac{\tilde{\alpha}(\epsilon,\omega)}{\omega_{\bar{s}'\bar{s}}-\omega} 
\right) \nonumber\\
&&\hspace*{-2.2cm}+ P_{s\bar{s}}(\epsilon,t) \left( -\pi 
\tilde{\alpha}(\epsilon,\omega_{\bar{s}\bar{s}'}) +i 
\int d\omega {\cal P} \frac{\tilde{\alpha}(\epsilon,\omega)}{\omega_{\bar{s}\bar{s}'}-\omega} \right)
S_{\bar{s}\bar{s}'} S_{\bar{s}'s'} \nonumber\\
&&\hspace*{-2.2cm}+ S_{s\bar{s}} 
\left( \pi P_{\bar{s}\bar{s}'}(\epsilon-\hbar \omega_{\bar{s}'s'},t) 
\tilde{\alpha}(\epsilon-\hbar \omega_{\bar{s}'s'},\omega_{\bar{s}'s'}) 
-i \int d\omega {\cal P} \frac{P_{\bar{s}\bar{s}'}(\epsilon-\hbar \omega,t) 
\tilde{\alpha}(\epsilon-\hbar \omega,\omega)}{\omega_{\bar{s}'s'}-\omega} \right)
S_{\bar{s}'s'} \nonumber\\
&&\hspace*{-2.2cm}+ S_{s\bar{s}}  
\left( \pi P_{\bar{s}\bar{s}'}(\epsilon-\hbar \omega_{\bar{s}s},t) 
\tilde{\alpha}(\epsilon-\hbar \omega_{\bar{s}s},\omega_{\bar{s}s}) 
+i \int d\omega {\cal P} \frac{P_{\bar{s}\bar{s}'}(\epsilon-\hbar \omega,t) 
\tilde{\alpha}(\epsilon-\hbar \omega,\omega)}{\omega_{\bar{s}s}-\omega}
\right) S_{\bar{s}'s'} \; \}. \nonumber \\                                                                \label{2Bbaab}
\end{eqnarray}

\subsection{Population dynamics} \label{popdynneweq}

To study the population dynamics, we now neglect the contribution of the coherences to the population dynamics.
We get 
\begin{eqnarray}            
\dot{P}_{ss}(\epsilon,t)&=&                                                                                                                          
- 2 \pi \frac{\lambda^2}{\hbar^2} \sum_{\bar{s}} \vert S_{s\bar{s}} \vert^2  
\tilde{\alpha}(\epsilon,-\omega_{\bar{s}s}) P_{ss}(\epsilon,t)                                            \label{2Bbaac} \\
&&+ 2 \pi \frac{\lambda^2}{\hbar^2} \sum_{\bar{s}} \vert S_{s\bar{s}} \vert^2  
\tilde{\alpha}(\epsilon-\hbar \omega_{\bar{s}s},\omega_{\bar{s}s}) 
P_{\bar{s}\bar{s}}(\epsilon-\hbar \omega_{\bar{s}s},t) \nonumber       
\end{eqnarray}
\begin{figure}[h]
\centering
\rotatebox{0}{\scalebox{0.6}{\includegraphics{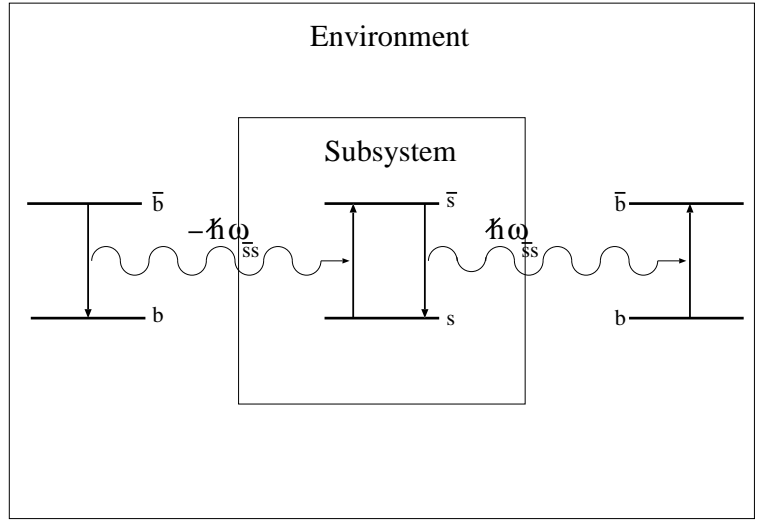}}}
\caption{The two processes of emission (on the right) and of absorption (on the left) of energy from the environment 
to the subsystem are schematically represented. $b$ and $\bar{b}$ are the states of the environment 
which can be either energy in the microcanonical case or temperature in the canonical case.} \label{imnewequation}
\end{figure}
This master equation describes the population dynamics of the subsystem and can be physically interpreted 
in a similar way as the Redfield population equation (\ref{1Gbaag}). \\
The quantity $\tilde{\alpha}(\epsilon,-\omega_{\bar{s}s})$ is proportional to the Fermi golden rule transition rate 
for the emission of an energy $\hbar \omega_{\bar{s}s}$ by the environment at an energy $\epsilon$. 
This later energy is therefore absorbed by the subsystem in the state $s$ which jumps to the state $\bar{s}$.
Because this transition depopulates the state $s$, this term appears as a negative contribution to the 
dynamics of $P_{ss}(\epsilon,t)$.
This term is similar to the negative contribution term of the Redfield population equation (\ref{1Gbaag})
if one applies it to microcanonical environment correlation functions.
The quantity $\tilde{\alpha}(\epsilon-\hbar \omega_{\bar{s}s},\omega_{\bar{s}s})$ is proportional to the  
transition rate for the absorption of an energy $\hbar \omega_{\bar{s}s}$ by the environment at an energy 
$\epsilon-\hbar \omega_{\bar{s}s}$.
This energy is thus emitted by the subsystem in the state $\bar{s}$ and allows it to jump to the state $s$.
This process contributes therefore positively to the dynamics of $P_{ss}(\epsilon,t)$.
Notice that this term is different from the positive contribution term of the Redfield population equation 
(\ref{1Gbaag}) applied to microcanonical environment correlation functions. 
In fact, in our equation, this terms takes into account the fact that the environment has an energy 
$\epsilon-\hbar \omega_{\bar{s}s}$ before absorbing the quantum $\hbar \omega_{\bar{s}s}$.
This is consistent with the principle of conservation of the total system energy which states that the energy 
gained (respectively lost) by the environment has to be lost (respectively gained) by the subsystem.
The Redfield population equation, if applied to microcanonical environments, assumes that the environment is always 
at the same energy $\epsilon$ and is therefore in contradiction with the total system energy conservation. 
However, this violation can sometimes be neglected as we will see in the next section.
The absorption and emission processes are schematically represented in figure \ref{imnewequation}.\\
We finally notice that equation (\ref{2Bbaac}) can satisfy detailed balance. 
If $P_{ss}^{eq}(\epsilon)$ denotes the stationary distribution, the detail balance condition reads
\begin{eqnarray}
\tilde{\alpha}(\epsilon,-\omega_{\bar{s}s}) P_{ss}^{eq}(\epsilon)=
\tilde{\alpha}(\epsilon-\hbar \omega_{\bar{s}s},\omega_{\bar{s}s}) 
P_{\bar{s}\bar{s}}^{eq}(\epsilon-\hbar \omega_{\bar{s}s}).                                              \label{2Bbabc}
\end{eqnarray}
Therefore, if the detailed balance condition is satisfied, we get [using the property (\ref{AppAaaav})] 
\begin{eqnarray}
\frac{P_{ss}^{eq}(\epsilon)}{P_{\bar{s}\bar{s}}^{eq}(\epsilon-\hbar \omega_{\bar{s}s})}= 
\frac{n(\epsilon)}{n(\epsilon-\hbar \omega_{\bar{s}s})} .                                                     \label{2Bbacc}
\end{eqnarray}
We will see in the next section that this relation leads to the subsystem Boltzmann distribution for 
large heat capacity environments.  

\subsection{Coherence dynamics} \label{cohdynneweq}

To study the coherence dynamics, we isolate one coherence and neglect its coupling to the other 
coherences and populations in equation (\ref{2Bbaab}). 
We get an equation of the form
\begin{eqnarray}  
\dot{P}_{ss'}(\epsilon,t) &\approx& (-i \tilde{\omega}_{ss'} - \gamma_{ss'} )  
P_{ss'}(\epsilon,t)                                                                                     \label{2Bbaad}
\end{eqnarray}
with the frequency
\begin{eqnarray} 
\tilde{\omega}_{ss'} = \frac{(\tilde{E}_s-\tilde{E}_{s'})}{\hbar}  ,                                          \label{2Bbaae}
\end{eqnarray}
where
\begin{eqnarray} 
\tilde{E}_s = E_s + \frac{\lambda}{\hbar} \bra{\epsilon} \hat{B} \ket{\epsilon} S_{ss}
+ \frac{\lambda^2}{\hbar^2} \sum_{\bar{s}} \vert S_{s\bar{s}} \vert^2 
\int d\omega {\cal P} \frac{\tilde{\alpha}(\epsilon,\omega)}{\omega_{s\bar{s}}-\omega}                       \label{2Bbaaf}
\end{eqnarray}
and with the damping rate
\begin{eqnarray} 
\gamma_{s,s'} &=&- \pi \frac{\lambda^2}{\hbar^2} (S_{ss}-S_{s's'})^2 
\tilde{\alpha}(\epsilon,0)                                                                                   \label{2Bbaag}\\ 
&&+ \pi \frac{\lambda^2}{\hbar^2} \sum_{\bar{s} \neq s} 
( \vert S_{s\bar{s}} \vert^2 \tilde{\alpha}(\epsilon,\omega_{s\bar{s}})                                               
+ \vert S_{s'\bar{s}} \vert^2 \tilde{\alpha}(\epsilon,\omega_{s'\bar{s}}) ) . \nonumber
\end{eqnarray}
The environment interaction has two kind of effects on the coherence dynamics.
A Lamb shift effect which modifies the free coherences oscillations of the subsystem 
and a decoherence effect which irreversibly damps the coherences.
We can notice that the approximate coherences equation (\ref{2Bbaad}) obtained from our new equation 
and the one (\ref{1Gbaah}) obtained from the Redfield equation are the same.
However, the approximations performed to derive these two equations are crude and the real coherences 
dynamics predicted by both equation are not the same.

\section{The infinite-environment limit} \label{infinitelimit}

In this section, we will investigate the limit of large environments of our new theory and
we will show that our new equation simplifies to the Redfield equation in this limit. 
We will also discuss the greater generality of our new theory
with respect to the Redfield theory.\\

Let us start by integrating our new equation (\ref{2Baaav}) over energy in order to obtain the reduced density 
matrix of the subsystem by using the fact that $\hat{\rho}_S(t)=\int d\epsilon \hat{P}(\epsilon,t)$.
Making the change of variable of the energy $\epsilon \to \epsilon'-\hbar \omega$ for the two last 
order $\lambda^2$ terms, the energy integrated equation becomes
\begin{eqnarray}
\dot{\hat{\rho}}_S (t)
&=&\hat{\hat{{\cal L}}}_S \hat{\rho}_S (t)                                                                    
- \frac{i}{\hbar} \lambda \int d\epsilon \sum_{\kappa}  \lbrack \hat{S}_{\kappa} , \hat{P}(\epsilon,t) \rbrack                        
\bra{\epsilon} \hat{B}_{\kappa} \ket{\epsilon}                                                                                                  
+ \frac{\lambda^2}{\hbar^2} \sum_{\kappa,\kappa'} \int^{t}_{0} d\tau \int d\omega \; \{ \nonumber \\ 
&&\hspace*{0.6cm}- \hat{S}_{\kappa} \hat{S}_{\kappa'}(-\tau) \int d\epsilon \hat{P}(\epsilon,t) 
\tilde{\alpha}_{\kappa \kappa'}(\epsilon,\omega)
e^{-i \omega \tau} \nonumber\\
&&\hspace*{0.6cm}- \int d\epsilon \hat{P}(\epsilon,t) 
\tilde{\alpha}^{*}_{\kappa \kappa'}(\epsilon,\omega)
e^{i \omega \tau} \hat{S}_{\kappa'}(-\tau) \hat{S}_{\kappa} \nonumber\\
&&\hspace*{0.6cm}+ \hat{S}_{\kappa} \int d\epsilon' \hat{P}(\epsilon',t) 
\tilde{\alpha}^{*}_{\kappa \kappa'}(\epsilon',\omega)
e^{i \omega \tau} \hat{S}_{\kappa'}(-\tau) \nonumber\\
&&\hspace*{0.6cm}+ \hat{S}_{\kappa'}(-\tau) \int d\epsilon' \hat{P}(\epsilon',t) 
\tilde{\alpha}_{\kappa \kappa'}(\epsilon',\omega)
e^{-i \omega \tau} \hat{S}_{\kappa} \; \} \label{2Bcaaa}.
\end{eqnarray} 
No assumption has been made till now.
If the environment is large compared to the subsystem, on can assume that the environment quantities 
$\bra{\epsilon} \hat{B}_{\kappa} \ket{\epsilon}$ and $\tilde{\alpha}_{\kappa \kappa'}(\epsilon,\omega)$
do not vary significantly on energy scales of the order of the subsystem energy:
\begin{eqnarray}
\bra{\epsilon \pm \Delta} \hat{B}_{\kappa} \ket{\epsilon \pm \Delta} &\approx& 
\bra{\epsilon} \hat{B}_{\kappa} \ket{\epsilon}                                                            \label{2Bcaba}\\
\tilde{\alpha}_{\kappa \kappa'}(\epsilon \pm \Delta,\omega) &\approx&
\tilde{\alpha}_{\kappa \kappa'}(\epsilon,\omega), \nonumber
\end{eqnarray} 
where $\Delta$ represents the typical energy scale implied in the subsystem dynamics.
Therefore, if the environment is initially in a microcanonical state at energy $\epsilon$, 
the assumption (\ref{2Bcaba}) means that it will stay in this state without being affected by the 
subsystem emitted or absorbed quanta. 
Equation (\ref{2Bcaaa}) can therefore be written as
\begin{eqnarray}
\dot{\hat{\rho}}_S (t)
&=&\hat{\hat{{\cal L}}}_S \hat{\rho}_S (t)                                                                    
- \frac{i}{\hbar} \lambda \bra{\epsilon} \hat{B}_{\kappa} \ket{\epsilon}  
\sum_{\kappa}  \lbrack \hat{S}_{\kappa} , \hat{\rho}_S (t) \rbrack                                                                                                                         
+ \frac{\lambda^2}{\hbar^2} \sum_{\kappa,\kappa'} \int^{t}_{0} d\tau \int d\omega \; \{ \nonumber \\ 
&&\hspace*{0.6cm}- \tilde{\alpha}_{\kappa \kappa'}(\epsilon,\omega) e^{-i \omega \tau}
\hat{S}_{\kappa} \hat{S}_{\kappa'}(-\tau) \hat{\rho}_S (t) \nonumber\\
&&\hspace*{0.6cm}- \tilde{\alpha}^{*}_{\kappa \kappa'}(\epsilon,\omega) e^{i \omega \tau} 
\hat{\rho}_S (t) \hat{S}_{\kappa'}(-\tau) \hat{S}_{\kappa} \nonumber\\
&&\hspace*{0.6cm}+ \tilde{\alpha}^{*}_{\kappa \kappa'}(\epsilon,\omega) e^{i \omega \tau} 
\hat{S}_{\kappa} \hat{\rho}_S (t) \hat{S}_{\kappa'}(-\tau) \nonumber\\
&&\hspace*{0.6cm}+ \tilde{\alpha}_{\kappa \kappa'}(\epsilon,\omega) e^{-i \omega \tau}
\hat{S}_{\kappa'}(-\tau) \hat{\rho}_S (t) \hat{S}_{\kappa} \; \}                                            \label{2Bcaab}.
\end{eqnarray}   
We see that this equation is the Redfield equation for a microcanonical environment.\\

The greater generality of our new equation compared to the Redfield equation can now be physically understood.
\begin{figure}[p]
\centering
\vspace*{0cm}
\hspace*{-1cm}
\rotatebox{0}{\scalebox{0.7}{\includegraphics{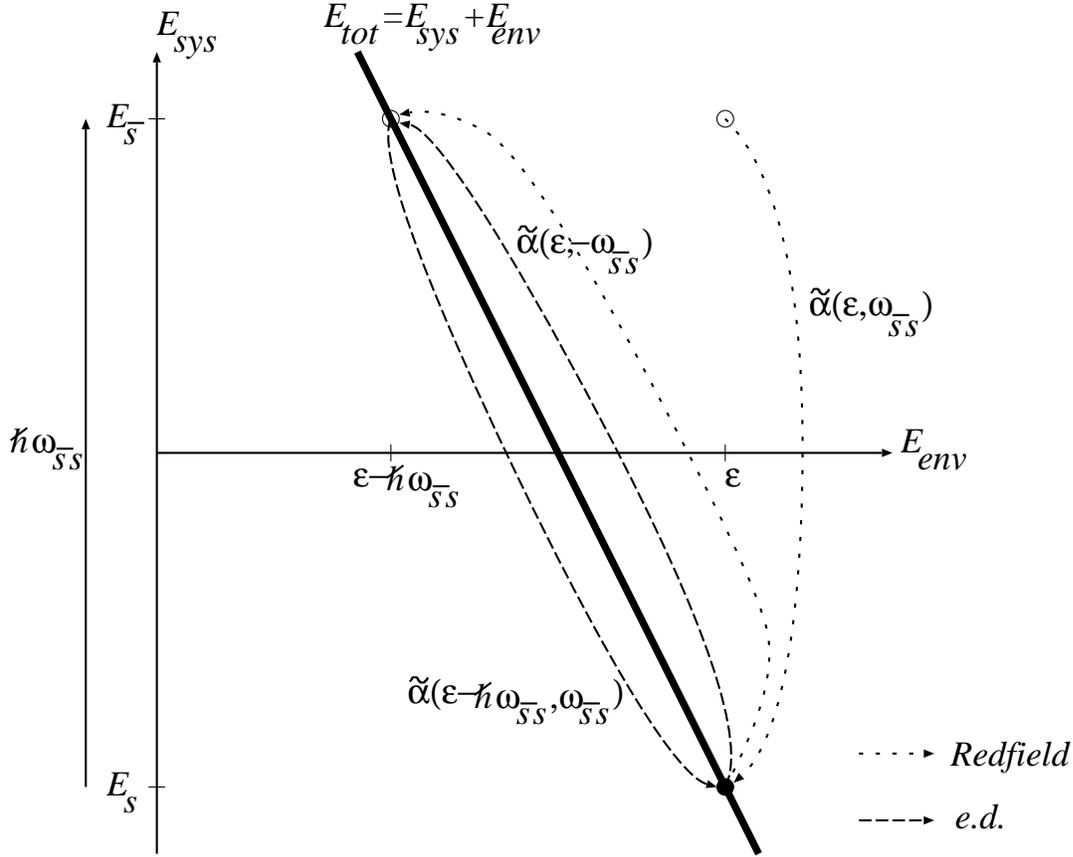}}}
\caption{Representation of the energy exchanges between the environment and the subsystem described 
respectively by the Redfield population equation (\ref{1Gbaag}) and by our new energy distributed (e.d.)
population equation (\ref{2Bbaac}) in the case of a two-level system. 
The subsystem energy $E_{sys}$ increases along the $y$-axis and the environment energy $E_{env}$ along the $x$-axis.  
The total energy of the system is represented by $E_{tot}$. 
The initial condition corresponding to the subsystem at energy $E_s$ and the environment at energy $\epsilon$ is 
denoted by the full black dot in the $(x,y)$-plane. 
The transitions that preserve the energy of the total system have to occur along the
total energy line $E=E_s+E_b$ (the thick full line) in the $(x,y)$-plane.
One can see that only the transition described by our energy distributed (e.d.) equation (dashed line) satisfies 
this condition. 
The Redfield equation describes transitions (dotted line) that occur along a constant environment 
energy line and is therefore erroneous when the subsystem energies are of the order or larger than the typical 
energy scale on which the Fourier transform of the correlation function of the environment varies.}
\label{energycons}
\end{figure}
In our new equation, the state of the environment is correlated to the state of the subsystem. 
The quanta which are emitted from the subsystem to the environment an vice versa have an effect on the subsystem 
as well as on the environment. 
In the Redfield equation, the Fourier transform of the environment microcanonical correlation functions which appears 
in the equation is always evaluated at the same energy.
This means that the emitted or absorbed quanta have no effect on the environment in the Redfield equation.
The only way to justify this assumption is to suppose that the environment is large so that its energy-dependent 
quantities vary on energy scales which are much larger than the typical energies emitted or absorbed by the subsystem.
This is precisely the meaning of the approximation (\ref{2Bcaba}) that we made to reduce our new equation to the Redfield equation.  
This point is particularly explicit if one compares the population dynamics obtained from our new equation 
(\ref{2Bbaac}) to the one obtained from the Redfield equation (\ref{1Gbaag}).
If one applies both equations to a subsystem interacting with a small environment, our new equation takes into account the 
energy conservation of the total system (subsystem plus environment) although the Redfield equation does not 
(see figure \ref{energycons}).
This shows that our new equation is more general than the Redfield equation and 
more pertinent to describe small total systems such as nanosystems.\\

Another way to see the problem of energy conservation if the Redfield equation is applied to a finite environment, 
is to consider the expectation value of the total energy.
When the total system evolves, the energy of the subsystem, the energy of the environment and the interaction energy
vary in time while keeping the total energy constant
\begin{eqnarray}
\mean{\hat{H}} = \mean{\hat{H}_{S}}_t + \mean{\hat{H}_{B}}_t + \lambda \mean{\hat{S} \hat{B}}_t                  \label{2Bcabc}
\end{eqnarray}
If the coupling parameter is small, the contribution of the interaction energy to the total energy can be neglected
\begin{eqnarray}
\mean{\hat{H}} \stackrel{\lambda \to 0}{=} \mean{\hat{H}_{S}}_t + \mean{\hat{H}_{B}}_t  .                        \label{2Bcacc}
\end{eqnarray}
In the Redfield theory, the environment does not evolve and the total density matrix is assumed to have the form 
\begin{eqnarray} 
\hat{\rho}_{T}(t) = \hat{\rho}_{S}(t) \hat{\rho}_{B}^{eq}   .                                                      \label{2Bcadc}
\end{eqnarray}
Therefore the subsystem energy evolves, but the environment energy does not. 
\begin{eqnarray}
\mean{\hat{H}} \stackrel{\lambda \to 0}{=} \sum_{s} E_s \bra{s} \hat{\rho}_{S}(t) \ket{s} 
+ \int d\epsilon \; n(\epsilon) \; \epsilon \; \bra{\epsilon} \hat{\rho}_{B}^{eq} \ket{\epsilon}                   \label{2Bcaec}
\end{eqnarray}
In order to avoid a violation of the total energy conservation, one has to suppose
\begin{eqnarray}
\mean{\hat{H}} \approx \mean{\hat{H}_{B}} \gg \mean{\hat{H}_{S}}_t , \; \lambda \mean{\hat{S} \hat{B}}_t.       \label{2Bcafc}
\end{eqnarray}
In our new theory, the total density matrix is assumed to have the form 
\begin{eqnarray} 
\hat{\rho}_{T}(t) = \frac{\hat{P}(\hat{H}_B,t)}{n(\hat{H}_B)}.                                                     \label{2Bcagc}
\end{eqnarray}
Therefore the total system energy is kept constant but energy is transferred between 
the subsystem and the environment 
\begin{eqnarray}
\mean{\hat{H}} \stackrel{\lambda \to 0}{=} \sum_{s} E_s \int d\epsilon P_{ss}(\epsilon,t) 
+ \int d\epsilon \; \epsilon \sum_{s} P_{ss}(\epsilon,t) .                                                         \label{2Bcahc}
\end{eqnarray}
This theory is valid if 
\begin{eqnarray}
\mean{\hat{H}} \gg \lambda \mean{\hat{S} \hat{B}}_t,                                                            \label{2Bcaic}
\end{eqnarray}
which is much less restrictive than the condition (\ref{2Bcafc}).
This again shows the larger range of validity of our theory compared to the Redfield theory.\\

One can also show that our new theory allows a consistent understanding of the emergence of a
canonical Boltzmann equilibrium distribution of the subsystem when the environment in a microcanonical state
becomes large. 
The temperature of the Boltzmann distribution is then given by the microcanonical temperature of the environment.
In fact, in our new theory, the equilibrium distribution (which is obtained by using the detailed balance condition) 
is given by equation (\ref{2Bbacc}) and has a simple form because of the property found in equation (\ref{AppAaaav}). 
Saying that the environment is large enough so that its energy-dependent quantities do not vary on 
energy scales of the order of the typical subsystem energies is equivalent to saying that the environment heat 
capacity has to be large.
Using the results of appendix \ref{AppB} for systems with a large heat capacity, equation (\ref{AppBaaag}) 
can be written as 
\begin{eqnarray}
n(\epsilon-\hbar \omega_{s\bar{s}}) \approx 
n(\epsilon) e^{- \beta_{mic}(\epsilon) \hbar \omega_{\bar{s}s}}.                                                \label{2Bcaag}
\end{eqnarray}
where $\beta_{mic}(\epsilon)=1/(k_B T_{mic}(\epsilon))$.
Therefore the detailed-balance condition given by equation (\ref{2Bbacc}) becomes
\begin{eqnarray}
\frac{P_{ss}^{eq}(\epsilon)}{P_{\bar{s}\bar{s}}^{eq}(\epsilon-\hbar \omega_{\bar{s}s})} \approx  
e^{ \beta_{mic}(\epsilon) \hbar \omega_{\bar{s}s}} .                                                            \label{2Bcaah}
\end{eqnarray}
Integrating equation (\ref{2Bcaah}) over energy and using the property (\ref{AppBaaah}), we get
\begin{eqnarray}
\bra{s} \hat{\rho}_{S}^{eq} \ket{s} &=&  
\int d\epsilon \; e^{ \beta_{mic}(\epsilon) \hbar \omega_{\bar{s}s}} 
P_{\bar{s}\bar{s}}^{eq}(\epsilon-\hbar \omega_{\bar{s}s})                                                \label{2Bcaai}\\
&=&\int d\epsilon' \; e^{ \beta_{mic}(\epsilon'+\hbar \omega_{\bar{s}s}) \hbar \omega_{\bar{s}s}} 
P_{\bar{s}\bar{s}}^{eq}(\epsilon') \nonumber\\
&\approx& e^{ \beta_{mic}(\epsilon) \hbar \omega_{\bar{s}s}} \bra{\bar{s}} \hat{\rho}_{S}^{eq} \ket{\bar{s}}
\end{eqnarray}
Therefore, we find that the subsystem equilibrium distribution is given by the Boltzmann distribution 
at the microcanonical temperature of the environment
\begin{eqnarray}
\frac{\bra{s} \hat{\rho}_{S}^{eq} \ket{s}}{\bra{\bar{s}} \hat{\rho}_{S}^{eq} \ket{\bar{s}}} 
&\approx& e^{ \beta_{mic}(\epsilon) \hbar \omega_{\bar{s}s}}            .                                      \label{2Bcaaj}
\end{eqnarray}
Here, we have shown how, in our new kinetic theory, an equilibrium canonical Boltzmann distribution can arise on 
a subsystem in contact with an environment in a microcanonical state if the environment has a large heat capacity. 
This treatment cannot be made in the Redfield theory.
In fact, in the Redfield theory, the microcanonical condition (\ref{AppAaaav}) cannot be used to simplify 
the subsystem equilibrium distribution (\ref{1Gbacg}).
This indicates that, contrary to our new equation, the Redfield equation applied to a microcanonical environment 
contains non-physical features which make the microcanonical environment description inconsistent with the 
canonical equilibrium distribution of the subsystem.
If the total system (subsystem plus environment) is an isolated finite system, our new theory becomes essential. 
When the environment becomes very large, our theory reduces to the Redfield theory.
Numerical evidence for this will be obtained in chapter \ref{ch4} devoted to the spin-GORM model. \\  

\begin{figure}[p]
\centering
\vspace*{-0.2cm}
\hspace*{-1.2cm}
\rotatebox{0}{\scalebox{0.7}{\includegraphics{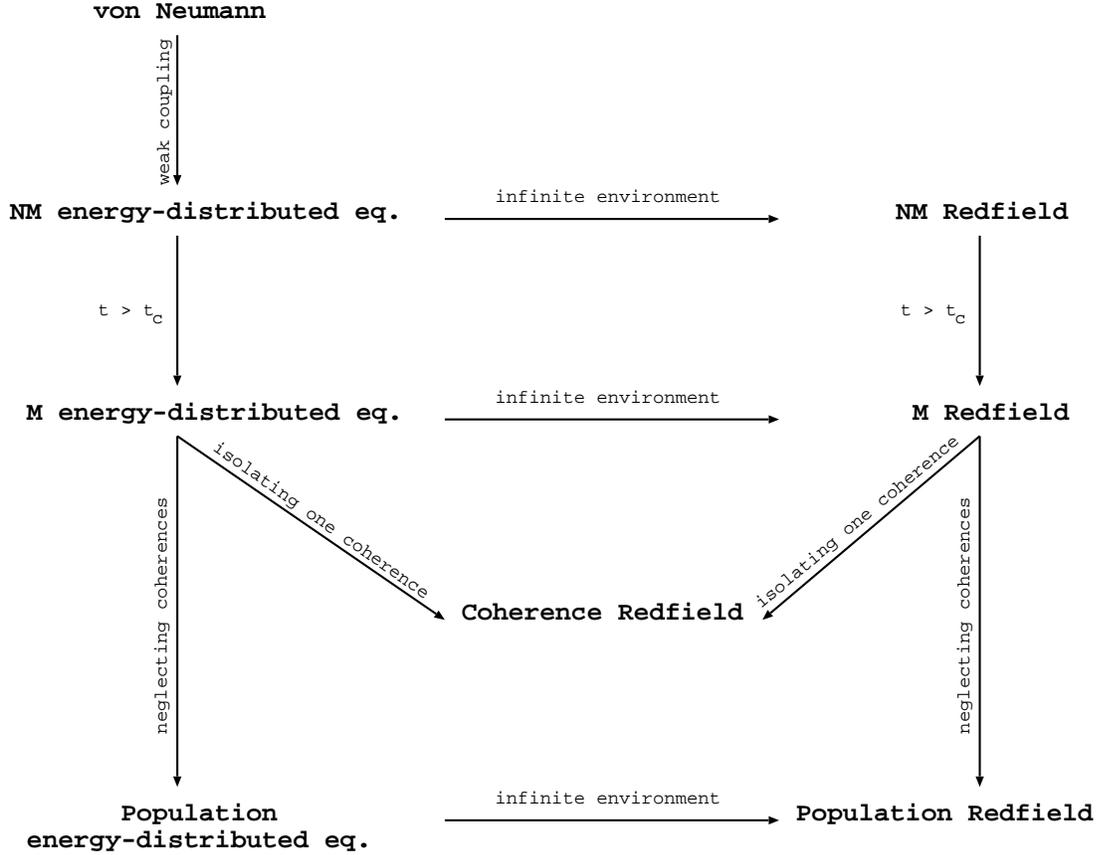}}}
\vspace*{-0.5cm}
\caption{Tree of possible approximations starting from the von Neumann equation to derive a quantum 
kinetic equation for the dynamics of a subsystem weakly interacting with an environment. 
The most general of these kinetic equations is the non-Markovian version of our new energy distributed equation 
(\ref{2Baaav}) ("NM energy distributed eq.").
By performing the Markovian approximation on this equation, one gets the Markovian version of it 
("M energy distributed eq.") which is valid on time scales longer than $t_c$ (the typical environment 
correlation function time scale).
If instead, the infinite environment limit is performed on the non-Markovian version of our new equation, 
we get the non-Markovian Redfield equation (\ref{1Gaaap}) ("NM Redfield").
If one further performs the Markovian approximation on this last equation, one gets the Markovian Redfield 
equation ("M Redfield"), which is equivalently obtained by performing the infinite-environment limit on 
the Markovian version of our new equation.
The energy distributed population equation (\ref{2Bbaac}) ("Population energy distributed eq.") can be derived 
from the Markovian version of our new equation by neglecting the coherences contribution to the population dynamics.
The Redfield population equation (\ref{1Gbaag}) ("Population Redfield") is obtained either by the same procedure
from the Markovian Redfield equation or by performing the infinite environment limit on the energy 
distributed population equation. 
Finally, an approximated coherence equation [(\ref{2Bbaag}) or (\ref{1Gbaah})] ("Coherence Redfield") is obtained 
from the Markovian version of our new equation, as well as from the Markovian Redfield equation, 
by neglecting the other coherences and the population contribution to the selected coherence dynamics.}
\label{allequation}
\end{figure}

A summary of the different equations that we discussed is compiled in figure \ref{allequation}.

\section{Application to a two-level subsystem} \label{twolevelgen}

We now apply our new theory to describe the dynamics of a two-level subsystem interacting in a non-diagonal way 
with a general environment.
Thanks to a simplification, the equations of our theory take in this case a sufficiently simple form to allow 
analytical solutions and an easy identification of the physical processes in play.
This simplification consists in neglecting the contribution of the diagonal elements of the environment coupling
operator in the basis diagonalizing the Hamiltonian of the environment.
It should be noticed that this simplification can be the feature of many systems.
For instance, it is exactly satisfied for any random matrix model in which the environment coupling operator 
$\hat{B}$ has a vanishing ensemble average.
In this case, the average of the matrix elements $\bra{\epsilon} \hat{B} \ket{\epsilon}$ over the ensemble of 
random matrices $\hat{B}$ also vanishes for all eigenvectors $\ket{\epsilon}$ of $\hat{H}_{B}$.\\

The Hamiltonian of a two-level subsystem interacting in a non-diagonal way with an environment is given by 
\begin{eqnarray}
\hat{H} = \frac{\Delta}{2} \hat{\sigma}_{z} + \hat{H}_{B} + \lambda \hat{\sigma}_x \hat{B} ,          \label{4Aaaaa} 
\end{eqnarray}
where \\ 
$\bullet$ $\hat{\sigma}_{x}$, $\hat{\sigma}_{y}$, and $\hat{\sigma}_{z}$ are the 
$2 \times 2$ Pauli matrices,\\ 
$\bullet$ $\frac{\Delta}{2} \hat{\sigma}_{z}$ is the Hamiltonian of the two-level subsystem,\\ 
$\bullet$ $\Delta$ is the energy spacing between the two levels of the subsystem,\\ 
$\bullet$ $\hat{H}_{B}$ is the Hamiltonian of the environment,\\ 
$\bullet$ $\hat{\sigma}_x$ is the coupling operator of the subsystem,\\ 
$\bullet$ $\hat{B}$ is the coupling operator of the environment,\\ 
$\bullet$ $\lambda$ is the coupling parameter between the subsystem and the environment. \\
The subsystem eigenvalue equation reads
\begin{eqnarray}
\hat H_S \vert s \rangle = \frac{\Delta}{2} \hat{\sigma}_{z} \vert s
\rangle = s \frac{\Delta}{2} \vert s \rangle ,                                                             \label{4Aaaab}
\end{eqnarray}
where $s=\pm 1$. \\

The two-level subsystem density matrix is, in our new theory, distributed in energy
as follows
\begin{equation}
\hat{\rho}_S(t)=\int d\epsilon
\left(\begin{array}{cc}
P_{++}(\epsilon,t) & P_{+-}(\epsilon,t) \\
P_{-+}(\epsilon,t) & P_{--}(\epsilon,t) \\
\end{array} \right).                                                                                       \label{4Aaaac}
\end{equation}
The energy-distributed elements of the subsystem density matrix evolve according to our 
new non-Markovian equation (\ref{2Baaav}). 
In the subsystem eigenbasis, equation (\ref{2Bbaaa}) is given in the present case by
\begin{eqnarray}
\dot{P}_{\pm \pm}(\epsilon,t)
&=&- \frac{i}{\hbar} \lambda \bra{\epsilon} \hat{B} \ket{\epsilon} 
(P_{\mp \pm}(\epsilon,t) - P_{\pm \mp}(\epsilon,t))                                                       \label{4Aaaad}\\                      
&&+ \frac{\lambda^2}{\hbar^2} \int^{t}_{0} d\tau \int d\omega \; \{ \nonumber \\                                  
&&\hspace*{0.5cm}- P_{\pm \pm}(\epsilon,t) 
\tilde{\alpha}(\epsilon,-\omega) (e^{\frac{i}{\hbar} (\pm \Delta+\hbar \omega) \tau} 
+e^{-\frac{i}{\hbar} (\pm \Delta+\hbar \omega) \tau}) \nonumber\\
&&\hspace*{0.5cm}+ P_{\mp \mp}(\epsilon-\hbar \omega,t) 
\tilde{\alpha}(\epsilon-\hbar \omega,\omega)
(e^{\frac{i}{\hbar} (\pm \Delta+\hbar \omega) \tau}
+e^{-\frac{i}{\hbar} (\pm \Delta+\hbar \omega) \tau}) \; \} \nonumber
\end{eqnarray}
for the distributed population and by
\begin{eqnarray}
\dot{P}_{\pm \mp}(\epsilon,t)
&=&\mp \frac{i}{\hbar} \Delta P_{\pm \mp}(\epsilon,t)
- \frac{i}{\hbar} \lambda \bra{\epsilon} \hat{B} \ket{\epsilon} 
(P_{\mp \mp}(\epsilon,t) - P_{\pm \pm}(\epsilon,t))                                                        \label{4Aaaae}\\              
&&+ \frac{\lambda^2}{\hbar^2} \int^{t}_{0} d\tau \int d\omega \; \{ \nonumber \\
&&\hspace*{0.5cm}- P_{\pm \mp}(\epsilon,t) 
\tilde{\alpha}(\epsilon,\omega) (e^{\frac{i}{\hbar} (\pm \Delta-\hbar \omega) \tau} 
+e^{\frac{i}{\hbar} (\pm \Delta+\hbar \omega) \tau})   \nonumber \\                                                   
&&\hspace*{0.5cm}+ P_{\mp \pm}(\epsilon-\hbar \omega,t) 
\tilde{\alpha}(\epsilon-\hbar \omega,\omega)
(e^{-\frac{i}{\hbar} (\pm \Delta-\hbar \omega) \tau}
+e^{-\frac{i}{\hbar} (\pm \Delta+\hbar \omega) \tau}) \; \} \nonumber
\end{eqnarray}
for the coherences.
Performing the Markovian approximation on equation (\ref{4Aaaad}) and (\ref{4Aaaae}) or, equivalently, 
using the Markovian equation (\ref{2Bbaab}), we get
\begin{eqnarray}
\dot{P}_{\pm \pm}(\epsilon,t)
&=&- \frac{i}{\hbar} \lambda \bra{\epsilon} \hat{B} \ket{\epsilon} 
(P_{\mp \pm}(\epsilon,t) - P_{\pm \mp}(\epsilon,t))                                                       \label{4Aaaaf} \\ 
&&- 2 \pi \frac{\lambda^2}{\hbar^2}
\tilde{\alpha}(\epsilon,\pm \Delta / \hbar) P_{\pm \pm}(\epsilon,t) \nonumber\\
&&+ 2 \pi \frac{\lambda^2}{\hbar^2}  
\tilde{\alpha}(\epsilon \pm \Delta,\mp \Delta / \hbar) P_{\mp \mp}(\epsilon \pm \Delta,t) \nonumber
\end{eqnarray}
for the population and 
\begin{eqnarray}
\dot{P}_{\pm \mp}(\epsilon,t)
&=&\mp \frac{i}{\hbar} \Delta P_{\pm \mp}(\epsilon,t)
- \frac{i}{\hbar} \lambda \bra{\epsilon} \hat{B} \ket{\epsilon} 
(P_{\mp \mp}(\epsilon,t) - P_{\pm \pm}(\epsilon,t))                                                       \label{4Aaaag}\\ 
&&\mp i 2 \Delta \frac{\lambda^2}{\hbar} P_{\pm \mp}(\epsilon,t) \int d\omega {\cal P}
\frac{\tilde{\alpha}(\epsilon,\omega)}{\Delta^2 - (\hbar \omega)^2}  \nonumber \\
&&\mp i 2 \Delta \frac{\lambda^2}{\hbar} \int d\omega {\cal P} \frac{ 
\tilde{\alpha}(\epsilon-\hbar \omega,\omega)}{\Delta^2 -(\hbar \omega)^2} 
P_{\mp \pm}(\epsilon-\hbar \omega,t) \nonumber \\
&&- \pi \frac{\lambda^2}{\hbar^2} P_{\pm \mp}(\epsilon,t) 
(\tilde{\alpha}(\epsilon,\pm \Delta)+\tilde{\alpha}(\epsilon,\mp \Delta)) \nonumber \\
&&+ \pi \frac{\lambda^2}{\hbar^2} ( P_{\mp \pm}(\epsilon \mp \Delta,t) 
\tilde{\alpha}(\epsilon \mp \Delta, \pm  \Delta / \hbar) \nonumber \\
&&\hspace*{1.0cm}+ P_{\mp \pm}(\epsilon \pm \Delta,t) 
\tilde{\alpha}(\epsilon \pm \Delta, \mp  \Delta / \hbar)) \nonumber
\end{eqnarray}
for the coherences.
We see that the population and the coherence dynamics are coupled together by the first-order term in 
$\lambda$ in equation (\ref{4Aaaaf}) and (\ref{4Aaaag}).
As mentioned before, to simplify the analysis, we will decouple them by assuming that the coupling 
operator of the environment is diagonal in the eigenbasis representation of the environment Hamiltonian :
$\bra{\epsilon} \hat{B} \ket{\epsilon}=0$ for all $\epsilon$.
Doing this, all the first-order terms in $\lambda$ in equation (\ref{4Aaaaf}) and (\ref{4Aaaag}) disappear.
Let us now define
\begin{eqnarray}
z(t)&\equiv&\textrm{Tr} \hat{\rho}(t) \hat{\sigma}_{z}= \int d\epsilon' Z(\epsilon',t)                   \label{4Aaaah}\\
x(t)&\equiv&\textrm{Tr} \hat{\rho}(t) \hat{\sigma}_{x}= \int d\epsilon' X(\epsilon',t) \nonumber \\
y(t)&\equiv&\textrm{Tr} \hat{\rho}(t) \hat{\sigma}_{y}= \int d\epsilon' Y(\epsilon',t) \nonumber ,
\end{eqnarray}
where
\begin{eqnarray}
C(\epsilon,t) &\equiv& P_{++}(\epsilon,t)+P_{--}(\epsilon+\Delta,t),                                      \label{4Aaaai}\\
Z(\epsilon,t) &\equiv& P_{++}(\epsilon,t)-P_{--}(\epsilon+\Delta,t), \nonumber \\
X(\epsilon,t) &\equiv& P_{+-}(\epsilon,t)+P_{-+}(\epsilon,t), \nonumber \\
Y(\epsilon,t) &\equiv& iP_{+-}(\epsilon,t)-iP_{-+}(\epsilon,t).\nonumber
\end{eqnarray}
Using equation (\ref{4Aaaaf}) and (\ref{4Aaaag}), we find that the population dynamics can be 
studied with the variable $Z(\epsilon,t)$ 
\begin{eqnarray}
\dot{Z}(\epsilon,t)
&=&- 2 \pi \frac{\lambda^2}{\hbar^2} 
(\tilde{\alpha}(\epsilon, \Delta / \hbar) - \tilde{\alpha}(\epsilon + \Delta,- \Delta / \hbar)) 
C(\epsilon,t)                                                                                             \label{4Aaaaj}\\
&&- 2 \pi \frac{\lambda^2}{\hbar^2}  
(\tilde{\alpha}(\epsilon, \Delta / \hbar) + \tilde{\alpha}(\epsilon + \Delta,- \Delta / \hbar))
Z(\epsilon,t) \nonumber,
\end{eqnarray}
and the coherence dynamics with $X(\epsilon,t)$ 
\begin{eqnarray}
\dot{X}(\epsilon,t)
&=&- \frac{\Delta}{\hbar} \left( 1 + 2 \lambda^2 \int d\omega {\cal P}
\frac{\tilde{\alpha}(\epsilon,\omega)}{\Delta^2 - (\hbar \omega)^2} \right) Y(\epsilon,t)           \label{4Aaaak}\\
&&+ 2 \frac{\Delta \lambda^2}{\hbar} 
\int d\omega {\cal P} \frac{ \tilde{\alpha}(\epsilon-\hbar \omega,\omega) 
}{\Delta^2 -(\hbar \omega)^2} Y(\epsilon-\hbar \omega,t) \nonumber \\
&&- \pi \frac{\lambda^2}{\hbar^2} 
(\tilde{\alpha}(\epsilon,\Delta / \hbar)+\tilde{\alpha}(\epsilon,-\Delta / \hbar)) X(\epsilon,t) \nonumber \\
&&+ \pi \frac{\lambda^2}{\hbar^2} 
\tilde{\alpha}(\epsilon-\Delta,\Delta / \hbar) X(\epsilon - \Delta,t)  \nonumber \\
&&+ \pi \frac{\lambda^2}{\hbar^2}  
\tilde{\alpha}(\epsilon + \Delta, -\Delta / \hbar) X(\epsilon + \Delta,t) \nonumber
\end{eqnarray}
and $Y(\epsilon,t)$
\begin{eqnarray}
\dot{Y}(\epsilon,t)
&=&\frac{\Delta}{\hbar} \left( 1 + 2 \lambda^2 \int d\omega {\cal P}
\frac{\tilde{\alpha}(\epsilon,\omega)}{\Delta^2 - (\hbar \omega)^2} \right) X(\epsilon,t)            \label{4Aaaal}\\
&&+ 2 \frac{\Delta \lambda^2}{\hbar} 
\int d\omega {\cal P} \frac{ \tilde{\alpha}(\epsilon-\hbar \omega,\omega)}
{\Delta^2 -(\hbar \omega)^2} X(\epsilon-\hbar \omega,t) \nonumber \\
&&- \pi \frac{\lambda^2}{\hbar^2}  
(\tilde{\alpha}(\epsilon,\Delta / \hbar)+\tilde{\alpha}(\epsilon,-\Delta / \hbar)) Y(\epsilon,t) \nonumber \\
&&- \pi \frac{\lambda^2}{\hbar^2}  
\tilde{\alpha}(\epsilon-\Delta,\Delta / \hbar) Y(\epsilon - \Delta,t) \nonumber \\
&&- \pi \frac{\lambda^2}{\hbar^2}  
\tilde{\alpha}(\epsilon + \Delta, -\Delta / \hbar) Y(\epsilon + \Delta,t) \nonumber .
\end{eqnarray}
We also find that $\dot{C}(\epsilon,t)=0$ is a constant of motion.
If one chooses the initial condition of the total system to be the product of a general subsystem 
density matrix with a microcanonical distribution at energy $\epsilon$ for the reservoir, we have that
\begin{eqnarray} 
\hat{\rho}(0)&=&\hat{\rho}_{S}(0) \frac{\delta(\epsilon-\hat{H}_B)}{n(\hat{H}_B)}                          \label{4Aaaam}\\
&=&\frac{\hat{P}(\hat{H}_B,0)}{n(\hat{H}_B)}  \nonumber                                                               
\end{eqnarray}
and, therefore, the reduced density matrix of the subsystem reads
\begin{eqnarray} 
\hat{\rho}_{S}(0)&=& \trace_B \hat{\rho}(0)                                                                \label{4Aaaan}\\
&=& \int d\epsilon' \hat{\rho}_{S}(0) \delta(\epsilon-\epsilon') \nonumber \\
&=& \int d\epsilon' \hat{P}(\epsilon',0) \delta(\epsilon-\epsilon'). \nonumber                     
\end{eqnarray}
We conclude that $\hat{\rho}_{S}(0)=\hat{P}(\epsilon,0)$. 
This means that
\begin{eqnarray}
Z(\epsilon',0)&=&C(\epsilon',0)=\bra{+} \hat{\rho}_{S}(0) \ket{+} \delta(\epsilon-\epsilon')               \label{4Aaaao}\\
&=&\frac{1+z(0)}{2} \delta(\epsilon-\epsilon')   \nonumber \\                                                            
Z(\epsilon'-\Delta,0)&=&-C(\epsilon'-\Delta,0)
=-\bra{-} \hat{\rho}_{S}(0) \ket{-} \delta(\epsilon-\epsilon')\nonumber \\ 
&=&-\frac{1-z(0)}{2} \delta(\epsilon-\epsilon') \nonumber \\
X(\epsilon',0) &=& (\bra{+} \hat{\rho}_{S}(0) \ket{-}+\bra{-} \hat{\rho}_{S}(0) \ket{+}) 
\delta(\epsilon-\epsilon') \nonumber \\
&=& x(0) \delta(\epsilon-\epsilon') \nonumber \\
Y(\epsilon',0) &=& i(\bra{+} \hat{\rho}_{S}(0) \ket{-}-\bra{-} \hat{\rho}_{S}(0) \ket{+}) 
\delta(\epsilon-\epsilon') \nonumber \\
&=& y(0) \delta(\epsilon-\epsilon') . \nonumber 
\end{eqnarray}
If the initial condition of the coherences is strictly "local" in the energy of the environment 
(i.e. a true delta distribution) and if the Markovian approximation is exact, 
the non-local terms in the coherence evolution equations (\ref{4Aaaak}) and (\ref{4Aaaal}) do not contribute 
to the coherence dynamics and can be neglected.  
However, the initial condition is never strictly "local" in the real world. 
Furthermore, the non-Markovian initial effects also contribute to give a width to the distribution.
Therefore, even if these non-local effects are generally small, neglecting these terms in the coherence 
evolution has to be considered as an approximation in order to get analytical solutions for the subsystem dynamics. 
Notice that the terms we neglect are non-local only if $\Delta$ is not too small compared to the typical energy scale 
of variation of the microcanonical correlation function $\tilde{\alpha}(\epsilon \pm \Delta, \omega) \approx 
\tilde{\alpha} (\epsilon, \omega)$. 
If $\Delta$ is not too small, our assumption is not valid and our new equation reduces to the Redfield equation 
(see appendix \ref{AppC}). 
However, for the population, the non-local character of the dynamics is strong and cannot be neglected. 
This is precisely the improvement of our new theory with respect to the Redfield theory. 
Performing the local approximation on the coherence dynamics (\ref{4Aaaak}) and (\ref{4Aaaal}), the dynamics of 
the populations and of the coherences can be rewritten
\begin{eqnarray}
\dot{Z}(\epsilon,t)
&=&\gamma_{\rm pop}(\epsilon) (Z(\epsilon,\infty)- Z(\epsilon,t))                                         \label{4Aaaap} \\ 
\dot{X}(\epsilon,t)
&\approx&-\left( \Delta/\hbar + \Gamma(\epsilon) \right) Y(\epsilon,t)               
- \gamma_{\rm coh}(\epsilon) X(\epsilon,t) \nonumber \\
\dot{Y}(\epsilon,t)
&\approx&\left( \Delta/\hbar + \Gamma(\epsilon) \right) X(\epsilon,t)               
- \gamma_{\rm coh}(\epsilon) Y(\epsilon,t) \nonumber  ,                                                         
\end{eqnarray}
where
\begin{eqnarray}
\gamma_{\rm pop}(\epsilon) &=& 2 \pi \frac{\lambda^2}{\hbar^2}  
(\tilde{\alpha}(\epsilon, \Delta / \hbar) + \tilde{\alpha}(\epsilon + \Delta,- \Delta / \hbar))            \label{4Aaaaq} \\
\gamma_{\rm coh}(\epsilon) &=& \pi \frac{\lambda^2}{\hbar^2}  
(\tilde{\alpha}(\epsilon,\Delta / \hbar)+\tilde{\alpha}(\epsilon,-\Delta / \hbar)) \nonumber \\ 
\Gamma(\epsilon) &=& 2 \frac{\Delta \lambda^2}{\hbar} \int d\omega {\cal P}
\frac{\tilde{\alpha}(\epsilon,\omega)}{\Delta^2 - (\hbar \omega)^2} \nonumber \\
Z(\epsilon,\infty)&=&
\frac{\tilde{\alpha}(\epsilon + \Delta,- \Delta / \hbar)-\tilde{\alpha}(\epsilon, \Delta / \hbar)}
{\tilde{\alpha}(\epsilon + \Delta,- \Delta / \hbar) 
+ \tilde{\alpha}(\epsilon, \Delta / \hbar)} C(\epsilon,0)  \nonumber\\
&=&\frac{n(\epsilon)-n(\epsilon+\Delta)} {n(\epsilon)+n(\epsilon+\Delta)} C(\epsilon,0) \nonumber .
\end{eqnarray}
The solutions of the equation (\ref{4Aaaap}) are given by
\begin{eqnarray}
C(\epsilon,t)&=& C(\epsilon,0)                                                                             \label{4Aaaaq}\\ 
Z(\epsilon,t) &=& Z(\epsilon,\infty) + \left[Z(\epsilon,0) - Z(\epsilon,\infty)
\right] e^{- \gamma_{\rm pop}(\epsilon) t} \nonumber \\ 
X(\epsilon,t) &\approx& \{X(\epsilon,0) \cos(\Delta/\hbar+\Gamma(\epsilon))t 
- Y(\epsilon,0) \sin(\Delta/\hbar+\Gamma(\epsilon))t \} 
e^{-\gamma_{\rm coh}(\epsilon) t} \nonumber \\ 
Y(\epsilon,t) &\approx& \{ X(\epsilon,0) \sin(\Delta/\hbar+\Gamma(\epsilon))t 
+ Y(\epsilon,0) \cos(\Delta/\hbar+\Gamma(\epsilon))t \} 
e^{-\gamma_{\rm coh}(\epsilon) t} \nonumber .
\end{eqnarray}
Using (\ref{4Aaaah}), we obtain the final result for the two-level subsystem reduced dynamics  
which reads
\begin{center} \fbox{\parbox{12.5cm}{
\begin{eqnarray}
z(t)&=&Z(\epsilon,t) + Z(\epsilon-\Delta,t)                                                                 \label{4Aaaar}\\
x(t)&=&X(\epsilon,t) \nonumber\\
y(t)&=&Y(\epsilon,t).\nonumber
\end{eqnarray} 
}} \end{center}
The two-level subsystem Markovian dynamics obeys a biexponential relaxation of the populations and an 
oscillatory exponential relaxation for the coherences.\\
Our results are consistent with the expected result at equilibrium for a large environment.
In fact, the coherences disappear on long time scales and the populations reach an equilibrium 
value which is given by the canonical Boltzmann distribution.
This is shown using the property (\ref{2Bcaag}) which becomes in case of the two-level subsystem 
\begin{eqnarray}
n(\epsilon-\Delta) \approx n(\epsilon) e^{-\beta_{mic}(\epsilon) \Delta}.                                   \label{4Aaaas}
\end{eqnarray}
Using now the fact that $C(\epsilon,0)+C(\epsilon-\Delta,0)=1$, we find that the equilibrium 
value of the population becomes
\begin{eqnarray}
z(\infty) &=& Z(\epsilon,\infty) + Z(\epsilon-\Delta,\infty)                                                \label{4Aaaat}\\
&=& \frac{e^{-\beta_{mic}(\epsilon) \Delta / 2}-e^{\beta_{mic}(\epsilon) \Delta / 2}}
{e^{-\beta_{mic}(\epsilon) \Delta / 2}+e^{\beta_{mic}(\epsilon) \Delta / 2}} \nonumber \\ 
&=& \coth  \frac{\beta_{mic}(\epsilon) \Delta}{2}.\nonumber
\end{eqnarray}
This is precisely the canonical expectation value of the population at a temperature given by the 
microcanonical temperature of the environment. 
Once again, we see that our new theory recovers the expected results 
from statistical mechanics when the environment becomes large enough.\\
\begin{figure}[p]
\centering
\vspace*{0cm}
\hspace*{-1cm}
\rotatebox{0}{\scalebox{0.7}{\includegraphics{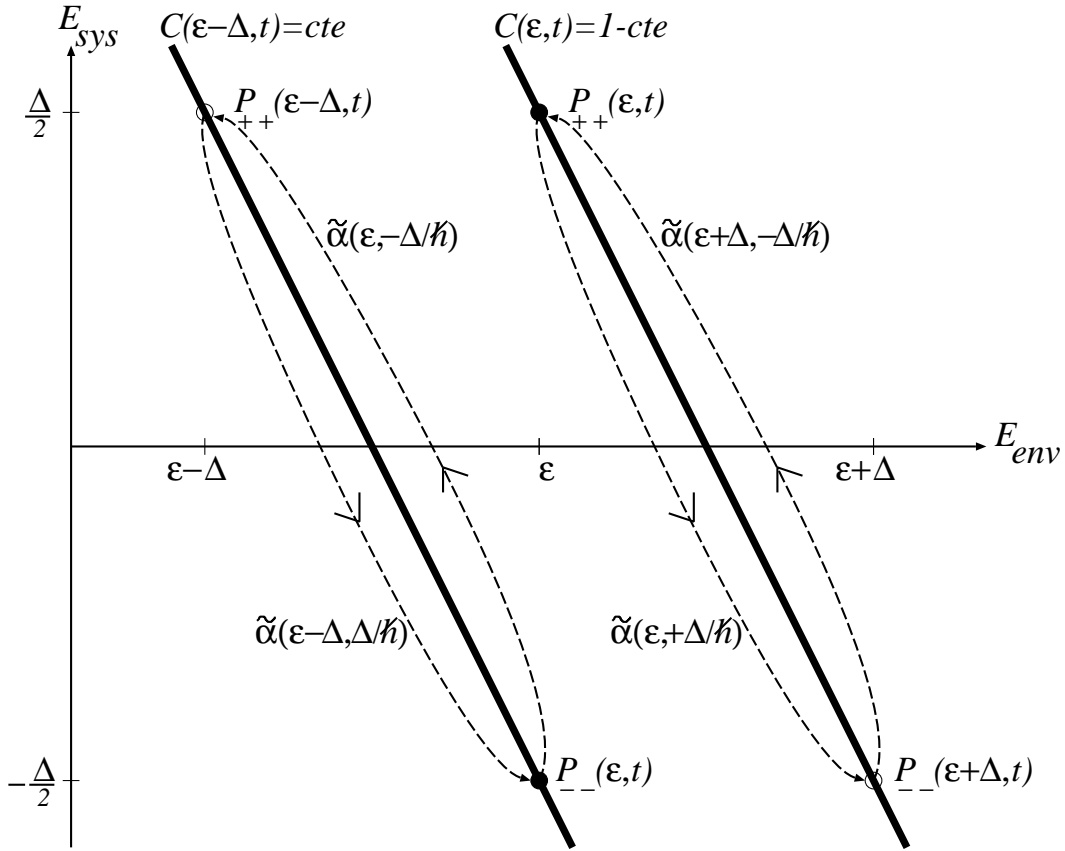}}}
\caption{Representation of the energy exchanges between the environment and the two-level subsystem described 
by the population equation (\ref{4Aaaar}). 
The subsystem energy $E_{sys}$ increases along the $y$-axis and the environment energy $E_{env}$ along the $x$-axis.  
The initial condition is denoted by the two full black dots in the $(x,y)$-plane and is the product of a general 
two-level subsystem density matrix with the environment in a microcanonical state at energy $\epsilon$.
The transitions due to the energy exchanges have to occur along the two thick black lines. 
These two lines correspond to the region of the plane preserving the quantity $C(\epsilon-\Delta,0)$ respectively 
$C(\epsilon,0)$ which are constants of motion. 
The dynamics of the population along these lines is described respectively by $Z(\epsilon-\Delta,t)$ and 
$Z(\epsilon,t)$ of equation (\ref{4Aaaar}).}
\label{energycons2level}
\end{figure}
The Redfield equation applied to the two-level system is derived in appendix \ref{AppC}.
One can notice that, in general, the solutions predicted by our new theory and those predicted by the Redfield theory 
are not the same.
The population dynamics is very different.
The relaxation rate as well as the equilibrium value is different.
But the coherence dynamics is similar.
Both theories predict an exponential damping of the oscillating coherences at the same rate but the frequency of the
oscillations differs only by the Lamb shift.  
One can assume for a very large environment that the microcanonical environment correlation function is almost constant 
on energies of the order of the subsystem typical energies around its microcanonical energy $\tilde{\alpha}(\epsilon \pm \Delta, 
\omega) \approx \tilde{\alpha} (\epsilon, \omega)$. 
As we have seen in section \ref{infinitelimit}, in this case our new equation reduces to the Redfield equation.
In fact our population equation [$z(t)$ in equation (\ref{4Aaaar})] reduces in this case to the Redfield population dynamics 
[$z(t)$ in equation(\ref{AppCaaag})].
However, we had to do the non-local approximation to derive an analytical expression for the coherence dynamics. 
But this approximation fails when the environment becomes too large.
Therefore, in this case, our approximated coherence equations [$x(t)$ and $y(t)$ in equation (\ref{4Aaaar})]
have to be replaced by the Redfield coherence equations [$x(t)$ and $y(t)$ in equation(\ref{AppCaaag})].

\chapter{The spin-GORM model} \label{ch4}

In the previous chapter, we have presented a weak-coupling kinetic theory extending the Redfield 
theory to subsystems interacting with a finite environment. 
The conditions of validity of these quantum kinetic equations remain however little known.
It is indeed very important to understand what are the conditions (in particular the minimum size) 
under which a quantum system can display an irreversible relaxation towards an equilibrium state.  
It is the purpose of this chapter to contribute to the clarification of these questions
by studying the simple model of a two-level system or spin\footnote{Spin means in what 
follows a spin $1/2$ and is therefore equivalent to a two-level system.} interacting with a 
random-matrix environment.  
Indeed, work done during the last decade has shown that typical quantum systems 
with a classical chaotic dynamics presents properties of random matrices on small energy scales
\cite{Bohigas84,Guhr98,HaakeB01,StockmannB99}. 
Here, we consider the Hamiltonian of the environment as well as the operator of coupling between 
the spin and the environment to be given by random matrices taken in a statistical ensemble of 
Gaussian orthogonal random matrices. 
This defines a model that we called the spin-GORM model \cite{Esposito2,Esposito3}. 
Its total Hamiltonian can be numerically diagonalized and, therefore, its physical properties can be 
computed numerically.
Furthermore, for this model many results can be obtained analytically.
It is therefore a prototype model to test and to compare the different kinetic theories.
Similar models using Gaussian orthogonal random matrices \cite{GelbartRice72,Lebowitz} or using banded random 
matrices \cite{Cohen1,Cohen2,Lutz98,MelloPereyra88,Pereyra91} have been studied. 
It has been shown that the dynamics of a subsystem interacting with an 
environment trough an interaction term modeled by random matrices present common features 
with other types of interaction terms found in the literature \cite{Lutz98}. 

The spin-GORM model differs from the spin-boson model by the density of states of the environment. 
Instead of monotonously increasing with energy as in the spin-boson model, the density of states of the 
environment obeys Wigner's semicircle law in our model and is thus limited to an interval of energy with 
a maximum density in between. 
We notice that such densities of states appear in systems where a spin is coupled to other (possibly 
dissimilar) spins as, for instance, in NMR, in which case the density of states of the other spins
forming the environment also present a maximum instead of a monotonous increase with energy.  
Our model may therefore constitute a simplification of such kind of interacting spin systems.  
Our main purpose is to understand the conditions under which a kinetic description can be used in order
to understand the relaxation of the spin under the effect of the coupling with the rest of the system, 
which we refer to as a complex environment.

Our study of the spin-GORM model is separated in four parts.
In section \ref{model}, we define the model. 
In section \ref{spectralanalysis}, we study the spectral properties of the total system. 
In section \ref{time}, we study the different regimes of the spin relaxation dynamics. 
Finally in section \ref{equilibrium}, we investigate the equilibrium distributions of the subsystem and of the total system.

\section{The model} \label{model}

We are interested in the study of a total system composed of a simple subsystem (with a few discrete levels) 
interacting with a complex environment (with many levels). 
A prototype of such system is given by a two-level subsystem interacting with an environment defined in terms 
of random matrices.
The von Neumann equation of this system reads
\begin{eqnarray}
\frac{d \hat{\rho}(\tilde{t})}{d\tilde{t}} = -\frac{i}{\hbar}
\lbrack \hat{\tilde{H}} , \hat{\rho}(\tilde{t}) \rbrack   ,                                                   \label{S-s-GOEvn}
\end{eqnarray}
where the total Hamiltonian is given by
\begin{eqnarray}
\hat{\tilde{H}} = \frac{\tilde{\Delta}}{2} \hat{\sigma}_{z} 
+ \hat{\tilde{H}}_{B} + \tilde{\lambda} \hat{\sigma}_x \hat{\tilde{B}}                                            \label{S-s-GOE}
\end{eqnarray}
where \\ 
$\bullet$ $\hat{\sigma}_{x}$, $\hat{\sigma}_{y}$, and $\hat{\sigma}_{z}$ are the $2 \times 2$ Pauli matrices,\\ 
$\bullet$ $\frac{\tilde{\Delta}}{2} \hat{\sigma}_{z}$ is the Hamiltonian of the two-level subsystem,\\ 
$\bullet$ $\tilde{\Delta}$ is the energy spacing between the two-levels of the subsystem,\\ 
$\bullet$ $\hat{\tilde{H}}_{B}$ is the Hamiltonian of the environment,\\ 
$\bullet$ $\hat{\sigma}_x$ is the coupling operator of the subsystem,\\ 
$\bullet$ $\hat{\tilde{B}}$ is the coupling operator of the environment,\\ 
$\bullet$ $\tilde{\lambda}$ is the coupling parameter between the subsystem and the environment. \\

The well-known spin-boson model \cite{GaspardRed99,Leggett94,Suarez92} is a particular case of this
total system where $\hat{\tilde{H}}_{B}$ corresponds to an infinite harmonic oscillator lattice and 
$\hat{\tilde{B}}$ is linear in the degrees of freedom of the environment. 
Here, we want to define a new model, the spin-GORM model, also described by the total Hamiltonian
(\ref{S-s-GOE}) and for which $\hat{\tilde{H}}_{B}$ and $\hat{\tilde{B}}$ are Gaussian orthogonal random 
matrices (GORM) (see appendix \ref{AppD} for some basic properties of GORM).\\

\begin{figure}[h]
\centerline{\rotatebox{0}{\scalebox{0.5}{\includegraphics{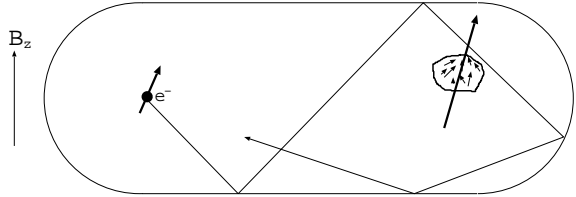}}}}
\caption{Schematic representation of an electronic spin in a chaotic billiard 
quantum dot with a magnetic field perpendicular to the plane of the dot interacting 
through magnetic dipole-dipole potential with a ferromagnetic impurity 
described by a classical spin. This physical system can be modeled 
by a Hamiltonian similar to the spin-GORM model.} \label{billiard}
\end{figure}

Such a model can represent for example an electronic spin traveling in a semiconducting quantum dot in the shape
of a chaotic billiard and, furthermore, submitted to a magnetic field perpendicular to the plane of the dot. 
The quantum dot contains a localized ferromagnetic impurity with a classical magnetic dipole. 
The two spin levels of the traveling electron are split by the magnetic field and interact through a magnetic 
dipole-dipole interaction in a non-diagonal way with the classical magnetic dipole of the localized ferromagnetic 
impurity. 
Since the position and momentum operators of the electron are associated with a classically chaotic dynamics, 
they can be represented by Gaussian orthogonal random matrices according to the Bohigas-Giannoni-Schmit conjecture 
\cite{Bohigas84}. 
This system is schematically depicted in figure \ref{billiard}.

\begin{figure}[h]
\centering
\rotatebox{0}{\scalebox{0.7}{\includegraphics{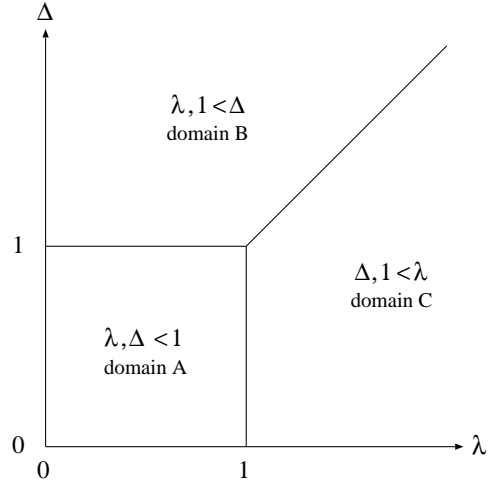}}} \\
\caption{Representation of the three different domains
in the space of the reduced parameters $\lambda$ and $\Delta$
of the model for a fixed number $N$ of states.}
\label{regimeschema}
\end{figure}

Let us now discuss the random matrix aspect of the spin-GORM model in more detail.
As we said, we want to model a two-level subsystem that interacts with an environment that has a complex dynamics. 
Here, complex is used in a generic way. 
The complexity can come, for example, from the fact that the corresponding classical system is chaotic 
\cite{Bohigas84,Guhr98,HaakeB01,StockmannB99} like in a quantum billiard or for the hydrogen atom in a strong magnetic field. 
It can also come from large number of couplings in an interacting many-body system as in nuclear physics \cite{Guhr98} or in
interacting fermion systems such as quantum computers \cite{Guhr98,BenentiCasatiShepelyansky01}. 
In the sixties, Wigner \cite{Brody81,Mehta67,Porter65} was the first to develop random-matrix theory for the purpose of 
modeling spectral fluctuations of complex quantum systems containing many states interacting with each other. 
This tool has now become very common in many fields from nuclear physics to quantum chaos. 
This is the reason why we consider random matrices to characterize the complexity of the environment operators.
The environment operators of the spin-GORM model, $\hat{\tilde{H}}_B$ and $\hat{\tilde{B}}$, are defined by
\begin{eqnarray}
\hat{\tilde{H}}_B &=& \sigma_{ND}^{\hat{\tilde{H}}_B} \hat{X} \nonumber \\ 
\hat{\tilde{B}} &=& \sigma_{ND}^{\hat{\tilde{B}}} \hat{X}^{'},                                                     \label{bof1}
\end{eqnarray}
where $\hat{X}$ and $\hat{X}^{'}$ are two different $\frac{N}{2} \times \frac{N}{2}$ Gaussian orthogonal 
random matrices with mean zero.
Their non-diagonal (diagonal) elements have standard deviation $\sigma_{ND}^{\hat{\tilde{X}}}= 1$ 
($\sigma_{D}^{\hat{\tilde{X}}} = \sqrt{2}$). 
$\hat{X}$ and $\hat{X}^{'}$ are two different realizations of the same random-matrix ensemble and have
therefore the same statistical properties. 
$\sigma_{ND}^{\hat{\tilde{H}}_B}$ and $\sigma_{ND}^{\hat{\tilde{B}}}$ are the standard deviations of the
non-diagonal elements of $\hat{\tilde{H}}_B$ and $\hat{\tilde{B}}$, respectively. 
For these random matrices, the width of their averaged smoothed density of states is given by
\begin{eqnarray}
\mathcal{D}\!\tilde{H}_B = \sigma_{ND}^{\hat{\tilde{H}}_B} \sqrt{8 N}, \nonumber \\
\mathcal{D}\!\tilde{B} = \sigma_{ND}^{\hat{\tilde{B}}} \sqrt{8 N} 
\end{eqnarray}
(see Appendix \ref{AppD}).\\
It is interesting to define the model in such a way that, when $N$ is increased, the averaged smoothed density 
of states of the environment increases without changing its width $\mathcal{D}\!\tilde{H}_B$. 
The width can be fixed to unity. 
This is equivalent to fixing the characteristic time scale of the environment. 
For doing this, it is necessary to rescale the parameters as follows:
\begin{equation}
\left\lbrace
\begin{array}{l}
\alpha=\sigma_{ND}^{\hat{\tilde{H}}_B} \sqrt{8 N} \; , \\
\\
t=\alpha \tilde{t} \nonumber \; ,\\
\\
\Delta = \frac{\tilde{\Delta}}{\sigma_{ND}^{\hat{\tilde{H}}_B} \sqrt{8 N}} \; ,\\
\\
\lambda=\tilde{\lambda} \frac{\sigma_{ND}^{\hat{\tilde{B}}}}{\sigma_{ND}^{\hat{\tilde{H}}_B}} .
\end{array}
\right.
\end{equation}
\\
The von Neumann equation of the spin-GORM model becomes
\begin{equation}
\frac{d \hat{\rho}(t)}{dt} = -\frac{i}{\hbar} \lbrack \hat{H} , \hat{\rho}(t) \rbrack,                       \label{VNeqspinGORM}
\end{equation}
with the rescaled total Hamiltonian
\begin{eqnarray}
\hat{H} &=&  \hat{H}_{S}  +  \hat{H}_{B} +  \lambda \hat{\sigma}_x \hat{B} \nonumber \\
 &=&  \frac{\Delta}{2} \hat{\sigma}_{z}  +  \frac{1}{\sqrt{8N}} \hat{X}  +
 \lambda  \hat{\sigma}_x  \frac{1}{\sqrt{8N}} \hat{X}^{'} .                                                       \label{hamiltonien}
\end{eqnarray}
As announced, we have now $\mathcal{D}\!H_B=\mathcal{D}\!B=1$.\\
In what follows, without loss of generality, $\alpha$ will always be taken equal to unity. 
Notice that, to model an environment with a quasi-continuous spectrum, the random matrix must be very large 
($N \to \infty$).\\
In order to get ensemble averaged results, one has to perform averages over the different results obtained for each
realization of (\ref{hamiltonien}). 
When we use finite ensemble averages, the number of members of the ensemble average will be denoted by $\chi$.\\
We see that the Hamiltonian (\ref{hamiltonien}) is characterized by three different parameters: $\Delta$, $\lambda$, and $N$. 
We define three different domains in the reduced parameter space corresponding to a fixed $N$ in order to facilitate 
the following discussion. 
These three regimes are represented in figure \ref{regimeschema}: domain A with $1 > \lambda, \Delta$; 
domain B with $\Delta > 1,\lambda$; domain C with $\lambda > 1,\Delta$.

\section{Spectral analysis} \label{spectralanalysis}

In this section, we study the spectrum of the complete system for the different values of the parameters. 
This study is important in order to understand the different dynamical behaviors that we encounter in the model.\\

Let us begin by defining the notations in the simple case where there is no coupling between the two parts of the total 
system ($\lambda \to 0$). 
The isolated subsystem has two levels separated by the energy $\Delta$:
\begin{equation}
\hat{H}_{S} \vert s \rangle = s \frac{\Delta}{2} \vert s \rangle,
\end{equation}
where $s=\pm 1$. 
The environment has the standard spectrum of a GORM
\begin{equation}
\hat{H}_{B} \vert b \rangle = E_{b}^{B} \vert b \rangle,
\label{VPenviron}
\end{equation}
where $b=1,2,...,N/2$. 
The Hamiltonian of the total system without interaction between the subsystem and the environment is thus
\begin{equation}
\hat{H}_{0}=\hat{H}_{S}+\hat{H}_{B},
\end{equation}
and the spectrum is therefore given by
\begin{equation}
\hat{H}_{0} \vert n \rangle = E^{0}_{n} \vert n \rangle,
\label{eqVPHo}
\end{equation}
with $ n=1,2,...,N$ and 
\begin{equation}
E^{0}_{n} = s \frac{\Delta}{2} + E_{b}^{B}. \label{eqVPHobis}
\end{equation}
The eigenvectors are tensorial products of both the subsystem and environment eigenvectors:
\begin{equation}
\vert n \rangle = \vert s \rangle \vert b \rangle.
\end{equation}
Let us now define the notations in the opposite case, when the coupling term is so large that the Hamiltonian 
of the subsystem and of the environment can both be neglected ($\lambda \to \infty$). 
Using the unitary matrix $\hat{U}$ acting only on the subsystem degrees of freedom
\begin{equation*}
\hat{U}=
\begin{bmatrix}
\frac{1}{\sqrt{2}} & \frac{1}{\sqrt{2}} \\ \frac{1}{\sqrt{2}} &
-\frac{1}{\sqrt{2}} \\
\end{bmatrix},
\end{equation*}
the total Hamiltonian becomes
\begin{eqnarray}
\tilde{\hat{H}}_{0} = \lambda \hat{\sigma}_z \hat{B}.
\end{eqnarray}
$E_{\kappa \eta}$ and $\vert \kappa \eta \rangle = \vert \kappa
\rangle \otimes \vert \eta \rangle$ are, respectively, the eigenvalues and eigenvectors of the Hamiltonian
\begin{equation}
\tilde{\hat{H}}_{0} \vert \kappa \eta \rangle = \lambda E_{\kappa
\eta} \vert \kappa \eta \rangle = \lambda \kappa E_{\eta} \vert
\kappa \eta \rangle ,
\label{eqVPHoinv}
\end{equation}
where $\eta=1,...,\frac{N}{2}$ and $\kappa=\pm1$.\\
After having defined the notation in the two extreme cases $\lambda \to 0$ and $\lambda \to \infty$, we start the study
of the spectrum with interaction $\lambda \neq 0$.
The total spectrum is given by the eigenvalues $\lbrace E_{\alpha} \rbrace$ that are solutions of the eigenvalue problem:
\begin{equation}
\hat{H} \vert \alpha \rangle = E_{\alpha} \vert \alpha
\rangle, \label{eqVPtot}
\end{equation}
where $\alpha=1,2,...,N$. 
It is very difficult to obtain analytical results for this problem. 
We will therefore study the total spectrum using a method of numerical diagonalization of the total Hamiltonian.

\subsection{Smoothed density of states}

In order to have a quantitative understanding of the global aspect of the spectrum (on large energy scales), we 
will study the total perturbed averaged smoothed density of states.\\

\begin{figure}[h]
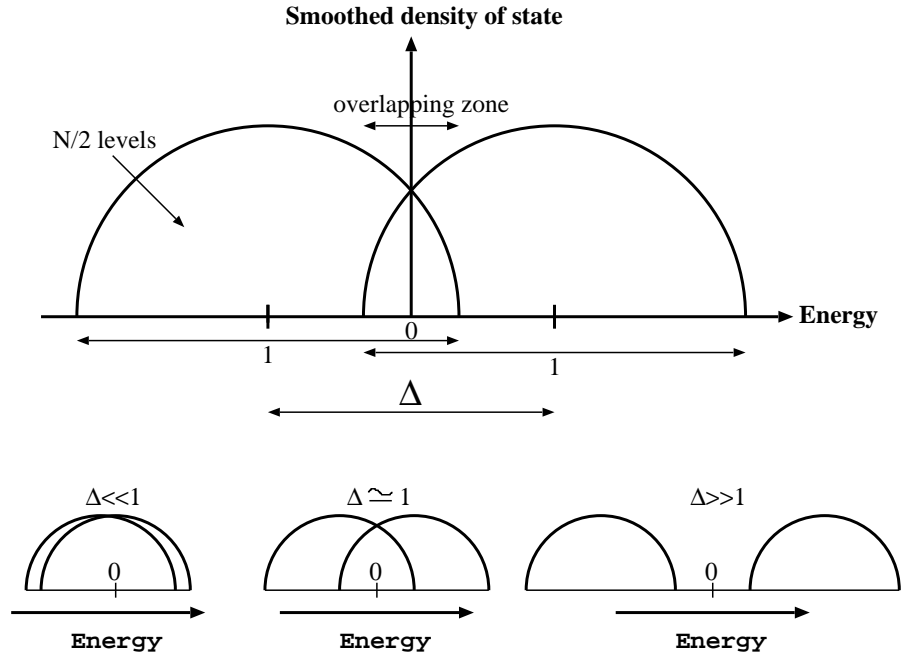

\centerline{\rotatebox{0}{\scalebox{1}{\includegraphics{fig2.eps}}}}
\vspace*{0.8cm}
\centerline{\rotatebox{0}{\scalebox{1}{\includegraphics{fig3.eps}}}}
\caption{Smoothed densities of states $n^w(\varepsilon-\frac{\Delta}{2})$ and
$n^w(\varepsilon+\frac{\Delta}{2})$ 
for different values of $\Delta$. The total smoothed averaged
density of states of the non-perturbed spectrum is obtained by the
sum of them (see equation (\ref{doubledistrib})).} \label{troisdelta}
\end{figure}

\begin{figure}[p]
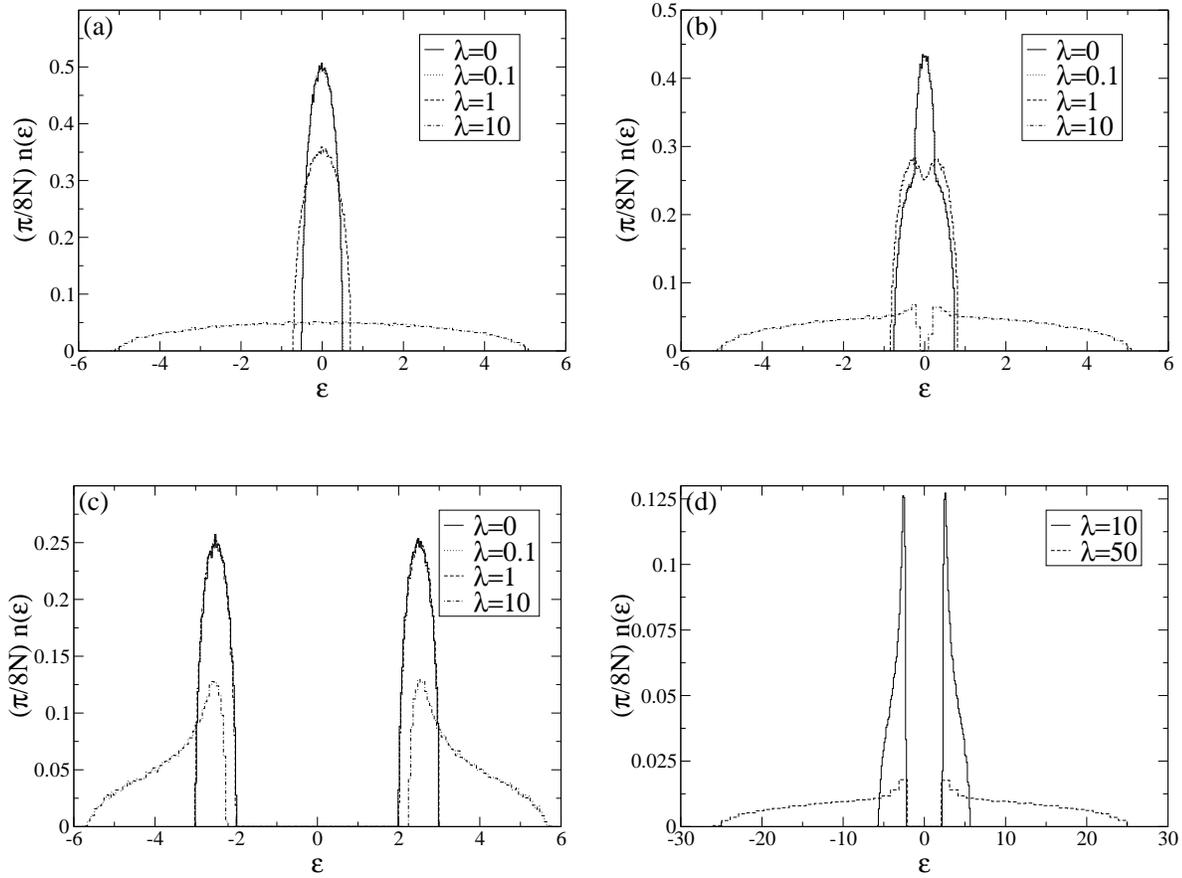

\centering
\begin{tabular}{c@{\hspace{0.5cm}}c}
\vspace*{1cm}
\hspace*{-2cm}
\rotatebox{0}{\scalebox{0.3}{\includegraphics{fig4.eps}}} &
\rotatebox{0}{\scalebox{0.3}{\includegraphics{fig5.eps}}} \\
\hspace*{-2cm}
\rotatebox{0}{\scalebox{0.3}{\includegraphics{fig6.eps}}} &
\rotatebox{0}{\scalebox{0.3}{\includegraphics{fig7.eps}}} \\
\end{tabular}
\caption{Total smoothed averaged
density of states obtained numerically for different values of
$\Delta$ and $\lambda$. (a) corresponds to $\Delta=0.01$, (b)
to $\Delta=0.5$, and (c) and (d) to $\Delta=5$. For all of them $N=500$
and $\chi=50$. Notice that the $\lambda=0$ and the $\lambda=0.1$ 
curves are not distinguishable.} \label{smoothedplot}
\end{figure}

The environment-averaged smoothed density of states obeys the semicircle Wigner law (see equation 
\begin{eqnarray}
n^w(\epsilon) = 
\begin{cases} 
\frac{4N}{\pi} \sqrt{(\frac{1}{2})^2-\epsilon^2}
\ \ &\mbox{if} \ \ \vert \epsilon \vert < \frac{1}{2} \\ 
0 \ \ &\mbox{if} \ \ \vert \epsilon \vert \geq \frac{1}{2},                                                         
\end{cases} \label{wignerdistr}
\end{eqnarray}
where $\epsilon$ is the continuous variable corresponding to the environment energy $E_{b}^{B}$.
Therefore, when $\lambda=0$, the total averaged smoothed density of states is the sum of the two environment 
semicircular densities of states which correspond to both states of the two-level system (see equation (\ref{eqVPHo})):
\begin{eqnarray}
n(\varepsilon) &=& n^w(\varepsilon-\frac{\Delta}{2})+n^w(\varepsilon+\frac{\Delta}{2}) \nonumber \\ 
&=& 
\begin{cases}
&\frac{4N}{\pi} \sqrt{(\frac{1}{2})^2-(\varepsilon-\frac{\Delta}{2})^2} +
\frac{4N}{\pi} \sqrt{(\frac{1}{2})^2-(\varepsilon+\frac{\Delta}{2})^2} \\ 
&\hspace{3.5cm} \mbox{if} \ \ (\frac{1}{2}-\frac{\Delta}{2}) < \vert \varepsilon
\vert < (\frac{1}{2}+\frac{\Delta}{2}) \\ 
& 0 \ \ \hspace{3cm} \rm{elsewhere}                                                                                         
\end{cases}, \label{doubledistrib}
\end{eqnarray}
where $\varepsilon$ is the continuous variable corresponding to the total energy $E^{0}_{n}$. 
The semicircular densities $n^w(\varepsilon-\frac{\Delta}{2})$ and $n^w(\varepsilon+\frac{\Delta}{2})$ 
are schematically depicted in figure \ref{troisdelta} for different values of $\Delta$. 
The numerical density of states of the total system corresponding to $\lambda=0$ are depicted 
in figure \ref{smoothedplot}(a)-(c).\\
When $\lambda \to \infty$ (meaning that the coupling term becomes dominant in the Hamiltonian), 
the averaged smoothed density of states of the total system (see equation (\ref{eqVPHoinv})) is given by
\begin{eqnarray}
n(\varepsilon) = n^w(\frac{\varepsilon}{\lambda}) + n^w(- \frac{\varepsilon}{\lambda}) = 
\begin{cases} 
\frac{8N}{\lambda \pi} \sqrt{(\frac{\lambda}{2})^2-(\varepsilon)^2} \ \ &\mbox{if} \ \ \vert
\varepsilon \vert < \frac{\lambda}{2} \nonumber \\ 
0 \ \ &\mbox{if} \ \ \vert \varepsilon \vert \geq \frac{\lambda}{2}                                            
\end{cases}, \label{1surleg0}
\end{eqnarray}
where $\varepsilon$ is the continuous variable corresponding to the total energy $E_{\kappa \eta}$. 
This result can be observed in figure \ref{smoothedplot}(a) for $\lambda=10$ (because $\Delta,1 \ll \lambda$).\\
When $\lambda \neq 0$, the total averaged smoothed density of states is also plotted in figure \ref{smoothedplot}. 
The main observation is that there is a broadening of the complete spectrum when one increases $\lambda$. 
In figure \ref{smoothedplot}(a), we see that the averaged smoothed density of states changes in a smooth way from
(\ref{doubledistrib}) to (\ref{1surleg0}). But in figure \ref{smoothedplot}(b) and much
more in figures \ref{smoothedplot}(c) and (d), the two semicircular densities $n^w(\varepsilon-\frac{\Delta}{2})$ and
$n^w(\varepsilon+\frac{\Delta}{2})$ seem to repel each other as $\lambda$ increases. 
This is due to the fact that the levels of a given semicircular density do not interact with each other but only 
interact with the levels of the other semicircular density. 
This is a consequence of the non-diagonal form of the coupling. 
Therefore, having in mind the perturbative expression of the energies (see equation (\ref{pertcas2}) of 
Appendix \ref{AppF}), one understands that when $\Delta$ is nonzero, the eigenvalues that are repelling
each other with the most efficiency are the ones closest to the center of the total spectrum.

\subsection{Eigenvalue diagrams}

The global effect of the increase of $\lambda$ on the eigenvalues has been studied with the average smoothed density of states.
But in order to have an idea of what happens on a finer energy scale inside the spectrum, it is interesting to individually follow
each eigenvalue $E_{\alpha}$ as a function of $\lambda$ on an eigenvalue diagram.\\

\begin{figure}[p]
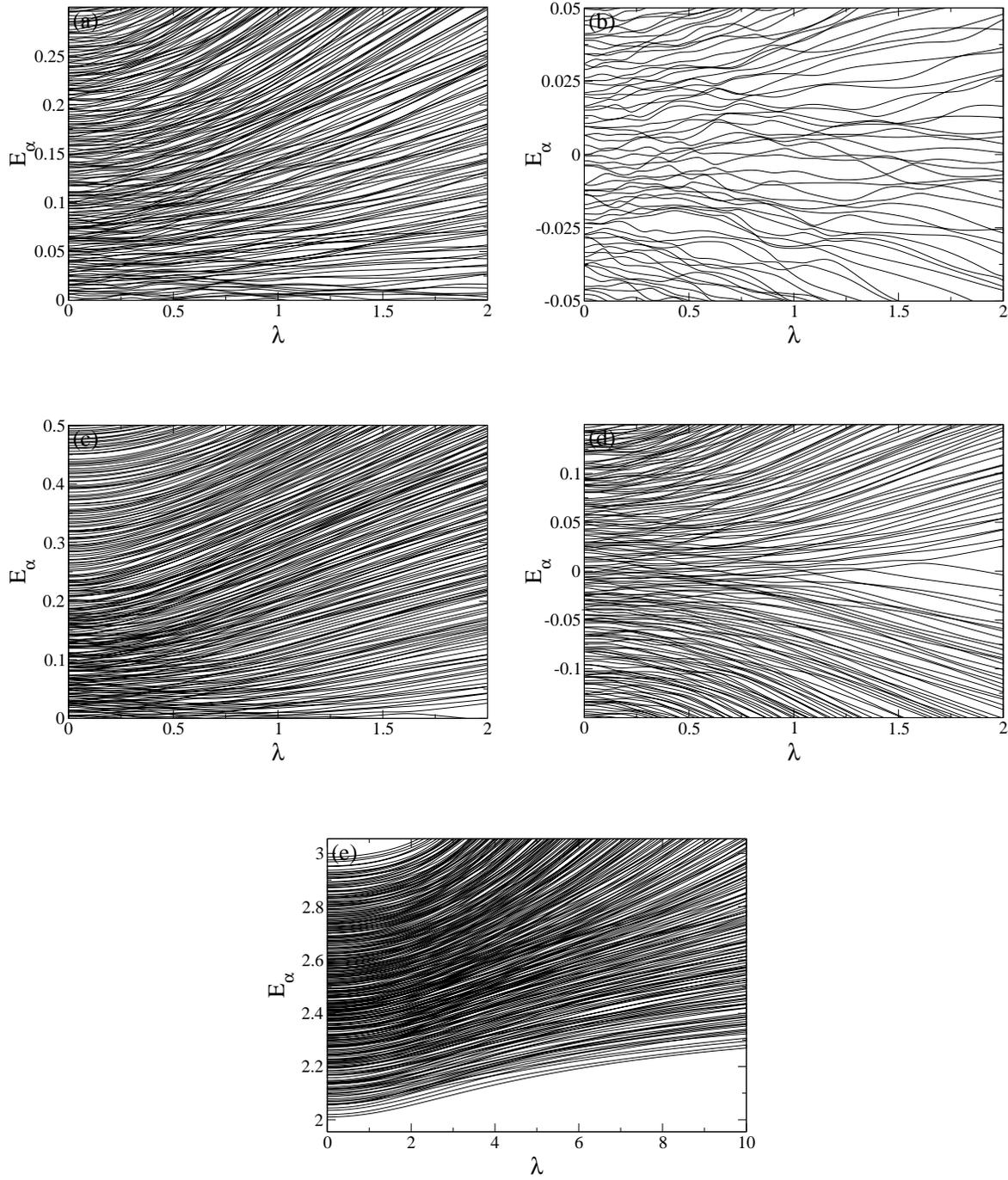

\vspace*{1cm}
\centering
\hspace*{-1cm}
\begin{tabular}{c@{\hspace{0.5cm}}c}
\vspace*{1cm}
\hspace*{-1cm}
\rotatebox{0}{\scalebox{0.3}{\includegraphics{fig8.eps}}}
&
\rotatebox{0}{\scalebox{0.3}{\includegraphics{fig9.eps}}}
\\
\vspace*{1cm}
\hspace*{-1cm}
\rotatebox{0}{\scalebox{0.3}{\includegraphics{fig10.eps}}}
&
\rotatebox{0}{\scalebox{0.3}{\includegraphics{fig11.eps}}}
\\
\end{tabular}
\centering
\rotatebox{0}{\scalebox{0.3}{\includegraphics{fig12.eps}}}
\caption{Different parts of the eigenvalue diagrams with $N=500$, corresponding
to (a) and (b) $\Delta=0.01$, (c) and (d) $\Delta=0.5$, and (e) $\Delta=5$. They represent the
eigenvalues of the total Hamiltonian (\ref{eqVPtot}) as a function
of $\lambda$.} \label{croisementd=0.01etd=0.5}
\end{figure}

The first thing to note (see figure \ref{croisementd=0.01etd=0.5}) is that the increase of the
coupling induces a repulsion between the eigenvalues. 
The result is the broadening of the total spectrum as we already noticed on the smoothed averaged density of states. 
If one looks closer inside the fine structure of the eigenvalue spectrum, we see that there
is no crossing between the eigenvalues. 
This is a consequence of the fact that there is no symmetry in the total system. 
Therefore, the non-crossing rule is always working. 
Each time two eigenvalues come close to each other, they repel each other and create an avoided crossing. 
One can notice that there is a large number of avoided crossings inside the region where the two
semicircles $n^w(\varepsilon-\frac{\Delta}{2})$ and $n^w(\varepsilon+\frac{\Delta}{2})$ overlap (see
figure \ref{troisdelta} in order to visualize the overlapping zone that extends from $\frac{\Delta}{2}-1$ to
$1-\frac{\Delta}{2}$). 
But outside this overlapping zone, there are very weak avoided crossings (and of course no crossing) and all
the eigenvalues appear to move in a regular and smooth way. 
The regions seen in figures \ref{croisementd=0.01etd=0.5}(a), (b) and (d) are inside 
the overlapping zone and the one in figure \ref{croisementd=0.01etd=0.5}(e) is outside. 
In figure \ref{croisementd=0.01etd=0.5}(c), the lower part of the
diagram is inside the overlapping zone and the upper part is outside. 
This phenomenon is due to the fact that the eigenvalues of a given semicircular density do not interact with 
the eigenvalues of their own semicircular density but only with those of the other semicircular density. 
It is the consequence of the non-diagonal nature of the coupling in the spin degrees of freedom. 
When the coupling becomes large enough ($\lambda > 1$), the avoided crossings also disappear inside the overlapping zone.
Around $\lambda=1$ and inside the overlapping zone, there is a smooth transition from an avoided-crossing regime 
that gives rise to a turbulent and complex $\lambda$ evolution to another regime without much interaction between 
the eigenvalues that gives rise to a smooth $\lambda$ evolution.

\subsection{Spacing distribution}

The eigenvalue diagrams only give us a qualitative understanding of the fine energy structure of the spectrum. 
For a more quantitative study, it is interesting to look at the spacing distribution of the spectrum.\\

\begin{figure}[p]
\centering 
\rotatebox{0}{\scalebox{0.4}{\includegraphics{fig13.eps}}} \\
\vspace*{1cm}
\rotatebox{0}{\scalebox{0.4}{\includegraphics{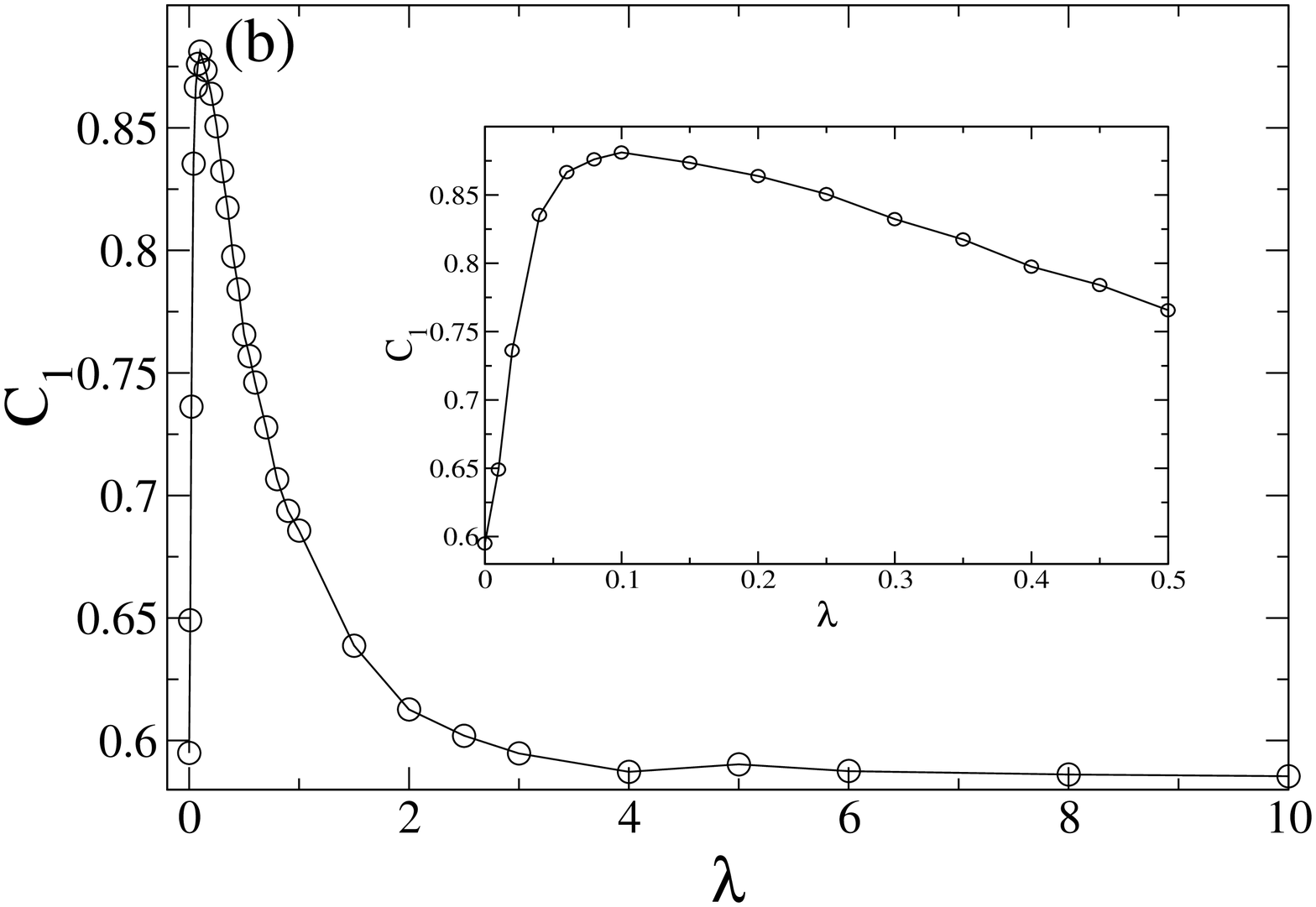}}}\\
\caption{(a) Spacing distribution for different
values of the coupling parameter $\lambda$. (b)
Fitted coefficient $C_1$ of equation (\ref{fit}). The closer is $C_1$ from
unity, the closer is the spacing distribution to the Wigner spacing
distribution. In the two figures, $\Delta=0.01$, $N=500$, and $\chi=50$.} 
\label{distribwigner}
\end{figure}

When $\lambda=0$, each eigenspace $s=\pm \frac{\Delta}{2}$, corresponding to a different subsystem level, has a Wignerian 
level spacing distribution
\begin{equation}
P^w(s) = \frac{\pi}{2} s e^{- \frac{\pi}{4} s^2}.
\end{equation}
But the total spectrum is obtained by the superposition with a shift $\Delta$ of two of such spectra. 
The superposition creates a Poissonian component to the total spacing distribution in the overlapping zone 
($\frac{\Delta}{2}-1$ to $-\frac{\Delta}{2}+1$). 
A Poissonian distribution is given by
\begin{equation}
P^p(s) = e^{-s}.
\end{equation}
Therefore, we choose to fit the total spacing distribution by the mixture
\begin{equation}
P^{fit}(s) = C_1 P^w(s) + (1-C_1) P^p(s). \label{fit}
\end{equation}
This choice of the form of the fit is empirical but reasonable because the correlation coefficient of the fit is 
always close to one (between $0.963$ and $0.982$).\\
We computed the spacing distribution and made the fit (\ref{fit}) in order to compute the mixing coefficient $C_1$ for
different values of $\lambda$ in the case where the overlapping zone covers almost the whole spectrum ($\Delta \ll 1$). 
The results are plotted in figure \ref{distribwigner}. 
We see that there is a specific region of $\lambda$ values where the total spacing distribution is close to a pure 
Wignerian one. 
This region corresponds to the situation where $\lambda^2 N =O(1)$, i.e., when the typical intensity of the interaction 
between the non-perturbed levels [ which is $O(\lambda^2)$ since the first order correction in perturbation theory is zero
in our model ] becomes of the order of the mean level spacing $O(\frac{1}{N})$ in 
the total system. 
In this region, the level repulsion is maximal and effective among almost all the states. 
In the limit $\lambda \to \infty$, the spectrum is again the superposition of two semicircle distributions, one for the 
$E_{1 \eta}$'s and one for the $E_{-1 \eta}$'s according to equation (\ref{eqVPHoinv}).
As a consequence, one gets a Poissonian type of spacing distribution.
The other $\Delta$ cases will have a Wignerian spacing distribution outside the overlapping zone and a mixed one 
(like for the case $\Delta \ll 1$) inside.
This can be visually seen on the eigenvalue diagram that we studied before.

\subsection{The shape of the eigenstates (SOE)}

We want now to have some information about the eigenstates of the total system inside the overlapping zone 
of the two semicircular densities.\\

\begin{figure}[p]
\centering
\rotatebox{0}{\scalebox{1}{\includegraphics{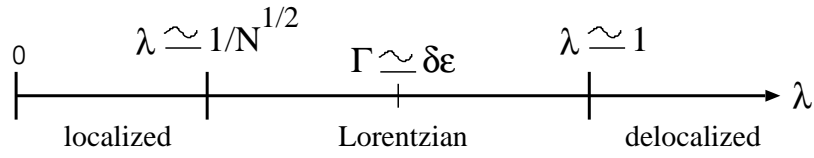}}} \\
\caption{Diagram of the different regimes as a function of the 
coupling parameter $\lambda$.}
\label{shemasoe}
\end{figure}

\begin{figure}[p]
\centering \rotatebox{0}{\scalebox{0.5}{\includegraphics{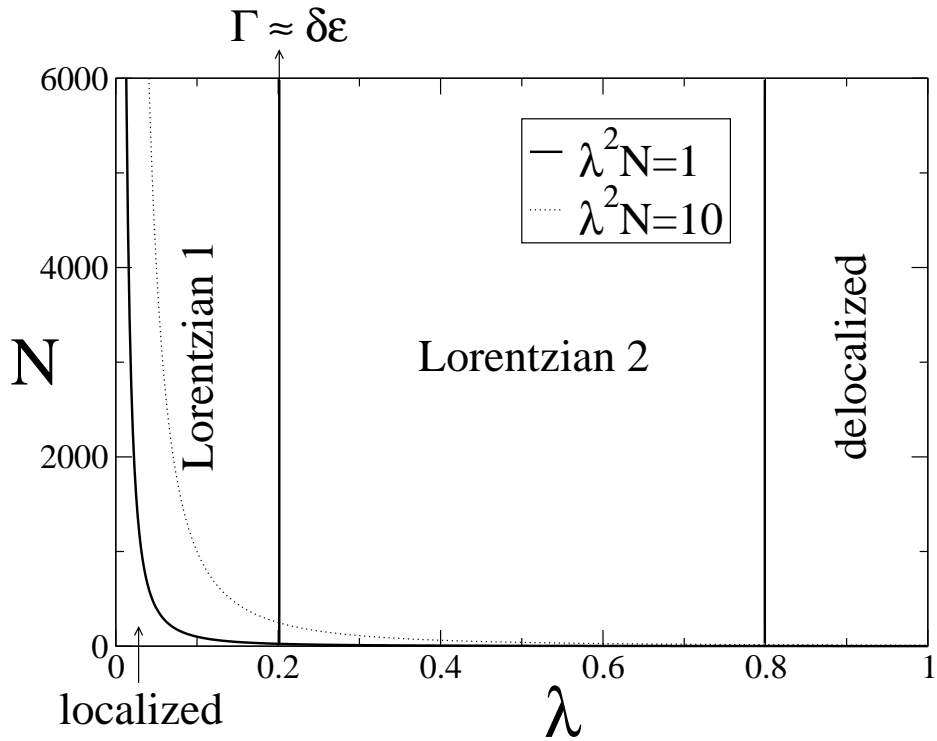}}}
\caption{Schematic representation of the different regimes in
the plane of the reduced parameter $\lambda$ and $N$. The Lorentzian $1$ 
regime corresponds to $\Gamma \lesssim \delta \epsilon$ and the Lorentzian 
$2$ regime to $\Gamma > \delta \epsilon$.}
\label{scheml2N}
\end{figure}

\begin{figure}[p]
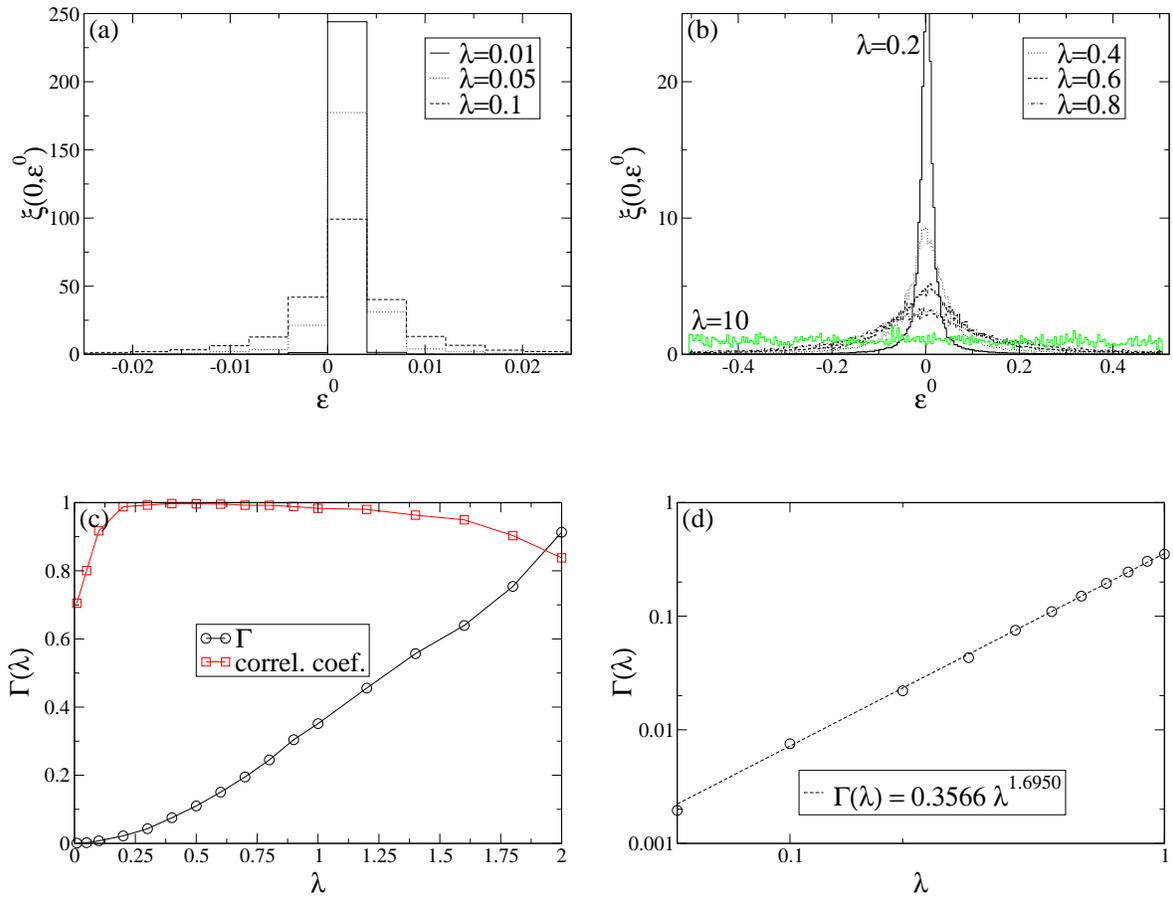

\centering
\begin{tabular}{c@{\hspace{0.5cm}}c}
\vspace*{1cm}
\hspace*{-2cm}
\rotatebox{0}{\scalebox{0.3}{\includegraphics{fig17.eps}}} &
\rotatebox{0}{\scalebox{0.3}{\includegraphics{fig18.eps}}} \\
\hspace*{-2cm}
\rotatebox{0}{\scalebox{0.3}{\includegraphics{fig19.eps}}} &
\rotatebox{0}{\scalebox{0.3}{\includegraphics{fig20.eps}}} \\
\end{tabular}
\caption{(a) SOE in the perturbative regime.
(b) SOE in the Lorentzian regime.
(c) The width and the correlation coefficient of the
fit of the SOE by a Lorentzian. (d) Power-law dependence of the width of
the Lorentzian SOE in the coupling
parameter. In all the figures: 
$\Delta=0.01$, $N=500$, $\chi=50$, and $\varepsilon=0$.} \label{SOEfigure}
\end{figure}

Following \cite{Cohen1}, we define the quantity
\begin{eqnarray}
\xi(\varepsilon,\varepsilon^0)=\sum_{\alpha,n} \vert \langle \alpha \vert n \rangle \vert^{2}
\delta(E_{\alpha}-\varepsilon) \delta(E_{n}-\varepsilon^{0})
\end{eqnarray}
We call local density of states (LDOS) the function $\xi(\varepsilon, \varepsilon^0)$ for fixed $\varepsilon^{0}$ 
and varying $\varepsilon$. \\ 
We call shape of the eigenstates (SOE) the function $\xi(\varepsilon,\varepsilon^0)$ for fixed $\varepsilon$
and varying $\varepsilon^{0}$.\\
We here focus on the SOE. 
The SOE tells us how close a perturbed eigenstate ($\lambda \neq 0$) at energy $\varepsilon$ is from the non-perturbed 
eigenstate ($\lambda = 0$) at energy $\varepsilon^0$. 
If the SOE is a very narrow function centered around $\varepsilon = \varepsilon^0$, the concept of a non-perturbed 
eigenstate is still useful. 
This regime corresponds to very small coupling, for which the interaction intensity is lower than the mean level 
spacing $0 < \lambda^2 N \lesssim 1$, and will be called to the \textit{localized} regime. 
In the limit $N \to \infty $, this regime disappears. 
If one increases the coupling, the interaction intensity between the non-perturbed states begins to be larger than 
the mean level spacing between the states: $\lambda^2 N > 1$. 
The non-perturbed levels start to be "mixed" by the interaction and the SOE starts then to have a Lorentzian shape 
with a finite width $\Gamma$, centered around $\varepsilon = \varepsilon^0$. 
This regime is called the \textit{Lorentzian} regime. 
If one further increases the coupling parameter, the SOE begins to spread over almost the whole spectrum.
This regime is called the \textit{delocalized} regime. \\
In the banded random matrix model of \cite{Cohen1}, the regimes are classified according to a different 
terminology and there is an additional regime corresponding to a spreading that goes beyond the energy
range where the coupling acts (due to the finite coupling range of the banded matrices). 
The motivation for our change of terminology will become clear in the study of the dynamics.\\
The Lorentzian regime can be separated into two parts. 
For small coupling, the width of the Lorentzian $\Gamma$ is smaller or of the order of magnitude of the 
typical energy scale of variation of the averaged smoothed density of states of the environment 
$\delta \epsilon$: $n(\epsilon + \delta \epsilon) \approx n(\epsilon)$. 
For larger coupling, the Lorentzian width extends on an energy scale larger than the typical energy
scale of variation of the density of states of the environment. 
Therefore, one has $\Gamma \lesssim \delta \epsilon $ in the former case and $\Gamma > \delta \epsilon$  
in the latter case. 
The different regimes are represented in figures \ref{shemasoe} and \ref{scheml2N}.
To represent the SOE of the spin-GORM model, we discretize the energy axis in small cells of the order of the 
mean level spacing $\frac{1}{N}$ and we average the SOE over $\chi$ realizations of the random-matrix ensemble. 
We see in figure \ref{SOEfigure}(a) the typical shape of the SOE going from the perturbative regime 
($\lambda=0.01,0.05$) to the beginning of the Lorentzian one ($\lambda=0.1$). 
In figure \ref{SOEfigure}(b), we see the SOE across the Lorentzian regime ($0.2 \leq \lambda \leq 0.8$). 
We also see the delocalized regime, when $\lambda \to \infty$ and the SOE gets completely flat ($\lambda=10$). 
Figure \ref{SOEfigure}(c) shows the width of the Lorentzian obtained from a fit made on the SOE curve. 
The correlation coefficient of the fit helps us to determine the region of the Lorentzian regime where the SOE is 
very well fitted by a Lorentzian. 
It has been verified that the width of the Lorentzian is independent of $N$ in the Lorentzian regime. 
Figure \ref{SOEfigure}(d) (log-log) shows that the width of the Lorentzian has a power-law dependence
in the coupling parameter close to two in the Lorentzian regime.

\subsection{Asymptotic transition probability kernel (ATPK)}

An interesting quantity, which is close to the SOE, but which has a nice physical interpretation, 
is the asymptotic transition probability kernel (ATPK).\\

\begin{figure}[p]
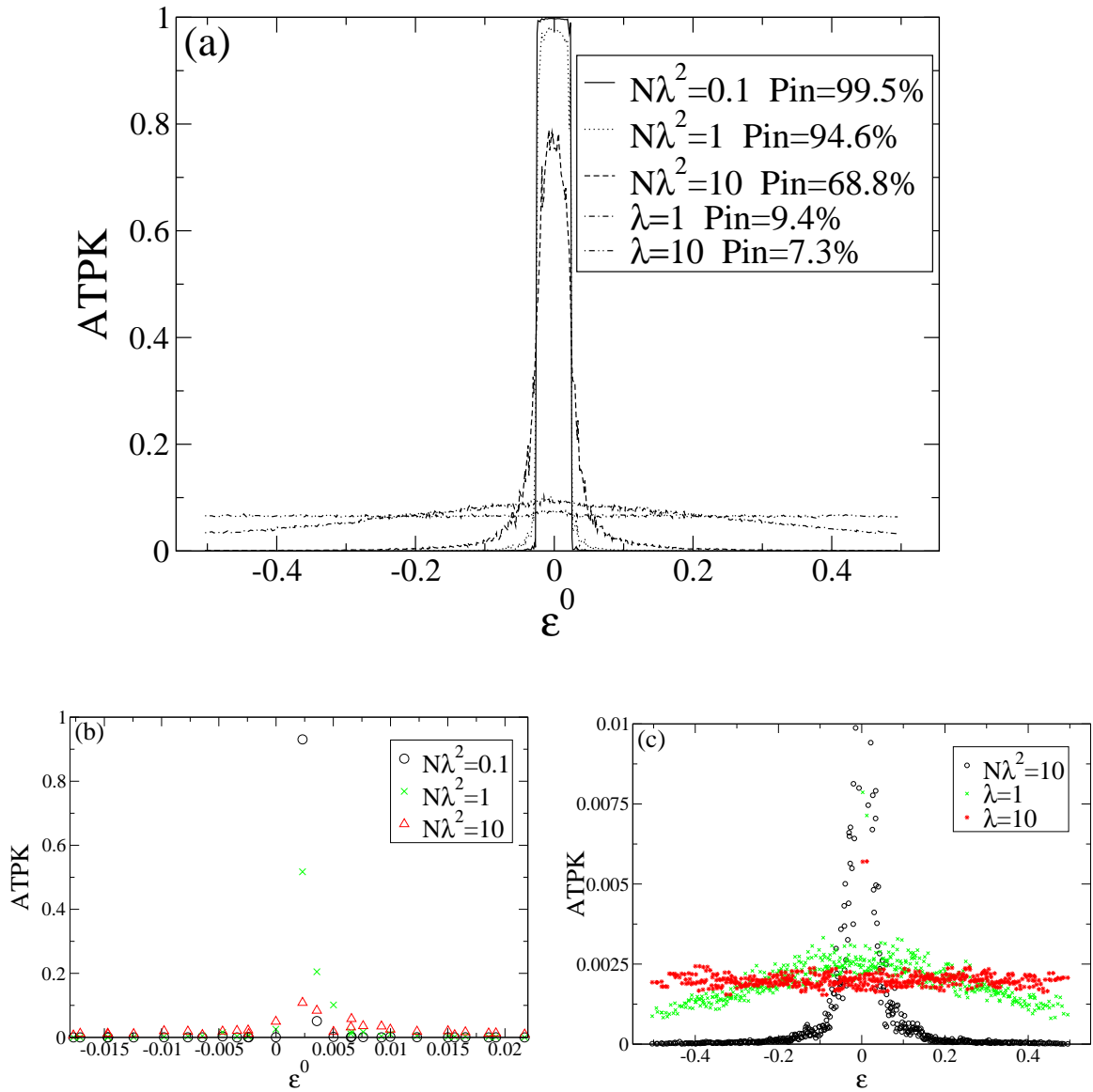

\centering
\hspace*{-2cm}
\rotatebox{0}{\scalebox{0.5}{\includegraphics{fig21.eps}}}\\
\centering \vspace*{1cm}
\begin{tabular}{c@{\hspace{0.5cm}}c}
\hspace*{-2cm}
\rotatebox{0}{\scalebox{0.3}{\includegraphics{fig22.eps}}} &
\rotatebox{0}{\scalebox{0.3}{\includegraphics{fig23.eps}}} \\
\end{tabular}
\caption{(a) ATPK from the localized
regime, through the Lorentzian regime, to the delocalized regime. The
ATPK is microcanonically averaged over the different initial
conditions corresponding to the levels inside the energy shell
centered at $\varepsilon^{0'}=0$ with width $\delta \varepsilon^{0'}=0.05$. 
"Pin" denotes the proportion of the ATPK that stays inside the initial energy 
shell after an infinite time. (b) and (c) ATPK for a single level as
an initial condition, without any average. All the figures are obtained
for very small system energy $\Delta=0.01$ with no random-matrix ensemble 
average $\chi=1$ and for $N=500$.} \label{ATPKdessin}
\end{figure}

The transition probability kernel (TPK) gives the probability at time $t$ to be in the level 
$\vert n \rangle$ if starting from $\hat{\rho}(0)$. 
It is defined as
\begin{eqnarray}
\Pi_{t}(n\vert\rho(0))=\langle n \vert  e^{-i \hat{H} t}
\hat{\rho}(0) e^{i \hat{H} t} \vert n \rangle
\end{eqnarray}
The ATPK is the time average of the TPK
\begin{eqnarray}
\Pi_{\infty}(n\vert\rho(0))&=& \lim_{T\to\infty} \frac{1}{T}
\int_{0}^{T} dt \Pi_{t}(n\vert\rho(0)) \nonumber \\ &=&
\sum_{\alpha} \vert \langle \alpha \vert n \rangle \vert^{2}
\langle \alpha \vert \hat{\rho}(0) \vert \alpha \rangle.
\end{eqnarray}
The distribution of the ATPK in energy is given by
\begin{eqnarray}
\Pi_{\infty}(\varepsilon^{0} \vert \rho(0)) = \sum_{n}
\Pi_{\infty}(n\vert\rho(0)) \delta(E_n-\varepsilon^{0}).
\end{eqnarray}
The ATPK has an intuitive physical interpretation. 
It represents the probability after a very long time to end up in a non-perturbed state $\vert n \rangle$, 
having started from the initial condition $\hat{\rho}(0)$. 
The ATPK is the convolution
\begin{eqnarray}
\Pi_{\infty}(\varepsilon^{0} \vert \varepsilon^{0'}) 
= \sum_{n,m} \sum_{\alpha} \vert \langle \alpha \vert n \rangle \vert^{2} \vert \langle \alpha 
\vert m \rangle \vert^{2} \delta(E_n-\varepsilon^{0}) \delta(E_m-\varepsilon^{0'}),
\end{eqnarray}
with $\hat{\rho}(0) = \sum_m \vert m \rangle \langle m \vert \delta(E_m-\varepsilon^{0'})$.\\
Since we are interested in a random matrix model and for the purpose of studying the dynamics, we will
perform averages of two different kinds. 
The first kind of average is a microcanonical average over states belonging to the same given energy shell 
of width $\delta \varepsilon$ for a given realization of the random-matrix ensemble used in our total Hamiltonian. 
The second kind of average is an ensemble average over the $\chi$ different realizations of the random matrices ensemble. 
For the microcanonical average, a choice of the width of the energy shell $\delta \varepsilon$ has to be
done in such a way that it is large enough to contain many levels (to get a good statistics) and small enough to be 
smaller than or equal to the typical energy scale of variation of the averaged smoothed density of states of the
environment $\delta \epsilon$. 
Therefore, the adequate choice of the width of the energy shell corresponds to 
$\frac{1}{N} < \delta \varepsilon < \delta \epsilon$. \\
Different ATPK are depicted in figure \ref{ATPKdessin} where there is no random-matrix ensemble average $\chi=1$. 
In figure \ref{ATPKdessin}(a), the ATPK is a microcanonical average inside the energy shell at energy $\varepsilon^{0'}$ 
and of width $\delta \varepsilon^{0'}$. 
"Pin" denotes the total probability of staying inside the energy shell after a very long time. 
$N \lambda^2=1$ is on the border between the localized and the Lorentzian regime and $\lambda=1,10$ in the delocalized regime. 
We see that, until $N \lambda^2=1$, the main probability stays in the initial energy shell. 
But when the Lorentzian regime starts, the probability spreads over energies larger than $\delta \epsilon$ 
(here $\delta \epsilon \approx \delta \varepsilon^{0'}$). 
In figures \ref{ATPKdessin}(b) and (c), no average has been done. 
The initial condition is a non-perturbed pure state (corresponding to an energy close to zero) and one can see the individual
probability of being on another non-perturbed state for the different regimes. 
We can see that in the localized regime the probability of staying on the initial state is much more important than the 
probability of leaving it. 
We also see that the Lorentzian regime starts when the initial state loses its privileged position containing the main 
probability and, therefore, when the neighboring levels begin to have an important fraction of the total probability.

\section{Spin dynamics} \label{time}

In this section, we study the different relaxation regimes of the spin in the spin-GORM model. 
This study relies on a detailed comparison between the numerical exact dynamics 
(subsection \ref{numresspinGORM}) and the theoretical predictions in the weak-coupling limit 
(subsection \ref{weakcoupling}) and in the strong-coupling limit (subsection \ref{strongcoupling}).
The important question of the dynamics of a statistical ensemble versus the dynamics of a single element
of the ensemble is investigated in subsection \ref{statvsind}.\\

The exact spin dynamics in the spin-GORM model is obtained from the von Neumann equation (\ref{VNeqspinGORM}) 
using the reduced density matrix $\hat{\rho}_S(t)=\textrm{Tr}_{B} \hat{\rho}(t)$. 
In this study, the total system has always a given finite and constant energy. 
We also always consider initial conditions for which the subsystem is in any state and the environment 
is in a microcanonical state at a given fixed energy $\epsilon$ corresponding to a uniform distribution 
inside an energy shell centered at $\epsilon$ and of width $\delta \varepsilon$. 
$\delta \varepsilon$ is chosen in such a way that it is large enough to contains many levels (to get a good statistics)
and small enough to be smaller than or equal to the typical energy scale of variation of the averaged smoothed density 
of states of the environment $\delta \epsilon$. 
Therefore, the adequate choice for the width of the energy shell corresponds to 
$\frac{1}{N} < \delta \varepsilon < \delta \epsilon$. 
The dynamics is then averaged over the $\chi$ realizations of the random-matrix ensemble.

\subsection{The weak-coupling regime ($\lambda \ll 1$)} \label{weakcoupling}

In this subsection, we make the theoretical study of the dynamics of the spin, weakly interacting with its 
random-matrix environment.
This means that we apply the weak-coupling theory that we developed in chapter \ref{ch3} to the spin-GORM model.\\  

We derived in section \ref{derivation} of chapter \ref{ch3} a kinetic equation to describe the evolution of a subsystem 
(with a discrete spectrum) weakly interacting with its environment (with a quasi-continuous spectrum) and
applied this equation in section \ref{twolevelgen} to a two-level subsystem interacting in a non-diagonal way with any
environment.
The spin-GORM model is in fact a specific case of the two-level subsystem model studied in section \ref{twolevelgen} 
where the environment operators are Gaussian orthogonal random matrices as defined in section \ref{model}.
The central quantity characterizing the environment operators in the weak-coupling kinetic theories is the correlation
function of the environment.
If one is able to obtain a simple expression for the correlation function of the environment in the spin-GORM model, all the results 
of section \ref{twolevelgen} apply to this chapter by replacing the correlation functions by their random-matrix expressions.\\
We therefore start by evaluating the correlation function of the environment and its Fourier transform in the spin-GORM model.
Using the random-matrix environment operators that we defined in section \ref{model} to write the spin-GORM Hamiltonian 
(\ref{hamiltonien}), we find that the Fourier transform of the environment correlation function defined by equation 
(\ref{AppAaaaj}) of appendix \ref{AppA} is given by
\begin{eqnarray}
\tilde{\alpha}(E_{b},\omega) 
&=& \sum_{b'} \delta(\omega + \omega_{bb'}) \vert B_{bb'} \vert^2                                             \label{4Baaaa}\\
&=& \delta(\omega) \vert B_{bb} \vert^2 + \sum_{b' \neq b} \delta(\omega + \omega_{bb'}) 
\vert B_{bb'} \vert^2 \nonumber 
\end{eqnarray}
where $\omega_{bb'}=(E_b-E_{b'})/\hbar$.
Averaging the Fourier transform of the environment correlation function over the random-matrix ensemble, we get
\begin{eqnarray}
\tilde{\alpha}(E_{b},\omega) =
\delta(\omega) \frac{1}{4 N} + \sum_{b' \neq b} \hbar \delta(\hbar \omega + E_b - E_{b'}) \frac{1}{8 N} .      \label{4Baaab}
\end{eqnarray}
To simplify the notations, we will use the convention
\begin{equation}
\sqrt{x} \equiv \left\{
\begin{array}{ll}
\sqrt{x} & {\rm if} \quad 0 < x \\
0        & {\rm if} \quad x \leq 0 \; .
\end{array}
\right.                                                                                                         \label{sqrt}
\end{equation}
Assuming that the environment has a spectrum that is dense enough to perform the quasi-continuous limit
and using the smoothed density of states given by (\ref{wignerdistr}), we get
\begin{eqnarray}
\tilde{\alpha}(\epsilon,\omega) 
&=& \delta(\omega) \frac{1}{4 N} + \int d\epsilon' \frac{\hbar}{2 \pi} \sqrt{\frac{1}{4}-(\epsilon')^2} 
\delta(\hbar \omega + \epsilon - \epsilon')                                                                    \label{4Baaac}\\               
&=& \delta(\omega) \frac{1}{4 N} + \frac{\hbar}{2 \pi} \sqrt{\frac{1}{4}-(\epsilon+\hbar \omega)^2} .\nonumber        
\end{eqnarray}
This means that
\begin{center} \fbox{\parbox{12.5cm}{
\begin{eqnarray}
\tilde{\alpha}(\epsilon,\omega) 
\stackrel{N \to \infty}{=} \frac{\hbar}{2 \pi} \sqrt{\frac{1}{4}-(\epsilon+\hbar \omega)^2}  .                  \label{4Baaad}
\end{eqnarray}
}} \end{center}
The environment correlation function can now be obtained by taking the inverse Fourier transform of (\ref{4Baaad})
\begin{eqnarray}
\alpha(\epsilon,t)                         
&=& \int d\omega e^{- i \omega t}  \tilde{\alpha}(\epsilon,\omega)                                               \label{4Baaae}\\
&\stackrel{N \to \infty}{=}& \int d\epsilon' \frac{1}{2 \pi} \sqrt{\frac{1}{4}-(\epsilon')^2} 
e^{- i \epsilon' t / \hbar} e^{i \epsilon t / \hbar} \nonumber .
\end{eqnarray}
Performing the integral we get 
\begin{center} \fbox{\parbox{12.5cm}{
\begin{eqnarray}
\alpha(\epsilon,t) 
\stackrel{N \to \infty}{=} \frac{J_1(t/(2 \hbar))}{4 t/\hbar} e^{i \epsilon t / \hbar} .                        \label{4Baaaf}
\end{eqnarray}
}} \end{center}
$J_1(z)$ is the Bessel function of the first kind.
We insist on the fact that these two results suppose that the environment is quasi-continuous ($N \to \infty$)
and that the random-matrix ensemble average has been performed ($\chi \to \infty$).\\
Given these fundamental quantities, the results of section \ref{twolevelgen} directly apply to our
present study of the spin-GORM model.
The non-Markovian dynamics of the subsystem is given by (\ref{4Aaaah}) using (\ref{4Aaaad}) and (\ref{4Aaaae}).
The order $\lambda$ term is neglected because in the spin-GORM model the random-matrix ensemble average of 
$\bra{\epsilon} \hat{B} \ket{\epsilon}$ vanishes. 
The non-Markovian population dynamics become
\begin{center} \fbox{\parbox{12.5cm}{
\begin{eqnarray}
z(t) &=& \int d\epsilon' \left[ P_{++}(\epsilon',t)-P_{--}(\epsilon'+\Delta,t) \right] ,                 \label{zdefpauligen1}\\
x(t) &=& \int d\epsilon' \left[ P_{+-}(\epsilon',t)+P_{-+}(\epsilon',t) \right] ,\nonumber \\
y(t) &=& i \int d\epsilon' \left[ P_{+-}(\epsilon',t)-P_{-+}(\epsilon',t) \right] ,\nonumber
\end{eqnarray}
}} \end{center}
where the quantities $P_{s s'}(\epsilon,t)$ obey the equations
\begin{eqnarray}
\dot{P}_{\pm \pm}(\epsilon,t) &=& \frac{\lambda^2}{\pi \hbar} \int \;
d\omega \frac{\sin(\pm \Delta+\hbar \omega)t}{(\pm \Delta+\hbar \omega)}                                 \label{paulispingoeNM11}\\
&&\hspace*{1cm} \left[ P_{\mp \mp}(\epsilon-\hbar \omega,t) \sqrt{\frac{1}{4}-{\epsilon}^2} - 
P_{\pm \pm}(\epsilon,t) \sqrt{\frac{1}{4}-(\epsilon-\hbar \omega)^2} \right] \nonumber                                           
\end{eqnarray}
and
\begin{eqnarray}
\dot{P}_{\pm \mp}(\epsilon,t)
&=&\mp \frac{i}{\hbar} \Delta P_{\pm \mp}(\epsilon,t)                    
+ \frac{\lambda^2}{2 \pi \hbar} \int^{t}_{0} d\tau \; \{ \;                                             \label{paulispingoeNMcoh}\\
&&\hspace*{-0.2cm}- \int d\omega P_{\pm \mp}(\epsilon,t) 
\sqrt{\frac{1}{4}-(\epsilon-\hbar \omega)^2} (e^{\frac{i}{\hbar} (\pm \Delta-\hbar \omega) \tau} 
+e^{\frac{i}{\hbar} (\pm \Delta+\hbar \omega) \tau}) \nonumber \\                        
&&\hspace*{-0.2cm}+ \int d\omega P_{\mp \pm}(\epsilon-\hbar \omega,t) 
\sqrt{\frac{1}{4}-{\epsilon}^2}
(e^{-\frac{i}{\hbar} (\pm \Delta-\hbar \omega) \tau}
+e^{-\frac{i}{\hbar} (\pm \Delta+\hbar \omega) \tau}) \; \} \nonumber .
\end{eqnarray}
This non-Markovian spin dynamics can be computed numerically. 
In the following we will refer to this equation [(\ref{zdefpauligen1}) with (\ref{paulispingoeNM11}) and 
(\ref{paulispingoeNMcoh})] as "{\bf new NM}" which stands for "our new non-Markovian equation".\\
Notice that the application of the non-Markovian version of the Redfield equation 
[(\ref{AppCaaac}) with (\ref{AppCaaaa}) and (\ref{AppCaaab}) in appendix \ref{AppC}] to the 
spin-GORM model is performed in the same way by the replacement of the Fourier tranform of the 
environment correlation function by its expression (\ref{4Baaac}) and by neglecting the 
order-$\lambda$ term which disappears due to the random-matrix average. 
In the following we will refer to this equation as "{\bf Redfield NM}" which stands for 
"non-Markovian Redfield equation".\\

We now focus on the Markovian spin dynamics. 
We can again simply apply the results of section \ref{twolevelgen} to 
the spin-GORM model.
The analytical expression (\ref{4Aaaar}) can be used if one evaluates the spin-GORM expression 
of the different environment-dependent quantities $\gamma_{\rm pop}(\epsilon)$, 
$\gamma_{\rm coh}(\epsilon)$, $\Gamma(\epsilon)$ and $Z(\epsilon,\infty)$.
This can be done easily because all these quantities depend on the Fourier transform of the environment 
correlation function which is now given by (\ref{4Baaad}).
Therefore, using (\ref{4Aaaar}) with (\ref{4Aaaap}) and (\ref{4Aaaao}), the final solution of our new 
Markovian equation, to which we later will refer to as "{\bf new M}", can be written  
\begin{center} \fbox{\parbox{12.5cm}{
\begin{eqnarray}                                                                            
z(t)&=& \frac{z(0)+1}{2} \left[ z(\epsilon,\infty)+(1 + z(\epsilon,\infty)) 
e^{- \gamma_{\rm pop}(\epsilon) t} \right]                                                               \label{eqnewM}\\ 
&&+\frac{z(0)-1}{2} \left[ z(\epsilon-\Delta,\infty)
+(1 + z(\epsilon-\Delta,\infty))  
e^{- \gamma_{\rm pop}(\epsilon-\Delta) t} \right] \nonumber \\ 
x(t) &\approx& \left[ x(0) \cos(\Delta/\hbar+\Gamma(\epsilon))t 
- y(0) \sin(\Delta/\hbar+\Gamma(\epsilon))t \right]
e^{-\gamma_{\rm coh}(\epsilon) t} \nonumber \\ 
y(t) &\approx& \left[ x(0) \sin(\Delta/\hbar+\Gamma(\epsilon))t 
+ y(0) \cos(\Delta/\hbar+\Gamma(\epsilon))t \right]
e^{-\gamma_{\rm coh}(\epsilon) t} \nonumber ,
\end{eqnarray}
}} \end{center}
where the two population relaxation rates are given by 
\begin{eqnarray}
\gamma_{\rm pop}(\epsilon) &=& 2 \pi \frac{\lambda^2}{\hbar^2}  
(\tilde{\alpha}(\epsilon, \Delta / \hbar) + 
\tilde{\alpha}(\epsilon + \Delta,- \Delta / \hbar)) \nonumber \\
&=& \frac{\lambda^2}{\hbar} \left[ \sqrt{\frac{1}{4}-(\epsilon+\Delta)^2} +
\sqrt{\frac{1}{4}-\epsilon^2} \right]                                                                     \label{Ch4aaaaa}
\end{eqnarray}
and
\begin{eqnarray}
\gamma_{\rm pop}(\epsilon-\Delta) &=& 2 \pi \frac{\lambda^2}{\hbar^2}  
(\tilde{\alpha}(\epsilon-\Delta, \Delta / \hbar) + 
\tilde{\alpha}(\epsilon,- \Delta / \hbar)) \nonumber \\
&=& \frac{\lambda^2}{\hbar} \left[ \sqrt{\frac{1}{4}-\epsilon^2} +
\sqrt{\frac{1}{4}-(\epsilon-\Delta)^2} \right],                                                          \label{Ch4aaaab}
\end{eqnarray}
while the coherence relaxation rate is
\begin{eqnarray}
\gamma_{\rm coh}(\epsilon) &=& \pi \frac{\lambda^2}{\hbar^2}  
(\tilde{\alpha}(\epsilon,-\Delta / \hbar)+\tilde{\alpha}(\epsilon,\Delta / \hbar)) \nonumber \\ 
&=& \frac{\lambda^2}{2 \hbar} \left[ \sqrt{\frac{1}{4}-(\epsilon-\Delta)^2} +
\sqrt{\frac{1}{4}-(\epsilon+\Delta)^2} \right] ,                                                         \label{Ch4aaaac}
\end{eqnarray}
the Lamb shift affecting the coherence frequencies
\begin{eqnarray}
\Gamma(\epsilon) &=& 2 \frac{\Delta \lambda^2}{\hbar} \int d\omega {\cal P} 
\frac{\tilde{\alpha}(\epsilon,\omega)}{\Delta^2-(\hbar \omega)^2} \nonumber \\
&=& \frac{\Delta \lambda^2}{\hbar \pi} \int d\epsilon' {\cal P}
\frac{\sqrt{\frac{1}{4}-{\epsilon'}^2}}{\Delta^2-(\epsilon'-\epsilon)^2} ,                               \label{Ch4aaaad}
\end{eqnarray}
which after solving the integral gives
\begin{eqnarray} 
\Gamma(\epsilon) = \lambda^2 \Delta 
&-& \lambda^2 \frac{\sqrt{(\epsilon+\Delta)^2-\frac{1}{4}}}{\pi} 
\arctan \left( \frac{(\epsilon+\Delta)+\frac{1}{2}}
{\sqrt{(\epsilon+\Delta)^2-\frac{1}{4}}} \right) \nonumber \\
&-& \lambda^2 \frac{\sqrt{(\epsilon+\Delta)^2-\frac{1}{4}}}{\pi}
\arctan \left( \frac{(\epsilon+\Delta)-\frac{1}{2}}
{\sqrt{(\epsilon+\Delta)^2-\frac{1}{4}}} \right) \nonumber \\ 
&+& \lambda^2 \frac{\sqrt{(\epsilon-\Delta)^2-\frac{1}{4}}}{\pi}
\arctan \left( \frac{(\epsilon-\Delta)+\frac{1}{2}}
{\sqrt{(\epsilon-\Delta)^2-\frac{1}{4}}} \right) \nonumber \\
&+& \lambda^2 \frac{\sqrt{(\epsilon-\Delta)^2-\frac{1}{4}}}{\pi}
\arctan \left( \frac{(\epsilon-\Delta)-\frac{1}{2}}
{\sqrt{(\epsilon-\Delta)^2-\frac{1}{4}}} \right).                                                        \label{Ch4aaaae}
\end{eqnarray}
The two asymptotic population values are given by
\begin{eqnarray} 
z(\epsilon,\infty) 
&=& \frac{Z(\epsilon,\infty)}{C(\epsilon,0)} 
= \frac{n(\epsilon)-n(\epsilon+\Delta)}{n(\epsilon)+n(\epsilon+\Delta)} \nonumber \\
&=& \frac{\sqrt{\frac{1}{4}-\epsilon^2}-\sqrt{\frac{1}{4}-(\epsilon+\Delta)^2}}
{\sqrt{\frac{1}{4}-\epsilon^2}+\sqrt{\frac{1}{4}-(\epsilon+\Delta)^2}}                                   \label{Ch4aaaaf}
\end{eqnarray}
and
\begin{eqnarray} 
z(\epsilon-\Delta,\infty) 
&=& \frac{Z(\epsilon-\Delta,\infty)}{C(\epsilon-\Delta,0)}
= \frac{n(\epsilon-\Delta)-n(\epsilon)}{n(\epsilon-\Delta)+n(\epsilon)} \nonumber \\
&=& \frac{\sqrt{\frac{1}{4}-(\epsilon-\Delta)^2}-\sqrt{\frac{1}{4}-\epsilon^2}}
{\sqrt{\frac{1}{4}-(\epsilon-\Delta)^2}+\sqrt{\frac{1}{4}-\epsilon^2}}.                                  \label{Ch4aaaag}
\end{eqnarray}
Notice that the Redfield Markovian equation for the spin dynamics, which we will refer 
to as "{\bf Redfield M}", can be obtained in a similar way using (\ref{AppCaaag}), where 
the population as well as half of the coherence relaxation rate is given by
\begin{eqnarray} 
\gamma 
&=& 2 \pi \frac{\lambda^2}{\hbar^2} (\tilde{\alpha}(\epsilon,-\Delta/\hbar)
+\tilde{\alpha}(\epsilon,\Delta/\hbar)) \nonumber \\
&=& \frac{\lambda^2}{\hbar} \left[ \sqrt{\frac{1}{4}-(\epsilon-\Delta)^2} 
+ \sqrt{\frac{1}{4}-(\epsilon+\Delta)^2} \right] ,                                                         \label{Ch4aaaah}
\end{eqnarray}
the asymptotic value of the populations by
\begin{eqnarray}
z(\infty) 
&=& \frac{\tilde{\alpha}(\epsilon,-\Delta/\hbar)-\tilde{\alpha}(\epsilon,\Delta/\hbar)}
{\tilde{\alpha}(\epsilon,-\Delta/\hbar)+\tilde{\alpha}(\epsilon,\Delta/\hbar)} \nonumber \\
&=& \frac{\sqrt{\frac{1}{4}-(\epsilon-\Delta)^2}-\sqrt{\frac{1}{4}-(\epsilon+\Delta)^2}}
{\sqrt{\frac{1}{4}-(\epsilon-\Delta)^2}+\sqrt{\frac{1}{4}-(\epsilon+\Delta)^2}},                           \label{Ch4aaaai}
\end{eqnarray}
and where the Lamb shift $\Gamma$ is the same as for our new equation (\ref{Ch4aaaae}).\\
We recall that our new equation has been shown to be equivalent to the Redfield equation when the typical energy
scale of the system (typical energy spacing between the system levels) can be considered small compared to 
the typical energy scale of variation of the environment correlation function 
\footnote{The only restriction to this statement concerns the analytical expression for the Markovian coherence
dynamics which has been derived by making the non-local approximation as explained in section \ref{twolevelgen}. 
Because of the fact that this assumption is not satisfied when $\Delta \to 0$, the Markovian Redfield analytical 
expression has to be used in this specific case.}. 
In the spin-GORM model, this condition is reduced to $n(\epsilon \pm \Delta) \approx n(\epsilon)$.


\subsection{The strong-coupling regime ($\lambda \gg 1$)} \label{strongcoupling}

We are now interested in making a theoretical prediction for the spin dynamics when the coupling parameter 
$\lambda$ is very large and, therefore, dominant in front of $1$ and $\Delta$. \\

The dynamics of the subsystem is fully described by
\begin{eqnarray}
\theta(t)
&=& \textrm{Tr} \; \hat{\rho}(t) \hat{\sigma}_\theta                                                          \label{strongAaaaa}\\
&=&\textrm{Tr} \; e^{i (\frac{\Delta}{2}\hat{\sigma}_z+\hat{H}_{B}+\lambda\hat{\sigma}_x\hat{B}) t} 
\hat{\rho}(0) e^{-i (\frac{\Delta}{2}\hat{\sigma}_z+\hat{H}_{B}+\lambda\hat{\sigma}_x\hat{B})t} 
\hat{\sigma}_\theta, \nonumber
\end{eqnarray}
where $\theta=x,y,z$.
Using the following unitary transformation acting on the degrees of freedom of the spin
\begin{equation}
\hat{U}=
\left(\begin{array}{cc}
\frac{1}{\sqrt{2}} & \frac{1}{\sqrt{2}} \\
\frac{1}{\sqrt{2}} & -\frac{1}{\sqrt{2}}
\end{array} \right)                                                                                            \label{rhoSexplicite}
\end{equation}
we have
\begin{equation}
\hat{U}^{\dag} \hat{\sigma}_z \hat{U} = \hat{\sigma}_x \  \ , \  \
\hat{U}^{\dag} \hat{\sigma}_x \hat{U} = \hat{\sigma}_z \  \ ,  \  \
\hat{U}^{\dag} \hat{\sigma}_y \hat{U} = -\hat{\sigma}_y .                                                      \label{strongAaaab}
\end{equation}
We can therefore rewrite (\ref{strongAaaaa}) as
\begin{eqnarray}
\theta(t)
&=&\textrm{Tr} \; e^{i(\frac{\Delta}{2}\hat{\sigma}_x+\hat{H}_{B}+\lambda\hat{\sigma}_z\hat{B})t} 
\hat{U}^{\dag} \hat{\rho}(0) \hat{U}
e^{-i(\frac{\Delta}{2}\hat{\sigma}_x+\hat{H}_{B}+\lambda \hat{\sigma}_z \hat{B}) t}  
\hat{U}^{\dag} \hat{\sigma}_\theta \hat{U}                                                                     \label{strongAaaac}\\ 
&=&\textrm{Tr} \; e^{i \lambda (\frac{1}{\lambda}\frac{\Delta}{2}\hat{\sigma}_x + 
\frac{1}{\lambda} \hat{H}_{B} + \hat{\sigma}_z \hat{B}) t} \hat{U}^{\dag} \hat{\rho}(0) \hat{U} 
e^{-i \lambda (\frac{1}{\lambda} \frac{\Delta}{2} \hat{\sigma}_x 
+ \frac{1}{\lambda}\hat{H}_{B} +\hat{\sigma}_z \hat{B}) t} \hat{U}^{\dag} \hat{\sigma}_\theta \hat{U} \nonumber.
\end{eqnarray}
It is easy to check that 
\begin{equation}
\hat{U}^{\dag} \hat{\rho}_{S}(0) \hat{U}=\frac{1}{2}
\left(\begin{array}{cc}
1 + x(0) & z(0) + i y(0)  \\
z(0) - i y(0) & 1 - x(0)
\end{array} \right) .                                                                                           \label{strongAaaaf}
\end{equation}
Using the following perturbative expansion of the evolution operator to order zero in $\frac{1}{\lambda}$
\begin{eqnarray}
e^{-i \lambda (\frac{1}{\lambda} \frac{\Delta}{2}\hat{\sigma}_x +
\frac{1}{\lambda}\hat{H}_{B}+ \hat{\sigma}_z \hat{B}) t}
&\stackrel{\frac{1}{\lambda} \to 0}{=}& e^{-i \lambda t
\hat{\sigma}_z \hat{B}} + O\left(\frac{1}{\lambda}\right)+ O\left(\frac{\Delta}{\lambda}\right)                 \label{strongAaaad}
\end{eqnarray}
and using the fact that $\hat{\rho}(0)=\hat{\rho}_{S}(0) \hat{\rho}_{B}^{eq}$, we get
\begin{eqnarray}
\theta(t) \stackrel{\frac{1}{\lambda} \to 0}{=}
\textrm{Tr} \; e^{i \lambda t \hat{\sigma}_z \hat{B}} \hat{U}^{\dag} \hat{\rho}_{S}(0) \hat{U}
\hat{\rho}_{B}^{eq} e^{-i \lambda t \hat{\sigma}_z \hat{B}} \hat{U}^{\dag} \hat{\sigma}_\theta \hat{U}
+ O \left(\frac{1}{\lambda}\right) + O\left(\frac{\Delta}{\lambda}\right).                                     \label{strongAaaae}
\end{eqnarray}
Using the following notation
\begin{eqnarray}
\hat{\sigma}_z \hat{B} \vert \kappa \eta \rangle = \hat{\sigma}_z
\vert \kappa \rangle \hat{B} \vert \eta \rangle = \kappa E_{\eta} \vert \kappa \eta \rangle                     \label{strongAaaag}
\end{eqnarray}
and
\begin{eqnarray}
\hat{H}_B  \vert b \rangle =  E_b^{B} \vert b \rangle,                                                          \label{strongAaaah}
\end{eqnarray}
we find that
\begin{eqnarray}
\theta(t) &\stackrel{\frac{1}{\lambda} \to 0}{=}& 
\sum_{\kappa,\kappa',\eta} e^{i \lambda t (\kappa-\kappa') E_{\eta} } 
\langle \eta \vert \hat{\rho}_B^{eq} \vert \eta \rangle 
\langle \kappa \vert \hat{U}^{\dag} \hat{\rho}_{S}(0) \hat{U} \vert \kappa' \rangle
\langle \kappa' \vert \hat{U}^{\dag} \hat{\sigma}_\theta \hat{U} \vert \kappa \rangle \nonumber\\
&&+ O\left(\frac{1}{\lambda}\right) + O\left(\frac{\Delta}{\lambda}\right).                                     \label{szprov1}
\end{eqnarray}
Because $\hat{\rho}_B$ is diagonal in the basis that diagonalizes $\hat{H}_{B}$, we have
\begin{eqnarray}
\langle \eta \vert \hat{\rho}_B(0) \vert \eta \rangle = 
\sum_{b} \vert \langle \eta \vert b \rangle \vert^2 \langle b \vert \hat{\rho}_B \vert b \rangle.               \label{strongAaaai}
\end{eqnarray}
If we perform an ensemble average over different realizations of $\hat{H}_{B}$ and use the statistics 
of the random-matrix eigenvectors \cite{Brody81,Mehta67,Porter65}, we find
\begin{eqnarray}
\overline{\langle \eta \vert \hat{\rho}_B(0) \vert \eta \rangle}^{(\hat{H}_{B})} 
&=& \sum_{b} \overline{\vert \langle \eta \vert b \rangle \vert^2}^{(\hat{H}_{B})} 
\overline{\langle b \vert \hat{\rho}_B \vert b \rangle}^{(\hat{H}_{B})} \nonumber \\
&=&\sum_{b} \frac{2}{N} \overline{\langle b \vert \hat{\rho}_B \vert b \rangle}^{(\hat{H}_{B})} = \frac{2}{N}.  \label{strongAaaaj}
\end{eqnarray}
Equation (\ref{szprov1}) therefore becomes
\begin{eqnarray}
\overline{\theta(t)}^{(\hat{H}_{B})} &\stackrel{\frac{1}{\lambda} \to 0}{=}&
\frac{2}{N} \sum_{\kappa,\kappa',\eta} e^{i \lambda t (\kappa-\kappa') E_{\eta} } 
\langle \kappa \vert \hat{U}^{\dag} \hat{\rho}_{S}(0) \hat{U} \vert \kappa' \rangle
\langle \kappa' \vert \hat{U}^{\dag} \hat{\sigma}_\theta \hat{U} \vert \kappa \rangle \nonumber\\
&&+ O\left(\frac{\Delta}{\lambda}\right) + O\left(\frac{1}{\lambda^2}\right).                                   \label{strongAaaak}
\end{eqnarray}
Using the relations (\ref{strongAaaab}), we therefore have that
\begin{eqnarray}
\overline{z(t)}^{(\hat{H}_{B})} &\stackrel{\frac{1}{\lambda} \to 0}{=}&
\frac{2}{N} \sum_{\kappa,\eta} e^{2 \kappa i \lambda t E_{\eta} } 
\langle \kappa \vert \hat{U}^{\dag} \hat{\rho}_{S}(0) \hat{U} \vert -\kappa \rangle 
+ O\left(\frac{\Delta}{\lambda}\right) + O\left(\frac{1}{\lambda^2}\right)                                     \label{strongAaaal}\\
\overline{x(t)}^{(\hat{H}_{B})} &\stackrel{\frac{1}{\lambda} \to 0}{=}&
\frac{2}{N} \sum_{\kappa,\eta} \kappa \;
\langle \kappa \vert \hat{U}^{\dag} \hat{\rho}_{S}(0) \hat{U} \vert \kappa \rangle 
+ O\left(\frac{\Delta}{\lambda}\right) + O\left(\frac{1}{\lambda^2}\right) \nonumber\\
\overline{y(t)}^{(\hat{H}_{B})} &\stackrel{\frac{1}{\lambda} \to 0}{=}&
- i \frac{2}{N} \sum_{\kappa,\eta} \kappa \; e^{2 \kappa i \lambda t E_{\eta} }
\langle \kappa \vert \hat{U}^{\dag} \hat{\rho}_{S}(0) \hat{U} \vert -\kappa \rangle 
+ O\left(\frac{\Delta}{\lambda}\right) + O\left(\frac{1}{\lambda^2}\right) .\nonumber
\end{eqnarray}
We now perform the following ensemble average over different realizations of $\hat{B}$
\begin{eqnarray}
\overline{\sum_{\eta} e^{\pm 2 i \lambda t E_{\eta}}}^{(\hat{B})} 
&=&\int_{-\frac{1}{2}}^{+\frac{1}{2}} d\epsilon \; n(\epsilon) e^{\pm 2 i \lambda t \epsilon} \nonumber \\ 
&=&\frac{4N}{\pi} \int_{-\frac{1}{2}}^{+\frac{1}{2}} d\epsilon \; \sqrt{\frac{1}{4}-\epsilon^2} \; 
e^{\pm 2 i \lambda t \epsilon} \nonumber\\ 
&=& N \frac{J_{1}(\lambda t)}{\lambda t},
\end{eqnarray}
where $J_{1}(t)$ is the Bessel function of the first kind and where we have used the fact that 
$J_1(t)$ is an odd function. 
Performing this ensemble average over (\ref{strongAaaal}) and using (\ref{strongAaaaf}), 
we finally find
\begin{center} \fbox{\parbox{12.5cm}{
\begin{eqnarray}
z(t) &\stackrel{\frac{1}{\lambda} \to 0}{=}& 2 \frac{J_{1}(\lambda t)}{\lambda t} z(0)                     \label{Besstlsurtl}\\
x(t) &\stackrel{\frac{1}{\lambda} \to 0}{=}& x(0) \nonumber\\
y(t) &\stackrel{\frac{1}{\lambda} \to 0}{=}& 2 \frac{J_{1}(\lambda t)}{\lambda t} y(0) . \nonumber
\end{eqnarray}
}} \end{center}
These equations are valid up to corrections $O\left(\frac{\Delta}{\lambda}\right)$ and $O\left(\frac{1}{\lambda^2}\right)$,
and are averaged over the $\hat{H}_{B}$ and the $\hat{B}$ realizations even if we have dropped the explicit notation.
We have therefore found a well-defined behavior of the system dynamics when the coupling parameter $\lambda$ is so large
that the coupling term can be considered to contribute alone to the dynamics.
We have again supposed that the environment is continuous ($N \to \infty$) and that the random-matrix 
ensemble averages have been performed ($\chi \to \infty$).

\subsection{Numerical results} \label{numresspinGORM}

We will now present the numerical results on the dynamical evolution of the spin in the spin-GORM model.
This will give us the opportunity to compare the theoretical predictions made in the two preceding subsections 
\ref{weakcoupling} and \ref{strongcoupling} with the exact dynamics obtained numerically.\\

\begin{figure}[p]
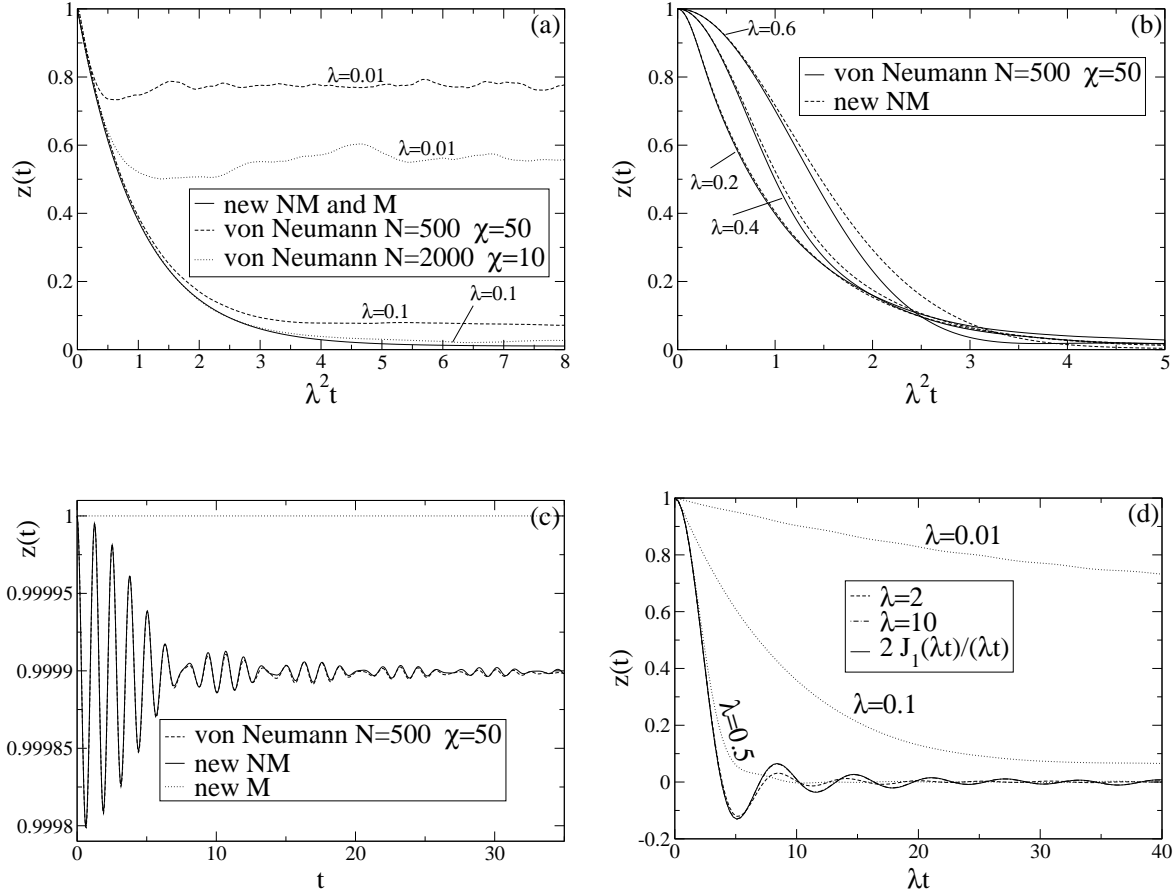

\centering
\hspace*{-2cm}
\begin{tabular}{c@{\hspace{0.5cm}}c}
\vspace*{1cm}
\rotatebox{0}{\scalebox{0.3}{\includegraphics{fig28.eps}}}
&
\rotatebox{0}{\scalebox{0.3}{\includegraphics{fig29.eps}}}
\\ 
\rotatebox{0}{\scalebox{0.3}{\includegraphics{fig30.eps}}}
&
\rotatebox{0}{\scalebox{0.3}{\includegraphics{fig31.eps}}} \\
\end{tabular}
\caption{Time evolution of the $z$ component of the spin. 
In all figures, $\chi=50$, $\epsilon=0$ and the von Neumann equations have been computed 
with $\delta \varepsilon=0.05$.
(a), (b), and (c) Comparison between the exact von Neumann equation and the non-Markovian 
version of our new equation ("new NM") for different values of the spin-GORM model parameters.
(a) and (b) correspond to $\Delta=0.1$ and (c) to $\Delta=5$ and $\lambda=0.1$.
(d) Comparison for $\Delta=0.01$ between the exact von Neumann equation with $N=500$ and 
$\chi=50$ and the Bessel strong-coupling result given by equation (\ref{Besstlsurtl}).}                 \label{differents regimes}
\end{figure}
The different parameter domains represented in figure \ref{regimeschema} will play an important 
role in our discussion on the dynamics.
\subsubsection{Weak-coupling (domain A)}
The most important point we want to investigate is the validity of our new weak-coupling kinetic theory. 
To do this, we will now compare the non-Markovian version of our new equation with the numerical exact von Neumann equation, 
staying in the parameter domain A (see figure \ref{regimeschema}) which corresponds to the weak-coupling regime. 
Figures \ref{differents regimes}(a) and \ref{differents regimes}(b) show the time evolution described by both 
equations at different $\lambda$ values using a $\lambda^2 t$ time scaling. 
The very important result is that the non-Markovian equation is valid not only for values of $\lambda$ below an 
upper bound, but also above a lower bound. 
The lower bound of the coupling parameter corresponds in our spectral analysis to the border between the localized 
and the Lorentzian regime of the SOE.
We recall that this bound corresponds to a value of the coupling parameter for which the first nonzero correction to 
the unperturbed energy levels of the total system becomes of the order of the mean level spacing 
between these coupled levels ($\lambda=\sqrt{1/N})$. 
This bound however disappears in the limit of a continuous environment spectrum ($N \to \infty$).
The upper bound corresponds to the transition between the Lorentzian $1$ regime and the Lorentzian $2$ regime 
(see figure \ref{scheml2N}).
This bound is located to a value of the coupling parameter for which the width of the SOE in the Lorentzian
regime becomes larger than the typical energy scale over which the density of states of the environment varies. 
In figure \ref{differents regimes}(a), we can see that when $\lambda$ is smaller than the upper bound 
(localized regime of the SOE), the non-Markovian version of our perturbative equation reproduces the initial dynamics 
but not the total relaxation process to the equilibrium value. 
One can see in figure \ref{differents regimes}(b) that our perturbative equation again looses its validity when $\lambda$ 
cannot be considered as a perturbation anymore. 
\subsubsection{Non-Markovian vs. Markovian dynamics}
We have often used the Markovian approximation in the various derivations of our new weak-coupling master 
equation presented in this thesis.
We argued that this approximation was justified on time scales longer than the characteristic time scale 
of the environment $\tau_c$ (which corresponds to the typical decay time of the environment correlation function), 
i.e., if $\tau_c$ is much smaller than the characteristic time scale of the subsystem $\tau_s$.
In the spin-GORM model, the characteristic time scale of the environment given by (\ref{4Baaaf}) is of 
the order $\tau_b = 2 \pi$ and the characteristic time scale of the spin dynamics is given by $\tau_s = 2 \pi / \Delta$.
It is interesting to test this prediction by numerically comparing the Markovian version ("new M") with 
the non-Markovian version of our equation ("new NM"). 
Notice that, using the $\lambda^2 t$ time scaling, the Markovian equation is independent of $\lambda$. 
The discrepancy between the non-Markovian and the Markovian equation is therefore a direct measure of the 
non-Markovian effects.
The curves of figures \ref{differents regimes}(a) and \ref{differents regimes}(b), for which $\tau_b < \tau_s$, 
confirm our theoretical predictions.
In fact, in figure \ref{differents regimes}(a), the Markovian and the non-Markovian versions of our new equation 
are so close that we only plotted the second one.
This is expected because the discrepancy is predicted to occur on a time scale $2 \pi \lambda^2$ which is here too 
short to be noticed.
However when $\lambda$ increases on the curves of figure \ref{differents regimes}(b), we see that the non-Markovian 
initial dynamics increases due to the increase of the predicted discrepancy region $2 \pi \lambda^2$.
\subsubsection{Pure non-Markovian dynamics (domain B)}
In the parameter domain B (see figure \ref{regimeschema}), we are in the situation where $\tau_b < \tau_s$.
Here, the Markovian version of our equation is predicted to be never valid and only the non-Markovian version can be used.
This situation is shown in figure \ref{differents regimes}(c), where we compare the Markovian and the non-Markovian versions 
of our perturbative equation with the numerical exact dynamics given by the von Neumann equation.
As predicted, the Markovian version of our equation does not reproduce at all the dynamics. 
But the non-Markovian equation reproduces this behavior with a very high accuracy. 
The frequency of the oscillations corresponds to the subsystem frequency $\tau_s = 0.4 \pi$ and the damping of these 
oscillations occurs on a time scale corresponding to the environment characteristic time scale $\tau_b = 2 \pi$. 
Notice that this pure non-Markovian dynamics scales in time according to $t$.
\subsubsection{Strong-coupling (domain C)}
We now want to verify our theoretical prediction of the spin dynamics in the strong-coupling regime corresponding to the 
parameter domain C (see figure \ref{regimeschema}).
We see in figure \ref{differents regimes}(d) that when $\lambda$ becomes large enough for the coupling term to dominate 
the total Hamiltonian, the dynamics obeys our predicted Bessel-function behavior derived in equation (\ref{Besstlsurtl}). 
It is important to notice that this behavior scales in time according to $\lambda t$.
\subsubsection{Summary of the different dynamical regimes}
\begin{figure}[p]
\centering
\rotatebox{0}{\scalebox{1.5}{\includegraphics{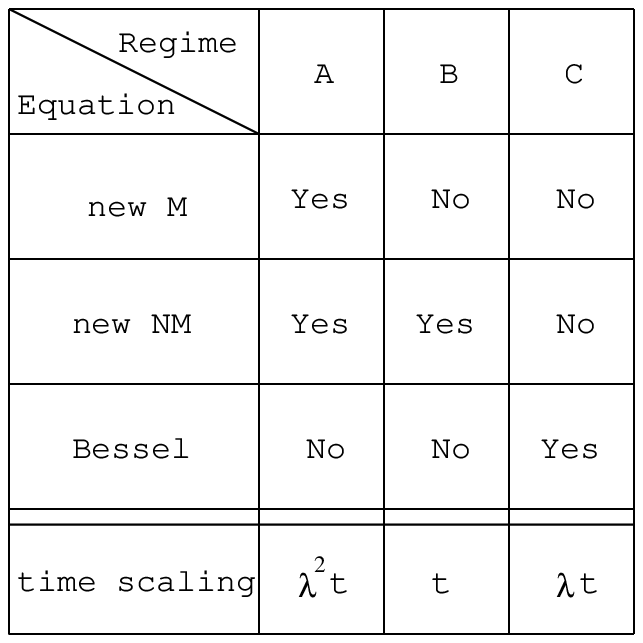}}} \\
\caption{Table summarizing the different types of spin relaxation that one can encounter in the spin-GORM model with 
their respective valid theory and time scaling.
The parameter domains are defined in figure \ref{regimeschema}. 
"new M" and "new NM" are the Markovian version and the non-Markovian version of our new weak-coupling kinetic equation,
respectively, and "Bessel" refers to the strong-coupling equation (\ref{Besstlsurtl}).} 
\label{valideq}
\end{figure}

We summarized in figure \ref{valideq} the different types of spin relaxation that one can encounter in the spin-GORM model. 
These different types of spin relaxation correspond to the different parameter domains A, B and C.
Each of these different relaxations has a characteristic time scaling. 
For each of them we have found a valid theoretical description. 
%
\subsubsection{Our new equation vs. Redfield equation} 
\begin{figure}[p]
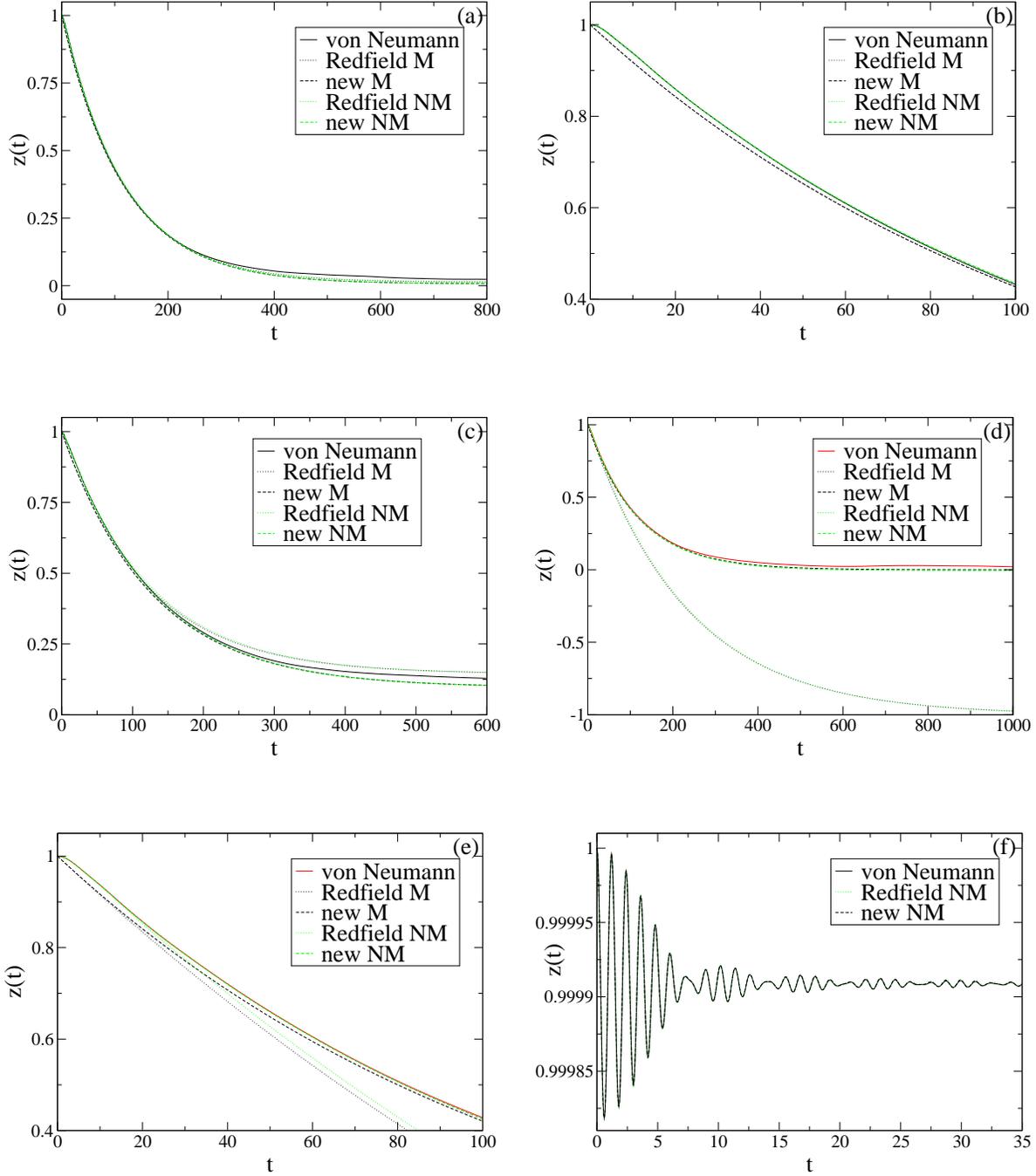

\vspace*{1cm}
\centering
\hspace*{-2cm}
\begin{tabular}{c@{\hspace{0.5cm}}c}
\vspace*{1cm}
\rotatebox{0}{\scalebox{0.3}{\includegraphics{Afig5.eps}}}
&
\rotatebox{0}{\scalebox{0.3}{\includegraphics{Afig6.eps}}}
\\ \vspace*{1cm}
\rotatebox{0}{\scalebox{0.3}{\includegraphics{Afig7.eps}}}
&
\rotatebox{0}{\scalebox{0.3}{\includegraphics{Afig8.eps}}}
\\ 
\rotatebox{0}{\scalebox{0.3}{\includegraphics{Afig9.eps}}}
&
\rotatebox{0}{\scalebox{0.3}{\includegraphics{Afig10.eps}}}
\\
\end{tabular}
\caption{Comparison between the exact von Neumann equation ("von Neumann"), the Markovian and the non-Markovian 
versions of our new equation ("new M" and "new NM",respectively) and the Markovian and the non-Markovian versions 
of the Redfield equation ("Redfield M" and "Refield NM",respectively).
Everywhere $\lambda=0.1$ and for the von Neumann equations $N=2000$, $\chi=10$ and $\delta \varepsilon=0.05$.
For (a) and (b): $\Delta=0.01$, $\epsilon=0.25$.
For (c): $\Delta=0.1$, $\epsilon=0.25$.
For (d) and (e): $\Delta=0.5$, $\epsilon=-0.25$.
For (f): $\Delta=5$, $\epsilon=0.25$.}                                                                         \label{redvspauli}
\end{figure}
We have motivated the extension of the Redfield theory to our new weak-coupling kinetic theory by the fact that when the
system energy becomes larger than the typical energy scale of variation of the density of states of the environment, the
energy conservation of the total system has to be taken into account. 
We now want to verify this statement by comparing the Redfield equation predictions (Markovian and non-Markovian) for the 
spin population relaxation in the spin-GORM model with the exact numerical results and with the prediction of our 
new master equation (Markovian and non-Markovian). 
We place ourself in the parameter domain A, where the weak-coupling applies, and choose an environment with a number 
of states such that the weak-coupling kinetic description is satisfied ($\lambda > \sqrt{1/N}$).
Figures \ref{redvspauli}(a), \ref{redvspauli}(c) and \ref{redvspauli}(d) depict the global relaxation of the spin 
population for increasing values of the energy splitting $\Delta$.
In accordance with what we argued on theoretical grounds in chapter \ref{ch3}, we see in these figures 
that the larger the energy splitting $\Delta$ of the system is, the bigger is the difference between the Redfield 
equation and our master equation. 
We also see that our equation always fits very well with the exact von Neumann equation, which is not the case for 
the Redfield equation. 
\begin{figure}[h]
\centering
\vspace*{+0.0cm}
\hspace*{-0.5cm}
\rotatebox{0}{\scalebox{0.54}{\includegraphics{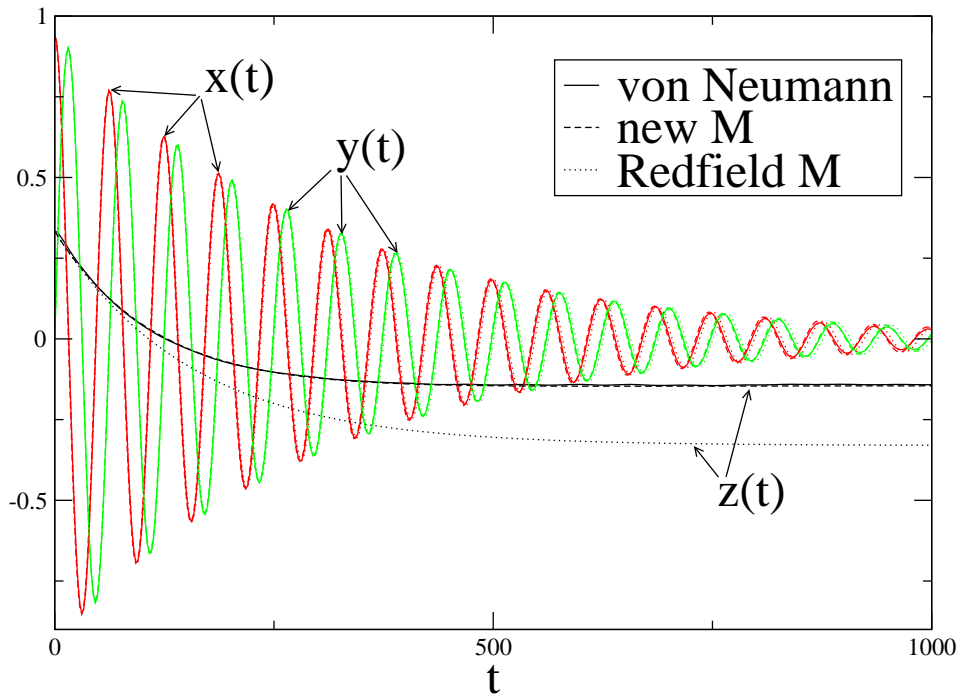}}} 
\caption{Population and coherence dynamics in the spin-GORM model for $\Delta=0.1$, $\epsilon=-0.35$, and $\lambda=0.1$.
We compare the exact von Neumann equation ("von Neumann") computed with $N=2000$, $\chi=10$, and $\varepsilon=0.05$ with 
the Markovian version of our new equation ("new M") as well as with the Markovian version of the Redfield equation 
("Redfield M").}                                                                                                 \label{cohpluspop}
\end{figure}
These results are a really strong support for the importance of our new theory. 
As expected and as already discussed, we see in figures \ref{redvspauli}(a) and \ref{redvspauli}(c)-\ref{redvspauli}(d) 
that the Markovian equations are, on time scales longer than $\tau_c=2\pi$, really close to the non-Markovian equations.
An interesting result can be deduced by comparing the non-Markovian results of the Redfield equation
with the non-Markovian results of our new equation in figures \ref{redvspauli}(b) and \ref{redvspauli}(e)-\ref{redvspauli}(f).
In fact, on a shorter time scale of the order $\tau_c = 2\pi$, there is no difference between the Redfield and our new equation,
and both give correct results.
The discrepancy between the Redfield equation and the exact dynamics when $\Delta$ is large enough, only occurs on longer time scales. 
This can be interpreted by the fact that, in quantum mechanics, energy conservation only takes place on long time scales
\footnote{This fact can be easily seen on the Pauli equation.}. 
This is particularly explicit in figure \ref{redvspauli}(f) where the relaxation of the spin population occurs in a pure non-Markovian 
regime (the Markovian version of the Redfield equation as well as our new equation predict no evolution in this case). 
We finish our comparison between the Redfield theory and our new weak-coupling kinetic theory by showing in figure \ref{cohpluspop}
the dynamics of the population as well as of the coherences of the spin.
As predicted in section \ref{weakcoupling}, the Redfield coherences dynamics is really close to the one predicted by our new equation.
They only slightly differ by their Lamb shift.
However, our new equation reproduces the exact dynamics with a higher accuracy than the Redfield equation. 
\subsubsection{Energy-distributed populations and coherences}
\begin{figure}[p]
\centering
\hspace*{-3.3cm}
\begin{tabular}{c@{\hspace{0.1cm}}c}
\vspace*{1cm}
\rotatebox{0}{\scalebox{0.9}{\includegraphics{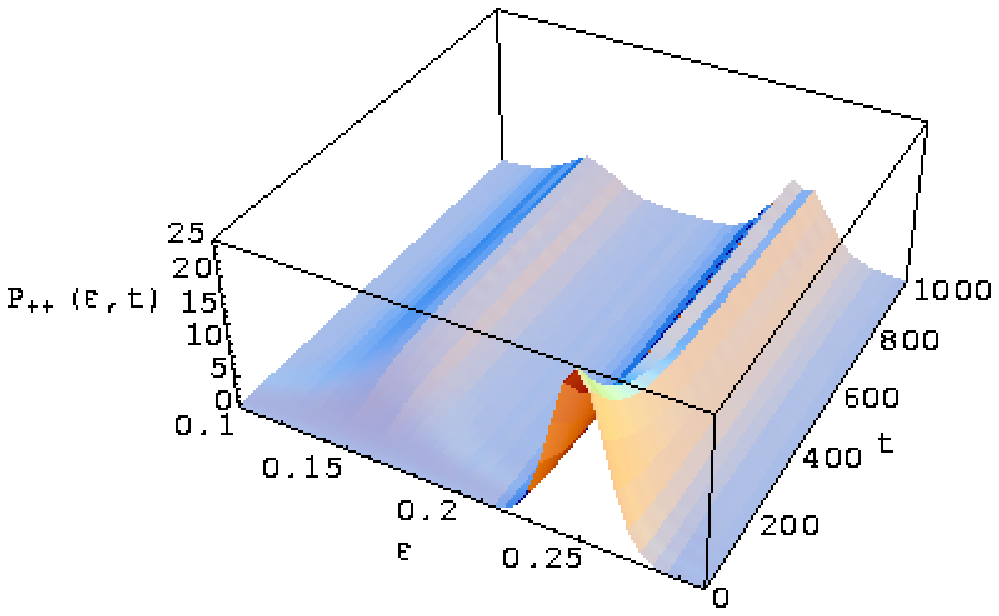}}}
&
\rotatebox{0}{\scalebox{0.9}{\includegraphics{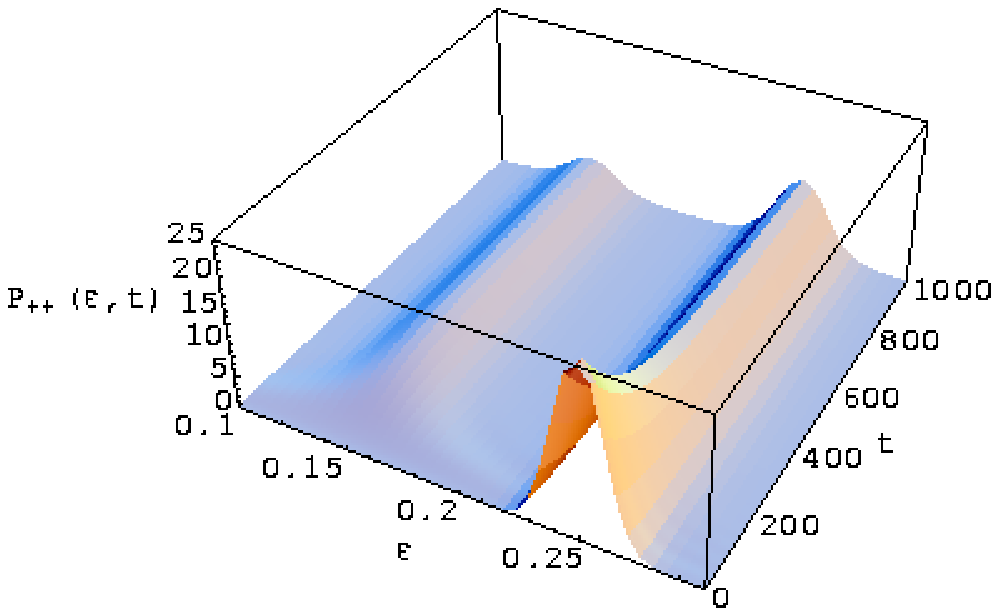}}}
\\
\rotatebox{0}{\scalebox{0.9}{\includegraphics{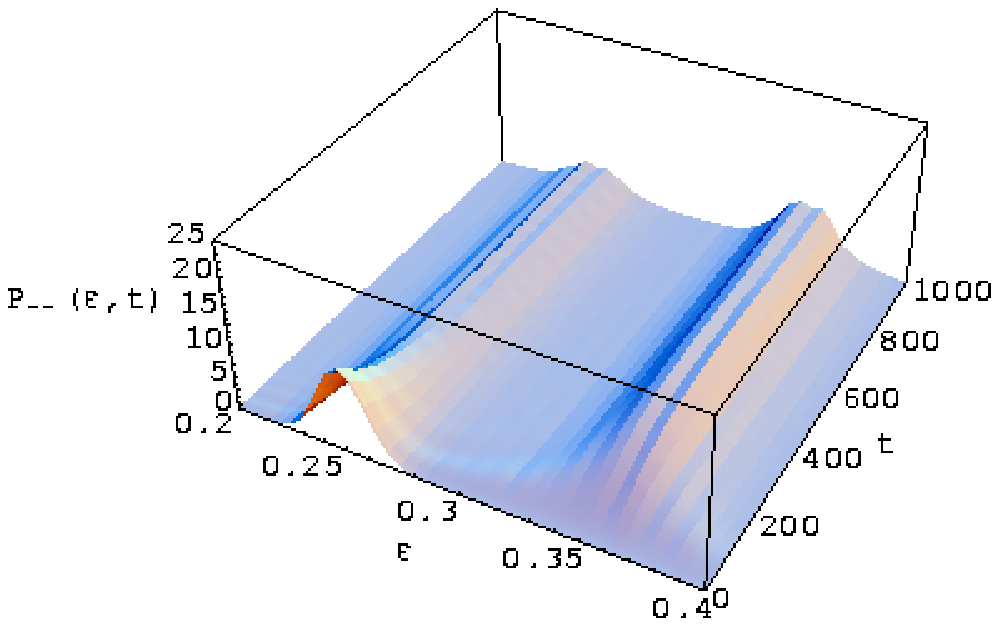}}}
&
\rotatebox{0}{\scalebox{0.9}{\includegraphics{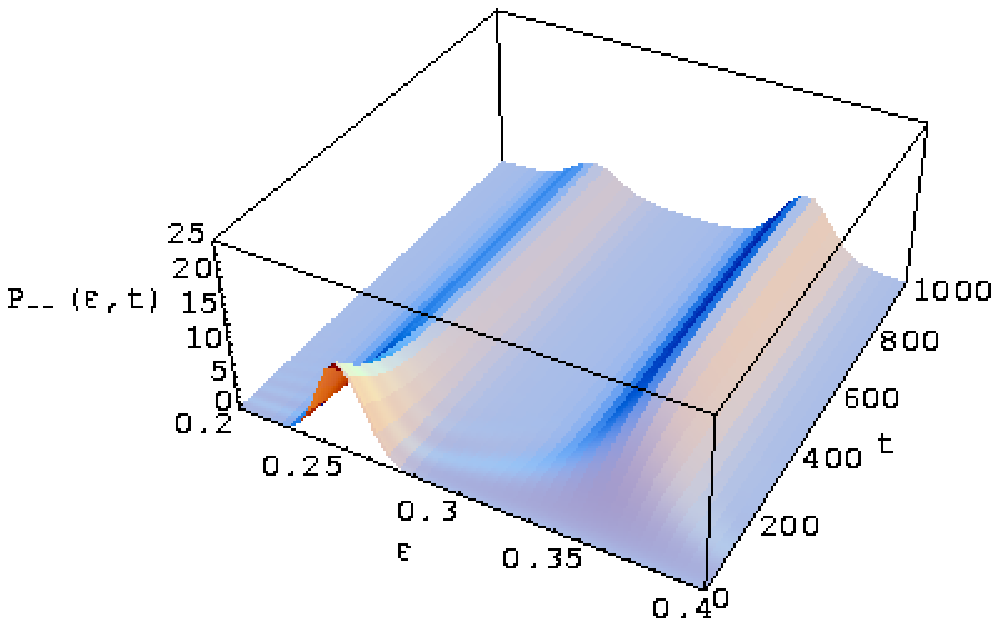}}}
\\
\end{tabular}
\caption{
Representation of the energy-distributed populations $P_{++}(\epsilon,t)$ and $P_{--}(\epsilon,t)$
which are central quantities in our new theory.
The plots on the right have been obtained from the exact von Neumann equation and the ones on 
the left from the non-Markovian version of our new equation.
The remarkable similarity between the right and left graphs is a strong support for the validity of 
our new theory.
We here have $\Delta=0.1025$, $\epsilon=0.25$, and $\lambda=0.1$. 
The exact results from the von Neumann equation have been obtained with $N=2000$ and $\chi=10$.
The initial condition of the subsystem is given by $\bra{1} \hat{\rho}_S(0) \ket{1} = 2/3$,
$\bra{-1} \hat{\rho}_S(0) \ket{-1} = 1/3$ and
$\bra{1} \hat{\rho}_S(0) \ket{-1} =  \bra{-1} \hat{\rho}_S(0) \ket{1} = \sqrt{2}/3$.
For a better graphical representation, the initial distribution of
the reservoir is not exactly defined as a delta but a Gaussian with standard deviation $0.01$.
}                                                                                                           \label{distribpop}
\end{figure}

\begin{figure}[p]
\centering
\hspace*{-3.3cm}
\begin{tabular}{c@{\hspace{0.1cm}}c}
\vspace*{1cm}
\rotatebox{0}{\scalebox{0.9}{\includegraphics{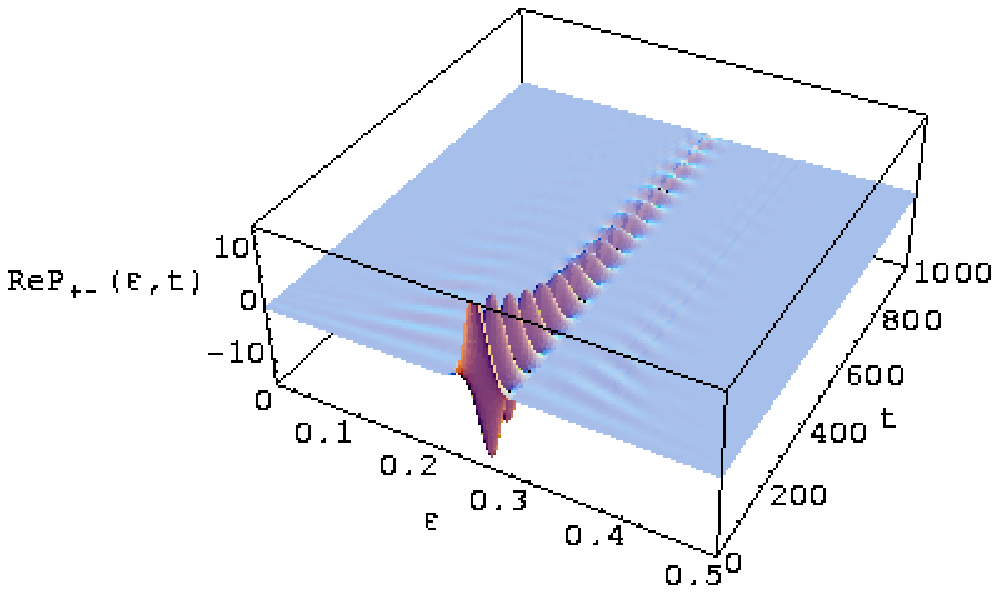}}}
&
\rotatebox{0}{\scalebox{0.9}{\includegraphics{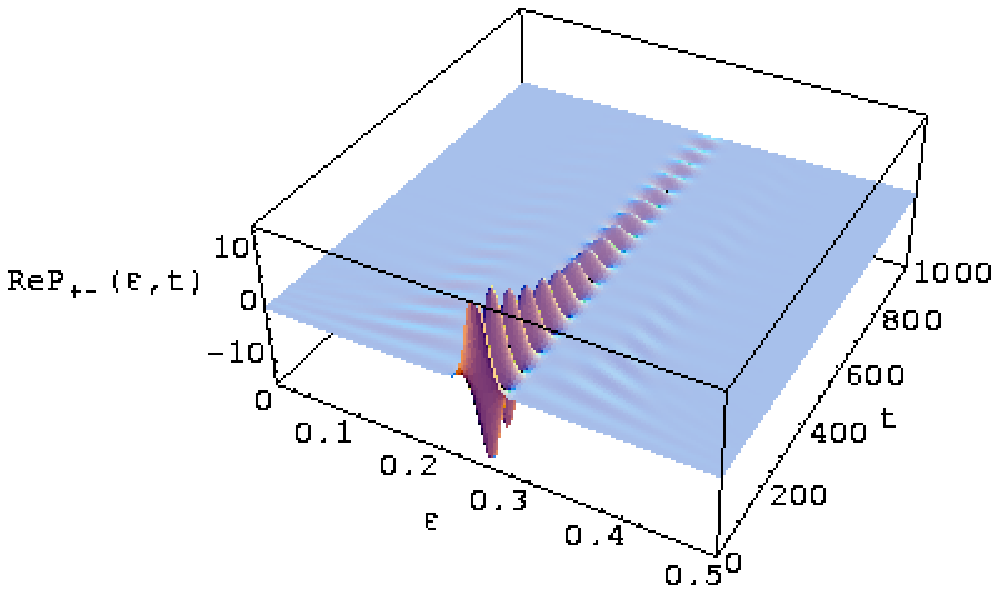}}}
\\
\rotatebox{0}{\scalebox{0.9}{\includegraphics{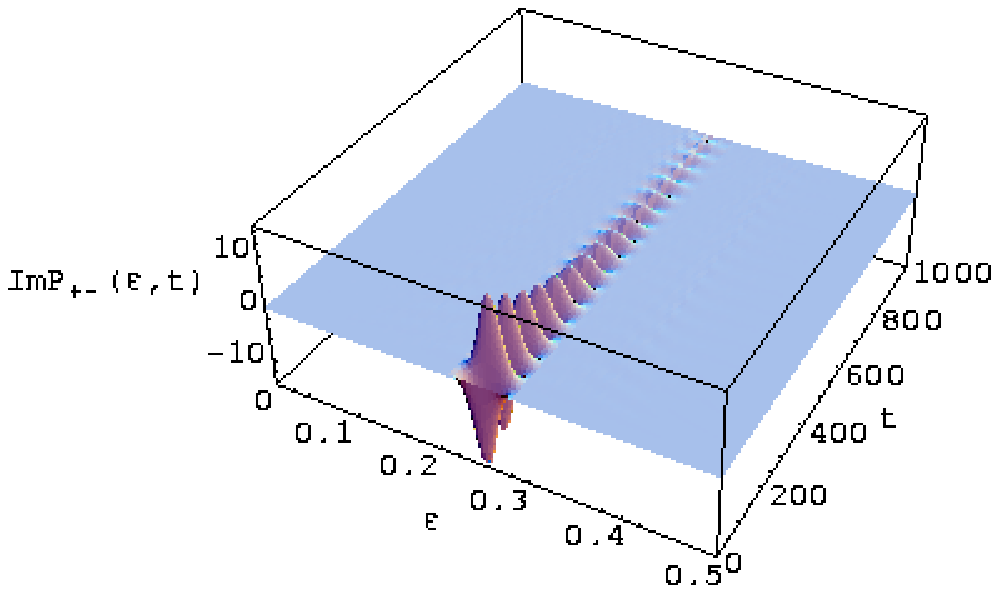}}}
&
\rotatebox{0}{\scalebox{0.9}{\includegraphics{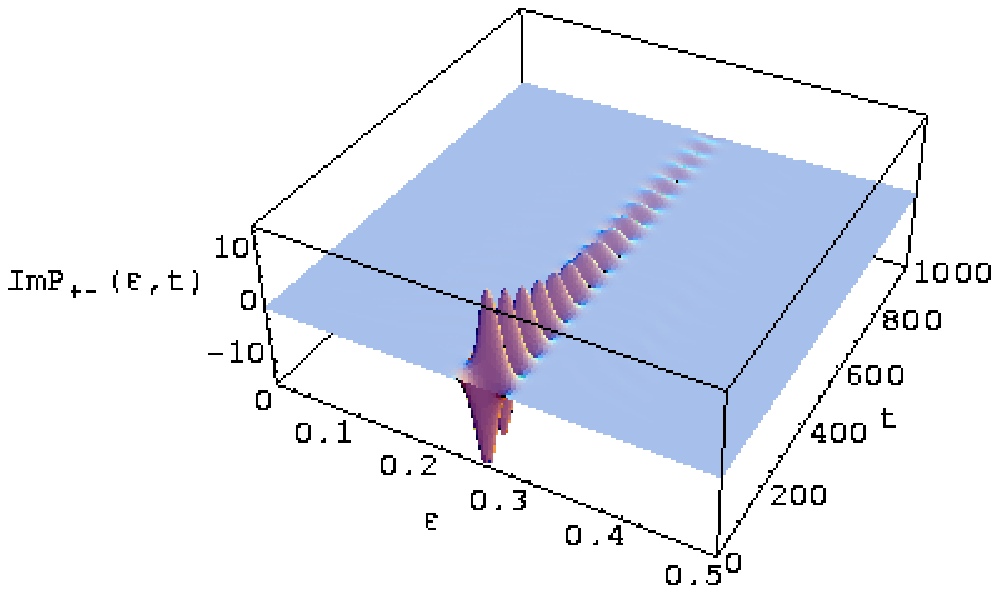}}}
\\
\end{tabular}
\caption{Representation of the real and imaginary parts of the energy-distributed coherence 
${\cal R} e \left[P_{+-}(\epsilon,t)\right]$ and ${\cal I} m \left[P_{+-}(\epsilon,t)\right]$ which are 
central quantities in our new theory.
The plots on the right have been obtained from the exact von Neumann equation and the ones on 
the left from the non-Markovian version of our new equation.
The remarkable similarity between the right and left graphs is again a strong support for the validity of 
our new theory.
The conditions are here the same as in figure \ref{distribpop}.
}                                                                                                           \label{distribcoh}
\end{figure}

We conclude the comparison of the exact spin dynamics with our new equation by representing the evolution of the 
energy-distributed elements of the spin density matrix.
These objects are the central elements of our new theory and are obtained, as we have seen in chapter \ref{ch3}, by a 
projection of the total density matrix. 
The very high similarity between the numerically exact dynamics of the energy-distributed spin density matrix element 
(obtained by the projection of the total density matrix evolving according to the exact von Neumann equation) and
this same dynamics predicted from our new theory (see figure \ref{distribpop} and \ref{distribcoh}) is once
again a really strong support for the validity of our new theory. 
The non-local aspect (in energy) of the population dynamics due to the energy conservation of the total system can be 
very well seen in figure \ref{distribpop}.
The almost local aspect of the coherence dynamics can be seen in figure \ref{distribcoh}.
One notices however the slight non-local effect at an energy of $\pm \Delta$ from the initial energy. 

\subsection{Average versus individual realizations} \label{statvsind}

An interesting point is the comparison, for the subsystem dynamics, between the averaged (random-matrix ensemble averaged 
or microcanonically averaged) dynamics and the dynamics of an individual realization within the statistical ensembles. 
This latter corresponds to a dynamics generated by an initial condition that corresponds to a pure state and without any
random-matrix average ($\chi=1$).\\

\begin{figure}[p]
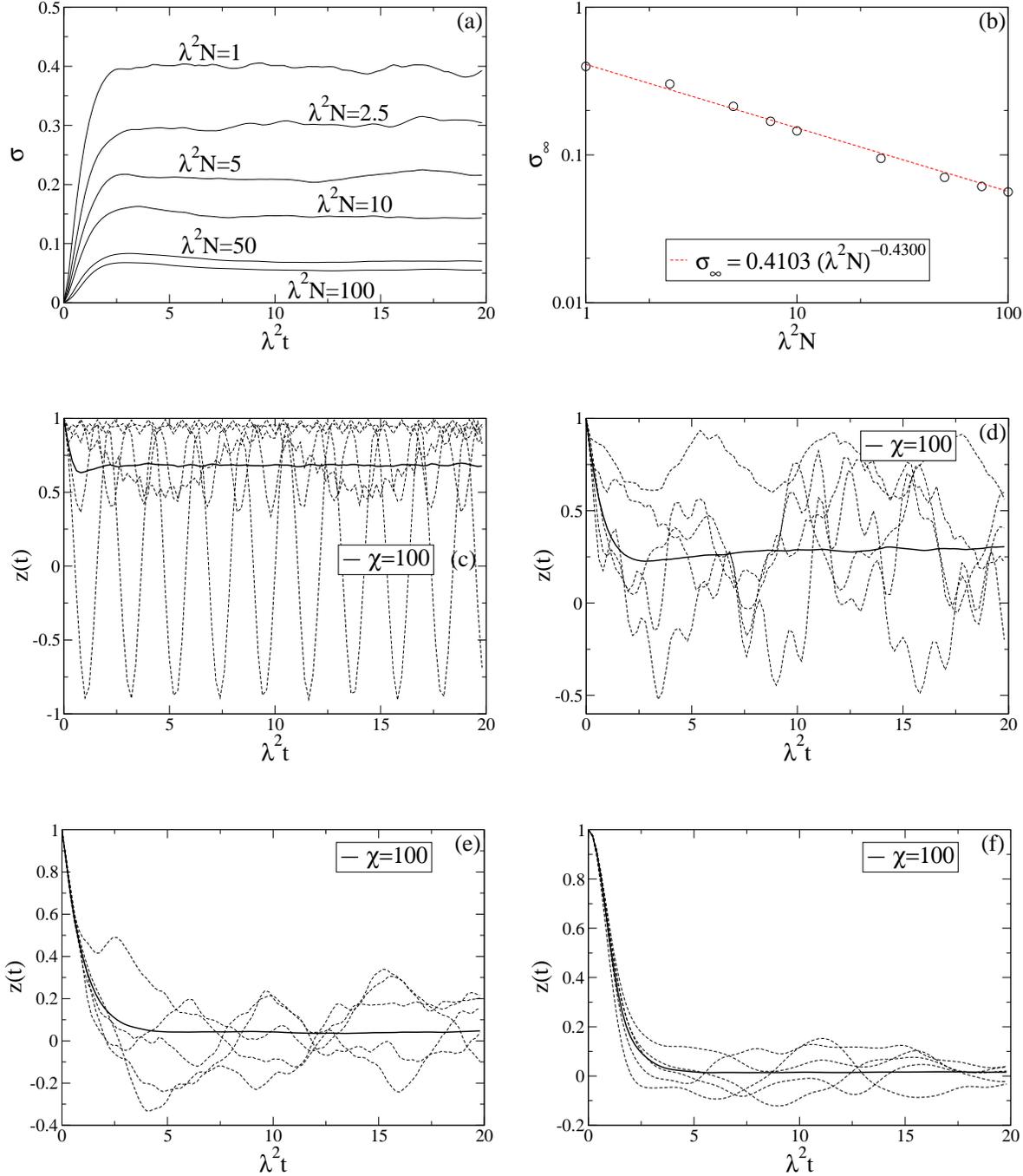

\vspace*{0.8cm}
\centering
\hspace*{-2cm}
\begin{tabular}{c@{\hspace{0.5cm}}c}
\vspace*{0.8cm}
\rotatebox{0}{\scalebox{0.3}{\includegraphics{fig33.eps}}}
&
\rotatebox{0}{\scalebox{0.3}{\includegraphics{fig34.eps}}}
\\ \vspace*{0.8cm}
\rotatebox{0}{\scalebox{0.3}{\includegraphics{fig35.eps}}}
&
\rotatebox{0}{\scalebox{0.3}{\includegraphics{fig36.eps}}}
\\ 
\rotatebox{0}{\scalebox{0.3}{\includegraphics{fig37.eps}}}
&
\rotatebox{0}{\scalebox{0.3}{\includegraphics{fig38.eps}}}
\\
\end{tabular}
\caption{In all the figures $\Delta=0.1$, $N=500$, $\epsilon=0$, 
and $\delta \varepsilon=0.05$. (a) Variance between individual
trajectories and the averaged one ($\chi=100$) as a function of 
time for different values of $\lambda^2 N$. (b) Power-law
dependence between the equilibrium value of this variance and
$\lambda^2 N$. (c)-(f) Individual
trajectories of the ensemble (dashed lines) and the ensemble
averaged trajectory (solid line). In (c) $\lambda^2 N=0.1$, in (d) 
$\lambda^2 N=1$, in (e) $\lambda^2 N=10$, and in (f) 
$\lambda^2 N=100$.} \label{Zfluct}
\end{figure}

We see in figure \ref{Zfluct}(c)-(f) the dynamics of a system with a small energy spacing $\Delta=0.1$ for different values 
of $\lambda^2 N$. 
The solid line represents the random-matrix and microcanonically averaged dynamics. 
The dashed lines depict some of the individual members of the random-matrix ensemble corresponding to an initial pure state 
inside the total unperturbed energy shell. 
We see that the larger $\lambda^2 N$ is, the closer the individual trajectories are from the averaged trajectories. 
We therefore have, when $\lambda^2 N$ is large enough, that the individual realizations are self-averaging in the random-matrix 
and microcanonical ensembles. 
In order to quantify this behavior, we plotted in figure \ref{Zfluct}(a) the variance between the individual trajectories and the 
averaged trajectories as a function of time for different values of $\lambda^2 N$. 
We observe that this variance decreases as $\lambda^2N\to\infty$. 
In figure \ref{Zfluct}(b), we show that the asymptotic value of the variance decreases with a power-law dependence with respect 
to $\lambda^2 N$.
We again see the connection with the SOE regimes. 
In the localized regime (figure \ref{Zfluct}(c)), the dynamics is governed by very different individual trajectories oscillating 
with a very few frequencies that differ from one individual trajectory to another.
This is a consequence of the fact that the perturbed levels are still close to the non-perturbed ones and are only slightly 
affected by neighboring non-perturbed levels. 
In the Lorentzian regime (figures \ref{Zfluct}(d)-(e)), each individual trajectory follows roughly the averaged trajectory 
and contains a very large number of different frequencies. 
This shows that the interaction "mixes" many of the non-perturbed levels, deleting the discrete structure of the spectrum.
Therefore, we can say that our new master equation derived in chapter \ref{ch3} holds with a given accuracy for a majority of 
individual trajectories if $\lambda$ is large enough to satisfy $\lambda \geq C N^{-\nu}$ with $\nu <\frac{1}{2}$  and a 
constant $C>0$, in the limit $N\to\infty$.

\section{Equilibrium distributions} \label{equilibrium}

In this section, we want to study the properties of the distribution reached on very long time in the spin-GORM model.
In other words, we want to study the spin-GORM equilibrium properties.
In subsection \ref{valequiobssys}, we study the equilibrium properties of the spin. 
In subsection \ref{thermsubsys}, we investigate the conditions under which the spin thermalizes.  
Finally, in subsection \ref{equisystot}, we study the thermalization of the total system.

\subsection{Equilibrium distribution of the subsystem} \label{valequiobssys}

\begin{figure}[p]
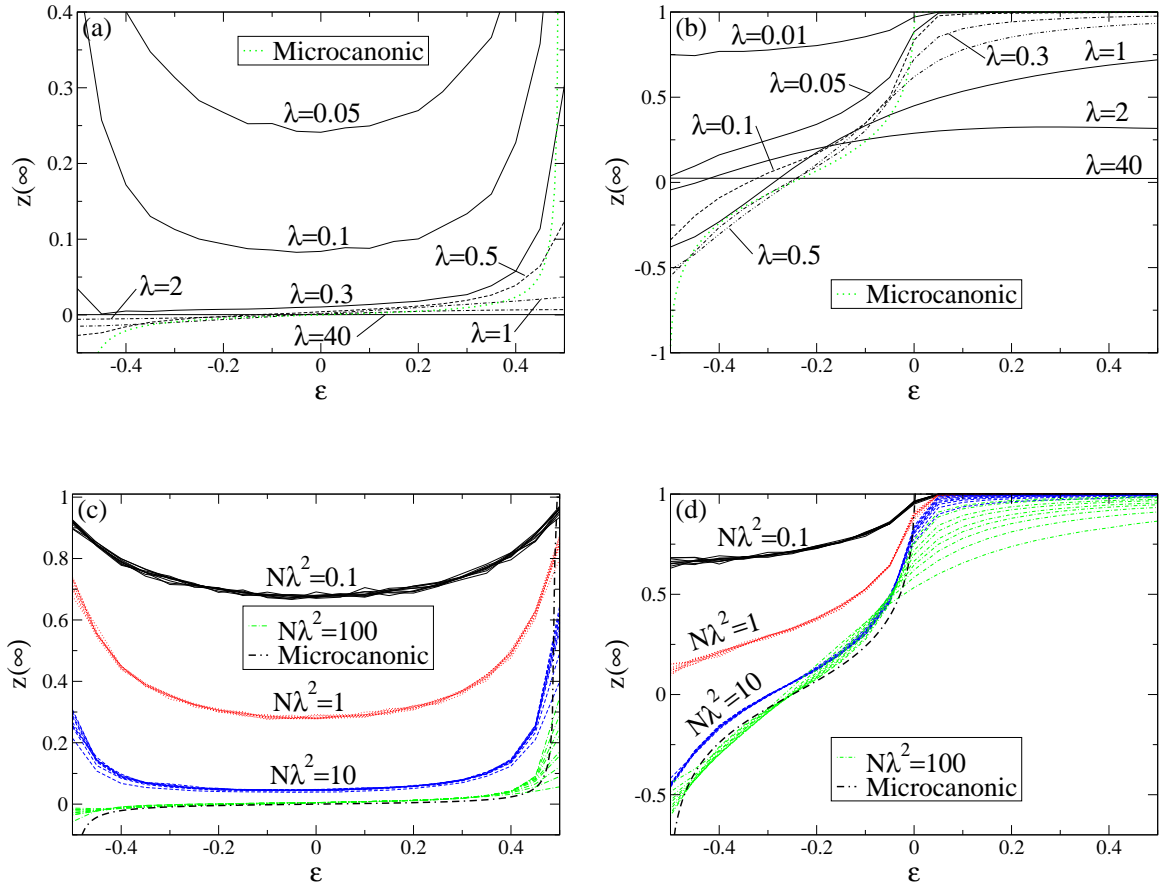

\centering
\hspace*{-2cm}
\begin{tabular}{c@{\hspace{0.5cm}}c}
\vspace*{1cm}
\rotatebox{0}{\scalebox{0.3}{\includegraphics{fig39.eps}}}
&
\rotatebox{0}{\scalebox{0.3}{\includegraphics{fig40.eps}}}
\\
\rotatebox{0}{\scalebox{0.3}{\includegraphics{fig41.eps}}}
&
\rotatebox{0}{\scalebox{0.3}{\includegraphics{fig42.eps}}}
\\
\end{tabular}
\caption{
Comparison between between the equilibrium value of $z(\infty)$ given by the Pauli equation which corresponds
to a microcanonical distribution on the total system and the exact values given by the time-averaged value 
for different values of $\lambda$. 
These equilibrium values are depicted as a function of the initial microcanonical energy $\epsilon$ of the environment. 
$\delta \epsilon=0.05$ in all figures. 
The parameter values are 
(a) $\Delta=0.01$ and $N=500$; 
(b) $\Delta=0.5$ and $N=500$; 
(c) $\Delta=0.01$ and $N=200-2000$; 
(d) $\Delta=0.5$ and $N=200-2000$.
}                                                                                                           \label{sigZinftavvspauli}
\end{figure}

As we have seen in the preceding sections, the spin coherences always decay to zero on long time scales due to 
the environment interaction.
Therefore, the only element left in the spin density matrix on these long time scales are the populations which 
are described by the spin observable $\hat{\sigma}_z$. 
This observable evolves in time according to
\begin{eqnarray}
z(t)
&=& \textrm{Tr} \; \hat{\rho}(0) e^{i \hat{H} t} \hat{\sigma}_z e^{-i \hat{H} t} \nonumber \\ 
&=& \sum_{\alpha,\alpha'} \langle \alpha \vert \hat{\rho}(0) \vert \alpha' \rangle \langle \alpha' \vert
\hat{\sigma}_z \vert \alpha \rangle e^{i (E_{\alpha}-E_{\alpha'})t}.                                                 \label{Ast}
\end{eqnarray}
The time-averaged value of $z(t)$ is obtained performing the following time average:
\begin{eqnarray}
z(\infty) &=& \lim_{T\to\infty} \frac{1}{T} \int_{0}^{T} dt \; z(t) \nonumber \\ 
&=& \sum_{\alpha} \langle \alpha \vert \hat{\rho}(0) \vert \alpha \rangle \langle \alpha 
\vert \hat{\sigma}_z \vert \alpha \rangle.                                                                           \label{Asinf}
\end{eqnarray}
We see that this time averaged value clearly depends on the initial condition.\\
An important result of our numerical simulation, which holds everywhere on the parameter space and for all kinds 
of initial conditions, is that the observed equilibrium value given by the exact von Neumann equation corresponds 
very well to the time-averaged value (\ref{Asinf}).
The study of the equilibrium properties of the subsystem is therefore equivalent to the study of the time 
averaged quantities (\ref{Asinf}).\\
We will use an initial condition for the total system which is the product of a microcanonical distribution at 
energy $\epsilon$ for the environment with a spin up state for the subsystem
\begin{eqnarray}
\hat{\rho}(0) = \vert 1 \rangle \langle 1 \vert \otimes \frac{\delta(\hat{H}_{B}-\epsilon)} {n(\epsilon)} 
= \sum_n \frac{\delta(E_n-\epsilon)} {n(\epsilon)} \vert 1 n \rangle \langle 1 n \vert.                          \label{rhozerostand}
\end{eqnarray}
We have therefore that
\begin{eqnarray}
\langle \alpha \vert \hat{\rho}(0) \vert \alpha \rangle 
= \sum_n \frac{\delta(E_n-\epsilon)} {n(\epsilon)} \vert \langle \alpha \vert 1 n \rangle \vert^2.                \label{arhoamiccan}
\end{eqnarray}
An interesting point is to understand when the Markovian version of our new weak-coupling kinetic equation 
predicts the correct population equilibrium value. 
We want therefore to compare the exact equilibrium distribution given by the time averaged populations $z(\infty)$ 
of (\ref{Asinf}) with $z(\epsilon,\infty)$ in (\ref{Ch4aaaaf}) (because for the initial condition 
(\ref{rhozerostand}), $z(0)=1$ and therefore the equilibrium distribution predicted by our kinetic equation 
in (\ref{eqnewM}) is such that $z(\infty)=z(\epsilon,\infty)$).\\ 
This comparison is done in figure \ref{sigZinftavvspauli} where we plotted the time averaged value $z(\infty)$ as
function of the initial energy of the environment $\epsilon$. 
The different curves in figures \ref{sigZinftavvspauli}(a) and \ref{sigZinftavvspauli}(b) correspond to different 
values of $\lambda$. 
We compare these curves to the $\lambda$ independent curve given by our new weak-coupling kinetic equation and 
corresponding to a microcanonical distribution inside the energy shell of the total system.  
As expected from our preceding study of the dynamics, we find again that, when $\lambda$ is too small, the "mixing" 
between the levels inside the total energy shell is not sufficient to reach equipartition in this shell and 
our kinetic equation, which assumes equipartition, overestimates the equilibrium values. 
If $\lambda$ is too large, our kinetic equation again gives bad results as expected from the restriction to
weak couplings of our equation. 
As already observed in the previous section, the prediction of our new kinetic equation are correct in a 
characteristic region of $\lambda$. 
The beginning of this region corresponds to $\lambda$ values such that $\lambda > \sqrt{1/N}$ 
at which the transition from the localized to the Lorentzian regimes occurs in the SOE.
Remember that this critical value goes to zero when $N \to \infty$. 
The end of this region corresponds to $\lambda$ values that cannot be considered anymore as perturbative
(when the width of the Lorentzian SOE becomes larger than $\delta \epsilon$ the typical energy scale of 
variation of the environment density of states). 
Figures \ref{sigZinftavvspauli}(c) and \ref{sigZinftavvspauli}(d) show very explicitly that $z(\infty)$
scales like $\lambda^2 N$. 
This again confirms our previous analysis which pointed out the importance of the quantity $\lambda^2 N$.

\subsection{Thermalization of the subsystem} \label{thermsubsys}

One of the important questions is to understand the conditions under which the subsystem thermalizes under 
the effect of a weak interaction with the environment or, in other words, under which conditions the subsystem 
relaxes to a canonical distribution corresponding to the microcanonical temperature of the environment.\\

We begin by recalling these conditions in the general case of a small subsystem weakly interacting 
with its environment. 
The isolated total system is composed of the subsystem and the environment and has the total energy
\begin{eqnarray}
E_{\rm tot} = \epsilon + e.
\end{eqnarray}
$\epsilon$ is the energy of the environment and $e$ the energy of the subsystem. 
We suppose that the contribution of the interaction energy between the environment and the subsystem is 
negligible compared to the total energy. 
The microcanonical environment entropy is defined as
\begin{eqnarray}
S_B(\epsilon)= k \ln \Omega_B(\epsilon), \label{Srho}
\end{eqnarray}
where $\Omega_B(\epsilon)$ is the number of states of the environment available at energy $\epsilon$. 
This number can be related to the density of states of the environment
$n_B(\epsilon)$ using the fact that $\Omega_B(\epsilon) = n_B(\epsilon) \delta\epsilon$, 
where $\delta\epsilon$ is a small energy interval but contains many states of the environment.
The microcanonical temperature of the environment is given by
\begin{eqnarray}
\frac{1}{T_B(\epsilon)} &=& \frac{d S_B(\epsilon)}{d \epsilon}  \label{defTmic}.
\end{eqnarray}
It can be expanded in the subsystem energy as
\begin{eqnarray}
T_B(\epsilon=E_{\rm tot}-e) &=& T_B(\epsilon=E_{\rm tot}) - e \frac{d
T_B(\epsilon=E_{\rm tot})}{d \epsilon} + \dots ,
\end{eqnarray}
because we suppose that the subsystem energy is much smaller than the environment energy. 
The specific heat capacity of the environment is
\begin{eqnarray}
\frac{1}{C_{vB}(\epsilon)} = \frac{d T_B(\epsilon)}{d \epsilon} .
\end{eqnarray}
If the condition
\begin{eqnarray}
\vert C_{vB}(\epsilon) \vert \gg \Big\vert \frac{e}{T_B(\epsilon)} \Big\vert \label{condenv2}
\end{eqnarray}
is satisfied, the temperature expansion can be truncated as
follows
\begin{eqnarray}
T_B(\epsilon=E_{\rm tot}-e) = T_B(\epsilon=E_{\rm tot}) \label{egaltemp}.
\end{eqnarray}
Therefore, we understand that if equation (\ref{condenv2}) is satisfied, the environment plays the role of 
a heat bath because its temperature is almost not affected by the subsystem energy.\\
Now, we suppose further that the interaction between the subsystem and the environment, even if small, is able 
to make the total probability distribution microcanonical on the total energy shell at energy $E_{\rm tot}$.
We suppose also that the energy levels are discrete and, therefore, that $E_{\rm tot}=E_s+E_b$ where $s$ 
and $b$ are discrete indices, respectively, for the subsystem and the environment. 
Therefore, the probability $P_S(E_s)$ for the subsystem being at energy $E_s$ is given by
\begin{eqnarray}
P_S(E_s)=\frac{\Omega_B(E_b=E_{\rm tot}-E_s)}{\Omega_{\rm tot}(E_{\rm tot})},
\end{eqnarray}
where $\Omega_B(E_b)$ is the number of states of the environment available at energy $E_b$, and 
$\Omega_{\rm tot}(E_{\rm tot})$ the number of states of the total system available at energy $E_{\rm tot}$.
Using equations (\ref{defTmic}) and (\ref{condenv2}), one gets that
\begin{eqnarray}
P_S(E_s)=\frac{e^{\frac{1}{k} S_B(E_b=E_{\rm tot}-E_s)}}{\Omega_{\rm tot}(E_{\rm tot})} \simeq
\frac{ e^{ \frac{1}{k}S_B(E_b=E_{\rm tot})-\frac{E_s}{kT_B(E_B=E_{\rm tot})}} }{\Omega_{\rm tot}(E_{\rm tot})} .
\end{eqnarray}
Using the normalization $ P_S(E_s)=1$, one finally gets the well-known canonical 
probability distribution for the subsystem
\begin{eqnarray}
P_S(E_s)=\frac{e^{-\frac{E_s}{k T_B(E_B=E_{\rm tot})}}}{Z},
\end{eqnarray}
where $Z=\sum_s e^{-\frac{E_s}{k T_B(E_B=E_{\rm tot})}}$.\\
We conclude that two conditions are necessary in order to thermalize the subsystem to a canonical probability 
distribution due to the interaction with the environment: a large heat capacity of the environment 
$\vert C_{vB}(\epsilon) \vert \gg \vert \frac{e}{T_B(\epsilon)} \vert$ and a microcanonical distribution
on the total energy shell.\\

\begin{figure}[h]
\vspace*{1cm}
\centering
\rotatebox{0}{\scalebox{0.4}{\includegraphics{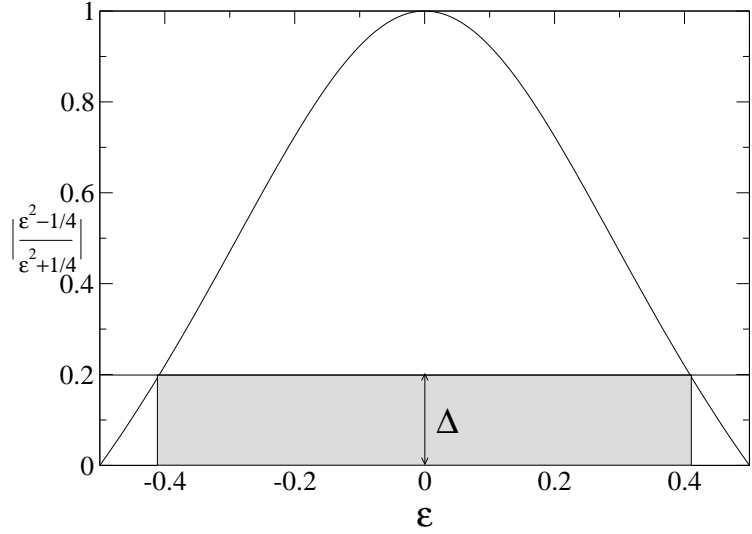}}}
\caption{
Representation of the region where the thermalization condition $\Big\vert \frac{\epsilon^2-\frac{1}{4}}
{\epsilon^2+\frac{1}{4}} \Big\vert > \Delta$ holds.
}                                                                                                               \label{CvT}
\end{figure}
\begin{figure}[p]
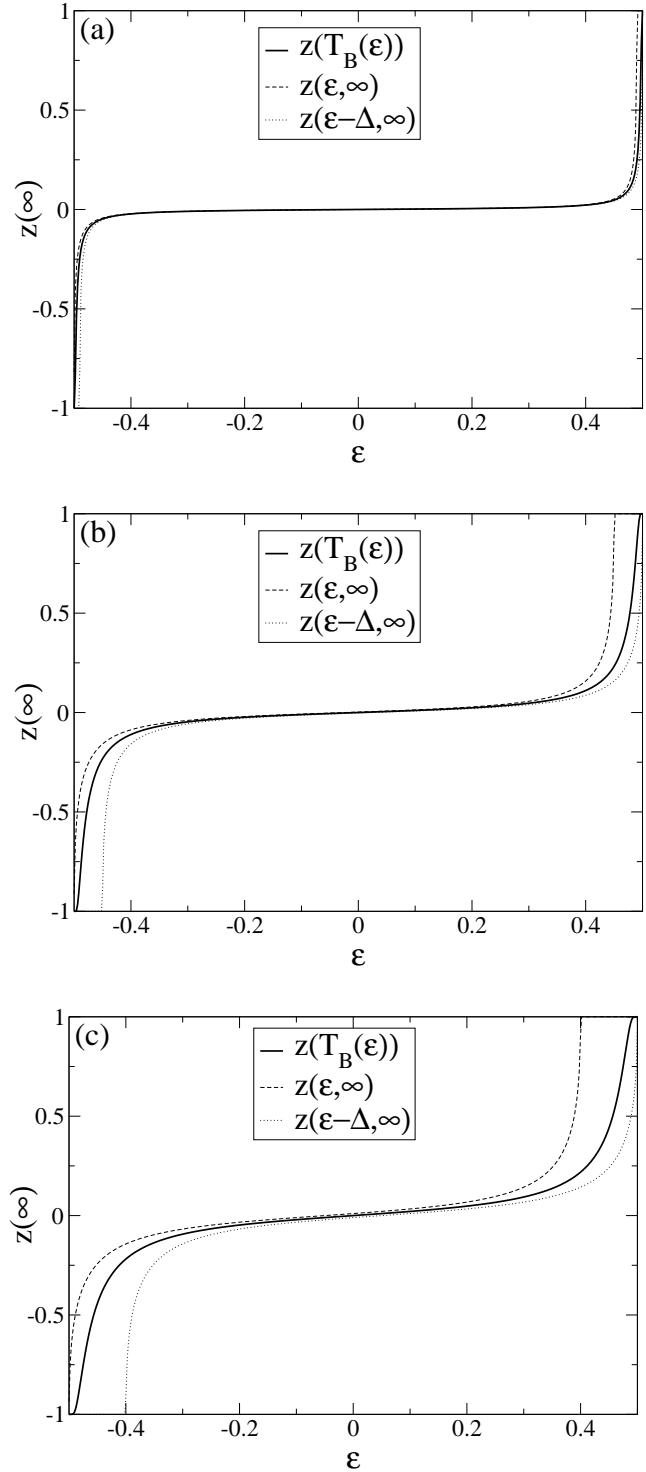

\vspace*{0.4cm}
\centering
\begin{tabular}{c@{\hspace{0.5cm}}c}
\vspace*{0.4cm}
\rotatebox{0}{\scalebox{0.35}{\includegraphics{fig44.eps}}} \\
\vspace*{0.4cm}
\rotatebox{0}{\scalebox{0.35}{\includegraphics{fig45.eps}}} \\
\rotatebox{0}{\scalebox{0.35}{\includegraphics{fig46.eps}}}
\end{tabular}
\caption{
Comparison between Eqs. (\ref{sigZcan}) and (\ref{sigZmic}) for different values of $\Delta$. 
The parameter values are (a) $\Delta=0.01$; (b) $\Delta=0.05$; (c) $\Delta=0.1$.
}                                                                                                       \label{miccancomp}
\end{figure}

Let us apply this result to the spin-GORM model.\\
The density of states of the environment is in this case
\begin{eqnarray}
n_B(\epsilon)=\frac{4N}{\pi} \sqrt{\frac{1}{4}-\epsilon^2}.
\end{eqnarray}
Therefore the microcanonical temperature of the environment is given by
\begin{eqnarray}
T_B(\epsilon)=\frac{\epsilon^2-\frac{1}{4}}{k \epsilon}
\end{eqnarray}
and the heat capacity is
\begin{eqnarray}
C_{vB}(\epsilon)=\frac{k \epsilon}{\epsilon^2+\frac{1}{4}}.
\end{eqnarray}
The first condition (\ref{condenv2}) in order to thermalize the spin to a canonical probability 
distribution becomes
\begin{eqnarray}
\vert C_{vB}(\epsilon) T_B(\epsilon) \vert 
= \Big\vert \frac{\epsilon^2-\frac{1}{4}}{\epsilon^2+\frac{1}{4}} \Big\vert \gg \Delta                   \label{condenv3}
\end{eqnarray}
and is depicted in Fig \ref{CvT}.
The second condition to have a microcanonical distribution on the total energy shell is satisfied 
(as we discussed in section \ref{valequiobssys}) when $\lambda^2 N > 1$. 
In this case, the populations of the spin obey the equilibrium distribution predicted by our new 
kinetic equation (see (\ref{eqnewM}))  
\begin{eqnarray}
z(\infty) = z(\epsilon,\infty) \frac{z(0)+1}{2} + z(\epsilon-\Delta,\infty) \frac{z(0)-1}{2}              \label{sigZmica}
\end{eqnarray}
where 
\begin{eqnarray}
z(\epsilon,\infty)=\frac{Z(\epsilon,\infty)}{C(\epsilon,0)}=
\frac{\sqrt{\frac{1}{4}-(\epsilon)^2}-\sqrt{\frac{1}{4}-(\epsilon+\Delta)^2} }
{\sqrt{\frac{1}{4}-(\epsilon)^2}+\sqrt{\frac{1}{4}-(\epsilon+\Delta)^2}}.                                 \label{sigZmic}
\end{eqnarray}
The canonical distribution of the spin populations, at the microcanonical temperature of the 
environment, is given by
\begin{eqnarray}
z(T_B(\epsilon)) = - {\rm tanh}\frac{\Delta}{2 k T_B(\epsilon)}
= - {\rm tanh}\frac{\Delta\epsilon}{2 (\epsilon^2-\frac{1}{4})} .                                         \label{sigZcan}
\end{eqnarray}
If the two conditions for the spin thermalization are satisfied, then $z(T_B(\epsilon))$ and $z(\infty)$ 
should be equal.\\ 
The comparison between these two quantities can be seen in figure \ref{miccancomp} for
different subsystem energies. 
We see, as expected, that the smaller the spin energy is the better the comparison is. \\

We can therefore conclude that under these two conditions ($\vert \frac{\epsilon^2-1/4}{\epsilon^2+
1/4} \vert \gg \Delta$ and $\lambda^2 N > 1$), the random matrices of the spin-GORM model can model 
an environment that behaves as a heat bath and that thermalizes the spin.
The particularity of this environment due to the semicircular density of states is that the 
microcanonical temperature can be negative (when the initial energy of the environment is positive).
Usual environments have positive microcanonical temperatures due to their monotonic increasing density 
of states. 

\subsection{Thermalization of the total system} \label{equisystot}

\begin{figure}[p]
\vspace*{0cm}
\centering
\hspace*{-1cm}
\vspace*{0.4cm}
\rotatebox{0}{\scalebox{0.35}{\includegraphics{fig47.eps}}} \\
\centering
\hspace*{-1cm}
\vspace*{0.4cm}
\rotatebox{0}{\scalebox{0.35}{\includegraphics{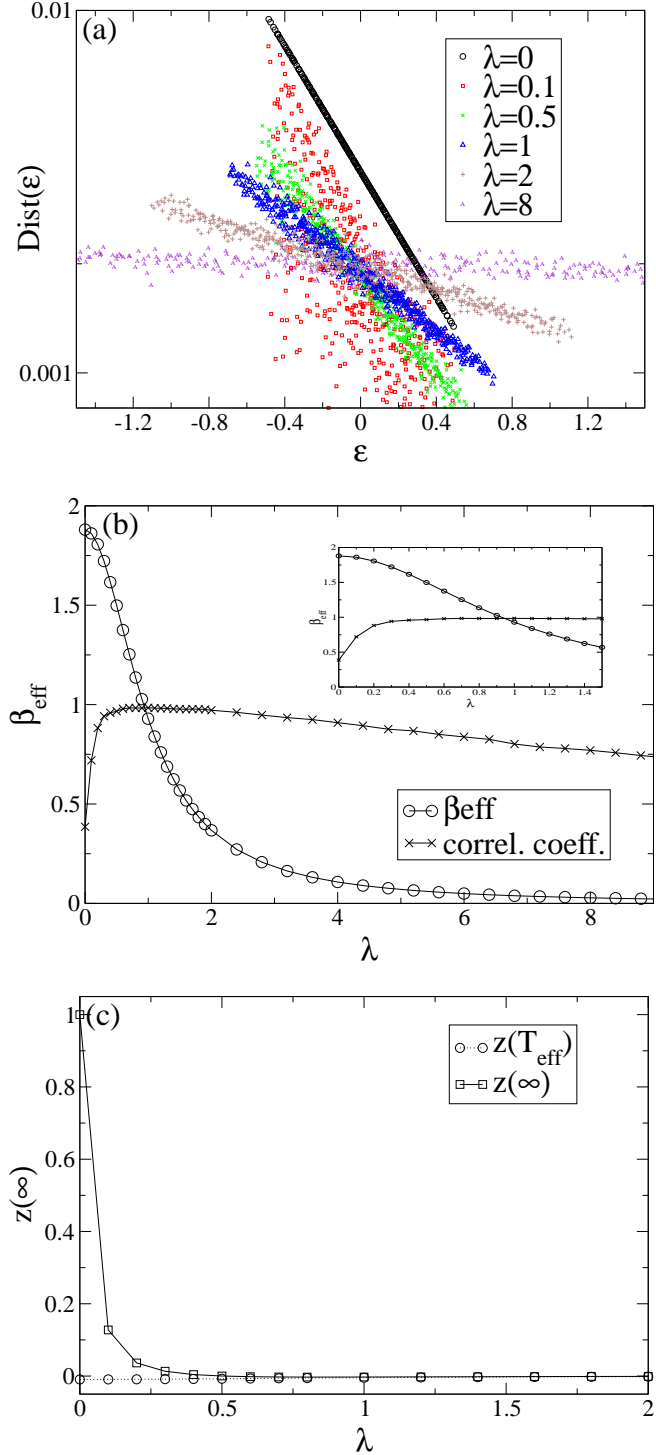}}} \\
\centering
\hspace*{-1cm}
\rotatebox{0}{\scalebox{0.35}{\includegraphics{fig49.eps}}}
\caption{
In all the figures $\Delta=0.01$, $N=500$, and $\beta=2$. 
(a) Probability at equilibrium $P(\varepsilon)={\rm Dist}(\varepsilon)$ of being at the total system 
energy $\varepsilon$ when starting from the initial condition $\hat{\rho}(0) = \vert 1 \rangle \langle 1 \vert 
\frac{e^{-\beta \hat{H}_B}}{Z_B}$ with $\frac{1}{\beta} = 0.5$. 
(b) Effective temperature of the equilibrium probability distribution obtained by fitting the data of (a)
with a canonical distribution. 
(c) Comparison between the exact equilibrium population value ((\ref{Asinf}) with (\ref{arhoacan})) and the 
canonical one ((\ref{Asinf}) with (\ref{TOTcanonicavAs})) computed with the effective temperature of plot (b).
}                                                                                                             \label{thermaB}
\end{figure}

Until now, we have always considered initial conditions of the environment corresponding to 
microcanonical distributions at a given energy like in (\ref{rhozerostand}). 
We now want to consider initial conditions of the environment corresponding to a canonical 
distribution at a given temperature.
We also suppose that the spin is up, so that the total initial condition reads  
\begin{eqnarray}
\hat{\rho}(0) 
= \vert 1 \rangle \langle 1 \vert \otimes \frac{e^{-\beta_{B} \hat{H}_{B}}}{Z_{B}} 
= \sum_n \frac{e^{-\beta_{B} E_{n}}}{Z_{B}} \vert 1 n \rangle \langle 1 n \vert.              
\end{eqnarray}
We therefore have that
\begin{eqnarray}
\langle \alpha \vert \hat{\rho}(0) \vert \alpha \rangle 
= \sum_n \frac{e^{-\beta_{B} E_{n}}}{Z_{B}} \vert \langle \alpha \vert 1 n \rangle \vert^2.                \label{arhoacan}
\end{eqnarray}
It is interesting to ask how the probability distribution looks like at equilibrium when the 
interaction is on.\\
One can clarify this point by plotting $\langle \alpha \vert \hat{\rho}(0) \vert \alpha \rangle$ versus energy. 
One uses the following energy representation
\begin{eqnarray}
P(\varepsilon)= \sum_{\alpha} \delta(E_{\alpha}-\varepsilon)
\langle \alpha \vert \hat{\rho}(0) \vert \alpha \rangle.
\end{eqnarray}
If the total system thermalizes and reaches a canonical distribution for the total system at an effective temperature 
$\beta_{\rm eff}^{-1}$, one would have that
\begin{eqnarray}
P(\varepsilon)= \frac{e^{-\beta_{\rm eff} \varepsilon}}{Z_{\rm tot}},
\end{eqnarray}
because
\begin{eqnarray}
\langle \alpha \vert \hat{\rho}(0) \vert \alpha \rangle =
\frac{e^{-\beta_{\rm eff} E_{\alpha}}}{Z_{\rm tot}}.
\label{TOTcanonicavAs}
\end{eqnarray}
As we shall see, it is the case if, again, $\lambda$ is large enough to induce "mixing" between the states: $\lambda^2 N > 1$.\\
Indeed, one sees in figure \ref{thermaB}(a) that, for $\lambda=0$, the states of the total system corresponding to the 
upper level of the spin are exponentially populated and the ones corresponding to the lower level are not. 
When the interaction is turned on and increased, one can notice that the probability distribution starts to accumulate around 
a mean effective canonical distribution. 
As expected, this accumulation becomes significant when $\lambda^2 N > 1$ and, in this case, the total system can be considered 
as having thermalized. 
One can calculate the final effective temperature that the total system has reached after interaction. 
This effective temperature is depicted in figure \ref{thermaB}(b) as a function of $\lambda$. 
The correlation coefficient indicates whether the exponential fit of the final effective temperature is good or not.
The effective temperature obeys the following law: $\beta_{\rm eff}=\frac{\beta_{i}}{1+\lambda^2}$. 
We also show in figure \ref{thermaB}(c) the comparison between the time-averaged value of $z(t)$ 
((\ref{Asinf}) with (\ref{arhoacan})) and its canonical average computed with the effective temperature 
((\ref{Asinf}) with (\ref{arhoacan})). 
One sees that, when $\lambda^2 N > 1$, both coincide.\\
It is important to realize that this thermalization is not statistical in the sense of the equivalence between the ensembles. 
It is an intrinsic thermalization due to the complexity of the interaction between the states. 
This thermalization appears at a critical value of the coupling parameter when the interaction term becomes 
of the order of or larger than the mean level spacing of the total system.

\chapter{Quantum diffusion} \label{ch5}

In this final chapter, we will focus on a particularly important quantum kinetic process: 
quantum diffusion.\\
Quantum diffusion is a quantum transport process where the transported quantity is the position of a 
particle on a spatially extended subsystem.
Understanding quantum diffusion is of huge importance, especially in nanosystems.
When the particles transported on the nanosystem have a charge, the transport properties of 
these particles determine the electric conduction properties of the nanosystem.
And one of the most common ways to study nanosystems such as quantum dots, nanotubes, or single linear 
molecules, is through their electric conduction.
Furthermore, the technological applications of these electronic devices are really promising. \\
If the subsystem is invariant under spatial translations, the Bloch theorem allows us to associate a Bloch 
number or wavenumber with each mode of the evolution superoperator. 
The dynamics of such a translationally invariant system is diffusive on long time scales if the dispersion relation, 
connecting the smaller real part of the mode eigenvalues (corresponding to the slower exponential relaxation 
modes) to the Bloch number, is parabolic around zero wavenumber. 
The proportionality coefficient between the eigenvalue and the squared wavenumber is then the 
diffusion coefficient.
Such a dispersion relation cannot be found in an isolated quantum subsystem where the real parts of the
mode eigenvalues are always zero. 
Therefore, we want to see if such a diffusive dispersion relation can occur in a translationally 
invariant subsystem weakly interacting with an environment. 
Our study is based on the Redfield equation (see section \ref{Redfield}) and the environment is therefore 
considered infinite.\\
Various attempts have been made in the literature to understand diffusion in quantum subsystems. 
Many of the early studies focused on polaron motion \cite{Flynn70,Holstein59}. 
Later studies used a kind of Quantum Brownian Motion model extended to periodic potentials 
\cite{ChenLebowitz89,ChenLebowitz92I,ChenLebowitz92II,Fisher85,Mak94,SassettiWeiss92,SassettiWeiss96,Weiss00}.   
However, it seems to us that none of them have avoided the presence of a non-translationally invariant 
term in the interaction Hamiltonian between the subsystems and its environment. 
Furthermore, they almost always use a path integral formalism, which has the advantage of sometimes
giving non-perturbative results but is not easy to use, especially for numerical simulations.
Our study, which relates to the full diagonalization of the evolution operator and therefore 
gives access to the complete dynamical information, seems to us to be a particularly interesting approach to
diffusion in quantum systems.\\

The plan of this chapter is the following.
Our translationally invariant model is defined in section \ref{Sec.system} by weakly coupling
a one-dimensional tight-binding Hamiltonian with a single energy band to a delta-correlated environment.  
The dynamics of this system is ruled by a Redfield quantum master equation.
The long-time evolution of the model can be studied in terms of the eigenvalues
and the associated eigenstates of the Redfield superoperator, as explained in section \ref{Sec.diag}.  
The eigenvalue problem is exactly solved for a finite chain in section \ref{Sec.finite}.  
We show in section \ref{Sec.dynamics} that diffusion dominates the long-time dynamics if the chain is large enough.  
The Liouvillian spectrum of the infinite chain as well as the temperature dependence of the diffusion coefficient 
are given in section \ref{Sec.infinite}.  
We finally summarize the results in section \ref{Sec.conclusions}.

\section{Defining the system} \label{Sec.system}

\subsection{Subsystem}

Let us consider a translationally invariant one-dimensional subsystem described by the following Hamiltonian
\begin{eqnarray}
\hat{H}_{S}=
\left(\begin{array}{ccccccc}
E_0    & -A    &   0     &  0     & \hdots & 0   &  -A    \\
-A     & E_0   &  -A     &  0     & \hdots & 0   &  0     \\
0      & -A    &  E_0    & -A     &        & 0   &  0     \\ 
\vdots &       &\ddots   &\ddots  & \ddots &     & \vdots \\
0      &  0    &         & -A     & E_0    & -A  &  0     \\
0      &  0    & \hdots  &  0     & -A     & E_0 & -A     \\
-A     &  0    & \hdots  &  0     &  0     & -A  & E_0    \\
\end{array} \right)_{N \times N}  \label{isolaaa}
\end{eqnarray}
represented in the site basis ${\vert l \rangle}$, where $l$ takes the values $l=0,1,\hdots,N-1$.
$N$ is the length of the chain.
We have here chosen periodic (Born-van Karman) boundary conditions.

Such a Hamiltonian, called tight-binding or H\"uckel Hamiltonian, is a simple model of 
spatially periodic subsystem and has therefore a wide range of applications. 
The tight-binding Hamiltonian is for instance commonly used in solid state physics to describe 
the electronic band structure of weakly interacting solids \cite{AshcroftB76}, in polymer physics 
to describe electronic conduction along the polymer backbone or polymer optical properties 
\cite{Heeger88,Heeger01}, as well as to understanding the conduction properties of carbon nanotubes 
\cite{SaitoB98}. 
This Hamiltonian describes a process of quantum tunneling from site to site.  
The parameter $A$ is given in terms of the overlap between the wavefunctions localized at the sites 
and is proportional to the quantum amplitude of tunneling.\\

The stationary Schr\"{o}dinger equation of the tight-binding Hamiltonian is given by
\begin{eqnarray}
\hat{H}_{S} \vert k \rangle = \epsilon_k \vert k \rangle ,\label{isolaab}
\end{eqnarray}
where the eigenvalues are
\begin{eqnarray}
\epsilon_k=E_0-2A \cos k \frac{2 \pi}{N} \label{isolaac}
\end{eqnarray}
and the eigenvectors
\begin{eqnarray}
\langle l \vert k \rangle = \frac{1}{\sqrt{N}} e^{i l k \frac{2 \pi}{N}} , \label{isolaad}
\end{eqnarray}
with $k=0,1,...,N-1$.
The closure relation is given by
\begin{eqnarray}
\frac{1}{N} \sum_{l=0}^{N-1} e^{i l (k-k') \frac{2 \pi}{N}}= \delta_{kk'} .\label{isolaae}
\end{eqnarray}
Accordingly, the energy spectrum of the Hamiltonian (\ref{isolaaa}) contains
a single energy band of width $2A$ and the motion of the particle
would be purely ballistic without coupling to a fluctuating environment.

\subsection{Coupling to the environment}

We now suppose that the subsystem is embedded in a large environment. \\

The Hamiltonian of the total system composed of the one-dimensional chain and its environment is
given by
\begin{eqnarray}
\hat{H}_{\rm tot} = \hat{H}_{S}  + \hat{H}_{B} +  \lambda \sum_{\kappa}
\hat{S}_{\kappa} \hat{B}_{\kappa} \; ,\label{isolaaf}
\end{eqnarray}
where $\hat{H}_{B}$ is the environment Hamiltonian, $\hat{S}_{\kappa}$ the subsystem coupling operators, 
$\hat{B}_{\kappa}$ the environment coupling operators, and $\lambda$ the coupling parameter which measures 
the intensity of the interaction between the subsystem and its environment.\\

The dynamics of the total system is described by the von Neumann equation 
\begin{eqnarray}
\frac{d \hat{\rho}_{\rm tot}(t)}{dt}=\hat{\hat{{\cal L}}}_{\rm tot} \hat{\rho}_{\rm tot}(t)
=-\frac{i}{\hbar} [\hat{H}_{\rm tot},\hat{\rho}_{\rm tot}(t)], \label{isolabf}
\end{eqnarray}
where $\hat{\hat{{\cal L}}}_{\rm tot}$ is the Liouvillian superoperator of the total system.
The reduced dynamics for the density matrix $\hat{\rho}(t)={\rm Tr}_{B} \hat{\rho}_{\rm tot}(t)$ 
of the subsystem is known to obey a Redfield quantum master equation for weak coupling to the environment. 
We have systematically derived this equation by second order perturbation theory in section 
\ref{Redfield} from the complete von Neumann equation for the total system (\ref{isolabf}). 
On time scales longer than the correlation time of the environment, the Redfield quantum master 
equation is Markovian  and reads     
\begin{eqnarray}
\frac{d \hat{\rho}}{dt} &=& \hat{\hat{{\cal L}}}_{\rm Red} \hat{\rho} \nonumber \\
&=& -\frac{i}{\hbar} \lbrack \hat{H}_{S} , \hat{\rho} \rbrack + \frac{\lambda^2}{\hbar^2} \sum_{\kappa} (
\hat{T}_{\kappa} \hat{\rho} \hat{S}_{\kappa} \nonumber \\ &&+
\hat{S}_{\kappa}^{\dagger} \hat{\rho}
\hat{T}_{\kappa}^{\dagger} - \hat{S}_{\kappa} \hat{T}_{\kappa}
\hat{\rho} - \hat{\rho} \hat{T}_{\kappa}^{\dagger}
\hat{S}_{\kappa}^{\dagger} ) + O(\lambda^3), \label{isolaag}
\end{eqnarray}
where $\hat{\hat{{\cal L}}}_{\rm Red}$ is called the Redfield superoperator and where
\begin{eqnarray}
\hat{T}_{\kappa} \equiv \sum_{\kappa'} \int_{0}^{\infty} d \tau \; 
\alpha_{\kappa \kappa'}(\tau) \; e^{-\frac{i}{\hbar} \hat{H}_{S} \tau} \hat{S}_{\kappa'} \; 
e^{\frac{i}{\hbar} \hat{H}_{S} \tau} . \label{isolaah}
\end{eqnarray}
The correlation function of the environment which contains all the necessary information to describe 
the coupling of the subsystem to its environment is given by
\begin{eqnarray}
\alpha_{\kappa \kappa'} (\tau) = {\rm Tr}_{B} \hat{\rho}_{B}^{eq} e^{\frac{i}{\hbar} \hat{H}_{B} \tau} 
\hat{B}_{\kappa} e^{-\frac{i}{\hbar} \hat{H}_{B} \tau} \hat{B}_{\kappa'} \label{isolaai}
\end{eqnarray}
where $\hat\rho_{B}^{eq}$ is the canonical equilibrium state of the environment.
Rigorous results show that the long-time dynamics of the subsystem is indeed ruled by
the Redfield superoperator according to
\begin{eqnarray}
\frac{d \hat{\rho}(t)}{dt} = {\rm Tr}_{B} \hat{\hat{{\cal L}}}_{\rm tot} \hat{\rho}_{\rm tot}(t) 
\stackrel{\lambda \ll 1}{=} \hat{\hat{{\cal L}}}_{\rm Red} \hat{\rho}(t) 
+ O(\lambda^3) \label{isolacf}
\end{eqnarray}
for weak coupling to its environment \cite{JaksicPillet97I,JaksicPillet97II}.\\

Let us now specify the interaction of the subsystem with its environment. 
We define the subsystem coupling operators as
\begin{eqnarray}
\langle l \vert \hat{S}_{\kappa} \vert l' \rangle=\delta_{\kappa l} \; \delta_{ll'} \; .\label{isolaal}
\end{eqnarray}
They are diagonal operators in the site basis of the subsystem taking the unit value if the
particle is located on the site $\kappa$ and zero otherwise.
These operators have the properties:
\begin{eqnarray}
\hat{S}_{\kappa}^n = \hat{S}_{\kappa} \; , 
\end{eqnarray}
for $n=2,3,...$ and
\begin{eqnarray}
\sum_{\kappa} \hat{S}_{\kappa} = \hat{I} \; .
\end{eqnarray}

We need now to specify the environment operators 
by the choice of the environment correlation functions.
We make two assumptions:\\
\underline{Assumption 1}: The environment dynamics has a very fast evolution compared to the subsystem dynamics. 
Therefore, the environment correlation functions decay in time so fast 
with respect to the characteristic time scales of the subsystem evolution,
that they can be assumed to be Dirac delta distributions in time.\\
\underline{Assumption 2}: The environment has very short range spatial correlations, 
much shorter than the distance between two adjacent sites of the subsystem. 
Accordingly, the environment correlation functions decay in space
so fast that they can be assumed to be given by Kronecker delta in space.\\
These two assumptions mean that each of the environmental fluctuations at the different subsystem sites 
are statistically independent. The environment correlation functions are thus given by
\begin{eqnarray}
\alpha_{\kappa \kappa'}(\tau)= 2\; Q \; \delta (\tau)\;  \delta_{\kappa \kappa'} \; , \label{isolaaj}
\end{eqnarray}
and the operators (\ref{isolaah}) in the Redfield equation (\ref{isolaag}) therefore become 
\begin{eqnarray}
\hat{T}_{\kappa} = Q \; \hat{S}_{\kappa} \; . \label{isolaak}
\end{eqnarray} 

Because of the fast decay of the temporal and the spatial correlations (\ref{isolaaj}) 
and the properties of the subsystem coupling operators (\ref{isolaal}), the Redfield equation we have to solve takes the form
\begin{eqnarray}
\frac{d \hat{\rho}}{dt} 
&=& \hat{\hat{{\cal L}}}_{\rm Red} \hat{\rho} \label{isolaam} \\ 
&=& -\frac{i}{\hbar} \lbrack \hat{H}_{S} ,
\hat{\rho} \rbrack + \frac{\lambda^2}{\hbar^2} Q \sum_{\kappa} (
2 \hat{S}_{\kappa} \hat{\rho} \hat{S}_{\kappa}
- \hat{S}_{\kappa}^2 \hat{\rho} - \hat{\rho} \hat{S}_{\kappa}^2) 
+ O(\lambda^3) \; . \nonumber 
\end{eqnarray}
It can easily be verified by projecting this equation onto the site basis that it is translationally invariant 
(shifting all the site indices appearing in the projected equation by a constant does not modify the equation). 
Furthermore, this equation preserves the complete positivity of the density matrix because it has the Lindblad 
form \cite{Lindblad76} which is the result of a coupling with delta correlation functions \cite{GaspardRed99}.

\section{Diagonalizing the Redfield superoperator} \label{Sec.diag}

The eigenvalues $s_{\nu}$ and associated eigenstates ${\hat{\rho}}^{\nu}$ of the  Redfield superoperator are defined by
\begin{eqnarray}
\hat{\hat{{\cal L}}}_{\rm Red}\;  {\hat{\rho}}^{\nu} = s_{\nu} \; {\hat{\rho}}^{\nu} , \label{blochaaa}
\end{eqnarray}
where $\nu$ is a set of parameters labeling the eigestates. Because the 
Redfield superoperator is not anti-Hermitian, its eigenvalues can be complex numbers 
with a nonzero real part.  The eigenvalue problem of the Redfield superoperator is important because 
the time evolution of the quantum master equation can then be decomposed onto the basis of the eigenstates as
\begin{eqnarray}
\hat{\rho}(t) = e^{\hat{\hat{{\cal L}}}_{\rm Red} t} \hat{\rho}(0) 
= \sum_{\nu=1}^{N^2} c_{\nu}(0) \; e^{s_{\nu} t} \hat{\rho}^{\nu}.
\label{isolaan}
\end{eqnarray}
The dynamics is therefore given by a linear superposition of exponential or oscillatory exponential
functions.  Since the reduced density matrix of the subsystem has $N^2$ elements, 
there is a total of $N^2$ eigenvalues and associated eigenstates.

\subsection{Bloch theorem}

Since the system is invariant under spatial translations, we can apply the Bloch theorem to
the eigenstates of the Redfied superoperator.  Thanks to this theorem, the state space of the
superoperator can be decomposed into independent superoperators 
acting onto decoupled sectors associated with a
Bloch number, also called wavenumber \cite{AshcroftB76}.\\

Let us define the superoperator $\hat{\hat{{\cal T}}}_a$ of the spatial translation by $a$ sites along 
the system ($a$ is an integer) as
\begin{eqnarray}
\sum_{m,m'} (\hat{\hat{{\cal T}}}_a)_{ll',mm'} \rho^{{\nu}}_{mm'} 
= \rho^{{\nu}}_{l+a,l'+a} \; , \label{blochaab}
\end{eqnarray}
where we use the notation
\begin{eqnarray}
\langle l \vert \hat{\rho} \vert l' \rangle = \rho_{ll'} \; .
\end{eqnarray}
This superoperator must have the group property
\begin{eqnarray}
\hat{\hat{{\cal T}}}_a \hat{\hat{{\cal T}}}_{a'} 
= \hat{\hat{{\cal T}}}_{a'} \hat{\hat{{\cal T}}}_a = \hat{\hat{{\cal T}}}_{a+a'} \; . \label{blochaac}
\end{eqnarray}
Because of the translational symmetry of the system, the translation superoperators commute 
with the Redfield superoperator
\begin{eqnarray}
\hat{\hat{{\cal T}}}_a \hat{\hat{{\cal L}}}_{\rm Red} {\hat{\rho}}(t) 
= \hat{\hat{{\cal L}}}_{\rm Red} \hat{\hat{{\cal T}}}_a {\hat{\rho}}(t) \; . \label{blochaad}
\end{eqnarray}
Therefore, the Redfield superoperator as well as the translation superoperators have a basis of
common eigenstates.
If $\tau(a)$ denotes the eigenvalues of the translation superoperator, we have that 
\begin{eqnarray}
\hat{\hat{{\cal T}}}_a \; {\hat{\rho}}^{{\nu}} = \tau(a) \; {\hat{\rho}}^{{\nu}} \; , \label{blochaae}
\end{eqnarray}
where, because of the unitarity of $\hat{\hat{{\cal T}}}_R$,
\begin{eqnarray}
\vert \tau(a) \vert^{2}=1 \; . \label{blochaaf}
\end{eqnarray}
Equation (\ref{blochaac}) implies
\begin{eqnarray}
\tau(a+a')=\tau(a) \; \tau(a') \; . \label{blochaag}
\end{eqnarray}
Because of equations (\ref{blochaaf}) and (\ref{blochaag}), we find that
\begin{eqnarray}
\tau(a)=e^{i q a} \; ,\label{blochaah}
\end{eqnarray}
where $q$ is the Bloch number, also called the wavenumber.
Finally, we find that
\begin{eqnarray}
\rho^{{\nu}}_{l+a,l'+a} = e^{iqa}  \rho^{{\nu}}_{ll'} \; . \label{blochaai}
\end{eqnarray}
A useful consequence is that
\begin{eqnarray}
\rho^{{\nu}}_{ll'} = e^{iql} \rho^{{\nu}}_{0,l'-l} \; .\label{blochaaj}
\end{eqnarray}
In order to determine the allowed values of the Bloch number, we write by using equation (\ref{isolaad}) that
\begin{eqnarray}
\rho^{\nu}_{ll'} = \frac{1}{N} \sum_{k,k'} \langle k \vert \hat{\rho}^{\nu} \vert k' \rangle 
e^{i (l k - l' k') \frac{2 \pi}{N}} \label{blochaak}
\end{eqnarray}
and
\begin{eqnarray}
\rho^{\nu}_{l+1,l'+1} &=& \frac{1}{N} \sum_{k,k'} \langle k \vert \hat{\rho}^{\nu} \vert k' \rangle 
e^{i (l k - l' k') \frac{2 \pi}{N}} e^{i (k - k') \frac{2 \pi}{N}} \; . \label{blochaal}
\end{eqnarray}
Due to equation (\ref{blochaai}), we also have
\begin{eqnarray}
\rho^{\nu}_{l+1,l'+1} &=& e^{iq} \; \rho^{\nu}_{ll'} \; .\label{blochabl}
\end{eqnarray}
Multiplying both sides of equations (\ref{blochaal}) and (\ref{blochabl}) by 
$\langle l' \vert k''' \rangle \langle k'' \vert l \rangle$, taking the sum
$\sum_{l,l'}$ of it, and identifying them, we get
\begin{eqnarray}
e^{iq} \langle k \vert \hat{\rho}^{\nu} \vert k' \rangle = e^{i (k - k') \frac{2 \pi}{N}} 
\langle k \vert \hat{\rho}^{\nu} \vert k' \rangle.  \label{blochaam}
\end{eqnarray}
We can now notice that if $q \neq (k-k')\frac{2 \pi}{N}$, then $\langle k \vert \hat{\rho}^{\nu} \vert k' \rangle=0$.
Using finally the periodicity 
\begin{eqnarray}
\rho_{l+N,l'}^{{\nu}}&=&\rho_{ll'}^{{\nu}} \; ,  \\
\rho_{l,l'+N}^{{\nu}}&=&\rho_{ll'}^{{\nu}} \; , \label{blochaao}
\end{eqnarray}
we find with (\ref{blochaai}) that the Bloch number takes the values
\begin{eqnarray}
q = j \frac{2 \pi}{N}, \ \ \text{where} \ \ j=0,1,\hdots,N-1 \; . \label{blochabo}
\end{eqnarray}
Accordingly, the Redfield superoperator can be block-diagonalized into $N$ 
independent blocks, which each contains $N$ eigenvalues as we shall see in the following.

\subsection{Simplifying the problem}

We will now formulate the eigenvalue problem of the Redfield superoperator
in each sector labeled by a given wavenumber $q$.  \\

For this purpose, we project equation (\ref{blochaaa}) onto the site basis using the explicit expression
(\ref{isolaam}) of the Redfield superoperator. We get
\begin{eqnarray}
s_{\nu}\rho^{\nu}_{ll'}&=& - i \frac{A}{\hbar} \left(- \rho^{\nu}_{l-1,l'} 
- \rho^{\nu}_{l+1,l'} + \rho^{\nu}_{l,l'-1} + \rho^{\nu}_{l,l'+1}\right) \nonumber \\
&&+ \frac{2 \lambda^2 Q}{\hbar^2} \left(\delta_{ll'} - 1\right) \rho^{\nu}_{ll'} \; . 
\label{spec1aaa}
\end{eqnarray}
Using equation (\ref{blochaaj}) and replacing $l'-l$ by $l$, we get
\begin{eqnarray}
\left( s_{\nu} + \frac{2 Q \lambda^2}{\hbar^2} \right) \rho^{{\nu}}_{0l} &=&  
\frac{2 A}{\hbar} \left(\sin\frac{q}{2}\right) \left( e^{-i \frac{q}{2}} \rho^{{\nu}}_{0,l+1} - 
e^{i \frac{q}{2}} \rho^{{\nu}}_{0,l-1}  \right)
\nonumber \\ & & +\frac{2 Q \lambda^2}{\hbar^2} \rho^{{\nu}}_{00} \; \delta_{0l} \; .
\label{spec1aac}
\end{eqnarray}
Making the change of variable 
\begin{eqnarray}
\rho^{{\nu}}_{0l}=i^{-l} e^{i \frac{q}{2} l} f_l \; ,
\end{eqnarray}
we obtain the simpler eigenvalue equation
\begin{eqnarray}
\mu_{\nu} f_l &=& \delta_{0l} f_l - i \beta (f_{l-1}+f_{l+1}) \; ,\label{spec1aad}
\label{eqvalproprefinale}
\end{eqnarray}
where 
\begin{eqnarray}
\mu_{\nu}=\frac{\hbar^2 s_{\nu}}{2Q \lambda^2} +1 \; ,\label{spec1baa}
\end{eqnarray}
and 
\begin{eqnarray}
\beta=\frac{A \hbar}{Q \lambda^2} \sin\frac{q}{2} \; . \label{spec1bab}
\end{eqnarray}

\section{Finite chain} \label{Sec.finite}

\subsection{The eigenvalue problem}

We can write the expression (\ref{eqvalproprefinale}) in matrix form (we no longer write the index ${\nu}$ to simplify the notation)
\begin{eqnarray}
\mu \vec{f} = \hat{W} \vec{f} \; , \label{spec1abe}
\end{eqnarray}
where $\mu$ are the eigenvalues, $\vec{f}=(f_{0},\hdots,f_{N-1})$ the eigenvectors of size
$N$, and $\hat{W}$ the following $N \times N$ matrix
\begin{eqnarray}
\left(\begin{array}{ccccccccc} 
1                                   & -i \beta &          &            &          &          & -i \beta i^{-N} e^{i N \frac{q}{2}}  \\
-i \beta                            & 0        & -i \beta &            &          &          &                                      \\
                                    & -i \beta & 0        & -i \beta   &          &          &                                      \\
                                    &          &  \ddots  & \ddots     & \ddots   &          &                                      \\
                                    &          &          & -i \beta   & 0        & -i \beta &                                      \\
                                    &          &          &            & -i \beta & 0        & -i \beta                             \\
-i \beta i^{N} e^{-i N \frac{q}{2}} &          &          &            &          & -i \beta & 0                                    \\                  
\end{array} \right)  \label{spec1aae}
\end{eqnarray}
We look for eigenstates of the form
\begin{eqnarray}
f_{l} = A e^{i \theta l} + B e^{-i \theta l}.\label{spec1aaf}
\end{eqnarray}
Solving equation (\ref{spec1abe}) with (\ref{spec1aaf}) gives
\begin{itemize}
\item for $0 < l < N-1$:
\begin{eqnarray}
\mu = -2i \; \beta \; \cos \theta \; , \label{spec1aag}
\end{eqnarray}
\item for $l=0$:
\begin{eqnarray}
&& (1-\mu)(A+B) - i \beta (A e^{i \theta}+B e^{-i \theta}) \nonumber \\  
&& - i \beta i^{-N} e^{i N \frac{q}{2}} (A e^{i \theta (N-1)}+B e^{-i \theta (N-1)}) =0 \; ,\label{spec1aah}
\end{eqnarray}
\item for $l=N-1$:
\begin{eqnarray}
A e^{i \theta N}+B e^{-i \theta N} - i^N e^{-i N\frac{q}{2}} (A+B) =0 \; .\label{spec1aai}
\end{eqnarray}
\end{itemize}
Solving the homogeneous linear system of equations (\ref{spec1aah}) and (\ref{spec1aai}) and replacing
$\mu$ by (\ref{spec1aag}), one gets the characteristic equation
\begin{eqnarray}
2 i \beta \sin \theta \left[ R(q) - \cos \theta N \right] = \sin \theta N, \label{spec1aaj} 
\end{eqnarray}
with
\begin{eqnarray}
R(q)=\frac{1}{2}\left(i^{N} e^{-i N \frac{q}{2}} + i^{-N} e^{i N \frac{q}{2}} \right). \label{spec1aatheta} 
\end{eqnarray}
Using equation (\ref{blochabo}), we find:
\begin{eqnarray}
&\mbox{for}& \; N \; \mbox{odd}: \qquad R(q_j)=0 \; , \\
&\mbox{for}& \; N=4I: \qquad R(q_j)=(-1)^j \; ,\\
&\mbox{for}& \; N=4I+2: \qquad R(q_j)=-(-1)^j \; , 
\end{eqnarray}
with $I$ integer.
From now on, we shall speak of even (respectively odd) $q$, if 
$q$ corresponds to an even (respectively odd) integer $j$ in equation (\ref{blochabo}).
Therefore, the characteristic equation (\ref{spec1aaj}) becomes
\begin{itemize}
\item for $N$ odd: 
\begin{eqnarray}
2 i \beta \sin \theta = - \tan \theta N \; ; \label{spec1aal} 
\end{eqnarray}
\item for $N=4I$ and $q$ even or for $N=4I+2$ and $q$ odd:\\ 
either
\begin{eqnarray}
2 i \beta \sin \theta  &=& {\rm cotan} \frac{\theta N}{2} \; , \label{spec1aam} 
\end{eqnarray}
or 
\begin{eqnarray}
\cos \theta  N &=& 1 \; ; \label{spec1abm} 
\end{eqnarray}
\item for $N=4I$ and $q$ odd or for $N=4I+2$ and $q$ even:\\ 
either
\begin{eqnarray}
2 i \beta \sin \theta  &=& - \tan \frac{\theta N}{2} \; , \label{spec1aan} 
\end{eqnarray}
or
\begin{eqnarray}
\cos \theta N &=& -1 \; .\label{spec1abn} 
\end{eqnarray}
\end{itemize}
We solve equations (\ref{spec1aal})-(\ref{spec1abn}) as follows.

\subsection{The diffusive eigenvalue $\mu^{(1)}$}

We first look for an eigenvalue $\mu$ which is real and should correspond to a monotonic
exponential decay.  With this goal, we suppose that the angle $\theta$ is complex
\begin{eqnarray}
\theta = \xi + i \eta \; , \label{theta}
\end{eqnarray}
so that the eigenvalue (\ref{spec1aag}) becomes
\begin{eqnarray}
\mu = -2i  \beta \cos \theta = - 2 \beta \sin\xi \sinh\eta - 2i\beta \cos\xi\cosh\eta \; .
\end{eqnarray}
This eigenvalue is real under the condition that $\cos\xi=0$
which is satisfied for
\begin{eqnarray}
\xi = - \frac{\pi}{2} \; , \label{xi}
\end{eqnarray}
in which case
\begin{eqnarray}
\mu =  - 2 \; \beta \; \sinh\eta  \; . \label{cmu}
\end{eqnarray}
We notice that the condition $\cos\xi=0$ is also satisfied for $\xi=\frac{\pi}{2}$
but it can be shown that this other case leads to the same eigenvalue as (\ref{xi}).
If we introduce the conditions (\ref{theta}) and (\ref{xi}) in equations (\ref{spec1aal}),
(\ref{spec1aam}), and (\ref{spec1aan}), we get
\begin{itemize}
\item for $N$ odd: 
\begin{eqnarray}
2 \beta \cosh \eta = \coth N\eta \; ; \label{cspec1aal} 
\end{eqnarray}
\item for $N=4I$ and $q$ even or for $N=4I+2$ and $q$ odd:\\ 
\begin{eqnarray}
2 \beta \cosh \eta  &=& \coth \frac{N\eta}{2} \; ; \label{cspec1aam} 
\end{eqnarray}
\item for $N=4I$ and $q$ odd or for $N=4I+2$ and $q$ even:\\ 
\begin{eqnarray}
2 \beta \cosh \eta  &=& \tanh \frac{N\eta}{2} \; . \label{cspec1aan} 
\end{eqnarray}
\end{itemize}

In the limit $N\to\infty$, the right-hand side of these equations tends to unity if
a non-vanishing solution $\eta \neq 0$ exists.  In this limit, this solution
is thus given by
\begin{eqnarray}
\eta_0 &=& {\rm arccosh} \; \frac{1}{2 \beta} \; ,\label{spec2abp}
\end{eqnarray}
which exists only if $\beta \leq \frac{1}{2}$.
Because
\begin{eqnarray}
\sinh \eta_0 &=& \sqrt{\left(\frac{1}{2 \beta}\right)^2 - 1} \; ,\label{spec2aar}
\end{eqnarray}
the corresponding eigenvalue should be
\begin{eqnarray}
\mu_0 &=& \sqrt{1-(2 \beta)^2} \; .\label{spec2aas}
\end{eqnarray}

However, for a finite chain with $N<\infty$, we expect a correction $\delta\eta$ to the solution
$\eta=\eta_0+\delta\eta$.  Replacing this correction in equations (\ref{cspec1aal}),
(\ref{cspec1aam}), and (\ref{cspec1aan}), we obtain by Taylor expansion that
\begin{itemize}
\item for $N$ odd:
\begin{eqnarray}
\delta \eta \simeq 2 \; \frac{e^{- 2 N {\rm arccosh} \; \frac{1}{2 \beta}}}{\sqrt{1-(2 \beta)^2}} \; ; \label{spec3abc}
\end{eqnarray}
\item for $N=4I$ and $q$ even or for $N=4I+2$ and $q$ odd:
\begin{eqnarray}
\delta \eta \simeq 2 \; \frac{e^{- N {\rm arccosh} \; \frac{1}{2 \beta}}}{\sqrt{1-(2 \beta)^2}} \; ;
\label{spec3acc}
\end{eqnarray}
\item for $N=4I$ and $q$ odd or for $N=4I+2$ and $q$ even:
\begin{eqnarray}
\delta \eta \simeq - 2 \; \frac{e^{- N {\rm arccosh} \; \frac{1}{2 \beta}}}{\sqrt{1-(2 \beta)^2}} \; ;
\label{spec3adc}
\end{eqnarray}
\end{itemize}
up to corrections of $O(\delta\eta^2)$.
Using the expression (\ref{cmu}), we finally obtain the eigenvalue
\begin{eqnarray}
\mu^{(1)} &=& 2\beta \sinh \eta_0 + 2 \beta \cosh \eta_0 \; \delta \eta + O(\delta \eta^2) \nonumber \\
&=& \sqrt{1-(2 \beta)^2} + \delta \eta + O(\delta\eta^2)\; ,\label{spec3aad}
\end{eqnarray}
where $\delta \eta$ is respectively given by equations (\ref{spec3abc}), (\ref{spec3acc}), and (\ref{spec3adc}).
Accordingly, the correction $\delta\eta$ to the eigenvalue decreases exponentially fast with the
size $N$ of the chain. \\

Using equations (\ref{spec1baa}) and (\ref{spec1bab}), we finally obtain the eigenvalue
\begin{eqnarray}
s^{(1)} &=& \frac{2 Q \lambda^2}{\hbar^2} (\mu^{(1)} - 1) \nonumber\\
&=& \frac{2 Q \lambda^2}{\hbar^2} \sqrt{1 - \left( \frac{2A \hbar}{Q\lambda^2} \sin \frac{q}{2}\right)^2} 
- \frac{2 Q \lambda^2}{\hbar^2} 
+ O(\delta\eta) \; .  \label{spec3aae} 
\end{eqnarray}
The eigenvalue $s^{(1)}=0$ corresponding to a vanishing wavenumber $q=0$ is always in the spectrum
of the Redfield superoperator.  The associated eigenstate describes the stationary equilibrium state.
At low wavenumbers $q,\beta\to 0$, we recover the dispersion relation of diffusion
\begin{eqnarray}
s^{(1)} = -D q^2 + O(q^4)\; , \label{spec3aaf} 
\end{eqnarray}
with the diffusion coefficient
\begin{eqnarray}
D = \frac{A^2}{Q \lambda^2} \; , \label{spec3aag}
\end{eqnarray}
which justifies calling $\mu^{(1)}$ or $s^{(1)}$ the diffusive eigenvalue.\\

We notice that the diffusive eigenvalue no longer exists beyond the critical value $\beta_{c}=\frac{1}{2}$. 
Since the matrix (\ref{spec1aae}) has a total of $N$ eigenvalues, we expect
further non-diffusive eigenvalues in a number of $N-1$ for $\beta< \frac{1}{2}$
and $N$ for $\beta > \frac{1}{2}$, as confirmed in the following subsections.

\subsection{The eigenvalues $\mu^{(2)}$}

Beside the diffusive eigenvalue, we expect eigenvalues corresponding to the chain-like
structure of the matrix (\ref{spec1aae}).  To obtain these eigenvalues, we set
$\tan \chi = 2i \beta \sin \theta $ so that equations (\ref{spec1aal}), (\ref{spec1aam}), and (\ref{spec1aan})
can be written respectively
\begin{eqnarray}
&& \sin(\theta  N+\chi)=0 \; , \\ && \cos\left(\frac{\theta  N}{2}+\chi\right)=0 \; , \\ 
&& \sin\left(\frac{\theta  N}{2}+\chi\right)=0 \; .
\end{eqnarray}
Therefore, we obtain
\begin{eqnarray}
i\left(n \pi - \theta  N\right)&=&{\rm arctanh}(-2 \beta \sin \theta  ) \; , \label{spec1aaq} \\
i\left(n\pi +\frac{\pi}{2} - \frac{\theta N}{2}\right)&=&{\rm arctanh}(-2 \beta \sin \theta  )\; , \label{spec1aar} \\
i\left(n \pi - \frac{\theta N}{2}\right)&=&{\rm arctanh}(-2 \beta \sin \theta  ) \; .\label{spec1abs}
\end{eqnarray}

We now expand ${\rm arctanh}(-2 \beta \sin \theta  )$ around $\beta=0$:
\begin{eqnarray}
{\rm arctanh}(-2 \beta \sin \theta  ) \stackrel{\beta \to 0}{=} 
-2\beta \sin \theta  - \frac{8}{3} \beta^3 \sin^3 \theta  + O(\beta^5) \; . \label{spec1aay} 
\end{eqnarray}

If $\beta=0$, the solutions of equations (\ref{spec1aaq}), (\ref{spec1aar}), and (\ref{spec1abs})
are respectively given by
\begin{eqnarray}
\theta _0&=&\frac{n \pi}{N}, \ \ \text{where} \ \ n=1,2,\hdots,N-1 , \label{spec1aas} \\
\theta _0&=&\frac{(2n+1) \pi}{N}, \ \ \text{where} \ \ n=0,1,\hdots,\frac{N}{2}-1 , \label{spec1aat} \\
\theta _0&=&\frac{2n \pi}{N}, \ \ \text{where} \ \ n=1,2,\hdots,\frac{N}{2}-1 , \label{spec1aau}
\end{eqnarray}
Notice that $n=0$ is rejected in equations (\ref{spec1aas}) and (\ref{spec1aau}).
It is due to the fact that $\theta =0$ does not correspond to an eigenvector 
because it can be seen that $f_{l}=A+B\neq 0$ in equation (\ref{spec1aaf})
cannot be an eigenvector of equation (\ref{spec1aae}).\\

Using the expansion (\ref{spec1aay}) in equations (\ref{spec1aaq}), (\ref{spec1aar}) and (\ref{spec1abs}) 
with $\theta =\theta_0 + \delta \theta $, we find
\begin{eqnarray}
\delta \theta \stackrel{\beta \to 0}{=} && - \frac{i 2 \beta}{M} \sin \theta_0 - \frac{i 8 \beta^3}{3 M} \sin^3 \theta_0 \nonumber \\
&& + O\left(i \frac{\beta^5}{M}\right) + O\left(\frac{\beta^2}{M^2}\right) ,
\label{spec1aaz} 
\end{eqnarray}
where $M=N$ for equation (\ref{spec1aaq}) and $M=\frac{N}{2}$ for equations (\ref{spec1aar}) and (\ref{spec1abs}).
The eigenvalue (\ref{spec1aag}) is now given by the expansion 
\begin{eqnarray}
\mu &=& - 2i \beta \cos(\theta_0 + \delta \theta) \nonumber \\
&\stackrel{\delta \theta \to 0}{=}& - 2i \beta \cos \theta_0 + 2 i \beta \sin \theta_0 \; \delta \theta \nonumber \\ 
&&+ i \beta \cos \theta_0 \;  \delta \theta^2 + O(\delta \theta^3) . \label{spec2aac}
\end{eqnarray}
Using equation (\ref{spec1aaz}) in (\ref{spec2aac}) gives
\begin{eqnarray}
\mu^{(2)} &\stackrel{\beta \to 0}{=}& - 2i \beta \cos \theta_0 + O\left(i \frac{\beta^3}{M^2}\right) 
\nonumber \\ 
&&+ \frac{4 \beta^2}{M} \sin^2 \theta_0 + \frac{16 \beta^4}{3 M} \sin^4 \theta_0 + 
O\left(\frac{\beta^6}{M}\right) \; .  \label{spec2aad} 
\end{eqnarray}
Consequently, we have
\begin{itemize}
\item for $N$ odd, using equation (\ref{spec2aad}) with (\ref{spec1aas}):
\begin{eqnarray}
\mu^{(2)} &\stackrel{\beta \to 0}{=}& - 2i \beta \cos \frac{n \pi}{N} + \frac{4 \beta^2}{N} \sin^2 \frac{n \pi}{N} 
+ \frac{16 \beta^4}{3 N} \sin^4 \frac{n \pi}{N} , \nonumber \\ && \text{where} \ \ 
n=1,2,\hdots,N-1 \; ; \label{spec2aaf} 
\end{eqnarray}
\item for $N=4I$ and $q$ even or for $N=4I+2$ and $q$ odd, using equation (\ref{spec2aad}) with (\ref{spec1aat}):
\begin{eqnarray}
\mu^{(2)} &\stackrel{\beta \to 0}{=}& - 2i \beta \cos \frac{(2n+1) \pi}{N} \nonumber \\ &&
+ \frac{8 \beta^2}{N} \sin^2 \frac{(2n+1) \pi}{N} 
+ \frac{32 \beta^4}{3 N} \sin^4 \frac{(2n+1) \pi}{N}, \nonumber \\ && \text{where} \ \ 
n=0,1,\hdots,\frac{N}{2}-1 \; ; \label{spec2aai} 
\end{eqnarray}
\item for $N=4I$ and $q$ odd or for $N=4I+2$ and $q$ even, using equation (\ref{spec2aad}) 
with (\ref{spec1aau}):
\begin{eqnarray}
\mu^{(2)} &\stackrel{\beta \to 0}{=}& - 2i \beta \cos \frac{2n \pi}{N} + \frac{8 \beta^2}{N} \sin^2 \frac{2n \pi}{N} 
+ \frac{32 \beta^4}{3 N} \sin^4 \frac{2n \pi}{N}, \nonumber \\ && \text{where} \ \ 
n=1,2,\hdots,\frac{N}{2}-1 \; .\label{spec2aal} 
\end{eqnarray}
\end{itemize}

\subsection{The eigenvalues $\mu^{(3)}$}

The solutions of equation (\ref{spec1abm}) are simply given by
\begin{eqnarray}
\theta =\frac{2 n \pi}{N}, \ \ \text{where} \ \ n=1,2,\hdots,\frac{N}{2}-1\;  . \label{spec1aao}
\end{eqnarray}
We reject $n=0$ and $n=\frac{N}{2}$ because the corresponding eigenvector does not exist in these cases. 
Similarly, the solutions of equation (\ref{spec1abn}) are given by
\begin{eqnarray}
\theta =\frac{(2n+1) \pi}{N}, \ \ \text{where} \ \ n=0,1,\hdots,\frac{N}{2}-1\; . \label{spec1aap}
\end{eqnarray}

Consequently, we have the further eigenvalues:\\
\begin{itemize}
\item for $N=4I$ and $q$ even or for $N=4I+2$ and $q$ odd, using equation (\ref{spec1aag}) with (\ref{spec1aao}):
\begin{eqnarray}
\mu^{(3)} &=& - 2i \beta \cos\left(\frac{2n \pi}{N}\right) \; , \ \ \text{where} \ \ n=1,2,\hdots,\frac{N}{2}-1 \; ;
\label{spec2aah} 
\end{eqnarray}
\item for $N=4I$ and $q$ odd or for $N=4I+2$ and $q$ even, using equation (\ref{spec1aag}) with (\ref{spec1aap}):
\begin{eqnarray}
\mu^{(3)} &=& - 2i \beta \cos\left(\frac{(2n+1) \pi}{N}\right), \nonumber \\
&& \text{where} \ \ n=0,1,\hdots,\frac{N}{2}-1 \; .
\label{spec2aak} 
\end{eqnarray}
\end{itemize}

\subsection{The eigenvalues $\mu^{(4)}$}

An important observation is that,
for $\beta < \frac{1}{2}$, the expansion (\ref{spec1aay}) which implies $2 \beta \sin \theta < 1$
is satisfied everywhere, i.e., for all the values of $\theta$ and therefore for all the eigenvalues. 
However, for $\beta > \frac{1}{2}$, the Taylor expansion around $\beta=0$ in equation (\ref{spec1aay}) 
is only valid if $2 \beta \sin \theta < 1$.
Therefore, a transition zone exists around  $\sin \theta \simeq 1/(2 \beta)$. 
According to equation (\ref{spec1aag}), this transition corresponds to the critical value of the eigenvalue
given by
\begin{eqnarray}
\mu_{c} &=& \pm i \sqrt{(2 \beta)^2-1} \; .  \label{spec2abc}
\end{eqnarray}
Therefore, for $\beta > \frac{1}{2}$, the expansion (\ref{spec1aay}) around $\beta=0$ is only valid if 
$\vert \mu \vert > \vert \mu_{c}\vert$. 
For $\vert \mu \vert < \vert \mu_{c}\vert$, we should instead consider the asymptotic expansion of 
${\rm arctanh}(-2 \beta \sin \theta  )$ around $\beta=\infty$:
\begin{eqnarray}
{\rm arctanh}(-2 \beta \sin \theta  ) &\stackrel{\beta \to \infty}{=}& 
i \frac{\pi}{2} -\frac{1}{2 \beta \sin \theta } -\frac{1}{24 \beta^3 \sin^3 \theta } \nonumber \\
&& -\frac{1}{160 \beta^5 \sin^5 \theta } + O\left(\frac{1}{\beta^7}\right) \; , \nonumber \\
\label{spec2aaa}
\end{eqnarray}
which leads to another family of eigenvalues existing for $\beta > \frac{1}{2}$.\\

If $\beta\to\infty$, the solutions of equations (\ref{spec1aaq}), (\ref{spec1aar}) and (\ref{spec1abs})
are respectively given by
\begin{eqnarray}
\theta _0&=&\frac{(n+\frac{1}{2}) \pi}{N}, \ \ \text{where} \ \ n=0,1,\hdots,N-1 \, , \label{spec1aav} \\
\theta _0&=&\frac{2n \pi}{N}, \ \ \text{where} \ \ n=1,2,\hdots,\frac{N}{2}-1 \, , \label{spec1aaw} \\
\theta _0&=&\frac{(2n+1) \pi}{N}, \ \ \text{where} \ \ n=0,1,\hdots,\frac{N}{2}-1 . \label{spec1aax} 
\end{eqnarray}
Because of the condition $\vert \mu \vert < \vert \mu_{c}\vert$ with the critical values (\ref{spec2abc}),
we should only consider the angles in the interval $\theta_{0,c}< \theta_0 < \pi - \theta_{0,c}$ with
\begin{eqnarray}
\theta_{0,c} &=& \arcsin \frac{1}{2\beta} \; ,  
\label{theta0c}
\end{eqnarray}
so that the integer $n$ in equations (\ref{spec1aav}), (\ref{spec1aaw}), and (\ref{spec1aax})
is restricted to take the intermediate values $n_{\rm min} < n < n_{\rm max}$ which do
not reach the extreme values.\\

Using the expansion (\ref{spec2aaa}) in equations (\ref{spec1aaq}), (\ref{spec1aar}), and (\ref{spec1abs}) 
with $\theta =\theta_0 + \delta \theta $, we find
\begin{eqnarray}
\delta \theta &\stackrel{\beta\to\infty}{=}& - \frac{i}{2 \beta M \sin \theta_0} 
- \frac{i}{24 \beta^3 M \sin^3 \theta_0} - \frac{i}{160 \beta^5 M \sin^5 \theta_0}\nonumber \\
&&+ O\left(\frac{i}{\beta^7 M}\right) + O\left(\frac{1}{\beta^2 M^2}\right) \; , \label{spec2aab}
\end{eqnarray}
where $M=N$ for equation (\ref{spec1aaq}) and $M=\frac{N}{2}$ for equations (\ref{spec1aar}) and (\ref{spec1abs}).\\

Using equation (\ref{spec2aab}) in (\ref{spec2aac}) gives
\begin{eqnarray}
\mu^{(4)} &\stackrel{\beta \to \infty}{=}& - 2i \beta \cos \theta_0 +  \frac{1}{M} \left(1+
\frac{1}{12 \beta^2 \sin^2 \theta_0 } + \frac{1}{80 \beta^4 \sin^4 \theta_0}\right) \nonumber \\ 
&&+ O\left(\frac{1}{\beta^6 M}\right) + O\left(\frac{i}{\beta M^2}\right) \; . \label{spec2aae}
\end{eqnarray}
Consequently, we have
\begin{itemize}
\item for $N$ odd, using equation (\ref{spec2aae}) with (\ref{spec1aav}):
\begin{eqnarray}
\mu^{(4)} &\stackrel{\beta \to \infty}{=}& - 2i \beta \cos \frac{(n+\frac{1}{2}) \pi}{N} 
+\frac{1}{N} \left(1+\frac{1}{12 \beta^2 \sin^2 \frac{(n+\frac{1}{2}) \pi}{N}} \right. \nonumber \\
&&\left.
+\frac{1}{80\beta^4\sin^4 \frac{(n+\frac{1}{2}) \pi}{N}}\right) 
, \nonumber \\ && \text{where} \ \ n=0,1,\hdots,N-1 \; ;\label{spec2aag} 
\end{eqnarray}
\item for $N=4I$ and $q$ even or for $N=4I+2$ and $q$ odd, using equation (\ref{spec2aae}) with (\ref{spec1aaw}):
\begin{eqnarray}
\mu^{(4)} &\stackrel{\beta \to \infty}{=}& - 2i \beta \cos \frac{2 n \pi}{N} \nonumber \\
&&+ \frac{2}{N} \left(1+\frac{1}{12 \beta^2 \sin^2 \frac{2 n \pi}{N}}
+\frac{1}{80 \beta^4 \sin^4 \frac{2 n \pi}{N}}\right) 
, \nonumber \\ && \text{where} \ \ n=1,2,\hdots,\frac{N}{2}-1 \; ;\label{spec2aaj} 
\end{eqnarray}
\item for $N=4I$ and $q$ odd or for $N=4I+2$ and $q$ even, using equation (\ref{spec2aae}) with (\ref{spec1aax}):
\begin{eqnarray}
\mu^{(4)} &\stackrel{\beta \to \infty}{=}& - 2i \beta \cos \frac{(2 n+1) \pi}{N} \nonumber \\
&&+ \frac{2}{N} \left(1+\frac{1}{12 \beta^2 \sin^2 \frac{(2 n+1) \pi}{N}} 
+\frac{1}{80 \beta^4 \sin^4 \frac{(2 n+1) \pi}{N}}\right)  
, \nonumber \\ && \text{where} \ \ n=0,1,\hdots,\frac{N}{2}-1 \; ;\label{spec2aam} 
\end{eqnarray}
\end{itemize}
with the aforementioned restriction on the values of the integer $n$.\\

\subsection{The eigenvalues $\mu^{(5)}$}

For $N=4I$ and $q$ even or $N=4I+2$ and $q$ odd,
two special eigenvalues exist in the limit $\beta\to\infty$ around $\theta_0=0$ and $\theta_0=\pi$.
They can be obtained by taking $\theta=\theta_0+\delta \theta$ and directly solving 
equation (\ref{spec1aam}) to get in both cases
\begin{eqnarray}
\delta \theta^2 \stackrel{\beta\to\infty}{=} \frac{1}{i \beta N} + O\left(\frac{1}{\beta^2}\right) \; . \label{spec1abo}
\end{eqnarray}
Inserting in equation (\ref{spec2aac}), we obtain 
\begin{eqnarray}
\mu^{(5)} &\stackrel{\beta \to \infty}{=}& \mp 2i \beta + \frac{1}{N} + O\left(\frac{1}{\beta}\right) \; .
\label{spec2abh} 
\end{eqnarray}

\subsection{Description of the spectrum} \label{spectradisuss}

Rewriting the eigenvalues (\ref{spec1aag}) of the Redfield superoperator with their explicit dependence 
in terms of equation (\ref{spec1bab}), we get 
\begin{eqnarray}
\mu_{\nu} = \mu_{q\theta} = -2 i \beta \cos \theta = 
-2 i \frac{A \hbar}{Q \lambda^2} \left(\sin \frac{q}{2}\right) \cos \theta \; ,\label{spec3abd} 
\end{eqnarray}
where $\nu$ can take $N^2$ different values because $q$ and $\theta$ take $N$ values each.
Remembering that according to equation (\ref{spec1baa}) 
\begin{eqnarray}
s_{\nu} = s_{q\theta} = \frac{2 Q \lambda^2}{\hbar^2} (\mu_{q\theta}- 1), \label{spec3acd} 
\end{eqnarray}
we conclude that we have found in this section all the eigenvalues of the Redfield superoperator. 
We list them in Tables \ref{table1} and \ref{table2}
according to the parameter regime in which they hold.\\

\begin{table}
\caption{{\bf For wavenumbers $q$ corresponding to $\beta < \frac{1}{2}$:} List of the eigenvalues $\mu_{q\theta}$ of 
the matrix (\ref{spec1aae}) given by equation (\ref{spec3abd}). 
These eigenvalues are directly related to the Redfield superoperator eigenvalues by equation (\ref{spec3acd}).}
\vspace*{0.5cm}
\scriptsize
\begin{tabular}{llll}
\hline
\hline
\\
$N$ odd                      & 
$N=4I$ and $q$ even          &                                                                          
$N=4I$ and $q$ odd      
\\
                                & 
or $N=4I+2$ and $q$ odd         &                                                                                
or $N=4I+2$ and $q$ even        
\\ 
\hline
\hline
\\ 
$\mu^{(1)}=\sqrt{1-(2 \beta)^2}$   & 
$\mu^{(1)}=\sqrt{1-(2 \beta)^2}$     &                       
$\mu^{(1)}=\sqrt{1-(2 \beta)^2}$  
\\
$\hspace{1.0cm} + 2 \frac{e^{- 2 N {\rm arccosh} \; 1/(2 \beta)}}{\sqrt{1-(2 \beta)^2}}$  &
$\hspace{1.0cm} + 2 \frac{e^{- N {\rm arccosh} \; 1/(2 \beta)}}{\sqrt{1-(2 \beta)^2}}$    &
$\hspace{1.0cm} - 2 \frac{e^{- N {\rm arccosh} \; 1/(2 \beta)}}{\sqrt{1-(2 \beta)^2}}$   
\vspace{0.5cm}
\\
\hline
\\
$\mu^{(2)} \stackrel{\beta \to 0}{=} - 2i \beta \cos \frac{n \pi}{N}$                &
$\mu^{(2)} \stackrel{\beta \to 0}{=} - 2i \beta \cos \frac{(2n+1) \pi}{N}$           &
$\mu^{(2)} \stackrel{\beta \to 0}{=} - 2i \beta \cos \frac{2n \pi}{N}$
\\
$\hspace{1.0cm} + \frac{4 \beta^2}{N} \sin^2 \frac{n \pi}{N} $            &
$\hspace{1.0cm} + \frac{8 \beta^2}{N} \sin^2 \frac{(2n+1) \pi}{N}$        &
$\hspace{1.0cm} + \frac{8 \beta^2}{N} \sin^2 \frac{2n \pi}{N}$
\\
$\hspace{1.0cm} + \frac{16 \beta^4}{3 N} \sin^4 \frac{n \pi}{N}$                &
$\hspace{1.0cm} + \frac{32 \beta^4}{3 N} \sin^4 \frac{(2n+1) \pi}{N}$           &
$\hspace{1.0cm} + \frac{32 \beta^4}{3 N} \sin^4 \frac{2n \pi}{N}$    
\\
$\hspace{1.0cm} + O(\frac{\beta^6}{N})+ O(\frac{i \beta^3}{N^2})$  &
$\hspace{1.0cm} + O(\frac{\beta^6}{N})+ O(\frac{i \beta^3}{N^2})$  &
$\hspace{1.0cm} + O(\frac{\beta^6}{N})+ O(\frac{i \beta^3}{N^2})$
\\
\hspace{1.0cm} for $n=1,2,\hdots,N-1$              &
\hspace{1.0cm} for $n=0,1,\hdots,\frac{N}{2}-1$    &
\hspace{1.0cm} for $n=1,2,\hdots,\frac{N}{2}-1$    
\vspace{0.5cm}
\\
\hline
\\
                                               &
$\mu^{(3)}=- 2i \beta \cos(\frac{2n \pi}{N})$      &
$\mu^{(3)}=- 2i \beta \cos(\frac{(2n+1) \pi}{N})$  
\\
                                                     &
\hspace{0.6cm} for $n=1,2,\hdots,\frac{N}{2}-1$      &
\hspace{0.6cm} for $n=0,1,\hdots,\frac{N}{2}-1$
\vspace{0.5cm}
\\
\hline
\hline
\end{tabular}
\normalsize
\label{table1}
\end{table}
\begin{table}
\caption{{\bf For wavenumbers $q$
corresponding to $\beta > \frac{1}{2}$:} List of the eigenvalues $\mu_{q\theta}$ of the matrix 
(\ref{spec1aae}) given by equation (\ref{spec3abd}). 
These eigenvalues are directly related to the Redfield superoperator eigenvalues by equation (\ref{spec3acd}).}
\vspace*{0.5cm}
\scriptsize
\begin{tabular}{llll}
\hline
\hline
\\
$N$ odd                      & 
$N=4I$ and $q$ even          &                                                                          
$N=4I$ and $q$ odd      
\\
                                & 
or $N=4I+2$ and $q$ odd         &                                                                                
or $N=4I+2$ and $q$ even        
\\ 
\hline
\hline
\\
\underline{If $\vert \mu^{(4)} \vert  < \vert \mu_c \vert$:}\footnotemark[1]
\vspace{0.5cm}
\\
$\mu^{(4)} \stackrel{\beta \to \infty}{=} - 2i \beta \cos \frac{(n+\frac{1}{2}) \pi}{N}$ &
$\mu^{(4)} \stackrel{\beta \to \infty}{=} - 2i \beta \cos \frac{2 n \pi}{N}$          &
$\mu^{(4)} \stackrel{\beta \to \infty}{=} - 2i \beta \cos \frac{(2 n+1) \pi}{N}$
\\
$\hspace{1.0cm}+ \frac{1}{N} (1+\frac{1}{12 \beta^2 \sin^2 \frac{(n+(1/2)) \pi}{N}}$  &
$\hspace{1.0cm}+ \frac{2}{N} (1+\frac{1}{12 \beta^2 \sin^2 \frac{2 n \pi}{N}}$        &
$\hspace{1.0cm}+ \frac{2}{N} (1+\frac{1}{12 \beta^2 \sin^2 \frac{(2 n+1) \pi}{N}}$ 
\\
$\hspace{2cm}+\frac{1}{32 \beta^4 \sin^4 \frac{(n+(1/2)) \pi}{N}})$ &
$\hspace{2cm}+\frac{1}{32 \beta^4 \sin^4 \frac{2 n \pi}{N}})$       &
$\hspace{2cm}+\frac{1}{32 \beta^4 \sin^4 \frac{(2 n+1) \pi}{N}})$
\\
$\hspace{1.0cm} + O(\frac{1}{\beta^6 N})+ O(\frac{i}{\beta N^2})$  &
$\hspace{1.0cm} + O(\frac{1}{\beta^6 N})+ O(\frac{i}{\beta N^2})$  &
$\hspace{1.0cm} + O(\frac{1}{\beta^6 N})+ O(\frac{i}{\beta N^2})$
\\
\hspace{1.0cm} for $n_{\rm min} < n <n_{\rm max}$              &
\hspace{1.0cm} for $n_{\rm min} < n <n_{\rm max}$    &
\hspace{1.0cm} for $n_{\rm min} < n <n_{\rm max}$    
\\
\\ 
                                                                                                             &
$\mu^{(5)} \stackrel{\beta \to \infty}{=} \mp 2i \beta + \frac{1}{N} + O(\frac{1}{\beta})$   &
\\
\vspace{0.5cm}
\\
\underline{If $\vert \mu^{(2)} \vert > \vert \mu_c \vert$:}\footnotemark[1]
\vspace{0.5cm}
\\
$\mu^{(2)} \stackrel{\beta \to 0}{=} - 2i \beta \cos \frac{n \pi}{N}$                &
$\mu^{(2)} \stackrel{\beta \to 0}{=} - 2i \beta \cos \frac{(2n+1) \pi}{N}$           &
$\mu^{(2)} \stackrel{\beta \to 0}{=} - 2i \beta \cos \frac{2n \pi}{N}$
\\
$\hspace{1.0cm} + \frac{4 \beta^2}{N} \sin^2 \frac{n \pi}{N} $            &
$\hspace{1.0cm} + \frac{8 \beta^2}{N} \sin^2 \frac{(2n+1) \pi}{N}$        &
$\hspace{1.0cm} + \frac{8 \beta^2}{N} \sin^2 \frac{2n \pi}{N}$
\\
$\hspace{1.0cm} + \frac{16 \beta^4}{3 N} \sin^4 \frac{n \pi}{N}$                &
$\hspace{1.0cm} + \frac{32 \beta^4}{3 N} \sin^4 \frac{(2n+1) \pi}{N}$           &
$\hspace{1.0cm} + \frac{32 \beta^4}{3 N} \sin^4 \frac{2n \pi}{N}$    
\\
$\hspace{1.0cm} + O(\frac{\beta^6}{N})+ O(\frac{i \beta^3}{N^2})$  &
$\hspace{1.0cm} + O(\frac{\beta^6}{N})+ O(\frac{i \beta^3}{N^2})$  &
$\hspace{1.0cm} + O(\frac{\beta^6}{N})+ O(\frac{i \beta^3}{N^2})$
\\
\hspace{1.0cm} for $n=1,2,\hdots,n_{\rm min}$ &
\hspace{1.0cm} for $n=0,1,\hdots,n_{\rm min}$ &
\hspace{1.0cm} for $n=1,2,\hdots,n_{\rm min}$   
\\
\hspace{1.0cm} and $n=n_{\rm max},...,N-1$              &
\hspace{1.0cm} and $n=n_{\rm max},...,\frac{N}{2}-1$    &
\hspace{1.0cm} and $n=n_{\rm max},...,\frac{N}{2}-1$  
\\
\vspace{0.5cm}
\\
\hline  
\\
                                               &
$\mu^{(3)}=- 2i \beta \cos(\frac{2n \pi}{N})$      &
$\mu^{(3)}=- 2i \beta \cos(\frac{(2n+1) \pi}{N})$  
\\
                                                     &
\hspace{0.6cm} for $n=1,2,\hdots,\frac{N}{2}-1$      &
\hspace{0.6cm} for $n=0,1,\hdots,\frac{N}{2}-1$
\vspace{0.5cm}
\\
\hline
\hline
\\
$^1 \mu_c = \pm i \sqrt{(2 \beta)^2-1}$
\end{tabular}
\normalsize
\label{table2}
\end{table}

We now discuss the main features of the spectrum when the different physical parameters are varied. 
This discussion is based on our analytical results for the eigenvalues 
and on the comparison between these 
results and the eigenvalues obtained by numerical diagonalization of the Redfield superoperator. 
Since the eigenvalues $\mu_{q\theta}$ are related to the Redfield 
superoperator eigenvalues $s_{q\theta}$ by equation (\ref{spec3acd}), we notice that
all the eigenvalues of the complete spectrum always satisfy $0\leq {\cal R}e \; \mu_{q\theta} \leq 1$ or, 
equivalently, $-2 Q \lambda^2 / \hbar^2 \leq {\cal R}e \; s_{q\theta} \leq 0$.
The imaginary part of $s_{q\theta}$ is simply proportional by a factor $2 Q \lambda^2 / \hbar^2$ 
to the imaginary part of $\mu_{q\theta}$.\\

We start by studying the $N$ eigenvalues $\mu_{q\theta}$ obtained by fixing the wavenumber 
$q$ (even or odd) and varying $\theta$. 
For given physical parameters ($A$, $\lambda$, $Q$, $N$), fixing $q$ is equivalent to fixing $\beta$.\\

\begin{figure}[p]
\centering
\hspace*{-1.5cm}
\begin{tabular}{c}
\vspace*{0.5cm}
\rotatebox{0}{\scalebox{0.35}{\includegraphics{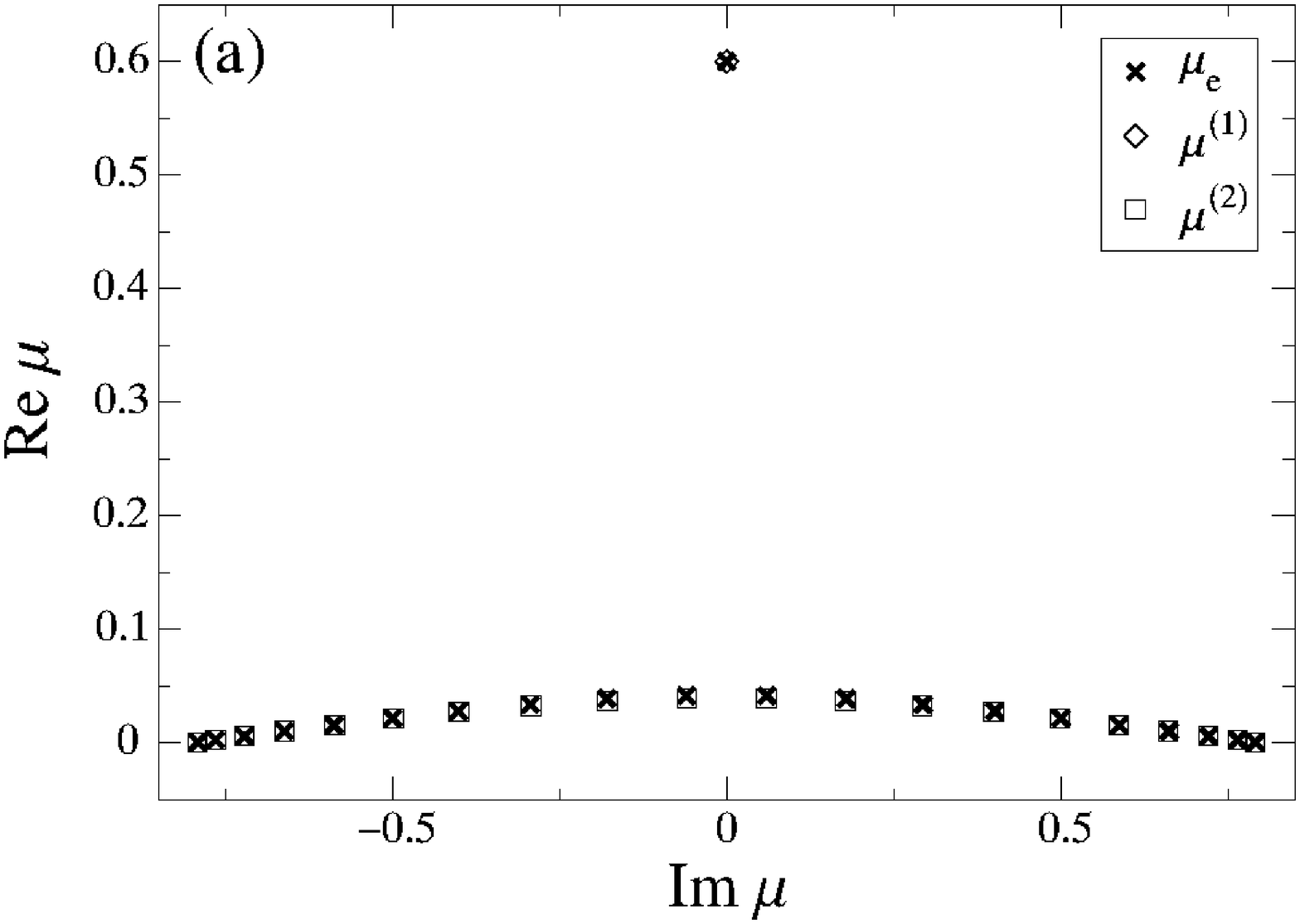}}} \\
\vspace*{0.5cm}
\rotatebox{0}{\scalebox{0.35}{\includegraphics{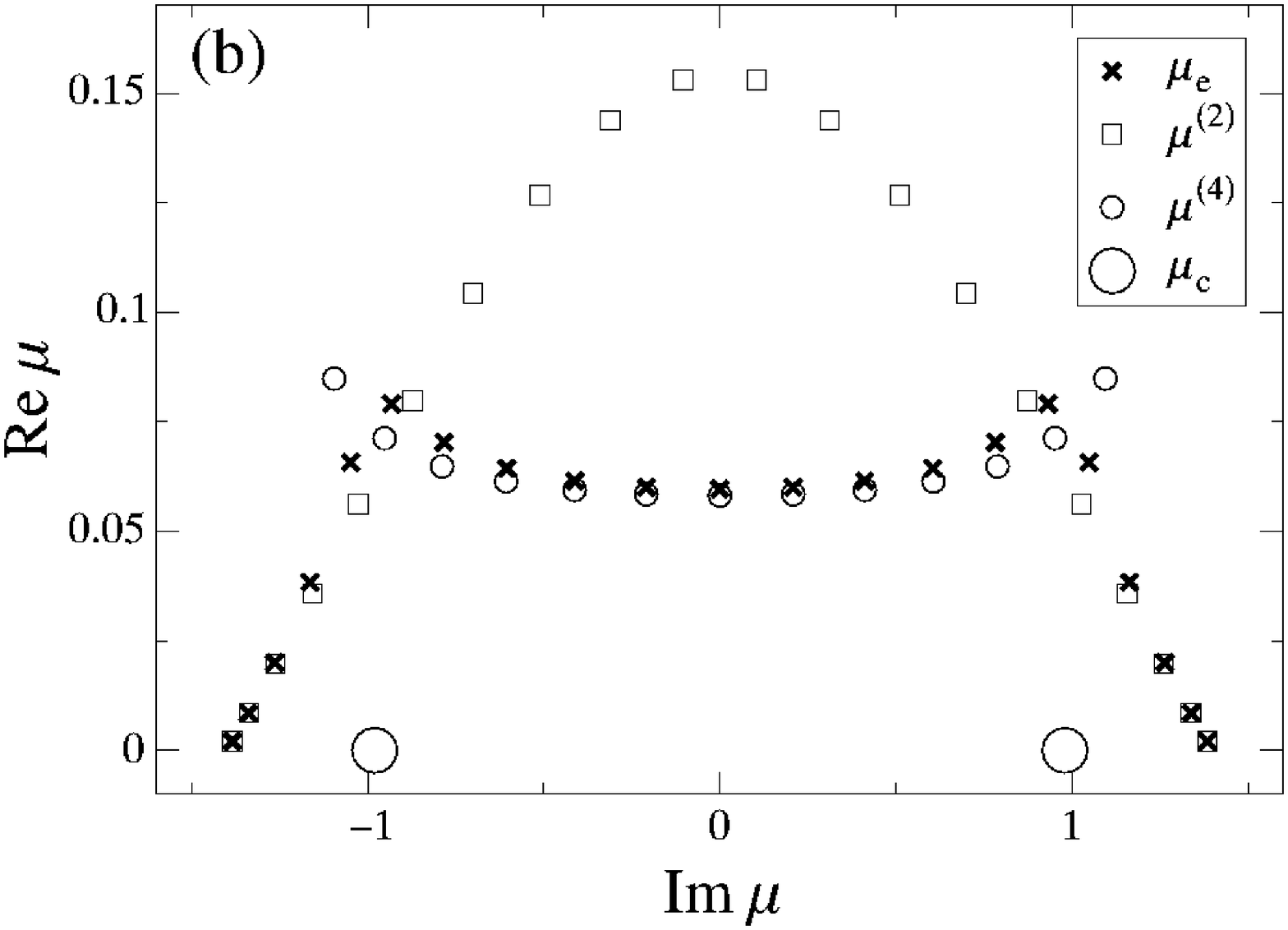}}} \\
\vspace*{0.5cm}
\rotatebox{0}{\scalebox{0.35}{\includegraphics{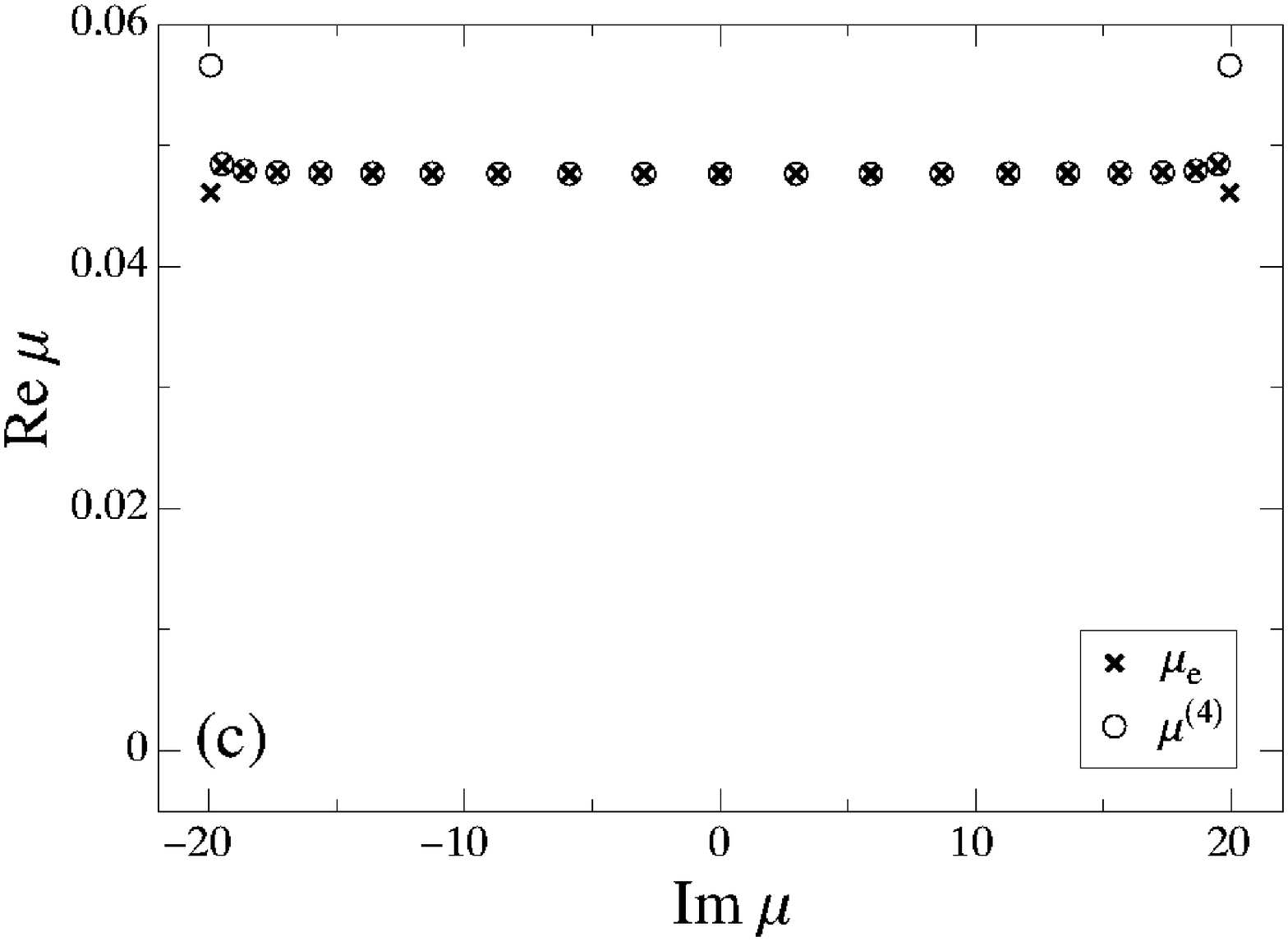}}}
\end{tabular}
\caption{
Eigenvalue spectrum for $N=21$ and given $q$: (a) $\beta=0.4$, (b) $\beta=0.7$, and (c) $\beta=10$. 
$\mu_{e}$ denotes the exact eigenvalues obtained by numerical diagonalisation and $\mu^{(i)}$ the 
eigenvalues of the different families given in Tables \ref{table1} and \ref{table2}.
} \label{spectre2D.Nodd}
\end{figure}

\begin{figure}[p]
\centering
\hspace*{-1.5cm}
\begin{tabular}{c}
\vspace*{0.5cm}
\rotatebox{0}{\scalebox{0.35}{\includegraphics{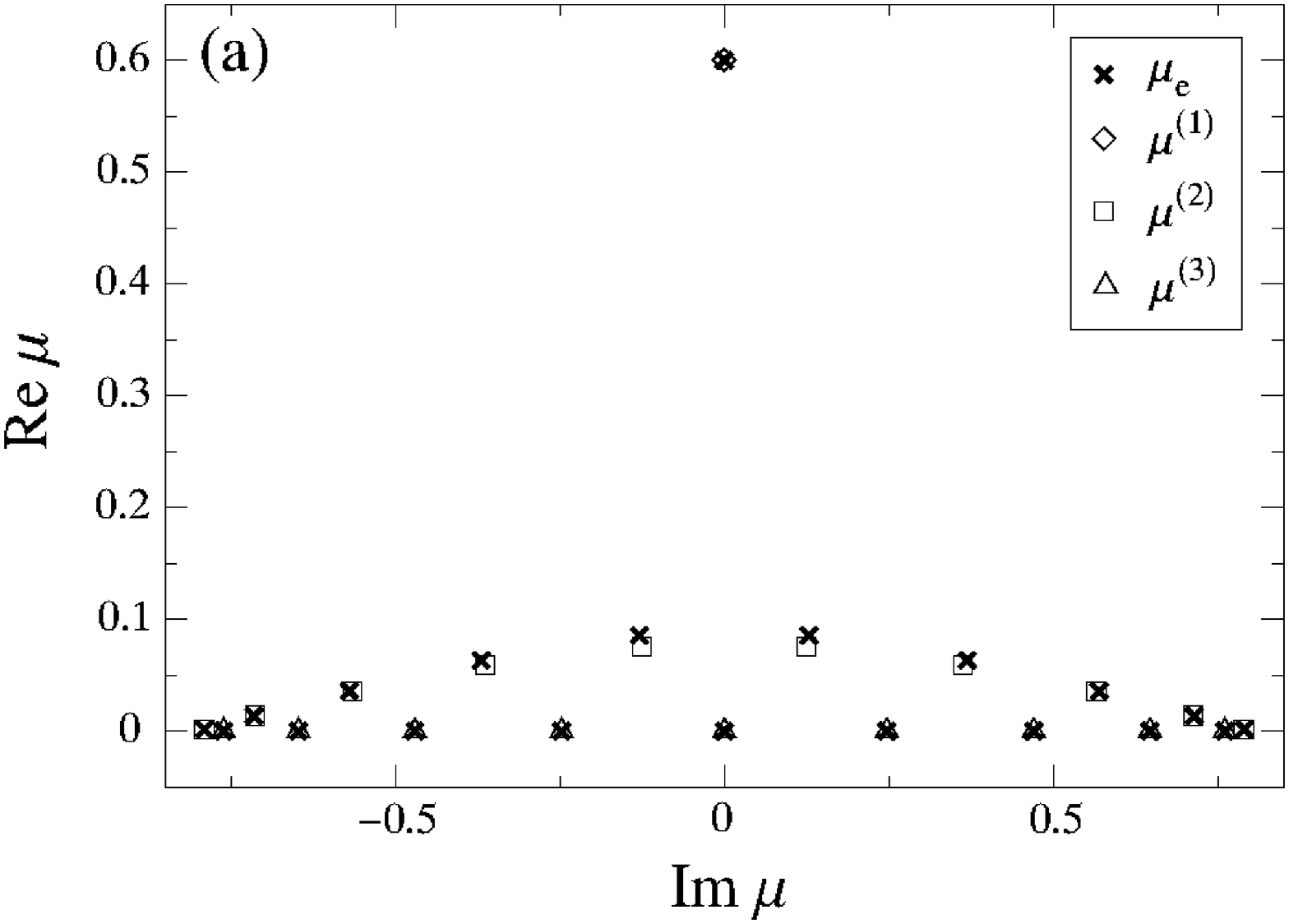}}} \\
\vspace*{0.5cm}
\rotatebox{0}{\scalebox{0.35}{\includegraphics{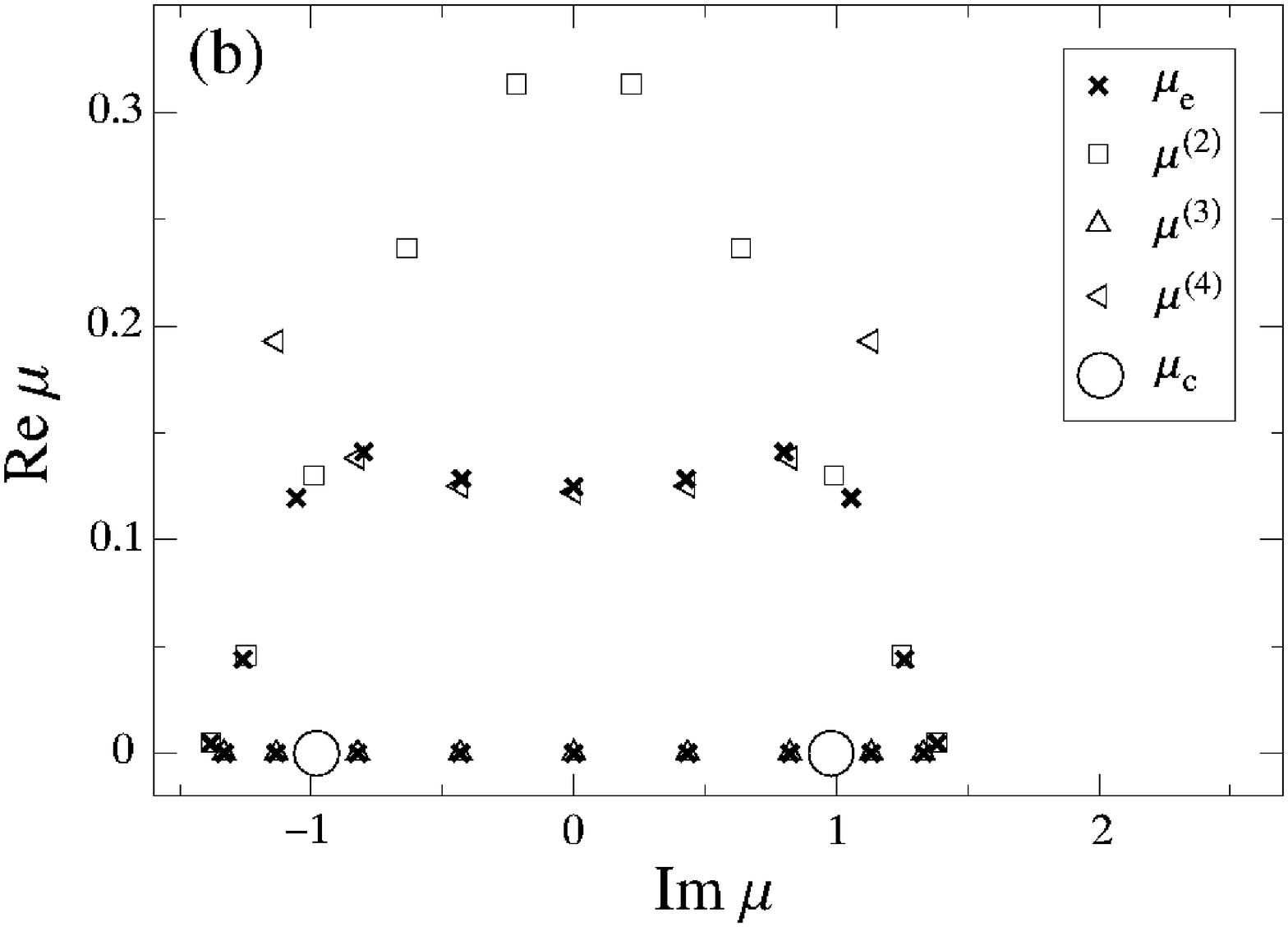}}}  \\
\vspace*{0.5cm}
\rotatebox{0}{\scalebox{0.35}{\includegraphics{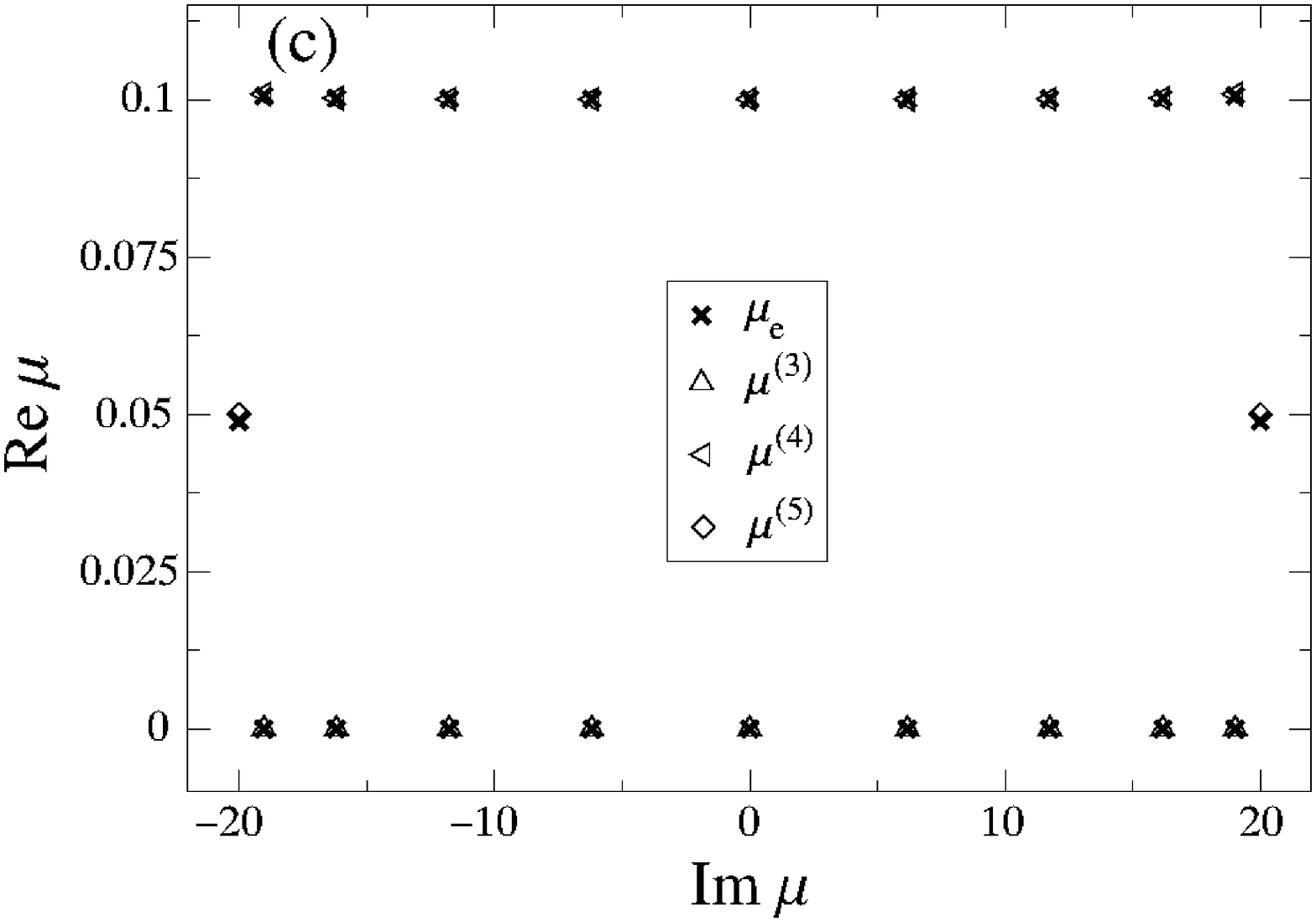}}} 
\end{tabular}
\caption{
Eigenvalue spectrum for $N=20$ and $q$ even: (a) $\beta=0.4$, (b) $\beta=0.7$, and (c) $\beta=10$. 
$\mu_{e}$ denotes the exact eigenvalues obtained by numerical diagonalisation and $\mu^{(i)}$ 
the eigenvalues of the different families given in Tables \ref{table1} and \ref{table2}.
} \label{spectre2D.Neven.qeven}
\end{figure}

\begin{figure}[p]
\centering
\hspace*{-1.5cm}
\begin{tabular}{c}
\vspace*{0.5cm}
\rotatebox{0}{\scalebox{0.35}{\includegraphics{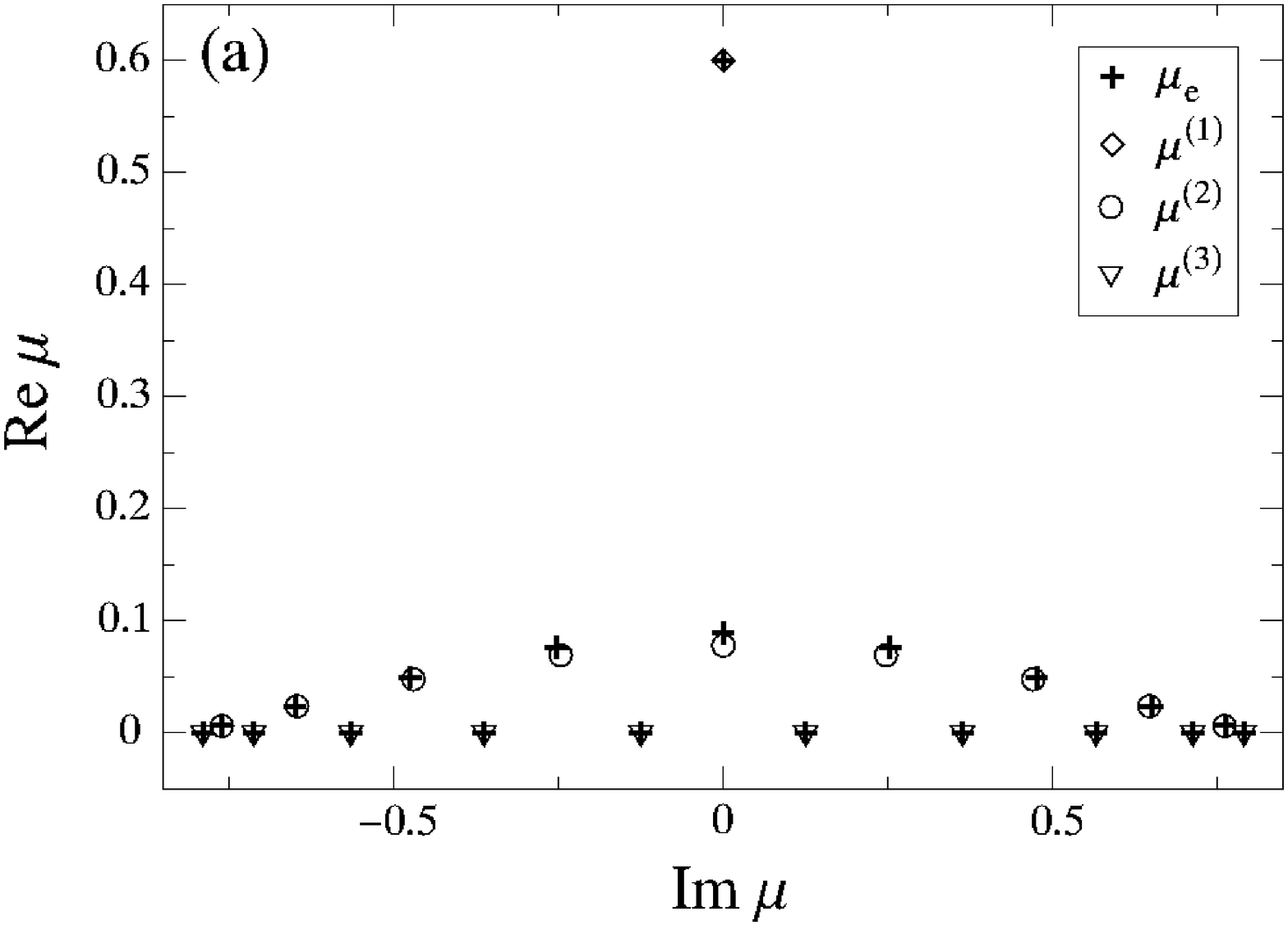}}} \\
\vspace*{0.5cm}
\rotatebox{0}{\scalebox{0.35}{\includegraphics{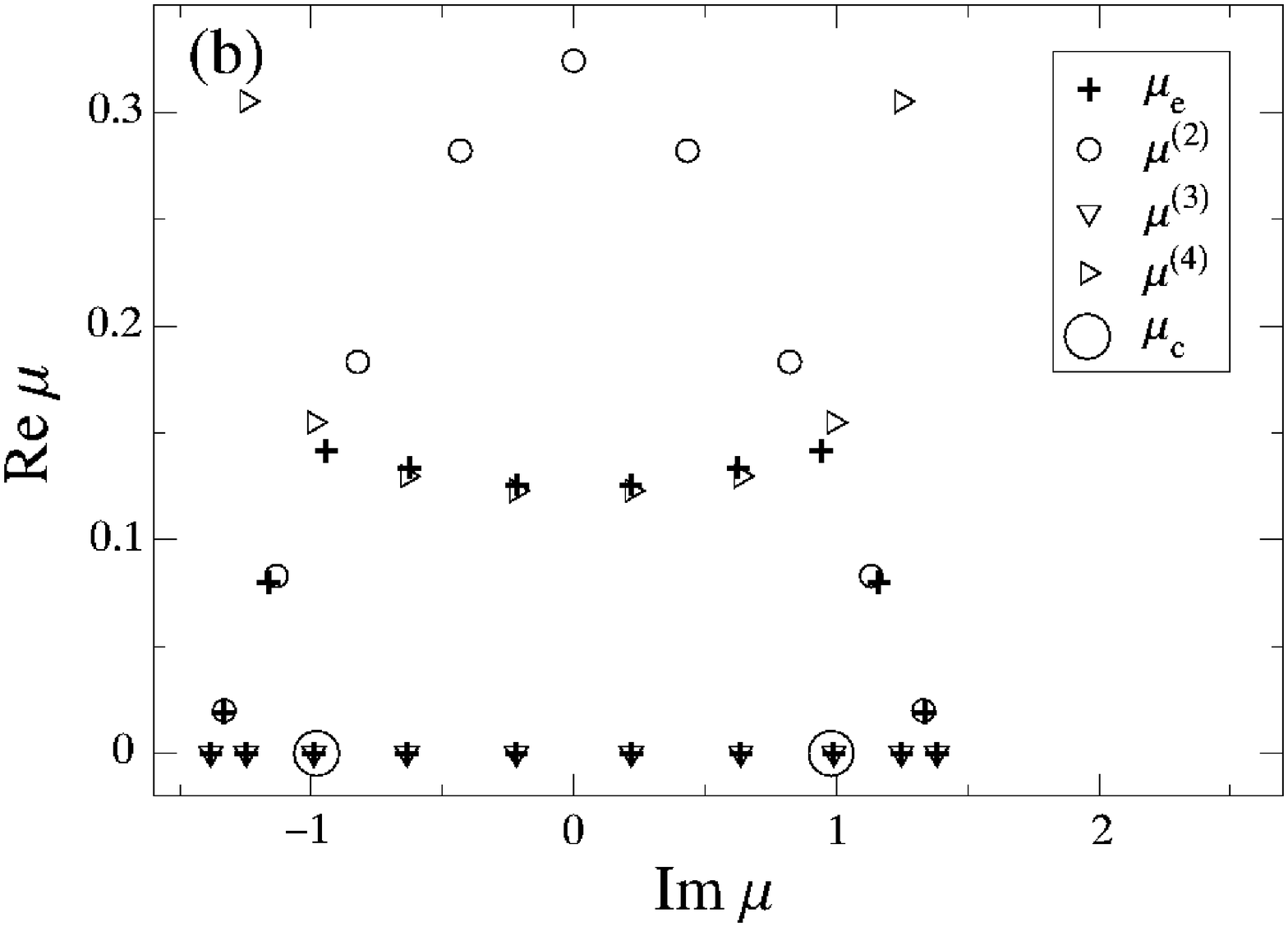}}}  \\
\vspace*{0.5cm}
\rotatebox{0}{\scalebox{0.35}{\includegraphics{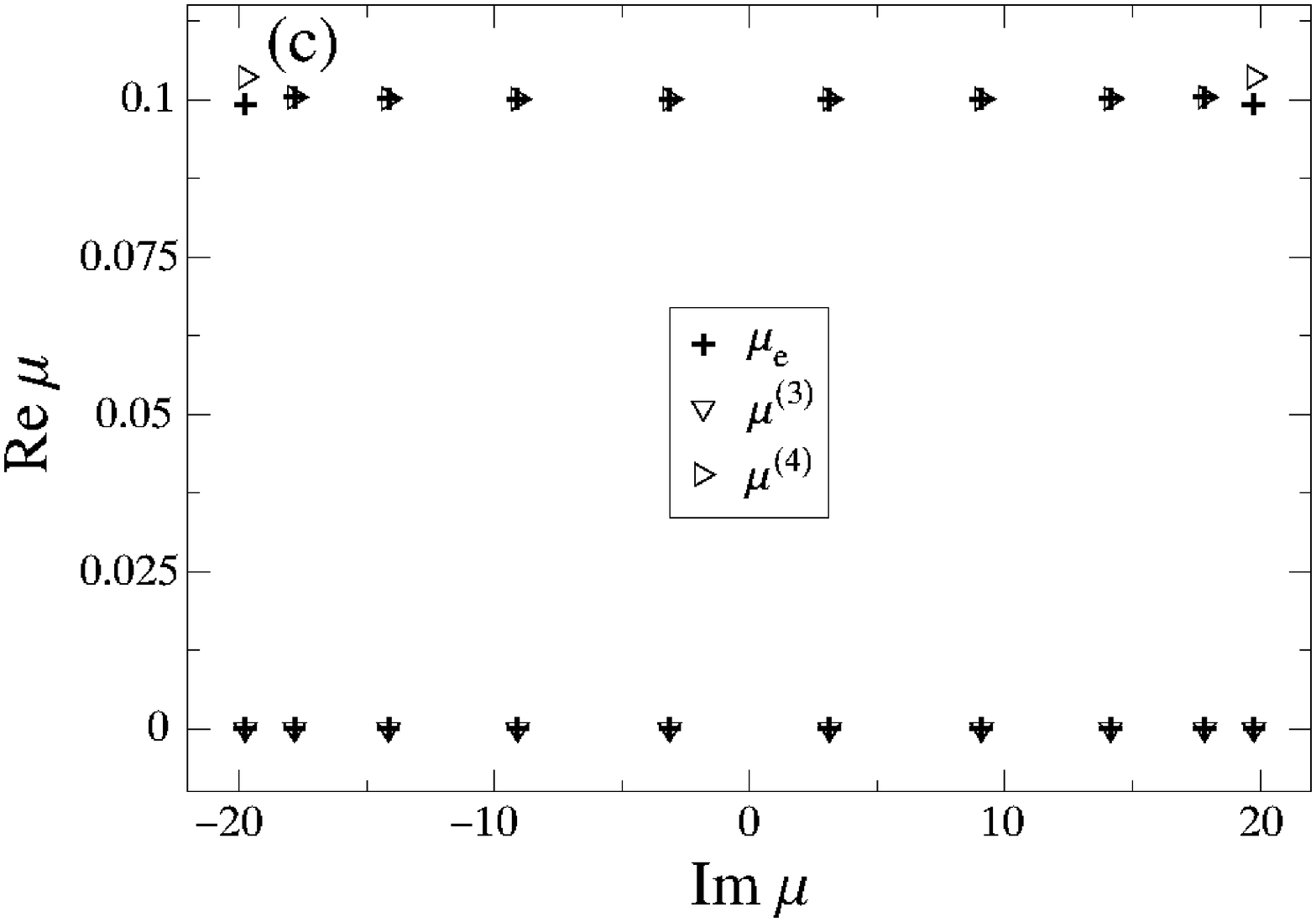}}} 
\end{tabular}
\caption{
Eigenvalue spectrum for $N=20$ and $q$ odd: (a) $\beta=0.4$, (b) $\beta=0.7$, and (c) $\beta=10$. 
$\mu_{e}$ denotes the exact eigenvalues obtained by numerical diagonalisation and $\mu^{(i)}$ 
the eigenvalues of the different families given in Tables \ref{table1} and \ref{table2}.
} \label{spectre2D.Neven.qodd}
\end{figure}

For $\beta <  \frac{1}{2}$, the analytical expressions of the eigenvalues which concern us are 
summarized in Table \ref{table1}. 
Two families of eigenvalues ($\mu^{(1)}$ and $\mu^{(2)}$) enter in the discussion for $N$ odd, and three families 
($\mu^{(1)}$, $\mu^{(2)}$, and $\mu^{(3)}$), for $N$ even.
The numerical eigenvalues are plotted in figures \ref{spectre2D.Nodd}(a),
\ref{spectre2D.Neven.qeven}(a), and \ref{spectre2D.Neven.qodd}(a)
and are in very good agreement with the analytical results.
The sole diffusive eigenvalue $\mu^{(1)}$ has a real part and no imaginary part.
The $N-1$ other eigenvalues, either belongs to the $\mu^{(2)}$ family for $N$ odd or to the $\mu^{(2)}$ 
and $\mu^{(3)}$ families for $N$ even. 
The eigenvalues $\mu^{(2)}$ and $\mu^{(3)}$ have an imaginary part which extends from $-2\beta$ to 
$2\beta$ and they generate oscillations in the dynamics as we shall see in the following section. 
The real part of the $\mu^{(2)}$ eigenvalues is small and tends to zero in the large $N$ limit.
The real part of the $\mu^{(3)}$ eigenvalues is always zero.\\

For $\beta >  \frac{1}{2}$, the diffusive eigenvalue $\mu^{(1)}$ has disappeared 
after merging with the other eigenvalues and
the situation is slightly more complicated.
The situation for a moderate value of $\beta >  \frac{1}{2}$ is depicted in figures \ref{spectre2D.Nodd}(b),
\ref{spectre2D.Neven.qeven}(b), and \ref{spectre2D.Neven.qodd}(b)
while the analytical expressions of the eigenvalues are given 
in Table \ref{table2}. 
Since $\mu^{(1)}$ no longer exists, we have
the two families of eigenvalues $\mu^{(2)}$ and $\mu^{(4)}$ if $N$ is odd, and
the three families  $\mu^{(2)}$, $\mu^{(3)}$ and $\mu^{(4)}$ if $N$ is even.
Two regions of the spectrum have to be distinguished. 
The eigenvalues $\mu^{(2)}$ exist in the region where $\vert \mu \vert > \vert \mu_c\vert$ while the eigenvalues 
$\mu^{(4)}$ exist in the region where $\vert \mu \vert <\vert \mu_c\vert$.
We observe that the extra family of eigenvalues $\mu^{(4)}$ has appeared because
of the collision with the diffusive eigenvalue $\mu^{(1)}$.
We can see in figures \ref{spectre2D.Nodd}(b),
\ref{spectre2D.Neven.qeven}(b), and \ref{spectre2D.Neven.qodd}(b) that the analytical results of Table \ref{table2}
reproduce very well the eigenvalues obtained by numerical diagonalization in the two regions.
Here again, the number of eigenvalues is equal to $N$ for a given wavenumber $q$,
the imaginary part of the eigenvalues extends from $-2 \beta$ to $2 \beta$, and
the real parts of all eigenvalues tends to zero in the large $N$ limit.\\

A special situation occurs when $\beta>  \frac{1}{2}$ is increased to large values.
This situation is depicted in figures \ref{spectre2D.Nodd}(c),
\ref{spectre2D.Neven.qeven}(c), and \ref{spectre2D.Neven.qodd}(c).
The situation is similar to the previous one but the region $\vert \mu \vert > \vert\mu_c\vert$ has disappeared 
so that the family of eigenvalues $\mu^{(2)}$ corresponding to the expansion $\beta\to 0$ no longer exists.
For $N$ odd and for $N$ even with $q$ odd, these eigenvalues are replaced by the eigenvalues $\mu^{(4)}$ eigenvalues.
For $N$ even and $q$ even, we find the two eigenvalues $\mu^{(5)}$ beside the family of eigenvalues  $\mu^{(4)}$. 
The agreement between the analytical and numerical results is very good here also.
As before, the imaginary part of all these eigenvalues extends from $-2 i \beta$ to $2 i \beta$ and
their real parts tends to zero in the large $N$ limit.\\

\begin{figure}[p]
\centering
\begin{tabular}{c}
\vspace*{0.2cm}
\rotatebox{0}{\scalebox{0.75}{\includegraphics{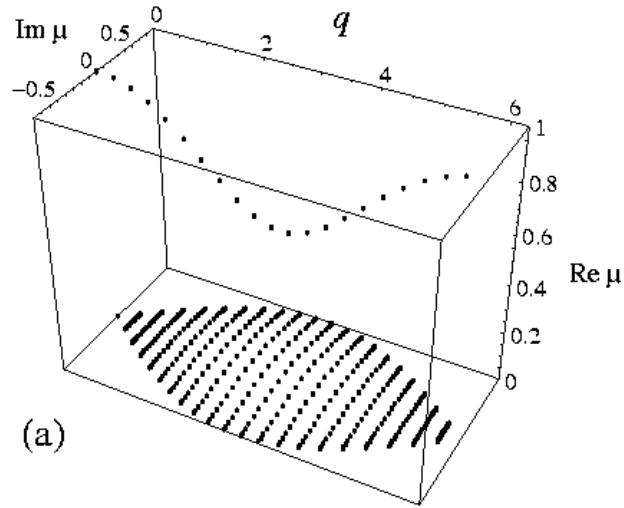}}} \\
\vspace*{0.2cm}
\rotatebox{0}{\scalebox{0.75}{\includegraphics{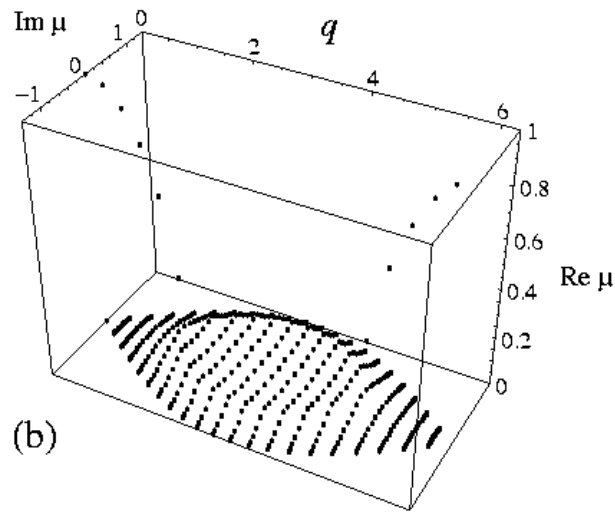}}} \\
\vspace*{0.2cm}
\rotatebox{0}{\scalebox{0.75}{\includegraphics{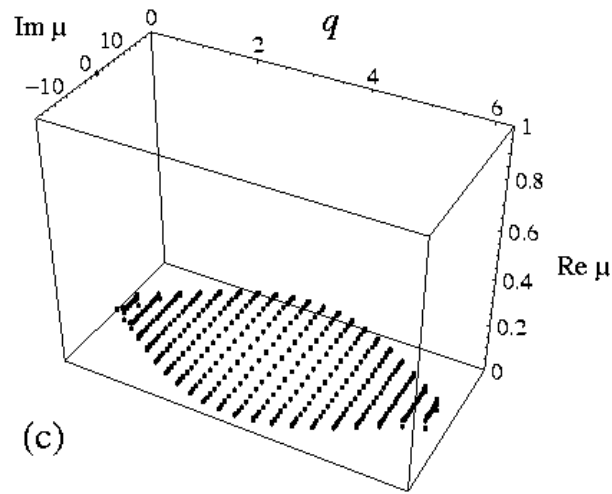}}} 
\end{tabular}
\caption{
Complete spectrum of the eigenvalues $\mu_{q\theta}$ related to the eigenvalues of the 
Redfield superoperator by equation (\ref{spec3acd}). 
The system size is here $N=21$. 
In (a), $\frac{A \hbar}{Q \lambda^2}=0.4$. 
In (b), $\frac{A \hbar}{Q \lambda^2}=0.7$. 
In (c), $\frac{A \hbar}{Q \lambda^2}=10$.
} \label{spectre3D}
\end{figure}

A global view of the complete spectrum of the $N^2$ eigenvalues of the Redfield superoperator
is depicted in figure \ref{spectre3D} by varying the wavenumber $q$ in a third dimension.
Here, we only consider for simplicity the case where $N$ is odd.
The relation between the wavenumber $q$ and the parameter $\beta$ is given by equation (\ref{spec1bab}).
The wavenumber $q$ varies in the first Brillouin zone or, equivalently, in the interval $0\leq q < 2\pi$.
We see in figure \ref{spectre3D}(a) that the diffusive eigenvalues $\mu^{(1)}$ exists for
all the values of the wavenumber in the case $\frac{A \hbar}{Q \lambda^2}< \frac{1}{2}$ 
which implies $\beta < \frac{1}{2}$.
However, if $\frac{A \hbar}{Q \lambda^2}> \frac{1}{2}$, the diffusive eigenvalue $\mu^{(1)}$ disappears as expected 
for some values of the wavenumber corresponding to $\beta > \frac{1}{2}$.
This situation is observed in figure \ref{spectre3D}(b).\\

For very large values of $\frac{A \hbar}{Q \lambda^2}> \frac{1}{2}$, the diffusive branch of the
spectrum is reduced to the sole eigenvalue at $q^{(1)}=0$, as seen in  figure \ref{spectre3D}(c).
In this case, diffusion has disappeared from the spectrum which only contains
eigenvalues associated with damped oscillatory behavior.
The diffusive branch can be supposed to have disappeared when its last nonzero eigenvalue disappears in equation (\ref{spec3aae}).
Therefore the diffusive branch disappears when the value of $\beta$ for the first nonzero eigenvalue 
corresponding to $q=\frac{2\pi}{N}$ is larger than the critical value $\beta_{c}=\frac{1}{2}$.
This happens when the coupling parameter exceeds the critical value given by
\begin{center} \fbox{\parbox{12.5cm}{
\begin{eqnarray}
\lambda_{c} = \sqrt{\frac{2 A \hbar}{Q} \sin \frac{\pi}{N}} \stackrel{N > 5}{\simeq} 
\sqrt{\frac{2 A \hbar \pi}{Q N}} \; . \label{spec3aah}
\end{eqnarray}
}} \end{center}
This disappearance of the diffusion branch can be observed in figure \ref{spectre3D}(c).
We notice that the diffusive branch always exists in the infinite-system limit ($N \to \infty$) 
in which case $\lambda_{c}$ can be arbitrarily small.

\section{Dynamical analysis} \label{Sec.dynamics}

We have described in the previous section the complete spectrum of the Redfield superoperator. 
We now need to discuss the dynamical implications of our spectral analysis.

\subsection{Theoretical predictions}

The spectrum of the Redfield superoperator determines the full dynamics of the subsystem 
interacting with its environment. 
In fact, projecting equation (\ref{isolaan}) in the site basis and using the Bloch property (\ref{blochaai}), 
we can write
\begin{eqnarray}
\rho_{ll'}(t) = \sum_{q,\theta} e^{s_{q\theta} t + i l q} c_{q\theta}(0) \rho_{0,l'-l}^{q\theta} \; ,
\label{diff1aaa}
\end{eqnarray}
We see that the imaginary part of the eigenvalues $s_{q\theta}$ will generate oscillations in the dynamics 
and that the real part of these eigenvalues will generate a damping. 
The contribution to the long-time dynamics from the modes of the Redfield superoperator 
corresponding to the eigenvalues $s_{q\theta}$ which have a large real part is very small. 
Only the modes corresponding to eigenvalues with a small real part will significantly contribute to 
the long-time dynamics and therefore to the transport property of the system. 
Here, we want mainly to focus on the long-time dynamics.\\

As pointed out in subsection \ref{spectradisuss}, the subsystem evolves in a non-diffusive
regime if the chain is small enough $\lambda < \lambda_{c}$.  In this regime, there is
no diffusive eigenvalue so that the long-time dynamics is dominated by
the eigenvalues $\mu^{(4)}$ describing the decay of the quantum coherences.
The lowest relaxation rate is of the order of magnitude of the real part of the eigenvalues 
belonging to the $\mu^{(4)}$ family of Table \ref{table2}. 
Using equation (\ref{spec3acd}), we find that this rate is given by
\begin{center} \fbox{\parbox{12.5cm}{
\begin{eqnarray}
s^{(4)} \simeq -\frac{2 Q \lambda^2}{\hbar^2} + O\left(\frac{1}{N}\right) \; . \label{spec4aad} 
\end{eqnarray}
}} \end{center}
We notice that these eigenvalues have a non-vanishing imaginary part so that the slowest modes relax 
exponentially but with oscillations. 
This means that the oscillating modes relax on the same time scales as the non-oscillating ones.
These damped oscillations are reminiscent of a similar behavior in the spin-boson model \cite{Leggett94}.
We can thus interpret these damped oscillations as due to the damping of the quantum coherences.\\

For a sufficiently large chain $\lambda > \lambda_{c}$, these damped oscillations disappear because
the eigenvalue of the Redfield superoperator with
the smaller real part is now the diffusive eigenvalue $\mu^{(1)}$.
In this case, the lowest relaxation rate is given according to equation (\ref{spec3aaf}) by
\begin{center} \fbox{\parbox{12.5cm}{
\begin{eqnarray}
s^{(1)} \simeq -D q^2 = - D \left(\frac{2\pi}{N}\right)^2 = - \frac{4 \pi^2 A^2 }{Q \lambda^2 N^2} \; , \label{spec4aac} 
\end{eqnarray}
}} \end{center}
because the wavenumber of the slowest nontrivial mode takes the value $q=2\pi/N$
and is inversely proportional to the length $N$ of the chain.
The signature of the diffusive regime is therefore that the relaxation rate scales as $N^{-2}$.\\

This behavior is consistent with the famous Einstein relation for the definition of the diffusion coefficient.
Indeed, we can define the position operator of the subsystem $\hat{x}$ by
\begin{eqnarray}
\langle l \vert \hat{x} \vert l' \rangle = l \; \delta_{ll'} \; .\label{diff1aba}
\end{eqnarray}
The variance of the position is therefore given by
\begin{eqnarray}
\langle \hat{x}^2 \rangle (t) = \sum_{l=0}^{N-1} l^2 \rho_{ll}(t) \; , \label{diff1aca}
\end{eqnarray}
where, using Eq (\ref{diff1aaa}),
\begin{eqnarray}
\rho_{ll}(t) = \sum_{q,\theta} a_{q\theta}(0) \; e^{s_{q\theta} t + i l q} \; ,
\label{diff1ada}
\end{eqnarray}
with the newly defined coefficients $a_{q\theta}(0)=c_{q\theta}(0) \rho_{00}^{q\theta}$.
In the long-time limit and for a sufficiently large chain, 
only the diffusive branch $s^{(1)}$ of eigenvalue spectrum will significantly contribute to the 
dynamics and one can write 
\begin{eqnarray}
\rho_{ll}(t) \stackrel{t \to \infty}{=} \sum_{q,\theta} a_{q\theta}(0) \; e^{s_{q\theta} t + i l q} \; ,
\label{diff1aea}
\end{eqnarray}
where $s_{q\theta} = - D q^2$.
Using $q_j=j \frac{2 \pi}{N}$, $\Delta q=q_{j+1}-q_{j}=\frac{2 \pi}{N}$ and 
$f(q_j)=N a_{q_{j}\theta}(0)$, we get
\begin{eqnarray}
\rho_{ll}(t) \stackrel{t \to \infty}{=} \frac{1}{2\pi} \sum_{j=0}^{N-1} \Delta q \; f(q_j) 
\; e^{-D q_j^2 t + i l q_j} \; ,
\label{diff1afa}
\end{eqnarray}
which becomes in the limit of an arbitrarily large chain 
\begin{eqnarray}
\rho_{ll}(t) \stackrel{N,t \to \infty}{=} \frac{1}{2\pi} \int_{-\infty}^{+\infty} d q \; f(q)\; 
e^{- D q^2 t + i l q} \; .   \label{diff1aga}
\end{eqnarray}
If we choose an initial condition of the density function centered on a given site, $f(q)=1$, and
\begin{eqnarray}
\rho_{ll}(t) \stackrel{N,t \to \infty}{=} \frac{e^{- \frac{l^2}{4 D t}}}{\sqrt{4 \pi D t}} \; .
\label{diff1aha}
\end{eqnarray}
Finally, we recover the well-known Einstein relation for the diffusion coefficient
\begin{eqnarray}
\langle \hat{x}^2 \rangle (t) \stackrel{N,t \to \infty}{=} 2 D t \; .
\label{diff1aia}
\end{eqnarray}
This demonstration is also a justification for calling $\mu^{(1)}$ or $s^{(1)}$
the diffusive eigenvalue.

\subsection{Numerical results}

The previous theoretical predictions are confirmed by the numerical integration 
of the Redfield equation. We always take $\hbar=1$.\\

Let us define the quantity we computed. Equation (\ref{diff1aaa}) can be written as
\begin{eqnarray}
\rho_{ll}(t) = \sum_{q,\theta} e^{s_{q\theta} t + i l q} \; a_{q\theta}(0)\; ,
\label{diff1baa}
\end{eqnarray}
where $a_{q\theta}(0)=c_{q\theta}(0) \rho_{00}^{{q\theta}}$.
If we introduce
\begin{eqnarray}
\tilde{\rho}(q',t) &=& \sum_{l=0}^{N-1} \rho_{ll} (t) \; e^{-i l q'} \; , \label{diff1bab}
\end{eqnarray}
using equation (\ref{isolaae}), we find that
\begin{eqnarray}
\tilde{\rho}(q',t) &=& N \sum_{\theta} a_{q'\theta}(0)  \; e^{s_{q'\theta} t} \; .\label{diff1bac}
\end{eqnarray}
The evolving quantity we have computed is 
\begin{eqnarray}
\vert X (t) \vert = \vert \tilde{\rho}(q_1,t) \vert \; ,
\label{spec4aab} 
\end{eqnarray}
where we recall that $q_1= 2 \pi / N$.
We consider that the subsystem has initial conditions given by
\begin{eqnarray}
\rho_{ll'} (0) = \delta_{ll'} \; \delta_{l0} \; . \label{spec4aaa} 
\end{eqnarray}
This means that $\sum_{\theta} a_{q\theta}(0)= 1 / N$. 
The quantity $X (t)$ is initially equal to unity ($X(0)=1$) and tends to zero after a long time
($X(\infty) = 0$) in an exponential way determined by the smallest non-vanishing real part of the eigenvalues $s_{q_1\theta}$.\\

\begin{figure}[p]
\vspace*{-0.3cm}
\centering
\hspace*{-1cm}
\begin{tabular}{c}
\vspace*{0.3cm}
\rotatebox{0}{\scalebox{0.5}{\includegraphics{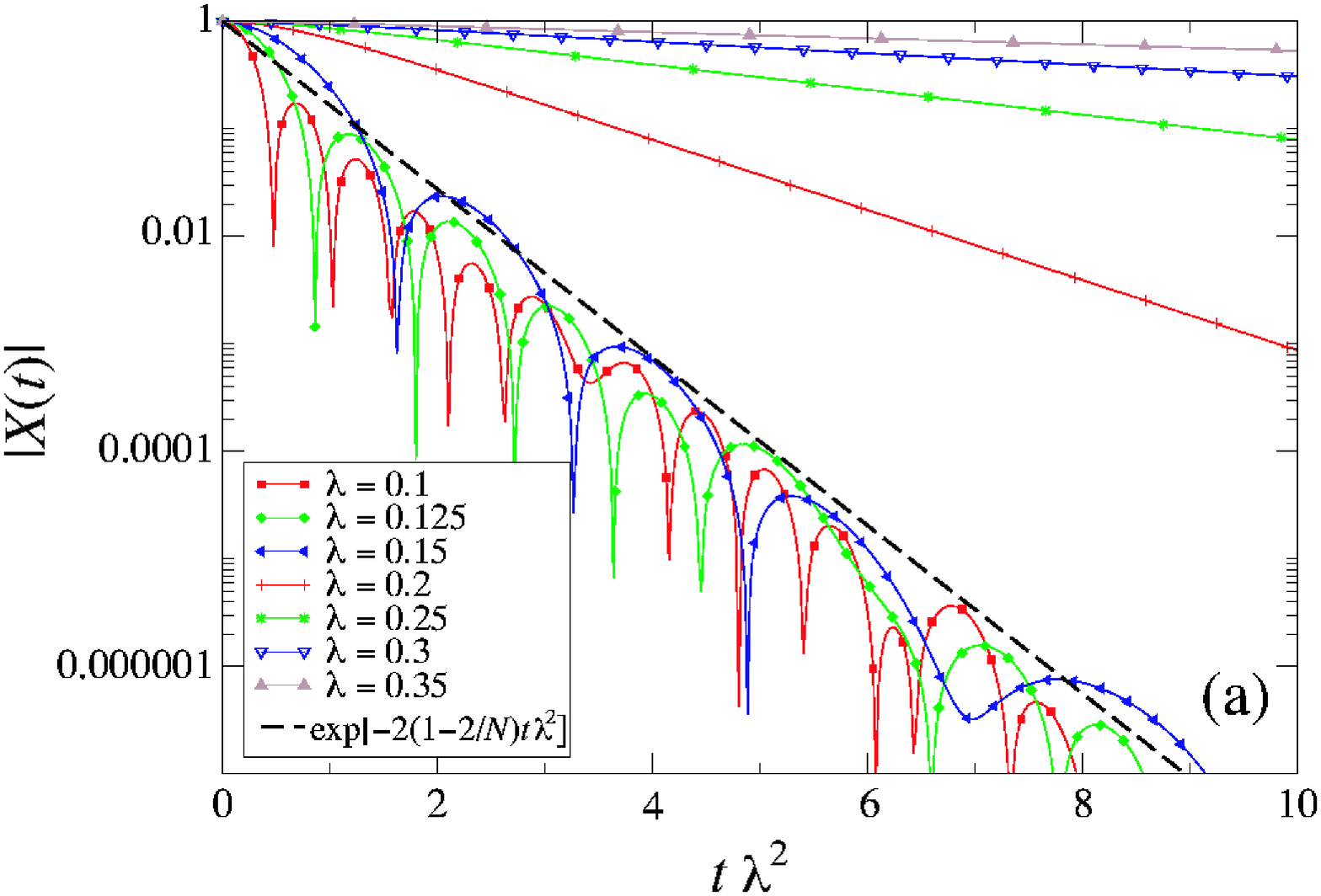}}} \\
\rotatebox{0}{\scalebox{0.5}{\includegraphics{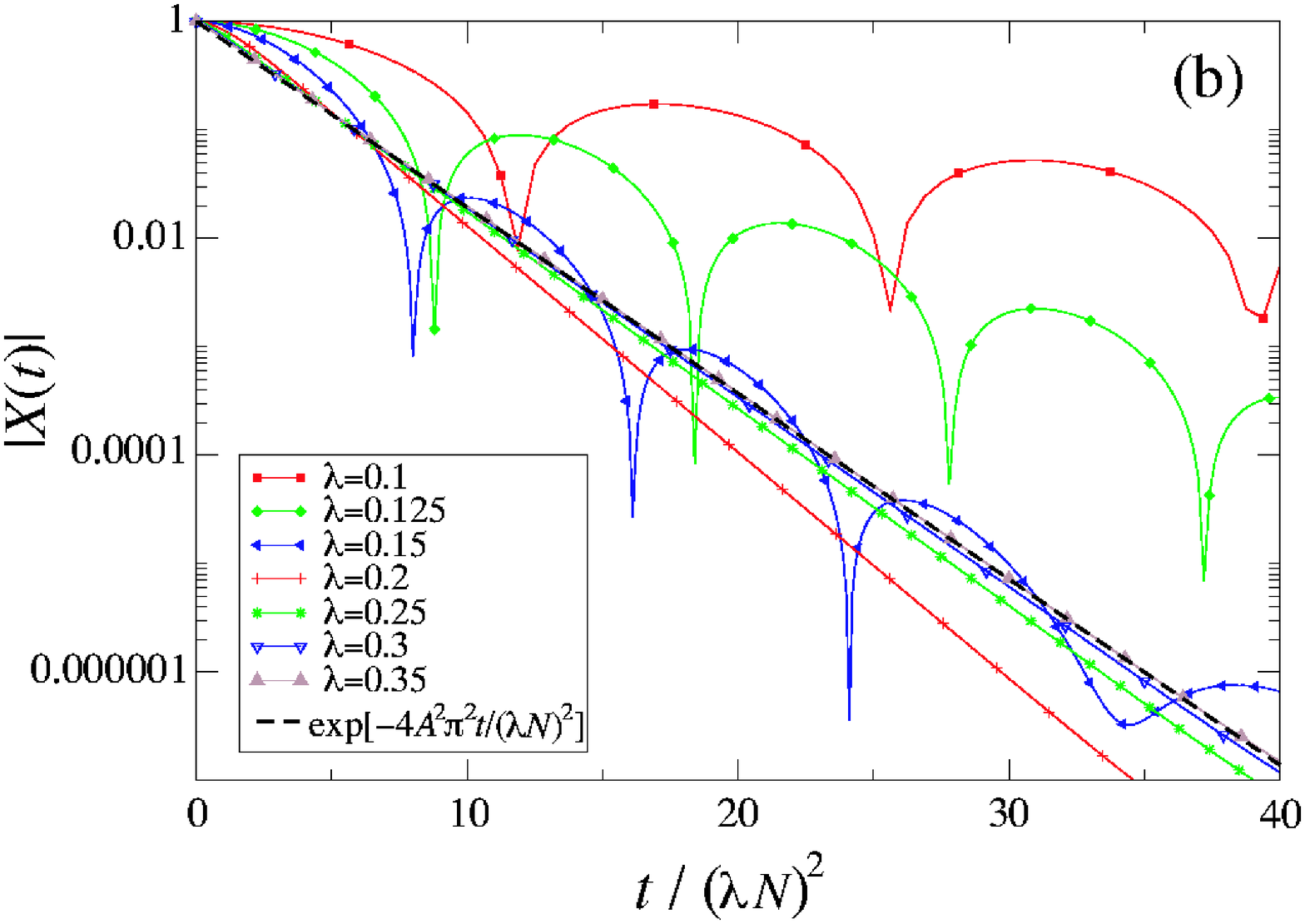}}}
\end{tabular}
\caption{
Representation of the transition between the perturbative and diffusive regimes
for a chain of size $N=20$ and parameters $E_0=0$, $A=0.1$, and $Q=1$.  The transition
take place at the critical value $\lambda_{c}=0.177$ given by equation (\ref{spec3aah}). 
(a) The quantity $\vert X(t)\vert$ versus the rescaled time $\lambda^2 t$ in order to identify the perturbative regime.
(b) The quantity $\vert X(t)\vert$ versus the rescaled time $t / (\lambda N)^2$ in order to identify the diffusive regime.
As predicted, we are in the perturbative regime if $\lambda < \lambda_{c}$ and
in the diffusive regime if $\lambda > \lambda_{c}$.
The quantity $\vert X(t) \vert$ is defined in equation (\ref{spec4aab}).
} \label{dynamicsfig1}
\end{figure}
Figures \ref{dynamicsfig1}(a) and \ref{dynamicsfig1}(b) show the same relaxation curves of $\vert X(t) \vert$ but for 
two different time scalings. 
The different curves correspond to different values of the coupling parameter $\lambda$, but to the same 
chain size $N$.
Figure \ref{dynamicsfig1}(a) has the scaling $\lambda^2 t$, which is characteristic of the non-diffusive regime and 
figure \ref{dynamicsfig1}(b) has the scaling $t/(\lambda N)^2$, which is characteristic of the diffusive regime.
Therefore, the curves in figure \ref{dynamicsfig1}(a) with the same time scaling correspond to values of
the coupling parameter such that $\lambda < \lambda_{c}$ (in the non-diffusive regime) and they relax with oscillations at 
rates given by equation (\ref{spec4aad}). On the other hand, the curves of figure \ref{dynamicsfig1}(a) with the same time 
scaling correspond to values of the coupling parameter such that $\lambda > \lambda_{c}$ (in the diffusive regime) and 
they relax exponentially without oscillations at a rate given by equation (\ref{spec4aac}).\\

\begin{figure}[h]
\centering
\hspace*{-1cm}
\rotatebox{0}{\scalebox{0.5}{\includegraphics{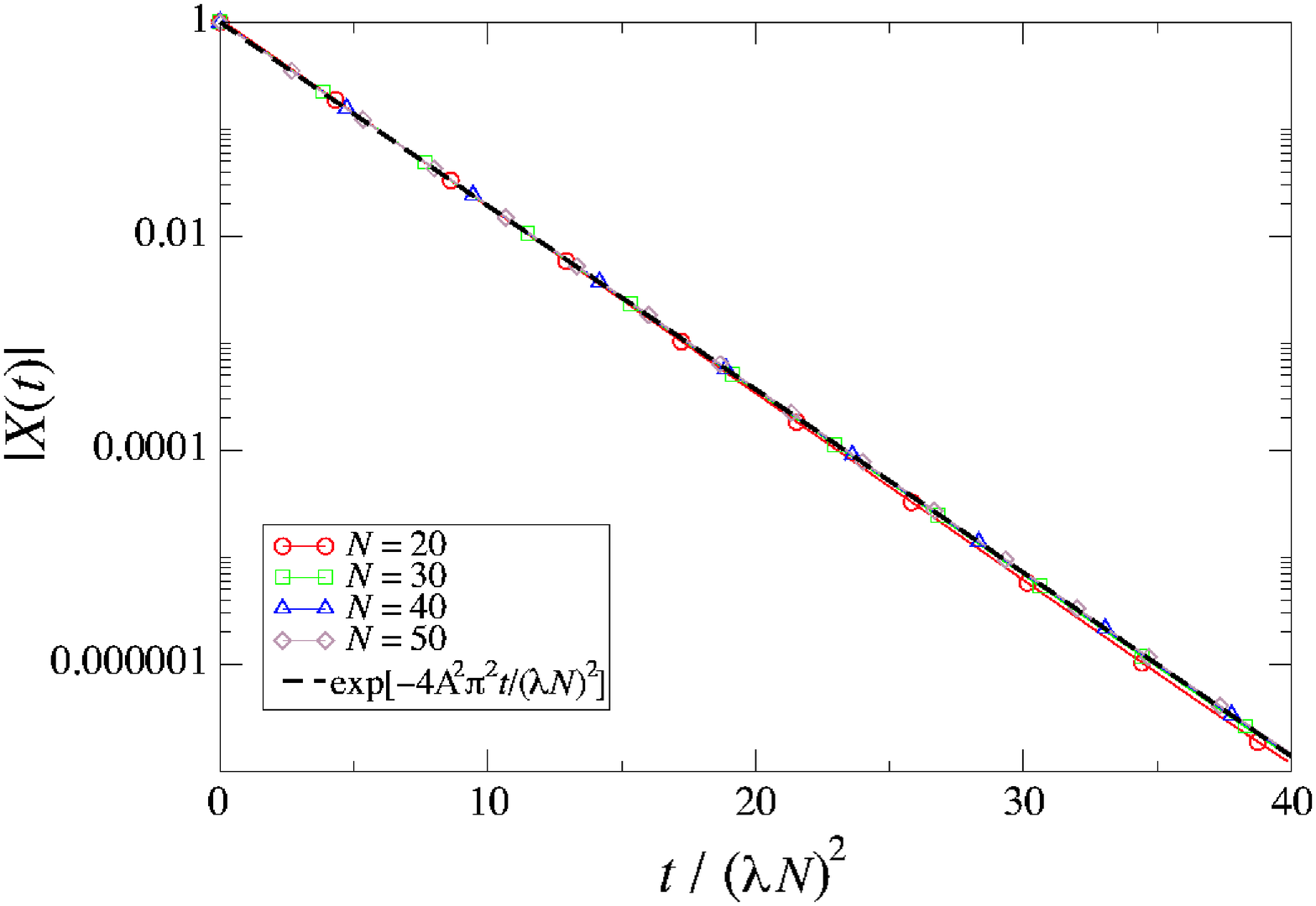}}}
\caption{
The quantity $\vert X(t)\vert$ defined by equation (\ref{spec4aab})
versus the rescaled time $t / (\lambda N)^2$
for various sizes $N=20-50$ of the chain.
The parameters are $E_0=0$, $A=0.1$, $Q=1$, and $\lambda=0.3$. 
Here, the system is always in the diffusive regime $\lambda > \lambda_{c}$.
} \label{dynamicsfig2}
\end{figure}
Figure \ref{dynamicsfig2} shows the scaling $t/(\lambda N)^2$ in the diffusive regime for different curves
at the same value of the coupling parameter but for different chain sizes $N$.  This figure numerically
demonstrates the property of diffusion in our model.\\

An important remark is that the time scaling of the non-diffusive regime is naturally expected
from perturbation theory. Indeed, according to perturbation theory, the relaxation rates are proportional to the square 
of the coupling constant $\lambda$.
We recall that the Redfield quantum master equation is obtained by second-order perturbation theory in the coupling parameter 
from the complete von Neumann equation for the subsystem interacting with its environment. 
It is quite remarkable that the time scaling of the relaxation in the diffusive regime is
completely different since the relaxation rate depends on the inverse of the square of the coupling constant.\\

\begin{figure}[p]
\centering
\hspace*{-1cm}
\begin{tabular}{c}
\vspace*{0.5cm}
\rotatebox{0}{\scalebox{1}{\includegraphics{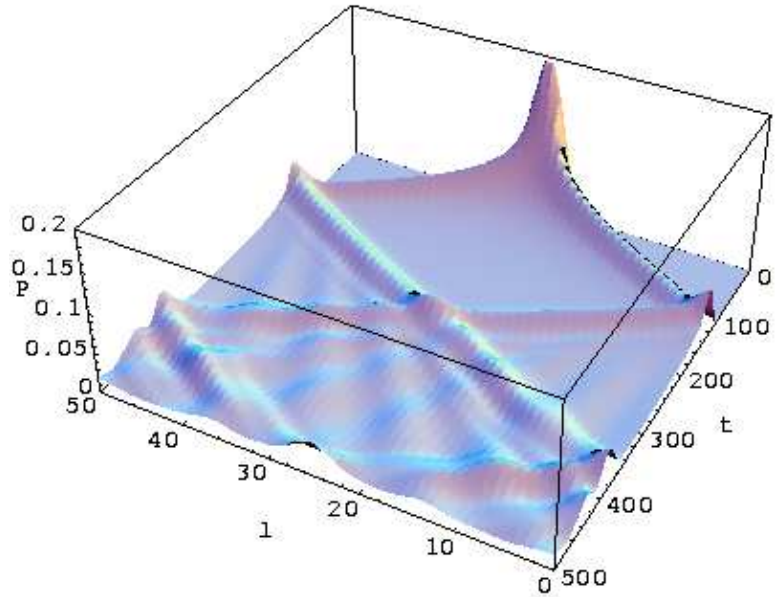}}} \\
\rotatebox{0}{\scalebox{1}{\includegraphics{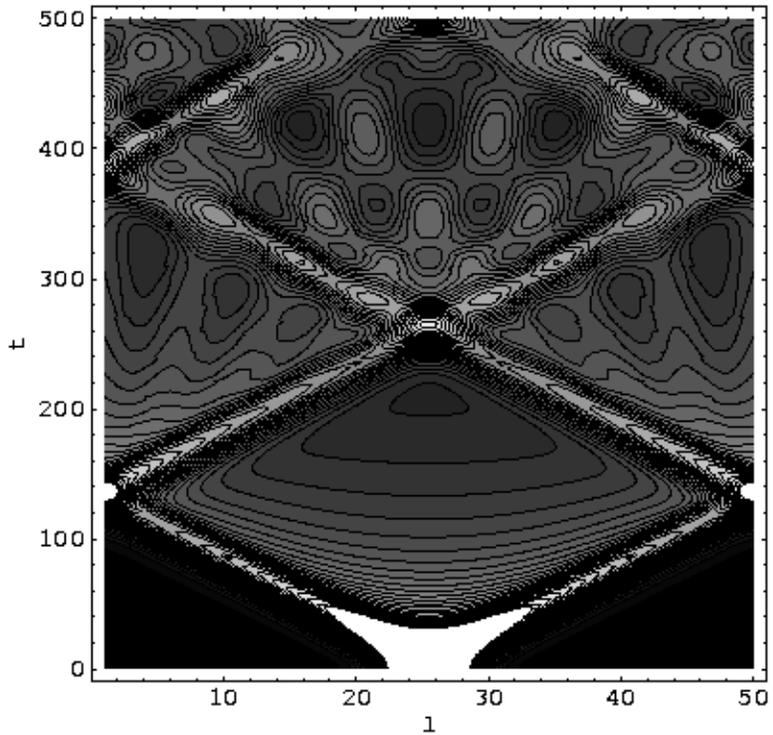}}}
\end{tabular}
\caption{Representation of the evolution of the probability $P=\rho_{ll}(t)$ to be on a given site of the system.
The initial condition corresponds to a Gaussian wave packet of standard deviation $2$ centered on the site $l=25$.
We always have: $N=50$, $E_o=0$, $A=0.1$, $Q=1$ and therefore $\lambda_c=0.1121$. 
The system is isolated ($\lambda=0$) and the evolution is therefore a pure Hamiltonian multiperiodic dynamics.
The center of the wave packet travels ballistically around the system.
} \label{dynamdiff3Da}
\end{figure}

\begin{figure}[p]
\centering
\hspace*{-1cm}
\begin{tabular}{c}
\vspace*{0.5cm}
\rotatebox{0}{\scalebox{1}{\includegraphics{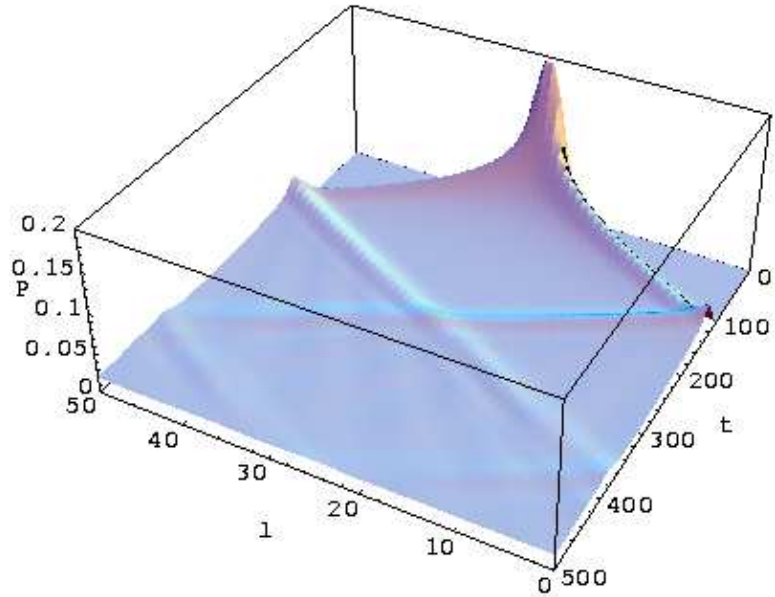}}} \\
\rotatebox{0}{\scalebox{1}{\includegraphics{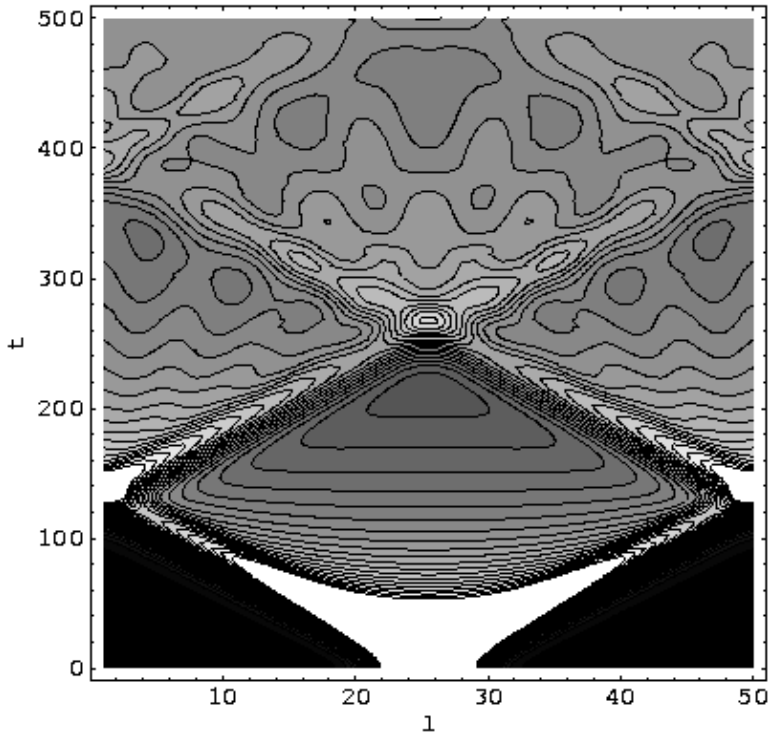}}}
\end{tabular}
\caption{Representation of the evolution of the probability $P=\rho_{ll}(t)$ to be on a given site of the system.
The situation is the same as in figure \ref{dynamdiff3Da}, except that the system is now interacting with its environment 
($\lambda=0.05$). 
The system therefore relaxes in the non-diffusive regime ($\lambda < \lambda_c$) and the evolution is a 
damped multiperiodic dynamics. 
However, the center of the wave packet keeps travelling ballistically around the system.
} \label{dynamdiff3Db}
\end{figure}

\begin{figure}[p]
\centering
\hspace*{-1cm}
\begin{tabular}{c}
\vspace*{0.5cm}
\rotatebox{0}{\scalebox{1}{\includegraphics{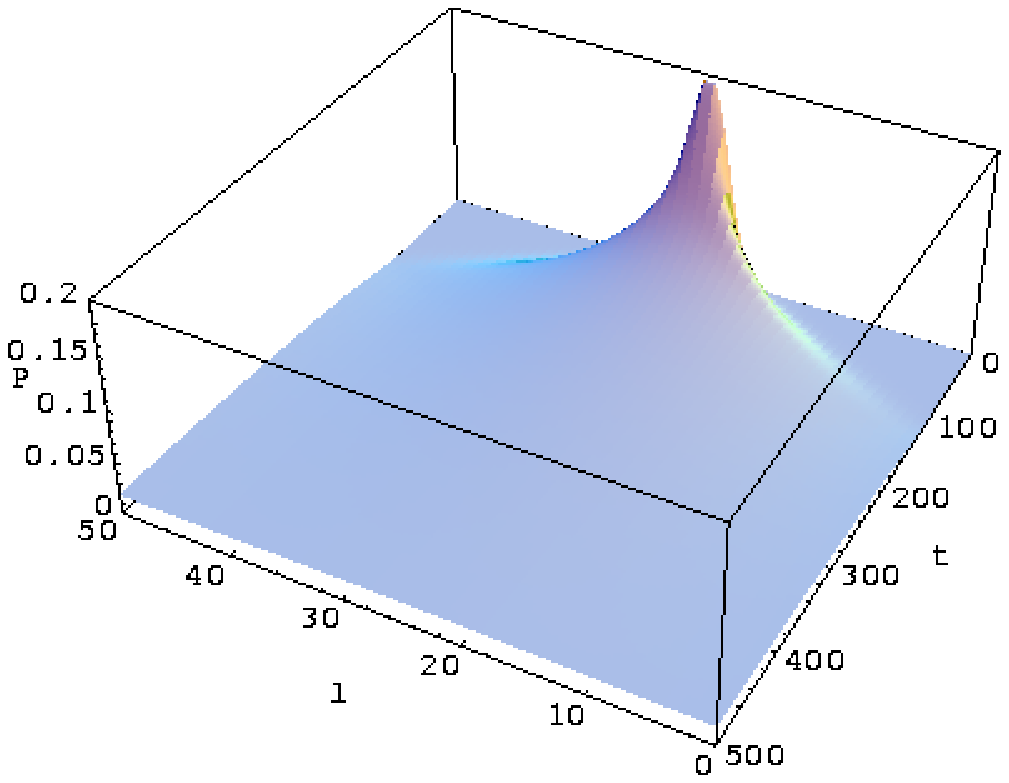}}} \\
\rotatebox{0}{\scalebox{1}{\includegraphics{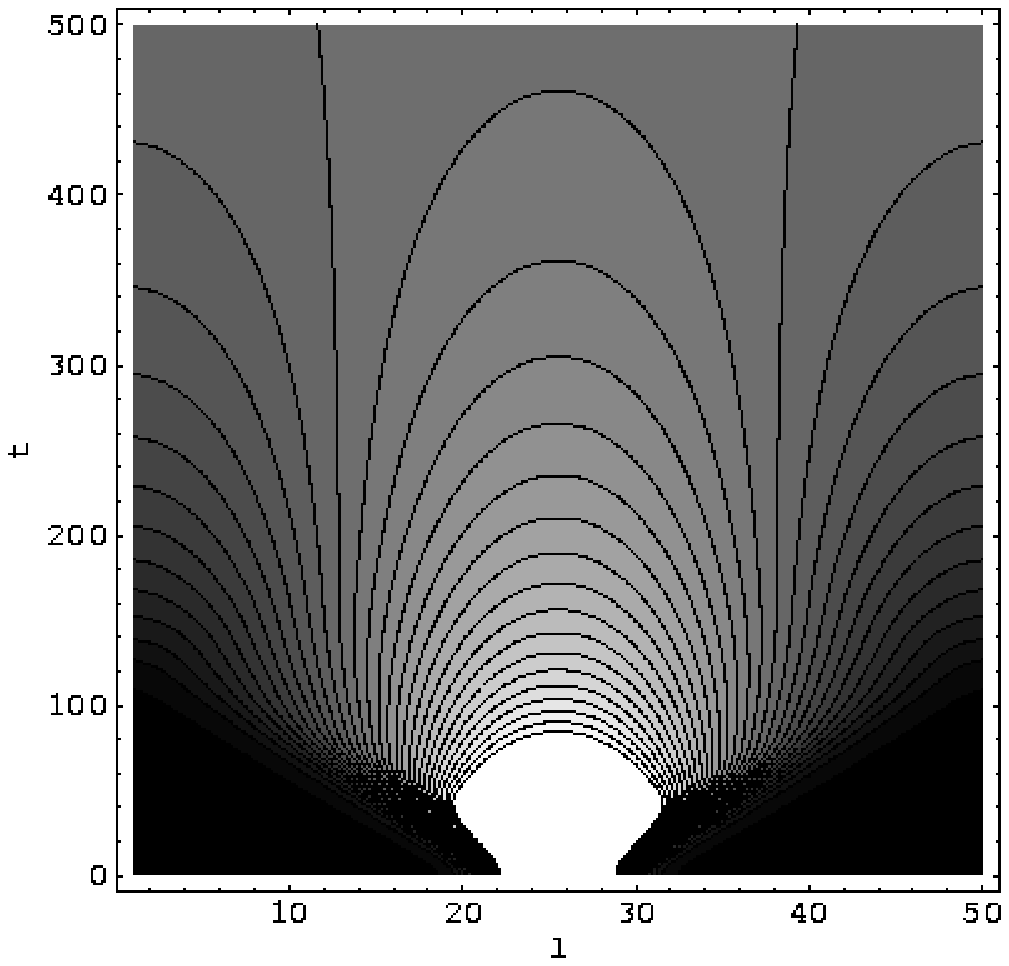}}}
\end{tabular}
\caption{Representation of the evolution of the probability $P=\rho_{ll}(t)$ to be on a given site of the system.
The situation is the same as in figure \ref{dynamdiff3Db}, but the coupling between the system and its environment 
is stronger ($\lambda=0.15$). 
The system now relaxes in the diffusive regime ($\lambda > \lambda_c$). 
The multiperiodicity has disappeared, the center of the wave packet does not move and a diffusive spreading of the 
wave packet can be observed.
} \label{dynamdiff3Dc}
\end{figure}

In figure \ref{dynamdiff3Da} to figure \ref{dynamdiff3Dc}, we have represented the evolution of the probability 
$P=\rho_{ll}(t)$ to be on a given site of the subsystem for tree different values of the coupling parameter.
In figure \ref{dynamdiff3Da}, the coupling parameter is zero and the subsystem evolution is Hamiltonian.
We can notice the interferences showing the coherent character of the dynamic.
In figure \ref{dynamdiff3Db}, the coupling parameter is such that the subsystem relaxes in the non-diffusive regime.
The dynamic still has interference patterns, but it is damped by the dissipation.
Let's notice too that the center of the wave packet travels along the subsystem.
Finally, in figure \ref{dynamdiff3Db}, the coupling parameter is now such that the subsystem relaxes in the diffusive regime.   
The subsystem dynamic is no longer coherent and the interference pattern has disappeared.
We also notice that the center of the wave packet no longer moves but that its width spreads diffusively.

\section{Infinite chain} \label{Sec.infinite}

\subsection{Spectrum}

The spectrum of the infinite chain coupled to its environment can be obtained from the
spectrum of the finite chain in the infinite size limit $N\to\infty$.
The wavenumber $q$ becomes a continuous parameter varying in the first Brillouin zone $-\pi \leq q < \pi$.\\

For given wavenumber, the diffusive eigenvalue $\mu^{(1)}$ or $s^{(1)}$ given by equation (\ref{spec3aae})
remains isolated.  Consequently, we obtain the result by which the dispersion relation of diffusion
is exactly given by the analytical expression
\begin{center} \fbox{\parbox{12.5cm}{
\begin{eqnarray}
s_q
&=& 2 \sqrt{\frac{Q^2 \lambda^4}{\hbar^4} - \left(\frac{2A}{\hbar} \sin \frac{q}{2}\right)^2} 
- \frac{2 Q \lambda^2}{\hbar^2}  = -Dq^2 + O(q^4) \; .  \label{diff.infinite} 
\end{eqnarray}
}} \end{center}
The diffusion coefficient
\begin{center} \fbox{\parbox{12.5cm}{
\begin{eqnarray}
D = \frac{A^2}{Q \lambda^2} \; , \label{diff.coeff}
\end{eqnarray}
}} \end{center}
is proportional to the square of the parameter $A$ of the tight-binding Hamiltonian 
and inversely proportional to the parameter $Q\lambda^2$ of the coupling to the environment.
The transport is therefore due to the tunneling from site to site, which is hindered by
the environmental fluctuations proportional to $Q\lambda^2$.
The eigenvalue (\ref{diff.infinite}) can be considered as a Liouvillian resonance \cite{JaksicPillet97I,JaksicPillet97II} 
similar to the Pollicott-Ruelle resonances describing diffusion in classical systems \cite{GaspardB98}.\\

\begin{figure}[p]
\centering
\begin{tabular}{c@{\hspace{0.5cm}}c}
\vspace*{1cm}
\rotatebox{0}{\scalebox{0.8}{\includegraphics{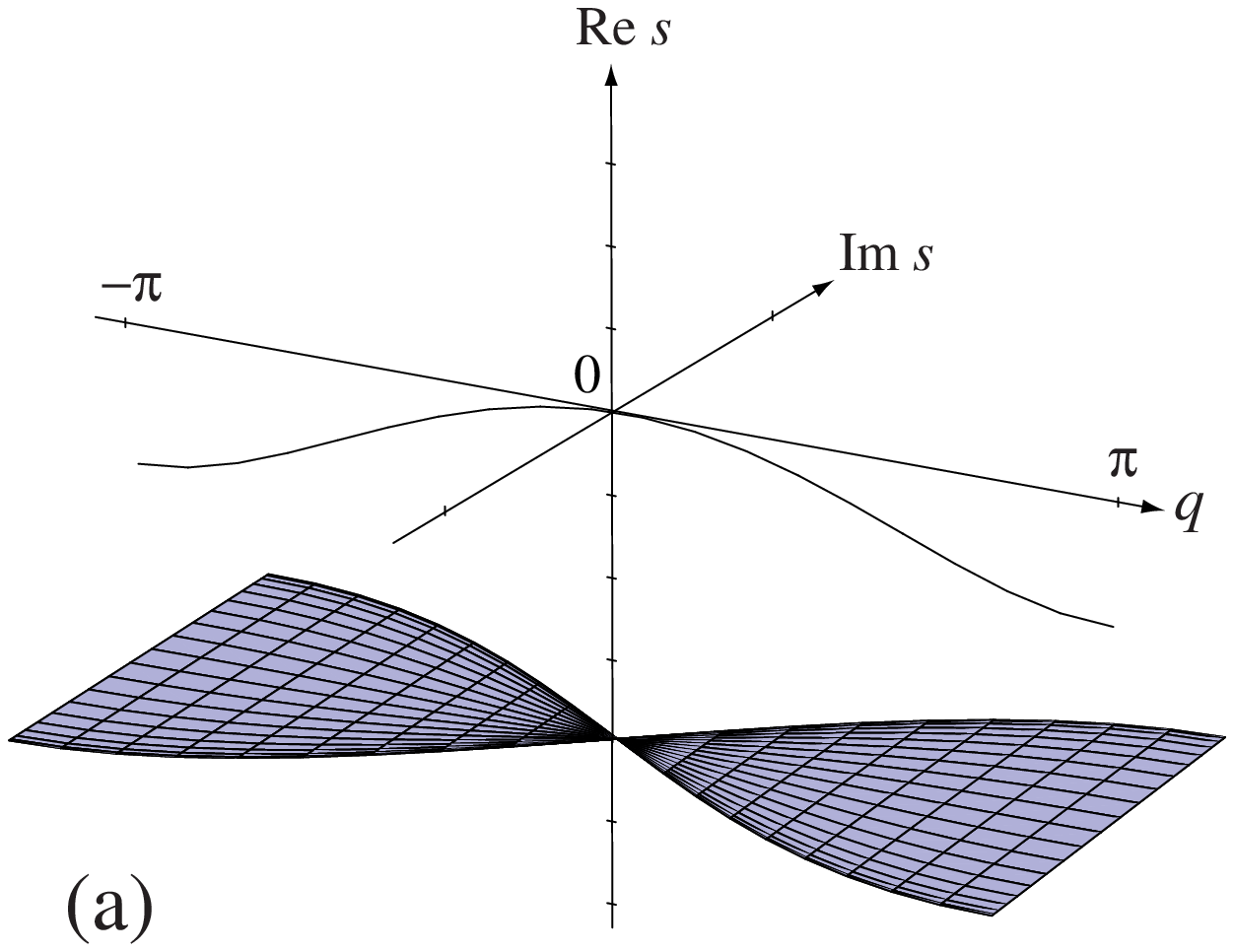}}} \\
\vspace*{1cm}
\rotatebox{0}{\scalebox{0.8}{\includegraphics{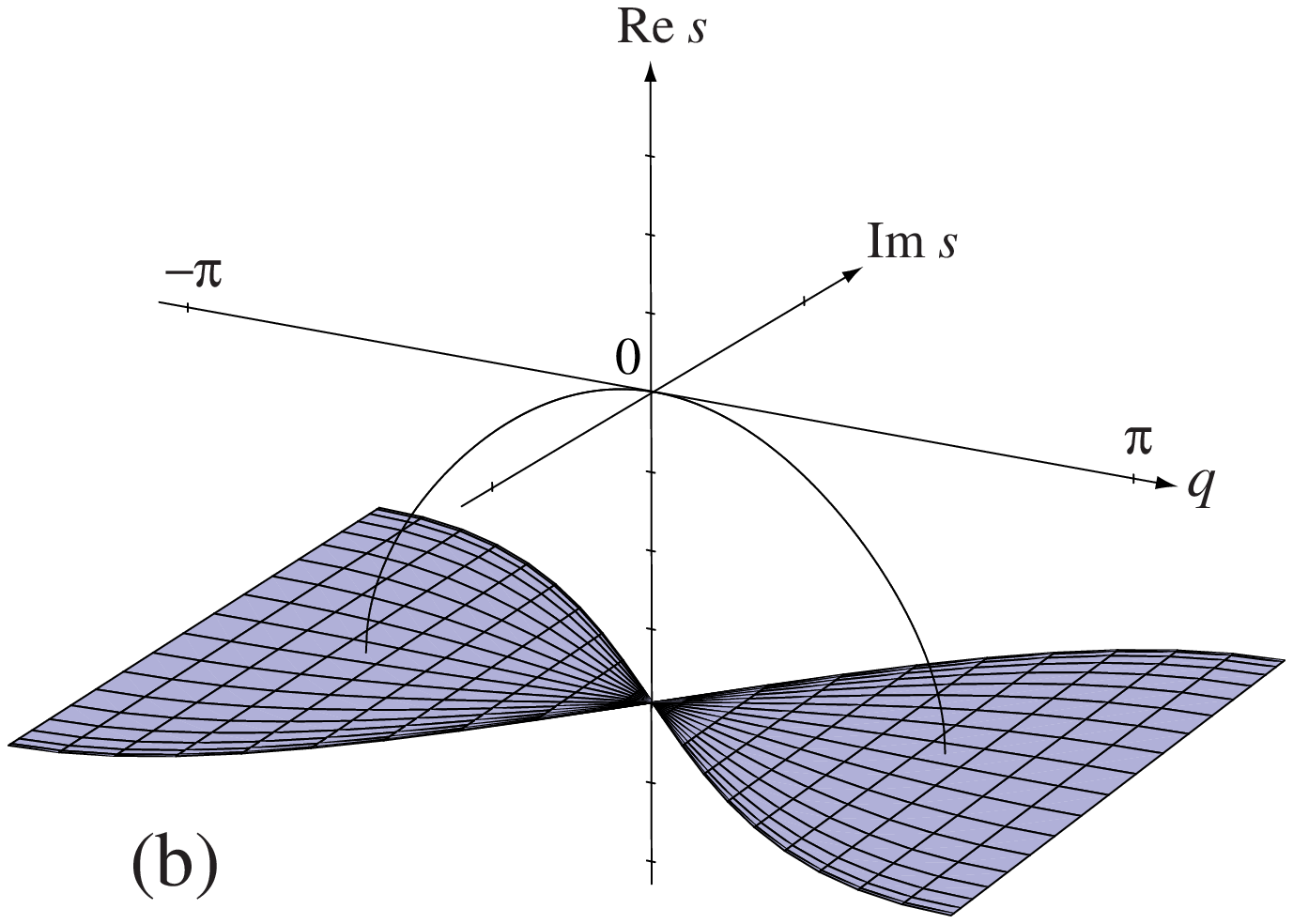}}}  \\
\end{tabular}
\caption{Spectrum of the infinite chain coupled to its environment:
(a) in the regime $2A \hbar <Q\lambda^2$ for $A \hbar =0.4$ and $Q\lambda^2=1$;
(b) in the regime $2A \hbar >Q\lambda^2$ for $A \hbar =0.6$ and $Q\lambda^2=1$
where the diffusive branch is limited to the wavenumbers $\vert q\vert < q_{c}=1.97022$.}
\label{spectre.infinite}
\end{figure}

For $2A \hbar \leq Q\lambda^2$, the diffusive eigenvalue exists for all the values of the wavenumber $-\pi \leq q < \pi$, 
as seen in figure \ref{spectre.infinite}(a).  \\

For $2A \hbar > Q\lambda^2$, the diffusive eigenvalue only exists for all the values of the wavenumber in the range
\begin{eqnarray}
-q_{c} \leq q \leq +q_{c} \; , \quad \mbox{with} \quad q_{c}= 2 \arcsin \frac{Q\lambda^2}{2A \hbar} \; .
\label{q_c}
\end{eqnarray}
as seen in figure \ref{spectre.infinite}(b). \\

Beside the isolated diffusive eigenvalue, the spectrum at given wavenumber $q$ contains a continuous part obtained 
by the accumulation of the eigenvalues $\mu^{(2)}$, $\mu^{(3)}$, $\mu^{(4)}$, and $\mu^{(5)}$ in the limit $N\to\infty$. 
Indeed, in this limit, all these eigenvalues accumulate into a segment of straight line given by
\begin{eqnarray}
s_{q\theta}
&=& - i \frac{4 A}{\hbar} \left( \sin \frac{q}{2}\right)\; \cos\theta - \frac{2 Q\lambda^2}{\hbar^2} \; , \label{non.diff.infinite} 
\end{eqnarray}
with $0\leq \theta\leq \pi$ and $-\pi \leq q < \pi$ [see equations (\ref{spec3abd}) and (\ref{spec3acd})]. 
This part of the spectrum is also depicted in figure \ref{spectre.infinite} and describes the time evolution of the 
quantum coherences which are damped at the exponential rate $-{\cal R}e \; s_{q\theta} = 2 Q \lambda^2 / \hbar^2$
with possible oscillations due to their non-vanishing imaginary part ${\cal I}m \; s_{q\theta}$.

\subsection{Temperature dependence of the diffusion coefficient}

In order to obtain the temperature dependence of the relaxation rates and the diffusion coefficient,
we need to specify the coupling of the subsystem to its environment.
In general, the coupling of a subsystem to its environment is described by the time correlation function. 
This correlation function is determined in terms of the spectral strength $J(\omega)$ according to
\begin{eqnarray}
\alpha(t) = \int_{0}^{\infty} d\omega \; J(\omega) \left(\hbar \coth \frac{\hbar \omega}{2 k_{B} T} 
\cos \omega t - i \; \hbar \sin \omega t \right) \; .
\label{cor1aaa}
\end{eqnarray}
The Fourier transform of the environment correlation function, which is given by
\begin{eqnarray}
\tilde{\alpha}(\omega) = \int_{-\infty}^{\infty} \frac{dt}{2\pi} e^{i \omega t} \alpha(t) \; , \label{cor1aab}
\end{eqnarray}
is therefore related to the spectral strength by
\begin{eqnarray}
\tilde{\alpha}(\omega) = \pm \frac{J( \pm \omega)}{2} \left(\hbar \coth \frac{\hbar \omega}{2 k_{B} T} + \hbar \right) \; ,
\label{cor1aac}
\end{eqnarray}
where the upper and lower signs respectively hold for $\omega > 0$ and $\omega < 0$.
Using equation (\ref{AppAaafb}), (\ref{1Haaay}) and (\ref{1Hbaaa}), one can notice that the spectral strength 
of the environment is proportional to the imaginary part of the environment susceptibility
\begin{eqnarray}
J(\omega) = 2 \; {\cal I}m \lbrack \tilde{\chi}(\omega) \rbrack .
\end{eqnarray}
Empirical forms for the spectral strength are often used in the literature \cite{Leggett94,Weiss00} such as
\begin{eqnarray}
J(\omega) = K \omega^{\gamma} e^{- \frac{\omega}{\omega_{c}}} \; , \label{cor1aad}
\end{eqnarray}
where $\omega_{c}$ is a cutoff frequency and $K$ a constant.
The environment is called ohmic if $\gamma=1$, subohmic if $\gamma<1$, and superohmic if $\gamma>1$.
Such empirical forms for the spectral strength are used to specify the dynamics of the subsystem over times longer 
than $1/\omega_{c}$, which is of the order of magnitude of the time scale of the environment.
Therefore, such spectral strengths are of primary importance to characterize the dependence  
of $\tilde{\alpha}(\omega)$ at small frequencies.
In the zero-frequency limit, the quantity
\begin{eqnarray}
\lim_{\omega \to 0} \tilde{\alpha}(\omega) = \lim_{\omega \to 0} J(\omega) \frac{k_{B} T}{\omega} 
= \lim_{\omega \to 0} K k_{B} T \omega^{\gamma-1} , \label{cor1aae}
\end{eqnarray}
is well defined and non-vanishing only in the ohmic case $\gamma=1$, which we consider here.\\

In our model [see equation (\ref{isolaaj})], the environmental correlation function is given by a Dirac delta 
distribution $\alpha(t)=2Q \delta(t)$. 
Therefore, the Fourier transform of the environmental correlation function is constant
\begin{eqnarray}
\tilde{\alpha}(\omega)=\frac{Q}{\pi} \, .\label{cor1abe}
\end{eqnarray}
This assumption is acceptable if the dynamics of the environment is much faster than the dynamics of the subsystem.  
Therefore, using equations (\ref{cor1aae}) and (\ref{cor1abe}), the following identification can be established
\begin{eqnarray}
Q = \pi K k_{B} T \, , \label{cor1aaf}
\end{eqnarray}
and the diffusion coefficient can now be written with an explicit temperature dependence
\begin{center} \fbox{\parbox{12.5cm}{
\begin{eqnarray}
D = \frac{A^2}{\pi K \lambda^2 k_{B} T} \; . \label{cor1aag}
\end{eqnarray}
}} \end{center}

The conductivity $\sigma$ can be obtained using the Einstein relation
\begin{eqnarray}
D = \frac{\sigma}{e^2} \left( \frac{\partial \mu}{\partial n}\right)_T \, , \label{Einstein}
\end{eqnarray}
where $e$ is the electric charge of the carriers, $n$ their density, and $\mu$ their chemical potential.
Assuming a low density of carriers, the chemical potential depends on the density according to
\begin{eqnarray}
\mu = \mu^0(T) + k_{B} T \; \ln n \; , \label{chem.pot}
\end{eqnarray}
so that the conductivity is given by
\begin{eqnarray}
\sigma = \frac{e^2 n}{k_{B} T} D  = \frac{e^2 A^2 n}{\pi  K \lambda^2 (k_{B} T)^2} \; . \label{conduct}
\end{eqnarray}
Therefore, the conductivity decreases as $T^{-2}$ with the temperature.
We notice that similar dependences are also obtained for the electric conductivity
in conducting polymers \cite{Heeger88} and in Fermi liquids \cite{MahanB90,PinesB89}.
This inverse power law is due to the existence of a single conduction band in our model.
The transport is therefore confined in this band and no thermally activated transport process
occurs in this model.  
Our diffusive transport phenomenon can be viewed as the result of the tunneling of the particle through the potential 
barriers of the system.  
This tunneling is more and more affected as the temperature increases.  \\

We expect that a crossover would occur to a regime where the diffusion coefficient has a temperature dependence 
of Arrhenius type if one considered a system with at least two conduction bands. 
In such systems, we should find that a crossover between the quantum tunneling regime and a thermally activated 
transport regime  when $k_{B} T$ becomes comparable to the energy spacing between the two bands 
\cite{DiFoggio82,Holstein59,Lauhon00,Poutier00,Storchak98,Weiss00}. 
From this viewpoint, the present model with a single energy band describes low-temperature behaviors.

\section{Summary} \label{Sec.conclusions}

In the present chapter, we have defined a simple translationally invariant subsystem which interacts with its 
environment by correlation functions which are delta-correlated in space and time. 
The reduced dynamics of the subsystem is described by a Redfield quantum master equation
which takes, for such environments, a Lindblad form.
Thanks to the invariance under spatial translations, we could apply the Bloch theorem to the subsystem density matrix.  
In this way, we succeeded in getting analytical expressions for all the eigenvalues of the Redfield superoperator. 
These eigenvalues control the time evolution of the subsystem and its relaxation to the thermodynamic equilibrium.
Two kinds of eigenvalues were obtained: the isolated eigenvalue (\ref{diff.infinite}) giving the dispersion
relation of diffusion along the one-dimensional subsystem and the other eigenvalues (\ref{non.diff.infinite}), 
which describe the decay of the quantum coherences, i.e., the process of decoherence in this subsystem.\\
The properties of the subsystem depend on the length $N$ of the one-dimensional chain, on the width $2A$ of the energy 
band of the unperturbed tight-binding Hamiltonian and on the intensity $Q$ of the environmental noise multiplied in 
the combination $Q\lambda^2$ with the square of the coupling parameter $\lambda$ of perturbation theory.\\
We discovered that for a finite chain, there are two regimes depending on
the chain length $N$ and the physical parameters $A$ and $Q\lambda^2$.\\
For a finite and sufficiently small chain, there is a non-diffusive regime characterized
by a time evolution with oscillations damped by decay rates proportional to $Q\lambda^2$.
The oscillations are the time evolution of the quantum coherences.
This non-diffusive regime exists if the coupling parameter is smaller
than a critical value which is inversely proportional to the square root of the chain size $N$:
$\lambda < \lambda_{c}=O(N^{-\frac{1}{2}})$.\\
For larger chains, we are in the diffusive regime with a monotonic decay on long times
at a rate controlled by the diffusion coefficient.
In this regime, the slower relaxation mode relaxes exponentially in time with the scaling $t/(\lambda N)^2$. \\
In the limit of an infinite chain $N \to \infty$ and for non-vanishing coupling parameter $Q\lambda^2$,
the non-diffusive regime disappears and the subsystem always diffuses.\\
The diffusion coefficient is proportional to the square of the width $2A$ of the energy band
and inversely proportional to the intensity $Q\lambda^2$ of the environmental noise.  
Accordingly, we are in the presence of a mechanism of diffusion in which the quantum tunneling of 
the particle from site to site is perturbed by the environmental fluctuations.
For an ohmic coupling to the environment, the diffusion coefficient is inversely proportional to the temperature.  
By using the Einstein relation between the diffusion coefficient and
the conductivity, this latter is inversely proportional to the square of the temperature.\\
Because the present model is very simple and shares some basic features of quantum diffusion and conductivity, 
we believe that it can easily be used to understand how quantum transport by diffusion can emerge out of quantum 
coherent behaviors as in a molecular wire or in a carbon nanotube in the form of a loop which becomes larger and larger.

\chapter{Conclusions} \label{conclusionstotal}

The main goal of this thesis has been to understand the emergence of kinetic processes such as transport, decoherence and 
relaxation to equilibrium in quantum nanosystems.
The understanding of these irreversible processes is of fundamental importance for the potential technological 
applications in the new field of nanosciences.

\section{Summary of the main results} \label{summarytotal}

In the two first chapters of this thesis, we have briefly reviewed the important results already known in the 
understanding of irreversible quantum processes.
In chapter \ref{ch1}, we have presented the important general concepts (quantum statistical ensemble, 
projected dynamics, Markovian evolution) necessary to understand how an irreversible evolution can arise from 
the reversible von Neumann equation. 
In chapter \ref{ch2}, we have reviewed the well-known and widely used theories of quantum irreversibility
(linear response theory, Pauli equation, Redfield equation) by deriving them from the same perturbative expansion 
of the von Neumann equation in order to show their similarities and differences. 
In particular, we pointed out the link that exists between the Fourier transform of the correlation functions 
(appearing in the linear response theory and the Redfield theory) and the Fermi golden rule (that is a central 
quantity for the Pauli equation).

\subsection{Emergence of relaxation and decoherence in nanosystems}   

In the traditional approaches like the Redfield theory, the study of the irreversible dynamics of a small quantum system
interacting with an environment relies on the assumption of an infinitely large environment.
Assuming that the environment is infinite means that it is not affected by the interaction with the small quantum system,
and, therefore, does not evolve and stays in its initial state.
Of course this hypothesis greatly simplifies the calculation but is not always valid, especially in the new context of 
nanotechnologies. 
Indeed, there are nanosystems in which a subsystem interacts with other degrees of freedom forming a finite environment.
In such cases, the environment is not sufficiently large to remain unaffected by the interaction with the small quantum subsystem. 
In this respect, an extension of the Redfield theory to finite environment, where the subsystem affects the environment
and of course vice versa, is necessary.
This first major achievement of this thesis is presented in chapter \ref{ch3}.\\

The central quantity of our new theory is the subsystem reduced density matrix, but distributed in the environment energy. 
This quantity is obtained by applying a projection operator acting on the full density matrix of the total system and 
which depends on the energy of the environment.
Let us notice that the projection operator used in the standard Redfield theory simply leads to the reduced density matrix of the 
subsystem without keeping information on the state of the environment.
Thanks to our projection operator, it is possible to monitor how the subsystem populations and coherences are 
distributed over the energy of the environment.
Our equation correctly takes into account the exchanges of energy between the subsystem and the environment satisfying 
the total system energy conservation, which is not the case if the Redfield equation is applied to a finite environment. 
Our new theory is derived from a second-order perturbative expansion of the total von Neumann equation and is therefore 
valid at weak couplings between the subsystem and the environment. 
Another condition is that the initial condition of the environment has to be an invariant state (i.e. it has to commute 
with the environment Hamiltonian).
Furthermore, only our equation gives a consistent understanding, based on detailed balance, of the way a canonical 
subsystem equilibrium distribution (at a temperature given by the microcanonical temperature of the environment) can arise 
when the initial state of the environment is a microcanonical distribution and when the environment has a large heat capacity. 
This indicates that applying the Redfield theory to systems where the environment is in an initial microcanonical state 
is not valid and that only our new theory is correct to describe kinetic processes within the microcanonical ensemble.
When the environment becomes very large compared to the subsystem, we showed that, as expected, our new equation reduces 
to the Redfield equation. \\
We have applied our new theory to a two-level subsystem interacting in a non-diagonal way with a general environment.
Assuming that the environment coupling operator in the environment eigenbasis representation vanishes,
we have been able to obtain analytical expressions for the bi-exponential relaxation of the subsystem population toward 
their equilibrium value and for the oscillating relaxation of the subsystem coherences toward zero.
This assumption is, in particular, exact in models where the environment coupling operators are given by random matrices
thanks to the average over the random-matrix ensemble. \\

The spin-GORM model studied in chapter \ref{ch4} of this thesis is a particular case of such models where
the environment operators (the environment Hamiltonian as well as the environment coupling operator) are given 
by random matrices taken in this case from the Gaussian orthogonal ensemble.\\
In the first part of our study of this model, we focused on various spectral quantities (density of states, 
mean level spacing, SOE, ATPK).
We classified the parameter space in three qualitatively distinct domains and we identified a crossover 
at which the various spectral quantities change qualitatively.
This crossover occurs at the value of the coupling parameter (between the spin and the environment) at which the 
interaction between the different coupled states becomes of the order of the mean-level spacing between these states.
For values of the coupling parameter below this critical point, the spectral quantities are almost not affected by the coupling 
and look basically as if there was no coupling.
But for values of the coupling parameter beyond this crossover, qualitative changes in the various spectral quantities occur.
It should be noticed that when the spectrum becomes continuous, the mean-level spacing and the crossover, both go 
to zero.\\
The second part of the study of the spin-GORM model focuses on the subsystem (spin) dynamics.
We identified three characteristic regimes of the spin relaxation corresponding to the three parameter domains identified 
in the spectral study.
The first domain corresponds to weak couplings (between the spin and the environment) and to situations where the energy
splitting between the two states of the spin is small compared to the energy width of the environment spectrum.
In this domain, the relaxation is exponential and scales in time as $\lambda^2 t$.
The second domain also corresponds to weak couplings, but the energy splitting between the two states of the spin is 
now large compared to the energy width of the environment spectrum.
In this domain, the relaxation is a purely non-Markovian dynamics (typically rapid oscillations at the spin frequency damped
on a time scale of the order of the environment correlation function time scale) and scales in time as $t$.
The last domain corresponds to very large couplings (larger than the spin energy splitting and than the environment spectral width).  
In this domain, the spin relaxation scales like $\lambda t$ and is governed by a Bessel function of the first kind 
$J_1(\lambda t)/ (\lambda t)$. \\ 
The study of the weak-coupling regime of this model was of special interest for us because it allowed a detailed comparison 
between the spin dynamics predicted by our theory, the Redfield theory, and the exact numerical results computed with the 
von Neumann equation for the total system. 
On one hand, this study confirmed the necessity of extending the Redfield theory to our new theory when the typical spin energy 
becomes non-negligible compared to the typical energy scale of variation of the environment density of states or, equivalently, 
when the heat capacity of the environment multiplied by the microcanonical temperature of the environment becomes of the order 
or smaller than the energy splitting between the two levels of the spin.
On the other hand, this study showed that both theories predict the same dynamics on non-Markovian short time scales.
Finally, and this is the second central result of this thesis, this study allowed us to find a very important criterion 
for the validity of kinetic equations in finite quantum systems.
In fact, for values of the coupling parameter $\lambda$ between the spin and the environment below the crossover that we 
identified in the spectral study, the exact dynamics does not follow the prediction of our kinetic theory.
The crossover take place at the critical value $\lambda_c=O(1/\sqrt{N})$, where $N$ is the number of energy 
levels in the environment. 
It is only beyond this crossover that the kinetic theory applies.
As expected, the critical value goes to zero in the thermodynamic limit and, consequently, the kinetic theory is always valid. 
Let us notice however that the coupling parameter cannot be too large, otherwise the weak-coupling assumption at the origin of the kinetic
theory fails.
It is conjectured that the existence of such a crossover to see kinetic processes is a general result and that kinetic processes only 
occur in finite quantum systems when the typical interaction terms between the coupled levels becomes larger than the mean level spacing 
between these coupled levels.
The interaction has to be large enough to somehow "mix" the coupled levels, so that the unperturbed structure of the spectrum disappears.\\     
This conclusion was strengthened by noting the fact that a self-averaging property arises for values of the coupling parameter 
beyond the crossover.
In fact, if the initial condition of the total system is chosen to be a pure state at a given energy of the environment, when the 
coupling parameter is below the crossover, the subsystem dynamics is strongly oscillating and does not follow the 
microcanonically averaged dynamics at this environment energy.
However, when the coupling parameter is above the crossover, the pure state dynamics starts to be smoother and to follow 
the microcanonically averaged one.
This self-averaging behavior is reminiscent of classical ergodicity and it is a very important point to understand the emergence 
of kinetic processes in quantum systems. \\
In the last part of the spin-GORM model study, we focused on the equilibrium properties of the subsystem (spin), but also of the 
total system.
Once again, we found that the equilibrium spin distributions expected from our kinetic theory are valid if the values of the 
coupling parameter are beyond the crossover.  
We also showed that when the heat capacity of the environment is large, the subsystem thermalizes at a temperature
given by the microcanonical temperature of the environment.
We finally showed that the complexity of internal couplings (given by random matrices) can have the effect of thermalizing 
the total system without the need of an external heat bath. 

\subsection{Transport in nanosystems}   

Finally, in chapter \ref{ch5}, we have defined a simple model in which a translationally invariant subsystem 
interacts with an environment by a coupling in terms of correlation functions which are delta-correlated in space and time.
The subsystem is a one-dimensional tight-binding subsystem made of $N$ sites and possessing a single energy band of width $2A$.
The intensity of the delta-correlated environmental noise is given by $Q$ and the intensity of the subsystem-environment
interaction by $\lambda$.
For weak coupling, the reduced subsystem dynamics is described by a Redfield quantum master equation which takes, for such 
correlation functions, a Lindblad form.
Thanks to the Bloch theorem, we succeeded in getting analytical expressions for all the eigenvalues of the Redfield superoperator.
By studying the Redfield superoperator spectrum we discovered that for a given system ($A$ and $N$ given) interacting 
with a given environment ($Q$ given), there are two subsystem relaxation regimes, depending on the intensity of the 
interaction between the subsystem and its environment ($\lambda$). 
These two relaxation regimes are separated from each other by a critical value of the coupling parameter 
$\lambda_c=\sqrt{2 A \hbar \pi/(Q N)}$.
In the diffusive regime corresponding to $\lambda > \lambda_c$, a diffusive branch exists in the Redfield superoperator spectrum 
which contains at least one nonzero eigenvalue. 
In this regime, the slower relaxation mode relaxes exponentially in time with a scaling $t/(\lambda N)^2$.
In the non-diffusive regime corresponding to $\lambda < \lambda_c$, the last nonzero eigenvalue of the diffusive branch has
disappeared.
The slower relaxation mode of this regime relaxes exponentially in time but with oscillations and with a scaling $\lambda^2 t$.\\
In the limit of an infinitely large number of sites on the subsystem $N \to \infty$, and for the nonzero value of the coupling 
parameter, the non-diffusive regime disappears ($\lambda_c \to 0$) and the subsystem always diffuses.
We showed that the existence of the diffusive branch in the Redfield superoperator spectrum implies an Einstein 
relation for the diffusion coefficient and that the inverse temperature dependence of the diffusion coefficient 
can be established if one considers our model of environment as the infinitely fast environment limit 
(fast compared to the subsystem dynamics) of an ohmic environment. 
Demonstrating the existence of quantum diffusion on a spatially extended system constitutes the third central 
result of this thesis.

\section{Perspectives} \label{perspectives}

The study of quantum kinetic equations started a long time ago and prestigious names are associated with this issue. 
The earlier results where first obtained by Pauli \cite{Pauli28} in the late twenties with the derivation of 
the Pauli equation and later (in the fifties and sixties) by Van Hove \cite{VanHove55}, Montroll \cite{Montroll61}, 
Zwanzig \cite{Zwanzig61} and others with more systematic derivations of the Pauli equation. 
These early studies only focus on the subsystem population dynamics.
Studies focusing on the full subsystem density matrix started at the end of the fifties with the first 
derivation of the Redfield equation by Redfield \cite{Redfield57} and have continued until now with various systematic derivation 
of this equations by Cohen-Tannoudji \cite{CohenTannoudjiB96}, Kubo \cite{KuboB2}, Gaspard \cite{GaspardRed99} and others.
In these studies of the Redfield equation, it is an external environment which generates the irreversible dynamics 
on the subsystem and one has to assume that this environment is large enough not to be affected by the subsystem.
Our kinetic equation is a new step in the theory of the quantum kinetic processes, because we further extend the
previous theories to include in our description some variables of the environment (here the environment energy) in 
order to study the irreversible dynamics of a subsystem interacting with an environment which can be affected by the 
subsystem dynamics.
The quantity on which our theory focuses is the density matrix of the subsystem distributed over the environment energy.\\

Our new approach to kinetic processes in quantum systems is certainly an important new tool to deepen our knowledge
of the foundation of statistical mechanics in quantum systems, especially because of the growing interest for 
the understanding of irreversible behavior such as decoherence in nanosystems. 
In these nanosystems, one often encounters "structured" environments (i.e. environments with a density of states which 
highly varies on short energy scales).
As soon as one wishes to study the dynamics of a subsystem interacting with such environments, one has to take into 
account the effects of the changes in the energy of the environment caused by the subsystem if the subsystem energy 
is equal or larger than the typical energy scales of the environment density of states. 
This requirement is fulfilled in our new theory.\\

Spintronics could provide situations in which models similar to the spin-GORM model could apply.
In spintronics, one studies the spin of an electron moving in a nanodevice.
If the energy difference between the two spin states of the electron (due for example to an applied magnetic field) is
not negligible compared to the typical energy scales of variation of the density of states of the nanodevice, again, a 
correct kinetic theory should take into account the effect of the spin on the nanodevice.
The correct description of the electronic spin relaxation and decoherence in such devices has a huge
importance for potential nanotechnologies such as quantum computers.\\

It should be possible to extend our new theory in order to obtain a mean field equation for many-body systems.
Such an extension could allow the description of the dynamics of an ensemble of interacting particles where energy conservation 
plays an important role.
In fact, we have seen that our equation is the only kinetic equation giving an adequate description of the irreversible
relaxation of a subsystem interacting with an environment described by a microcanonical state.
Such kinds of extension could possibly lead for example to quantum kinetic equations for Bose-Einstein condensates.\\

Our kinetic equation allows us to follow the dynamics of each element of the subsystem density matrix as a distribution in 
the energy of the environment. 
Our theory could therefore play an important role in order to understand the quantum measurement within a pure quantum scheme
where the measured system as well as the measurement apparatus are described by quantum mechanics. 
The density matrix of the measured system would be distributed over the energy of the measurement apparatus allowing
possible interesting interpretations.\\

Stochastic Schr\"odinger equations describing the wave function dynamics of a subsystem interacting with its environment 
have been intensely studied during this last decade \cite{EspositoMemoire00,GaspardStocSchrod99,Struntz01}.
These equations will probably play an important role in the future to describe individual elements of a quantum 
statistical ensemble.
It has been shown that if ensemble averaged, such equations obey the well-known master equations.
Stochastic Schr\"odinger equations corresponding in this way to the Redfield equation were also 
obtained \cite{GaspardStocSchrod99}.
An interesting issue would be to find the stochastic Schr\"odinger equations which, if ensemble averaged, gives
our new kinetic equation.\\

The results in the last chapter of this thesis open perspectives on transport in nanosystems. \\

Investigating quantum transport in a translationally invariant model with at least two conduction bands is an 
important extension of the single band model studied in this thesis.
In fact, in such a double band model, two kinds of transport are possible. 
At very low temperatures, only the lower conduction band is filled, and the transport occurs by the tunneling of the 
particle through the potential barriers. 
At high temperatures, the thermal fluctuations coming from the environment are sufficiently strong to allow the 
particle to jump from a conduction band to another and the transport occurs mainly by thermal hopping. 
This second type of transport becomes effective when the temperature creates energy fluctuations comparable to the 
energy gap between the conduction bands.
At low temperatures in the tunneling regime, we have seen that the transport is slowed down when the temperature is 
increased.
At high temperatures where the transport by hopping is dominant, one expects an Arrhenius temperature dependence of 
the transport coefficient.
Therefore, at a given temperature, a crossover between the two transport regimes should exist. \\

In our model of quantum diffusion we have used semi-qualitative arguments in order to assert that normal transport 
(transport with a well-defined transport coefficient) can only occur if the environment spectral strength is ohmic.
Here we may wonder if superohmic or subohmic environments generate abnormal transport on the subsystem, 
as suggested by possible extension of our ohmic model studied in chapter \ref{ch5} and also by Refs. 
\cite{Lutz01,Weiss00}.\\

In this thesis we have encountered the criterion $\lambda^2 N > Cte$ twice for the emergence of irreversible 
processes in nanosystems.
In the spin-GORM model, $N$ is the number of states of the environment. 
In the model of quantum diffusion, $N$ is the number of sites of the translationally invariant system.
Despite the different meanings of $N$ in both cases, it is worthwhile to make the analogy.
One can wonder if such a criterion for having kinetic processes in finite quantum systems will replace in the future the 
much stronger criterion of the thermodynamic limit (which requires $N \to \infty$).

\appendix

\chapter{Correlation functions of quantum systems} \label{AppA}

The correlation function between two operators $\hat{B}$ and $\hat{A}$ of a quantum system with Hamiltonian $\hat{H}$
is defined as
\begin{eqnarray}
\alpha_{AB}(t) = \trace \hat{\rho}^{eq} \hat{A}(t) \hat{B} 
= \trace \hat{\rho}^{eq} e^{\frac{i}{\hbar} \hat{H} t} \hat{A} e^{-\frac{i}{\hbar} \hat{H} t} \hat{B},              \label{AppAaaaa}
\end{eqnarray}
where $\hat{\rho}^{eq}$ is an invariant density matrix such that $\lbrack \hat{H}, \hat{\rho}^{eq} \rbrack=0$.
The correlation function has the property $\alpha_{AB}(t)=\alpha^{*}_{BA}(-t)$.\\
The Fourier transform of the correlation function is defined by
\begin{eqnarray}
\tilde{\alpha}_{AB}(\omega)=\int_{-\infty}^{\infty} dt \frac{e^{i \omega t}}{2 \pi} \alpha_{AB}(t).                 \label{AppAaaab}
\end{eqnarray}
The Fourier transform of the correlation function has the property 
$\tilde{\alpha}_{AB}(\omega)=\tilde{\alpha}^{*}_{BA}(\omega)$.

\subsubsection{General properties}

The correlation function can be written as
\begin{eqnarray}
\alpha_{AB}(t) = C_{AB}(t) + i D_{AB}(t),                                                                            \label{AppAaabb}
\end{eqnarray}
where
\begin{eqnarray}
C_{AB}(t)= \frac{1}{2} \trace \hat{\rho}^{eq} \lbrack \hat{A}(t) , \hat{B} \rbrack_{+}
=\frac{1}{2} (\alpha_{AB}(t)+\alpha_{AB}^{*}(t))                                                                     \label{AppAaacb}
\end{eqnarray}
and
\begin{eqnarray}
D_{AB}(t)= - \frac{i}{2} \trace \hat{\rho}^{eq} \lbrack \hat{A}(t) , \hat{B} \rbrack 
= -\frac{i}{2} (\alpha_{AB}(t)-\alpha_{AB}^{*}(t)) .                                                                 \label{AppAaadb}
\end{eqnarray}
Notice that $C_{AB}(t)$ and $D_{AB}(t)$ are real and have the property
\begin{eqnarray}
C_{AB}^{*}(t)&=&C_{AB}(t) \ \ , \; \ \  D_{AB}^{*}(t)=D_{AB}(t)                                                      \label{AppAaaeb}\\
C_{AB}(t)&=&C_{BA}(-t) \ \ , \; \ \  D_{AB}(t)=-D_{BA}(-t). \nonumber   
\end{eqnarray}
The Fourier transform of the correlation function can be written as
\begin{eqnarray}
\tilde{\alpha}_{AB}(\omega)=
\tilde{C}_{AB}(\omega) + i \tilde{D}_{AB}(\omega)=\tilde{\alpha}_{BA}^{*}(\omega),                                   \label{AppAaafb}
\end{eqnarray}
where
\begin{eqnarray}
\tilde{C}_{AB}^{*}(\omega)&=&\tilde{C}_{AB}(-\omega) \ \ , \; \ \  \tilde{D}_{AB}^{*}(\omega)=\tilde{D}_{AB}(-\omega) \label{AppAaagb}\\
\tilde{C}_{AB}(\omega)&=&\tilde{C}_{BA}(-\omega) \ \ , \; \ \  \tilde{D}_{AB}(\omega)=-\tilde{D}_{BA}(-\omega). \nonumber           
\end{eqnarray}
If we define
\begin{eqnarray}
\tilde{C}^{s}_{AB}(\omega)&\equiv&(\tilde{C}_{AB}(\omega)+\tilde{C}_{BA}(\omega))/2                                   \label{AppAaahb}\\
\tilde{C}^{a}_{AB}(\omega)&\equiv&(\tilde{C}_{AB}(\omega)-\tilde{C}_{BA}(\omega))/2 \nonumber \\
\tilde{D}^{s}_{AB}(\omega)&\equiv&(\tilde{D}_{AB}(\omega)+\tilde{D}_{BA}(\omega))/2 \nonumber \\
\tilde{D}^{a}_{AB}(\omega)&\equiv&(\tilde{D}_{AB}(\omega)-\tilde{D}_{BA}(\omega))/2 \nonumber ,
\end{eqnarray}
we can separate the real and the imaginary parts of the Fourier transform of the correlation function 
in the following way
\begin{eqnarray}
\tilde{\alpha}_{AB}(\omega)&=&{\cal R}e \lbrack \tilde{\alpha}_{AB}(\omega) \rbrack 
+ i {\cal I}m \lbrack \tilde{\alpha}_{AB}(\omega) \rbrack ,                                                          \label{AppAaaib}
\end{eqnarray}
where
\begin{eqnarray}
{\cal R}e \lbrack \tilde{\alpha}_{AB}(\omega) \rbrack &=& 
\tilde{C}^{s}_{AB}(\omega)+i \tilde{D}^{s}_{AB}(\omega)                                                              \label{AppAaajb}\\
{\cal I}m \lbrack \tilde{\alpha}_{AB}(\omega) \rbrack &=& 
\tilde{D}^{a}_{AB}(\omega) - i  \tilde{C}^{a}_{AB}(\omega) \nonumber .
\end{eqnarray}

\subsubsection{The case of a discrete spectrum}
 
If the spectrum of $\hat{H}$ is discrete ($\hat{H} \ket{n} = E_n \ket{n}$), the correlation function reads 
\begin{eqnarray}
\alpha_{AB}(t) = \sum_{nn'} \rho_{nn}^{eq} e^{\frac{i}{\hbar} (E_n-E_{n'}) t} A_{nn'} B_{n'n},                      \label{AppAaaac}
\end{eqnarray}
and the Fourier transform of the correlation function reads
\begin{eqnarray}
\tilde{\alpha}_{AB}(\omega) = 
\sum_{nn'} \rho_{nn}^{eq} \delta(\omega + \omega_{nn'}) A_{nn'} B_{n'n} ,                                          \label{AppAaaad}
\end{eqnarray}
where $\hbar \omega_{nn'}=E_n-E_{n'}$.\\
For the canonical case $\hat{\rho}^{eq}=e^{-\beta \hat{H}}/Z$ where $Z=\trace e^{-\beta \hat{H}_0}$, the correlation function reads
\begin{eqnarray}
\alpha_{AB}(\beta,t) = 
\sum_{nn'} \frac{e^{-\beta E_n}}{Z} e^{\frac{i}{\hbar} (E_n-E_{n'}) t} A_{nn'} B_{n'n},                             \label{AppAaaae}
\end{eqnarray}
and its Fourier transform
\begin{eqnarray}
\tilde{\alpha}_{AB}(\beta,\omega) = 
\sum_{nn'} \frac{e^{-\beta E_n}}{Z} \delta(\omega + \omega_{nn'}) A_{nn'} B_{n'n} .                                 \label{AppAaaaf}
\end{eqnarray}
For the microcanonical case where $\hat{\rho}^{eq}=\delta(\epsilon-\hat{H})/n(\epsilon)$ with 
$n(\epsilon)=\trace \delta(\epsilon-\hat{H}_0)$, the correlation function reads
\begin{eqnarray}
\alpha_{AB}(\epsilon,t) = 
\sum_{nn'} \frac{\delta(\epsilon-E_n)}{n(\epsilon)} e^{\frac{i}{\hbar} (E_n-E_{n'}) t} A_{nn'} B_{n'n},             \label{AppAaaag}
\end{eqnarray}
and its Fourier transform
\begin{eqnarray}
\tilde{\alpha}_{AB}(\epsilon,\omega) = 
\sum_{nn'} \frac{\delta(\epsilon-E_n)}{n(\epsilon)} \delta(\omega + \omega_{nn'}) A_{nn'} B_{n'n} .                \label{AppAaaah}
\end{eqnarray}
For the pure case $\hat{\rho}^{eq}=\ket{n} \bra{n}$, the correlation function reads
\begin{eqnarray}
\alpha_{AB}(E_{n},t) = 
\sum_{n'} e^{\frac{i}{\hbar} (E_{n}-E_{n'}) t} A_{nn'} B_{n'n} ,                                                  \label{AppAaaai}
\end{eqnarray}
and its Fourier transform
\begin{eqnarray}
\tilde{\alpha}_{AB}(E_{n},\omega) = 
\sum_{n'} \delta(\omega + \omega_{nn'}) A_{nn'} B_{n'n} .                                                          \label{AppAaaaj}
\end{eqnarray}

\subsubsection{The case of a quasi-continuous spectrum}

The spectrum of $\hat{H}$ is quasi-continuous if the discrete eigenvalues are so densely distributed in energy
that one can use a continuous description for the energy ($\hat{H} \ket{\epsilon} = \epsilon \ket{\epsilon}$)
and that the operator of interest can be considered as a smooth function of the energy in the eigenbasis 
representation of the system.\\
In this case, taking the quasi-continuous limit means that 
\begin{eqnarray}
E_n &\to& \epsilon                                                                                                   \label{AppAaaak}\\
\sum_{n} &\to& \int d\epsilon \; n(\epsilon) \nonumber \\
\rho_{nn}^{eq} &\to& \rho^{eq}(\epsilon) \nonumber \\
A_{nn'} &\to& A(\epsilon,\epsilon') \nonumber \\
B_{nn'} &\to& B(\epsilon,\epsilon') \nonumber
\end{eqnarray}
where $n(\epsilon)$ is the density of states of the system at energy $\epsilon$.\\
The correlation function of a quasi-continuous system reads
\begin{eqnarray}
\alpha_{AB}(t) = \int d\epsilon' \int d \epsilon'' n(\epsilon') n(\epsilon'') 
\rho^{eq}(\epsilon') A(\epsilon',\epsilon'') B(\epsilon'',\epsilon') e^{\frac{i}{\hbar}(\epsilon'-\epsilon'')t},     \label{AppAaaal}
\end{eqnarray}
and its Fourier transform
\begin{eqnarray}
\tilde{\alpha}_{AB}(\omega) &=& \hbar \int d\epsilon' \int d \epsilon'' n(\epsilon') n(\epsilon'')
\rho^{eq}(\epsilon) \delta(\hbar \omega + \epsilon' - \epsilon'') 
A(\epsilon',\epsilon'') B(\epsilon'',\epsilon') \nonumber \\
&=& \hbar \int d\epsilon' n(\epsilon') n(\hbar \omega + \epsilon')
\rho^{eq}(\epsilon') A(\epsilon',\hbar \omega + \epsilon') B(\hbar \omega + \epsilon',\epsilon') .                   \label{AppAaaam}
\end{eqnarray}
In the canonical case, the correlation function of a quasi-continuous system reads
\begin{eqnarray}
\alpha_{AB}(\beta,t) = \int d\epsilon' \int d \epsilon'' n(\epsilon') n(\epsilon'') 
\frac{e^{-\beta \epsilon'}}{Z} A(\epsilon',\epsilon'') B(\epsilon'',\epsilon') 
e^{\frac{i}{\hbar}(\epsilon'-\epsilon'')t},                                                                          \label{AppAaaan}
\end{eqnarray}
and its Fourier transform
\begin{eqnarray}
\tilde{\alpha}_{AB}(\beta,\omega)= \hbar \int d\epsilon' n(\epsilon') n(\hbar \omega + \epsilon')
\frac{e^{-\beta \epsilon'}}{Z} A(\epsilon',\hbar \omega + \epsilon') B(\hbar \omega + \epsilon',\epsilon') .         \label{AppAaaao}
\end{eqnarray}
In the microcanonical case, the correlation function of a quasi-continuous system reads
\begin{eqnarray}
\alpha_{AB}(\epsilon,t) &=& \int d\epsilon' \int d \epsilon'' n(\epsilon') n(\epsilon'') 
\frac{\delta(\epsilon-\epsilon')}{n(\epsilon)}
A(\epsilon',\epsilon'') B(\epsilon'',\epsilon') e^{\frac{i}{\hbar}(\epsilon'-\epsilon'')t} \nonumber \\                      
&=& \int d\epsilon' n(\epsilon') A(\epsilon,\epsilon') B(\epsilon',\epsilon) 
e^{\frac{i}{\hbar}(\epsilon-\epsilon')t},                                                                            \label{AppAaaap}                              
\end{eqnarray}
and its Fourier transform 
\begin{eqnarray}
\tilde{\alpha}_{AB}(\epsilon,\omega)&=& 
\hbar \int d\epsilon' n(\epsilon') A(\epsilon,\epsilon') B(\epsilon',\epsilon) 
\delta(\epsilon-\epsilon'+\hbar \omega) \nonumber \\
&=& \hbar n(\epsilon+\hbar \omega) A(\epsilon,\epsilon+\hbar \omega) B(\epsilon+\hbar \omega,\epsilon) .             \label{AppAaaaq}
\end{eqnarray}
We notice that
\begin{eqnarray}
\alpha_{AB}(\beta,t) &=& \int d\epsilon \frac{n(\epsilon) e^{-\beta \epsilon}}{Z} \alpha_{AB}(\epsilon,t),           \label{AppAaaar}                        
\end{eqnarray}
and therefore that
\begin{eqnarray}
\tilde{\alpha}_{AB}(\beta,\omega) 
= \int d\epsilon \frac{n(\epsilon) e^{-\beta \epsilon}}{Z} \tilde{\alpha}_{AB}(\epsilon,\omega).                     \label{AppAaaas}                          
\end{eqnarray}

\subsubsection{KMS property}

The usual KMS (Kubo-Martin-Schwinger) property, which is easy to verify, is that the 
canonical correlation function satisfies
\begin{eqnarray}
\alpha_{AB}(\beta;t)=\alpha_{BA}(\beta;-t - i \beta / \hbar)  .                                                      \label{AppAaaat}
\end{eqnarray}
Taking the Fourier transform, on gets that
\begin{eqnarray}
\tilde{\alpha}_{AB}(\beta;\omega)&=& e^{\beta \hbar \omega} \tilde{\alpha}^{*}_{AB}(\beta;-\omega)                   \label{AppAaaau}\\
&=&e^{\beta \hbar \omega} \tilde{\alpha}_{BA}(\beta;-\omega) \nonumber .
\end{eqnarray}
This implies that
\begin{eqnarray}
\tilde{C}_{AB}(\beta;\omega)= 2i \frac{E_{\beta}(\omega)}{\hbar \omega} \tilde{D}_{AB}(\beta;\omega)                \label{AppAaabu}
\end{eqnarray}
where we have defined 
\begin{eqnarray}
E_{\beta}(\omega)=\frac{\hbar \omega}{2} \coth{\frac{\beta \hbar \omega}{2}}   .                                     \label{AppAaacu}
\end{eqnarray}
The microcanonical analogue of the canonical KMS property is also easy to verify
and is given by the relation 
\begin{eqnarray}
\tilde{\alpha}_{AB}(\epsilon,\omega)  
&=& \frac{n(\epsilon+\hbar \omega)}{n(\epsilon)} \tilde{\alpha}^{*}_{AB}(\epsilon+\hbar \omega,-\omega)              \label{AppAaaav}\\
&=& \frac{n(\epsilon+\hbar \omega)}{n(\epsilon)} \tilde{\alpha}_{BA}(\epsilon+\hbar \omega,-\omega) .\nonumber                                   
\end{eqnarray}
In fact, by averaging (\ref{AppAaaav}) over energy with a canonical weight as in (\ref{AppAaaas}), 
one recovers (\ref{AppAaaau}).
However the KMS property is not as simple and therefore as useful in the microcanonical case as in canonical case.

\chapter{Systems with a large heat capacity} \label{AppB}

Here we consider an isolated system with a dense spectrum described by a density of states $n(\epsilon)$. 
The statistical state of the system is described by a microcanonical distribution at energy $\epsilon$.
For such a system, the microcanonical entropy is given by 
\begin{eqnarray}
S(\epsilon)=k_b \ln \Omega(\epsilon)=k_b \ln n(\epsilon) \delta \epsilon ,                                 \label{AppBaaaa}
\end{eqnarray}
where $\Omega(\epsilon)$ is the number of states in the energy shell which 
is given by the density of states $n(\epsilon)$ multiplied 
by the small width of the energy shell $\delta \epsilon$.  
The microcanonical temperature is defined as follows
\begin{eqnarray}
\frac{1}{T_{mic}(\epsilon)} = \frac{d S(\epsilon)}{d \epsilon}.                                            \label{AppBaaab}
\end{eqnarray}
The microcanonical heat capacity is defined as
\begin{eqnarray}
\frac{1}{C_{v}(\epsilon)} = \frac{d T_{mic}(\epsilon)}{d \epsilon}                                         \label{AppBaaac}                                                  
\end{eqnarray}
Using (\ref{AppBaaaa}), we can connect the density of states of the system at two different energies 
as follows
\begin{eqnarray}
\frac{n(\epsilon+\Delta)}{n(\epsilon)} &=& \frac{e^{S(\epsilon+\Delta)/k_b}}{e^{S(\epsilon)/k_b}} .        \label{AppBaaad}
\end{eqnarray}
Now, we suppose that the energy $\Delta$ by which the energy of the system has been increased is small 
enough to perform an expansion of the entropy $S(\epsilon+\Delta)$ in $\Delta$ around $\epsilon$.
Using (\ref{AppBaaab}) and (\ref{AppBaaac}) we can write
\begin{eqnarray}
S(\epsilon+\Delta) &=& S(\epsilon) + \frac{d S(\epsilon)}{d \epsilon} \Delta +
\frac{d^2 S(\epsilon)}{d \epsilon^2} \frac{\Delta^2}{2} + \dots  \nonumber \\
&=& S(\epsilon) + \frac{\Delta}{k_b T_{mic}(\epsilon)} -
\frac{1}{C_{v}(\epsilon) T_{mic}^2(\epsilon)} \frac{\Delta^2}{2} + \dots   .                               \label{AppBaaae}
\end{eqnarray}
This means that if the heat capacity of the system is large enough to neglect the second order 
term in (\ref{AppBaaae})
\begin{eqnarray}
C_{v}(\epsilon) \gg \frac{\Delta}{T_{mic}(\epsilon)},                                                      \label{AppBaaaf}
\end{eqnarray}
(\ref{AppBaaad}) can be written as
\begin{eqnarray}
\frac{n(\epsilon+\Delta)}{n(\epsilon)} \approx e^{\beta_{mic}(\epsilon) \Delta}                           \label{AppBaaag}
\end{eqnarray}
where $\beta_{mic}(\epsilon)=1/k_b T_{mic}(\epsilon)$.
This shows that a Boltzmann distribution, characterising a canonical ensemble, can naturally emerge 
from a microcanonical description. 
We also notice that the microcanonical temperature of a system satisfying 
(\ref{AppBaaaf}) can be considered constant on energy scales of order $\Delta$
\begin{eqnarray}
T_{mic}(\epsilon+\Delta)&=&T_{mic}(\epsilon) + \frac{d T_{mic}(\epsilon)}{d \epsilon} \Delta + \dots      \label{AppBaaah}\\
&=& T_{mic}(\epsilon) + \frac{1}{C_{v}(\epsilon)} \Delta + \dots \nonumber \\
&\approx& T_{mic}(\epsilon)\nonumber 
\end{eqnarray}

\chapter{Redfield equation for two-level subsystems} \label{AppC}

In this appendix we apply the Redfield equation (\ref{1Gaaap}) to a two-level subsystem interacting 
in a non-diagonal way with its environment. 
The Hamiltonian of the total system is given by (\ref{4Aaaaa}). 
The non-Markovian Redfield equation becomes
\begin{eqnarray}
\bra{\pm} \dot{\hat{\rho}}_S (t) \ket{\pm}
&=&- \frac{i}{\hbar} \lambda \mean{\hat{B}}
(\bra{\mp} \hat{\rho}_S (t) \ket{\pm} - \bra{\pm} \hat{\rho}_S (t) \ket{\mp})                           \label{AppCaaaa}\\
&&\hspace*{0.0cm} + \frac{\lambda^2}{\hbar^2} \int^{t}_{0} d\tau \int d\omega \; \nonumber \\ 
&&\hspace*{1.0cm} \{ \; - \bra{\pm} \hat{\rho}_S (t) \ket{\pm}
\tilde{\alpha}(\omega) (e^{\frac{i}{\hbar} (\pm \Delta-\hbar \omega) \tau} 
+e^{-\frac{i}{\hbar} (\pm \Delta-\hbar \omega) \tau}) \nonumber\\
&&\hspace*{1.0cm}+ \bra{\mp} \hat{\rho}_S (t) \ket{\mp} \tilde{\alpha}(\omega)
(e^{\frac{i}{\hbar} (\pm \Delta+\hbar \omega) \tau}
+e^{-\frac{i}{\hbar} (\pm \Delta+\hbar \omega) \tau}) \; \} \nonumber
\end{eqnarray}
for the population and 
\begin{eqnarray}
\bra{\pm} \dot{\hat{\rho}}_S (t) \ket{\mp}
&=&\mp \frac{i}{\hbar} \Delta \bra{\pm} \hat{\rho}_S (t) \ket{\mp}
- \frac{i}{\hbar} \lambda \mean{\hat{B}}
(\bra{\mp} \hat{\rho}_S (t) \ket{\mp} - \bra{\pm} \hat{\rho}_S (t) \ket{\pm}) \nonumber   \\                   
&&\hspace*{0.0cm} + \frac{\lambda^2}{\hbar^2} \int^{t}_{0} d\tau \int d\omega \;                        \label{AppCaaab} \\
&&\hspace*{1.0cm} \{ \; - \bra{\pm} \hat{\rho}_S (t) \ket{\mp}
\tilde{\alpha}(\omega) (e^{\frac{i}{\hbar} (\pm \Delta-\hbar \omega) \tau} 
+e^{\frac{i}{\hbar} (\pm \Delta+\hbar \omega) \tau})  \nonumber \\                                              
&&\hspace*{1.0cm} + \bra{\mp} \hat{\rho}_S (t) \ket{\pm}
\tilde{\alpha}(\omega)
(e^{-\frac{i}{\hbar} (\pm \Delta-\hbar \omega) \tau}
+e^{-\frac{i}{\hbar} (\pm \Delta+\hbar \omega) \tau}) \; \}  \nonumber                                         
\end{eqnarray}
for the coherences.
Performing the Markovian approximation, we get
\begin{eqnarray}
\bra{\pm} \dot{\hat{\rho}}_S (t) \ket{\pm}
&=&- \frac{i}{\hbar} \lambda \mean{\hat{B}}
(\bra{\mp} \hat{\rho}_S (t) \ket{\pm} - \bra{\pm} \hat{\rho}_S (t) \ket{\mp})                           \label{AppCaaba}\\
&& - 2 \pi \frac{\lambda^2}{\hbar^2} \tilde{\alpha}(\pm \Delta/\hbar) 
\bra{\pm} \hat{\rho}_S (t) \ket{\pm}  \nonumber \\
&&+ 2 \pi \frac{\lambda^2}{\hbar^2} \tilde{\alpha}(\mp \Delta/\hbar)
\bra{\mp} \hat{\rho}_S (t) \ket{\mp} \nonumber
\end{eqnarray}
for the population and 
\begin{eqnarray}
\bra{\pm} \dot{\hat{\rho}}_S (t) \ket{\mp}
&=&\mp \frac{i}{\hbar} \Delta \bra{\pm} \hat{\rho}_S (t) \ket{\mp}
- \frac{i}{\hbar} \lambda \mean{\hat{B}}
(\bra{\mp} \hat{\rho}_S (t) \ket{\mp} - \bra{\pm} \hat{\rho}_S (t) \ket{\pm})  \nonumber \\                   
&&- \frac{\lambda^2}{\hbar^2} \bra{\pm} \hat{\rho}_S (t) \ket{\mp}
\{ + \pi (\tilde{\alpha}(\Delta/\hbar)+\tilde{\alpha}(-\Delta/\hbar)) \nonumber \\
&&\hspace*{3cm} \pm i \frac{\Delta}{\hbar} \int d\omega {\cal P} \frac{\tilde{\alpha}(\omega)}
{(\Delta/\hbar)^2-\omega^2} \} \nonumber \\                                     
&&+ \frac{\lambda^2}{\hbar^2} \bra{\mp} \hat{\rho}_S (t) \ket{\pm}
\{ + \pi (\tilde{\alpha}(\Delta/\hbar)+\tilde{\alpha}(-\Delta/\hbar)) \nonumber \\
&&\hspace*{3cm} \mp i \frac{\Delta}{\hbar} \int d\omega {\cal P} \frac{\tilde{\alpha}(\omega)}
{(\Delta/\hbar)^2-\omega^2} \}                                                                         \label{AppCaabb}
\end{eqnarray}
for the coherences.
Now we use the Bloch variables given by
\begin{eqnarray}
z(t)&=&\textrm{Tr} \hat{\rho}(t) \hat{\sigma}_{z}
=\bra{+} \hat{\rho}_S (t) \ket{+} - \bra{-} \hat{\rho}_S (t) \ket{-}                                   \label{AppCaaac} \\
x(t)&=&\textrm{Tr} \hat{\rho}(t) \hat{\sigma}_{x}
=\bra{+} \hat{\rho}_S (t) \ket{-} + \bra{-} \hat{\rho}_S (t) \ket{+} \nonumber \\
y(t)&=&\textrm{Tr} \hat{\rho}(t) \hat{\sigma}_{y}=
i(\bra{+} \hat{\rho}_S (t) \ket{-} - \bra{-} \hat{\rho}_S (t) \ket{+}) \nonumber .
\end{eqnarray}
Assuming as we did in section \ref{twolevelgen} that $\mean{\hat{B}}=0$ 
\footnote{Notice that even if $\mean{\hat{B}} \neq 0$, the order $\lambda$ term in the Redfield 
equation can vanish using the redefinition (\ref{1Gaaas}) and (\ref{1Gaabs}).}, we get
\begin{eqnarray}
\dot{z}(t)
&=& \gamma (z(\infty) - z(t))                                                                          \label{AppCaaad}\\
\dot{x}(t) &=&-\frac{\Delta}{\hbar} y(t) \nonumber \\
\dot{y}(t) &=& (\frac{\Delta}{\hbar}+\Gamma) x(t) - \gamma  y(t) \nonumber,
\end{eqnarray}
where
\begin{eqnarray}
\gamma&=&
2 \pi \frac{\lambda^2}{\hbar^2} (\tilde{\alpha}(\Delta/\hbar)+\tilde{\alpha}(-\Delta/\hbar))          \label{AppCaaae}\\
\Gamma&=& 
2 \frac{\lambda^2}{\hbar} \Delta \int d\omega {\cal P} 
\frac{\tilde{\alpha}(\omega)}{\Delta^2-(\hbar \omega)^2} \nonumber \\
z(\infty)&=& \frac{\tilde{\alpha}(-\Delta/\hbar)-\tilde{\alpha}(\Delta/\hbar)}
{\tilde{\alpha}(-\Delta/\hbar)+\tilde{\alpha}(\Delta/\hbar)} .
\end{eqnarray}
The solutions of this equation are given by
\begin{eqnarray}
z(t)&=& z(\infty) + (z(0) - z(\infty)) e^{-\gamma t}                                                  \label{AppCaaag}\\
x(t)&=&\frac{x(0) \frac{\gamma}{2} - y(0) \frac{\Delta}{\hbar}}{\sqrt{\frac{\Delta}{\hbar}
(\frac{\Delta}{\hbar}+\Gamma)-(\frac{\gamma}{2})^2}}  
\sin \left( \sqrt{\frac{\Delta}{\hbar}(\frac{\Delta}{\hbar}+\Gamma)-(\frac{\gamma}{2})^2} \; t \right) 
e^{- \frac{\gamma}{2} t} \nonumber \\
&&+x(0) \cos \left( \sqrt{\frac{\Delta}{\hbar}(\frac{\Delta}{\hbar}+\Gamma)-(\frac{\gamma}{2})^2} \; t \right) 
e^{-\frac{\gamma}{2} t} \nonumber \\
y(t)&=&\frac{x(0) (\frac{\Delta}{\hbar}+\Gamma) - y(0) \frac{\gamma}{2}}{\sqrt{\frac{\Delta}{\hbar}
(\frac{\Delta}{\hbar}+\Gamma)-(\frac{\gamma}{2})^2}}  
\sin \left(\sqrt{\frac{\Delta}{\hbar}(\frac{\Delta}{\hbar}+\Gamma)-(\frac{\gamma}{2})^2} \; t \right) 
e^{-\frac{\gamma}{2} t} \nonumber \\
&&+ y(0) \cos \left(\sqrt{\frac{\Delta}{\hbar}(\frac{\Delta}{\hbar}+\Gamma)-(\frac{\gamma}{2})^2} \;t \right) 
e^{-\frac{\gamma}{2} t} \nonumber 
\end{eqnarray}
It can finally be noticed that
\begin{eqnarray}
\gamma&=& 4 \frac{\lambda^2}{\hbar^2} \int_{0}^{\infty} d\tau \cos (\Delta \tau / \hbar) \; C(\tau)     \label{AppCaaah}\\
\Gamma&=& 
4 \frac{\lambda^2}{\hbar^2} \int_{0}^{\infty} d\tau \sin (\Delta \tau / \hbar) \; C(\tau) \nonumber \\
\gamma \; z(\infty) &=& 
4 \frac{\lambda^2}{\hbar^2} \int_{0}^{\infty} d\tau \sin (\Delta \tau / \hbar) \; D(\tau) \nonumber ,
\end{eqnarray}
where we recall that $\alpha(t)=C(t)+iD(t)$ where $C(t)$ and $D(t)$ are real functions.

\chapter[Gaussian orthogonal random matrices]{Gaussian orthogonal random matrices (GORM)} \label{AppD}

A Gaussian orthogonal random matrix (GORM) $\hat{Y}$ is
characterized by $M$, the size of the matrix, and by the parameter
$a_{\hat{Y}}$, which enters the Gaussian probability distribution
$P(\hat{Y})=C e^{- \frac{a_{\hat{Y}}}{2} Tr(\hat{Y}^2)}$ of the
whole matrix. The statistical properties of a GORM are preserved
under orthogonal transformations. Because the matrix is
symmetric, each non-diagonal element $Y_{ij}$ is equal to its
transposed $Y_{ji}$. The $\frac{M(M+1)}{2}$ independent matrix
elements of $\hat{Y}$ are Gaussian random numbers of mean zero.
The standard deviation of the non-diagonal matrix elements
$\sigma_{ND}^{\hat{Y}}$ and the standard deviation of the diagonal
matrix element $\sigma_{D}^{\hat{Y}}$ are related to $a_{\hat{Y}}$
by
\begin{equation}
\sigma_{D}^{\hat{Y}}=\sqrt{2}
\sigma_{ND}^{\hat{Y}}=\sqrt{\frac{1}{a_{\hat{Y}}}}.
\end{equation}
The \textit{density of states} of the GORM $\hat{Y}$ is defined by
\begin{equation}
d(E)=\sum_{i=1}^M \delta(E-E_i),
\end{equation}
and the \textit{smoothed density of states} by
\begin{equation}
\bar{d}(E)= \lim_{\epsilon \to 0} \frac{1}{\epsilon}
\int_{E-\frac{\epsilon}{2}}^{E+\frac{\epsilon}{2}} d(E) \ \ dE,
\end{equation}
where $\epsilon$ is a small energy interval which is large enough to
contain many states in order for $\bar{d}(E)$ to be smooth. The
\textit{averaged smoothed density of states} is an ensemble
average of $\chi$ realizations of the GORM. Such an ensemble is
called \textit{Gaussian orthogonal ensemble} (GOE). It is well
known \cite{Mehta67,Porter65,Brody81} that the ensemble averaged smoothed density of
states $\langle \bar{d}(E) \rangle_{\chi}$ obey the
\textit{Wigner semicircle law} in the limit $\chi \to \infty$:
\begin{eqnarray}
\langle \bar{d}(E) \rangle_{\infty} =
\begin{cases} 
\frac{a_{\hat{Y}}}{\pi}
\sqrt{\frac{2M}{a_{\hat{Y}}}-E^2} \ \ &if \ \ \vert E \vert <
\sqrt{\frac{2M}{a_{\hat{Y}}}} \\
0 \ \ & if \ \ \vert E \vert \geq \sqrt{\frac{2M}{a_{\hat{Y}}}} 
\end{cases}. \label{semi-circ}
\end{eqnarray}
The domain of energy where the eigenvalues are distributed (i.e., the
width of the semi-circle) is
$\mathcal{D}\!Y=\sqrt{\frac{8M}{a_{\hat{Y}}}}$. Notice that when
$M \to \infty$, $\bar{d}(E) \to \langle \bar{d}(E)
\rangle_{\infty}$, and therefore $\bar{d}(E)$ follows the
semi-circle law. The following notation is used in the present
paper: $n^w(E)=\langle \bar{d}(E) \rangle_{\infty}$.

\chapter{Perturbation theory for the spin-GORM model} \label{AppF}

There is no analytical way of getting a general form of the
eigenvalues $E_{\alpha}$ of the total system, 
but the three terms in Eq. (\ref{hamiltonien}) have
different orders of magnitude, depending on the value of the
parameters $\Delta$ and $\lambda$. The system and the environment
Hamiltonians are of order $\Delta$ and $1$, respectively, while the
coupling term is of order $\lambda$. Therefore, we can examine the
different extreme cases that can be treated perturbatively.\\

\boldmath $\Delta,1 \gg \lambda$ \unboldmath :\\

When the system and the environment Hamiltonians are larger than
the interaction term in Eq. (\ref{hamiltonien}), we can treat the
interaction term in a perturbative way, taking the system and the
environment Hamiltonian as reference,
\begin{equation}
\hat{H}_0 \vert s b \rangle
= E_{s b}^{0} \vert s b \rangle \; ,
\end{equation}
where we replaced the index ${n}$ by the two indices ${s,b}$.
The perturbed energy is given to the second order by
\begin{equation}
E_{\alpha} = E_{s,b} = \frac{\Delta}{2} s + E_{b}^{B}
+ \lambda^2 \sum_{b' \neq b} \frac{\vert \langle b' \vert \hat{B}
\vert b \rangle \vert^2}{E_{b}^{B}-E_{b'}^{B}+ s \Delta} +O(\lambda^4).
\label{pertcas1}
\end{equation}
We notice that the first nonzero correction to the
non-perturbed eigenstate is of order $\lambda^2$.\\
\\

\boldmath $\lambda \gg 1,\Delta$ \unboldmath :\\

When $\lambda$ is large compared to $\Delta$ and $1$ in Eq.
(\ref{hamiltonien}), it is possible to consider the interaction
term as the reference Hamiltonian and to treat $\hat{H}_S$ and
$\hat{H}_B$ as small perturbation. Transforming
(\ref{hamiltonien}) by a unitary matrix acting only on the system
degree of freedom, we get
\begin{eqnarray}
\hat{H} = \frac{\Delta}{2} \hat{\sigma}_{x} + \hat{H}_B +
\lambda \hat{\sigma}_z \hat{B}.
\end{eqnarray}
The non-perturbed reference Hamiltonian is therefore,
\begin{eqnarray}
\hat{\tilde{H}}_{0} = \lambda \hat{\sigma}_z \hat{B}.
\end{eqnarray}
Let $E_{\kappa \eta}$ and $\vert \kappa \eta \rangle = \vert
\kappa \rangle \otimes \vert \eta \rangle$ be respectively the
eigenvalues and eigenvectors of $\hat{\tilde{H}}_0$:
\begin{equation}
\hat{\tilde{H}}_0 \vert \kappa \eta \rangle = \lambda \hat{B}
\sigma_z \vert \kappa \eta \rangle = \lambda E_{\kappa \eta} \vert
\kappa \eta \rangle = \lambda \kappa E_{\eta} \vert \kappa \eta
\rangle,
\end{equation}
where $\eta=1,...,\frac{N}{2}$ and $\kappa=\pm1$. The energy of
the perturbed Hamiltonian is thus given to the second order perturbation
in $\frac{1}{\lambda}$ by
\begin{equation}
\frac{E_{\alpha}}{\lambda} = \kappa E_{\eta} + \frac{1}{\lambda}
\langle \eta \vert \hat{H}_B \vert \eta \rangle +
\frac{1}{\lambda^2} \sum_{\stackrel{\kappa',\eta'}{\neq
\kappa,\eta}} \frac{\vert \frac{\Delta}{2} + \langle \eta \vert
\hat{H}_B \vert \eta \rangle \vert^2}{E^{0}_{\kappa \eta} -
E^{0}_{\kappa' \eta'}}+ O\left(\frac{1}{\lambda^3}\right).
\label{pertcas2}
\end{equation}

\boldmath $1 \gg \Delta,\lambda$ \unboldmath :\\

In this case, the environment Hamiltonian is large compared to the system Hamiltonian
and the interaction term so that they both can be considered as perturbations.
We get
\begin{equation}
E_{\alpha} = E_{s,b} = \frac{\Delta}{2} s + E_{b}^{B}
+ \lambda^2 \sum_{b' \neq b} \frac{\vert \langle b' \vert \hat{B}
\vert b \rangle \vert^2}{E_{b}^{B}-E_{b'}^{B}}
+ O(\Delta^2) + O(\lambda^2).
\label{pertcas3}
\end{equation}

\boldmath $\Delta \gg 1,\lambda$ \unboldmath :\\

We now suppose that the system Hamiltonian taken as reference is large compared
to the environment Hamiltonian and the interaction term, so that these last two terms can
be considered as perturbations.
We then get
\begin{equation}
E_{\alpha} = E_{s,b} = \frac{\Delta}{2} s + E_{b}^{B}
+ s \frac{\lambda^2}{\Delta} \sum_{b' \neq b} \vert \langle b' \vert \hat{B}
\vert b \rangle \vert^2 + O(1) + O(\lambda^2).
\label{pertcas4}
\end{equation}

Two additionals situations, \boldmath $1,\lambda \gg \Delta$ \unboldmath and
\boldmath $\Delta,\lambda \gg 1$ \unboldmath, could be considered but
cannot be treated perturbatively because no reference
basis exists in which $\hat{H}_B$ and $\hat{B}$ are simultaneously diagonal.

\end{document}